\RequirePackage[l2tabu, orthodox]{nag}
\documentclass[prd,aps,notitlepage,nofootinbib,floatfix,twocolumn,superscriptaddress,twoside]{revtex4-1}
\usepackage[T1]{fontenc}
\usepackage[utf8]{inputenc}
\usepackage{color} 
\usepackage{array}
\usepackage{calc}
\usepackage{units}
\usepackage{url}
\usepackage{multirow}
\usepackage{amsmath}
\usepackage{amssymb}
\usepackage{physics}
\usepackage{graphicx}
\usepackage[svgnames,dvipsnames]{xcolor} 
\usepackage[makeroom]{cancel}
\usepackage{booktabs}

\usepackage{siunitx}
\numberwithin{equation}{section}
\numberwithin{figure}{section}
\numberwithin{table}{section}
\usepackage{listings}
\usepackage{subcaption}
\usepackage{slashed}
\usepackage{titlesec}
\usepackage[bordercolor=white,backgroundcolor=gray!30,linecolor=black,colorinlistoftodos]{todonotes}

\usepackage[linewidth=1pt]{mdframed}
\usepackage{hhline}

\titleformat{\chapter}[display]{\Large\bfseries}{Chapter \Large\thechapter}{1ex}{}
\titleformat*{\section}{\large\bfseries}
\titleformat*{\subsection}{\normalsize\bfseries}
\titleformat*{\subsubsection}{\normalsize\bfseries}
\titleformat*{\paragraph}{\normalsize\bfseries}
\titleformat*{\subparagraph}{\normalsize\bfseries}

\usepackage{enumitem}		
\usepackage{footmisc}
\usepackage{natbib}
\usepackage[originalparameters]{ragged2e}
\usepackage{sidecap}

\usepackage{etoolbox}
\usepackage{soul}
\makeatletter
\patchcmd{\chapter}{\if@openright\cleardoublepage\else\clearpage\fi}{}{}{}
\makeatother

\makeatletter
\renewcommand{\l@section}{\@dottedtocline{1}{1.5em}{2.0em}}
\renewcommand{\l@subsection}{\@dottedtocline{2}{3.5em}{3.0em}}
\makeatother

\usepackage[unicode=true, bookmarks=true,bookmarksnumbered=true,bookmarksopen=false, 
breaklinks=false,pdfborder={0 0 1},backref=false,colorlinks=true] {hyperref}
\usepackage{breakurl}
\usepackage{bookmark}
\usepackage{cleveref}

\usepackage{pdfcomment}

\newcommand{\MODALLSS}{\texttt{MODAL-LSS}}
\newcommand{\LPICOLA}{\texttt{L-PICOLA}}
\newcommand{\GADGET}{\texttt{GADGET-3}}
\DeclareMathOperator\erfc{erfc}

\begin{document}

\title{Advancing the matter bispectrum estimation of large-scale structure:
  fast prescriptions for galaxy mock catalogues}
\author{Johnathan Hung}
\email{jmch2@damtp.cam.ac.uk}
\affiliation{Centre for Theoretical Cosmology, DAMTP, University of Cambridge, CB3 0WA, United Kingdom}
\author{Marc Manera}
\email{mmanera@ifae.es}
\affiliation{Institut de Física d’Altes Energies, The Barcelona Institute of Science and Technology,Campus UAB, 08193 Bellaterra (Barcelona), Spain}
\affiliation{Centre for Theoretical Cosmology, DAMTP, University of Cambridge, CB3 0WA, United Kingdom}\author{E.P.S. Shellard}
\email{E.P.S.Shellard@damtp.cam.ac.uk}
\affiliation{Centre for Theoretical Cosmology, DAMTP, University of Cambridge, CB3 0WA, United Kingdom}

\date{\today}

\begin{abstract}
 We investigate various phenomenological schemes for the rapid generation of 3D mock galaxy catalogues with a given power spectrum and bispectrum.  We apply the fast bispectrum estimator \MODALLSS{} to these mock galaxy catalogues and compare to $N$-body simulation data analysed with the halo-finder \texttt{ROCKSTAR} (our benchmark data).  We propose an assembly bias model  for populating parent halos with subhalos by using a joint lognormal-Gaussian probability distribution for the subhalo occupation number and the halo concentration.  This prescription enabled us to recover the benchmark power spectrum from $N$-body simulations to within 1\% and the bispectrum to within 4\% across the entire range of scales of the simulation.   A small further boost adding an extra galaxy to all parent halos above the mass threshold $M>2\times10^{14}\,h^{-1} M_\odot$ obtained a better than 1\% fit to both power spectrum and bispectrum in the range $K/3<1.1\,h\,\text{Mpc}^{-1}$, where $K=k_1+k_2+k_3$.   This statistical model should be applicable to fast dark matter codes, allowing rapid generation of mock catalogues which simultaneously reproduce the halo power spectrum and bispectrum obtained from $N$-body simulations.  We also investigate alternative schemes using the Halo Occupation Distribution (HOD) which depend only on halo mass, but these yield results deficient in both the power spectrum (2\%) and the bispectrum (>4\%) at $k,K/3 \approx 0.2\,h\,\text{Mpc}^{-1}$, with poor scaling for the latter.  Efforts to match the power spectrum by modifying the standard four-parameter HOD model result in overboosting the bispectrum (with a 10\% excess).  We also characterise the effect of changing the halo profile on the power spectrum and bispectrum.
\end{abstract}




\pagebreak

\maketitle

\section{Introduction}

One of the most active areas of cosmological research is to
understand the collapse of matter and the evolution of large scale structure
(LSS) in the Universe. This goal is facilitated by upcoming large
data sets offered by galaxy surveys such as the Dark Energy Survey (DES)
\citep{DES1, DES2}, the Large Synoptic Survey Telescope (LSST) \citep{LSST},
the ESA Euclid Satellite \citep{Euclid} and the Dark Energy Spectroscopic
Instrument (DESI) \citep{DESI}. In particular, the bispectrum has been shown
to be a crucial diagnostic in the mildly non-linear regime and,
combined with the galaxy power spectrum, can constrain parameters five times
better than the power spectrum alone \citep{forecast}, potentially offering
much tighter constraints for local-type primordial non-Gaussianities (PNG) than current
limits from Planck. The bispectrum also has a stronger dependence on
cosmological parameters so can provide tighter constraints than the power
spectrum for the same signal to noise and can help break degeneracies in
parameter space, notably those between $\sigma_8$ and bias \citep{bias}.
For this reason, our focus in this paper is on making the galaxy bispectrum a tractable diagnostic tool for analysing future galaxy surveys by deploying our efficient bispectrum estimator \MODALLSS{} on mock galaxy catalogues.  Previously we have used \MODALLSS{}
 to compare the dark matter bispectrum from $N$-body and fast dark matter codes\citep{DM}, but now we apply it to the halo (or galaxy) bispectrum.  This work builds on earlier efforts to estimate the full three-dimensional bispectrum
 from simulations (see, for example, \citep{wagner,regan,9param,glass,Andrei}) and direct measurements of the galaxy bispectrum using existing galaxy survey data from the Baryon Oscillation Spectroscopic Survey (BOSS) \citep{BOSS1, BOSS2, sdssi, sdssii, sdssiii}.
 
As we enter the age of precision cosmology we are ever more reliant
on cosmological simulations to understand the dynamics of dark matter
and baryons. Numerical simulations act as a buffer between theory and
observation: we test cosmological models by matching simulation results
to observational data, and hence obtain constraints on cosmological
parameters. On the other hand since we only observe one universe we must
turn to simulations to understand the statistical significance of our
measurements. This is especially important with large galaxy data sets
coming from current and near-future surveys such as DES, LSST, Euclid
and DESI. While it would be ideal to use full $N$-body simulations to
generate these so-called mock catalogues for statistical analysis, their
huge demand for computational resources is prohibitive for generating
the large number of simulations required for accurate estimates of
covariances \citep{l-picola}. Alternatively, compression
methods have also been developed to reduce the number of mocks required,
see e.g. \citep{gualdi1,gualdi2,gualdi3,heavens,alsing}.

Although dark matter simulations have given us a wealth of information
about the clustering of matter in the universe, ultimately we need to
map this information to the visible universe. Gravitational pull
induces the formation of bound dark matter halos, and these virialised
objects in turn create an environment in which baryons can collapse
and form bound objects such as galaxies. The galaxies we observe
in galaxy surveys, which live inside these halos, therefore act as biased
tracers to the underlying dark matter distribution, as the spatial
distribution of galaxies need not exactly mirror that of the dark matter
\citep{kaiser}. To take advantage of high resolution galaxy data from
future surveys we must therefore have a robust way to extract halo
and galaxy distributions from $N$-body dark matter simulations. Many
techniques for this process, known as halo finding, have been
developed over the years (e.g. \citep{halo1,halo2,halo3,halo4,halo5,
  halo6,halo7,ahf2,halo9,halo10,halo11,halo12,halo13,halo14,halo15,ahf,
  halo17,halo18,rockstar,halo20,halo21,halo22,halo23,halo24}), but
it remains a computationally intensive task, especially with the sheer 
number of simulations required for covariance matrix estimation.
Additionally, to put constraints on cosmological parameters halo
properties must be understood to percent level in order for
theoretical and statistical uncertainties to be at the same
level \citep{rockstar,percent1,percent2}. In this paper we present
fast phenomenological prescriptions for producing mock galaxy
catalogues that reproduce the power spectrum and bispectrum of
a reference catalogue to better than 1\% accuracy. In order to do
so we examine the effects of the spatial distribution of galaxies
within their host halos, the halo occupation number through the
Halo Occupation Distribution (HOD) model, as well as a more
sophisticated assembly bias model that jointly models the occupation
number and halo concentration. Previous work estimating the dark matter
bispectrum has shown its power in helping benchmark fast dark matter
codes \citep{DM}, and here we likewise validate these methods with
both the power spectrum and bispectrum.

The paper is outlined as follows: in \Cref{sec:methodology} we
detail our benchmark galaxy mock catalogue and the phenomenological
methods we use to reproduce the statistics of this catalogue.
Then in \Cref{sec:halo-polyspectra} we introduce the \MODALLSS{}
method for bispectrum estimation, as well as the phenomenological
3-shape model for the halo bispectrum. In \Cref{work}, we then present
the alternative prescriptions for generating mock catalogues as we investigate  
the effect of halo profiles and different HODs on the bispectrum, ultimately 
proposing a joint lognormal-Gaussian assembly bias model which is a key
outcome of this paper. Finally, we summarise the main 
results and conclude the paper in \Cref{sec:conclusions}.

\section{Halo catalogues\label{sec:methodology}}

There are many techniques that have been developed to identify
collapsed objects in dark matter simulations, but two methods
remain a core part of the halo finding process. These are the
Friends-of-Friends (FoF) algorithm \citep{FoF_Davis}, originally
proposed in 1985, and the Spherical Overdensity (SO) algorithm
\citep{SO_PS}, originally proposed in 1974. In its simplest form
the FoF algorithm simply links together particles that are separated
by a distance less than a given linking length $b$, resulting in
distinct connected regions that are identified as collapsed halos.
The SO algorithm on the other hand identifies peaks in the density
field as the candidate halo centres, then assuming a spherical
profile grows the halo until a density threshold is reached. There
are shortcomings associated with naive implementations of both of
these methods: the FoF algorithm is susceptible to erroneously
connecting two distinct halos to each other via \emph{linking
  bridges}, which are filaments between linked particles belonging
to the 2 distinct halos; whereas the spherical assumption in the SO
method does not reflect the true shape of
halos. A particular difficulty of these position-based finders, yet
crucial for the mapping of dark matter distribution to the galaxies
we observe, is the classification of halos within halos, or
\emph{subhalos}, i.e. virialised objects that sit inside and orbit
a larger, host halo. Many have introduced refinements to extend the
capabilities of FoF and SO, for example by changing the FoF linking
length or the SO density threshold as well as better taking
advantage of other information given to us by cosmological
simulations; for instance, see \citep{mad} for a comprehensive review.

A relatively recent and novel approach to this old problem is
the incorporation of velocity information of the particles, reducing
the ambiguity in determining particle membership between overlapping
halos. While this additional information is clearly useful for 
distinguishing subhalos from its host halos due to their relative
motion, working in phase-space necessitates the creation of a metric
that suitably weights the relative positions and velocities of the
particles. 
The 6D phase-space halo finder we adopt for this paper is \texttt{ROCKSTAR}
\citep{rockstar}, which further utilises \emph{temporal} data across
simulation time steps to ensure consistency of halo properties.
Furthermore the authors claim it to be the first grid- and
orientation-independent adaptive phase-space code, and possesses the
unprecedented ability to probe substructure masses down to the very
centres of host halos. Here we give a brief overview of the mechanics
of the \texttt{ROCKSTAR} algorithm.

The simulation box is first partitioned with a fast implementation
of position-based FoF and a large linking length of $b=0.28$ (in units
of the mean inter-particle distance). Likewise in the 3D case, an adaptive
metric must be used if one is to find substructures at all levels.
For each of these 3D FoF groups a hierarchy of 6D phase-space FoF
subgroups is built up by adapting the phase-space linking length at
every level so that only 70\% of the particles are linked together in
its subgroups, until the number of particles in the deepest level falls
under a predefined threshold (here set to 10). The phase-space metric
they adopt is weighted by the standard deviations in position,
$\sigma_x$, and velocity $\sigma_v$, of the particles within a (3D or
6D) FoF group, i.e. for two particles $p_1$ and $p_2$ the metric is:
\begin{align}
  \label{eq:rockstar_pp_metric}
  d(p_1,p_2)=\left(\frac{\left|\textbf{x}_1-\textbf{x}_2\right|^2}{\sigma_x^2}
  +\frac{\left|\textbf{v}_1-\textbf{v}_2\right|^2}{\sigma_v^2}\right)^{1/2}.
\end{align}

Once this phase-space hierarchy is built, the deepest levels in the
hierarchy are identified as seed halos, and all particles in the base
3D FoF group are assigned to these seed halos from the bottom-up. If
a seed halo is the only child of its parent then all the particles of
the parent will be assigned to that seed halo. Otherwise if a parent
has multiple subgroups then particle membership is determined by proximity
in phase-space. In this instance the metric (\Cref{eq:rockstar_pp_metric})
is modified to reflect halo and not particle properties; for a halo $h$
and particle $p$ the metric is
\begin{align}
  \label{eq:rockstar_hp_metric}
  d(h,p)=\left(\frac{\left|\textbf{x}_h-\textbf{x}_p\right|^2}{r_{vir}^2}
  +\frac{\left|\textbf{v}_h-\textbf{v}_p\right|^2}{\sigma_v^2}\right)^{1/2},
\end{align}
where $r_{vir}$ is the current virial radius of the halo and now 
$\sigma_v$ is the current velocity dispersion of the halo. This procedure
is repeated recursively along the hierarchical ladder until particle
assignment is complete. A significant advantage of this assignment scheme
is the assurance that particles that belong to the host halo will not be
mis-assigned to the subhalo, or vice versa, even if the subhalo sits close
to the host halo centre. This is because host halo particles and subhalo
particles should have different distributions in phase-space even if they
are close in position-space. 

Finally, host-subhalo relationships are determined based on phase-space
distances before halo masses are calculated to avoid ambiguity when multiple
halos are involved. At each level the halos are first ordered by the number of
assigned particles. Starting with the lowest one, each halo centre is
treated as a particle, and its distance to the other halos are calculated
with \Cref{eq:rockstar_hp_metric}. The halo being examined is then
assigned as a subhalo of the closest larger halo. These relationships are
checked against the previous time-step, if available, for consistency across
time-steps. After all assignments have been made, unbounded particles are
removed by a modified Barnes-Hut method from the halos, and halo properties
are calculated.

\subsection{Benchmark galaxy mock catalogue}

\begin{table}
  \begin{center}
    \begin{tabular}{c|c|c}
      Description & Symbol & Value \\[1ex] \hhline{=|=|=}
      \rule{0pt}{3ex}
      Hubble constant & $H_0$  & 67.74 $\text{km}\,\text{s}^{-1}$  \\
      Physical baryon density  & $\Omega_b h^2$  &  0.02230 \\
      Matter density  & $\Omega_m$   & 0.3089 \\
      Dark energy density  & $\Omega_{\Lambda}$   & 0.6911 \\
      Fluctuation amplitude at $8h^{-1}$ Mpc & $\sigma_8$  & 0.8196 \\ 
      Scalar spectral index & $n_s$  & 0.9667 \\
      Primordial amplitude & $10^9A_s$ & 2.142\\[1ex] \hhline{=|=|=}
      \rule{0pt}{3ex}
      Physical neutrino density  & $\Omega_{\nu} h^2$   & 0.000642 \\
      Number of effective neutrino species & $N_{eff}$   & 3.046 \\
      Curvature density  & $\Omega_{k}$   & 0.0000 \\
    \end{tabular}
    \caption{
      Planck 2015 cosmological parameters (Tables 4 and 5 in \citep{planck2015},
      rightmost columns), which we used to generate the input power spectrum
      from \texttt{CAMB}. The pivot scale for $n_s$ is 0.05 $\text{Mpc}^{-1}$.
      }
    \label{planck}
  \end{center}
\end{table}

\begin{table}
  \begin{center}
    \begin{tabular}{c|c|c}
      Name & Description & Value\\[1ex] \hhline{=|=|=}
      \rule{0pt}{3ex}
      MaxRMSDisplacementFac & \multirow{3}{2cm}{\centering Timestepping criteria}  & 0.1 \\
      ErrTolIntAccuracy &  & 0.01 \\
      MaxSizeTimestep &  & 0.01 \\[1ex] \hhline{=|=|=}
      \rule{0pt}{3ex}
      ErrTolTheta & \multirow{3}{2cm}{\centering Gravitational force criteria}  & 0.2 \\
      ErrTolForceAcc &  & 0.002 \\
      Smoothing length &  & $30\,h^{-1}$ kpc \\[1ex] \hhline{=|=|=}
      \rule{0pt}{3ex}
      Number of particles & \multirow{3}{2cm}{\centering Mass resolution}  & $2048^3$ \\
      Mass of particles &  & $2.1\times10^{10}\,h^{-1} M_\odot$ \\
      PM grid size &  & 2048
    \end{tabular}
    \caption{\GADGET{} parameters chosen in reference to
      \citep{transients,crocce2006} to ensure high
      numerical accuracy in our simulations.}
    \label{tab:gadget}
  \end{center}
\end{table}

Our benchmark dark matter simulation is a $N$-body simulation run 
with \GADGET{} code. We have chosen a cubical box of size 
$1280\,h^{-1}$ Mpc and run with $2048^3$ particles, obtaining 
a particle mass of $M_p=2.1\times10^{10}\,h^{-1} M_\odot$. We have
dark matter outputs at redshifts $z=0,0.5,1,2$. The Particle 
Mesh (PM) grid of the simulation is $2048^3$. 

We have generated the Gaussian initial conditions from second-order
Lagrangian Perturbation Theory (2LPT) displacements using
\LPICOLA{} \citep{l-picola,scoccimarro} at redshift $z_i=99$ to
ensure the suppression of transients in power spectra and bispectra
estimates of our simulations \citep{transients}. Our input linear power
spectrum at redshift $z=0$ was produced by \texttt{CAMB} \citep{CAMB}
using a flat $\Lambda$CDM cosmology with extended Planck 2015
cosmological parameters (TT,TE,EE+lowP+lensing+ext, see \Cref{planck}).
For neutrinos we had one massive neutrino species and two massless
neutrinos. The lack of radiation and neutrino evolution in
\LPICOLA{} and \GADGET{} has led us to define the matter power
spectrum to consist only of cold dark matter and baryons, which leads
us to recover the input power spectrum at $z=0$ to linear order.
This causes the raised value of $\sigma_8$ instead of the Planck
value of 0.8159. \Cref{tab:gadget} shows a number of \GADGET{}
parameter values we used guarantee high numerical precision in our
simulation.

To obtain a benchmark galaxy mock catalogue we first ran
\texttt{ROCKSTAR} on the \GADGET{} output. Since small halos are
unreliable we impose a mass threshold of
$M_{200b}>10^{12}\,h^{-1} M_\odot$ on the parent halos of the
\texttt{ROCKSTAR} output, where $M_{200b}$ means the mass enclosed
by the halo corresponds to a spherical overdensity of 200 times
the background density of the Universe. This cuts all parent halos
with fewer than 50 particles, which is roughly the same criterion
adopted in \citep{eisenstein1,eisenstein2}. The benchmark halo mock 
catalogue then consists of all parent halos that pass this threshold
alongside all subhalos they contain, if any. In this paper we use
the halos as proxies for galaxies, such that every parent halo
hosts a central galaxy at its core, and all the subhalos of the parent
hosts a satellite galaxy each. Our benchmark galaxy mock catalogue is
therefore identical to the benchmark halo mock catalogue, and we will
be using these terms interchangeably. 

The purpose of this paper is to investigate phenomenological methods
to reproduce the statistics of the benchmark galaxy mock catalogue
without detailed information given by the simulation. We restrict
ourselves to the mass, position, and halo concentration of the parent
halos, and build models that inform us of the number and positions
of the satellite galaxies in each parent halo. We define the benchmark
catalogue as above to examine these effects rather than reproduce a
realistic mock galaxy catalogue that matches observational data, e.g.
in \citep{eisenstein1}. We are also interested in first understanding
these effects in configuration space, and as such will not include
observational effects such as Redshift Space Distortions (RSD). This
is because the RSD signal will dominate in the bispectrum at small
scales and swamp the contributions that we are interested in here.
After we correctly model these effects in configuration space we
shall tackle RSD effects in a future paper. Additionally, both the
projected bispectrum \citep{projection} or bispectrum monopole
\citep{monopole} are rather insensitive to RSD effects, thus our
methods are well suited to the study of these observables. We note
here that our previous investigation of the dark matter bispectrum
using these simulations have uncovered problematic transient modes
that persist to late times \cite{DM}. However this should not interfere
with our work in this paper, as these modes only distort the bispectrum
signal at large scales, and their effects will cancel when we make
comparisons between different phenomenological methods. When calculating
statistics we follow the example of others, e.g. \citep{mock1,mock2},
and use the number density field where each object is weighted by 1
instead of their mass in the Cloud in Cell (CIC) assignment scheme,
which is on a $1024^3$ grid throughout the paper.

\subsection{Halo profile\label{subsec:profile}}

We tackle the distribution of galaxies within a halo by first
examining the relevance of the halo shape. It is well known in the
literature, particularly from dark matter simulations, that halos
are triaxial objects \citep{triaxial1,triaxial2,triaxial3},
and that their shape are complicated functions of time, halo mass,
and choice of halo radius. Halo shapes have been predicted analytically
as well within the ellipsoidal-collapse model in
\citep{triaxial4}. In principle one should take into account these
effects when building a halo mock catalogue, but as we shall see in
\Cref{work}, halo triaxiality only has a small effect compared to the
choice of halo profile in the power spectrum and bispectrum, and only
at small scales. Consequently, in this paper we only consider radially
symmetric profiles here and randomise the solid angle distribution of
each halo. We leave the inclusion of halo triaxiality for future work.

There are a number of radially symmetric halo profiles in the
literature that we can use to populate halos with satellite
galaxies. One popular choice is the NFW profile proposed by
Navarro, Frenk and White \citep{nfw}, which was adopted in the
generation of BOSS galaxy mock catalogues \citep{mock1}:
\begin{align}
  \label{eqn:nfw}
  \rho(r|r_s,\rho_s)=\frac{4\rho_s}{\frac{r}{r_s}\left(1+\frac{r}{r_s}\right)^2}.
\end{align}

The two parameters of the model are the scale radius $r_s$ and the
density at that radius $\rho_s=\rho(r_s)$. An alternative
parameterisation is with the concentration parameter $c=R_{vir}/r_s$,
and the virial mass of the halo $M_{vir}$; in \texttt{ROCKSTAR} the
virial radius $R_{vir}$ is defined such that the corresponding virial
mass $M_{vir}$ is consistent with the virial threshold in
\citep{virial}. Further imposing conservation of mass:
\begin{align}
  \label{eqn:nfw3}
  M_{vir}=\int^{R_{vir}}_0\rho(r|r_s,\rho_s) 4\pi r^2\,dr,
\end{align}
leads to
\begin{align}
  \label{eqn:nfw4}
  \rho_s=\frac{M_{vir}}{16\pi R_{vir}^3}\frac{c^3}{\log(1+c)-\frac{c}{1+c}}.
\end{align}
This allows us to write the radial density as
\begin{align}
  \label{eqn:nfw2}
  \rho(r|M_{vir},c)=\frac{M_{vir}}{4\pi rc(R_{vir}+rc)^2}
  \frac{c^3}{\log(1+c)-\frac{c}{1+c}}.
\end{align}

To populate the halos with the NFW profile we assume the radial
probability density function (PDF) of the mass distribution in a
halo is proportional to $\rho(r|M_{vir},c)$, and then obtain the
positions of the galaxies by inverse sampling. This first involves
calculating the cumulative distribution function (CDF) from the PDF:
\begin{align}
  \label{eqn:cdf}
  \text{CDF}_{\text{NFW}}(r|M_{vir},c)
  &=\frac{\int^{r}_0\rho(r'|M_{vir},c) 4\pi r'^2\,dr'}
    {\int^{R_{vir}}_0\rho(r'|M_{vir},c) 4\pi r'^2\,dr',}
    \nonumber \\
  &=\frac{\log(1+\frac{cr}{R_{vir}})-\frac{cr}{R_{vir}+cr}}
    {\log(1+c)-\frac{c}{1+c}}.
\end{align}
We then draw samples from the inverse of the CDF,
$\text{CDF}_{\text{NFW}}^{-1}$, with a uniform distribution
$u\sim U\in[0,1]$:
\begin{align}
  \label{eqn:inverse}
  r=\text{CDF}^{-1}_{\text{NFW}}(u|M_{vir},c).
\end{align}
Since the inversion of the CDF is numerically expensive we instead
calculate the desired $r$ by interpolating the tabulated CDF. Finally,
we model the concentration $c$ with this analytical fit as proposed in
\citep{concentration}:
\begin{align}
  \label{eqn:conc}
  \bar{c}(M,z)=\frac{9}{1+z}\left(\frac{M}{M_{NL}}\right)^{-0.13},
\end{align}
where $M_{NL}=\frac{4\pi}{3}\bar{\rho}(z)(\frac{2\pi}{k_{NL}})^3$
is the non-linear mass scale, and $k_{NL}$ is defined by the
linear power spectrum $P_L$ as $k^3_{NL}P_L(k_{NL},z)=2\pi^2$.

\begin{figure}
  \begin{center}    
    \includegraphics[width=\linewidth]{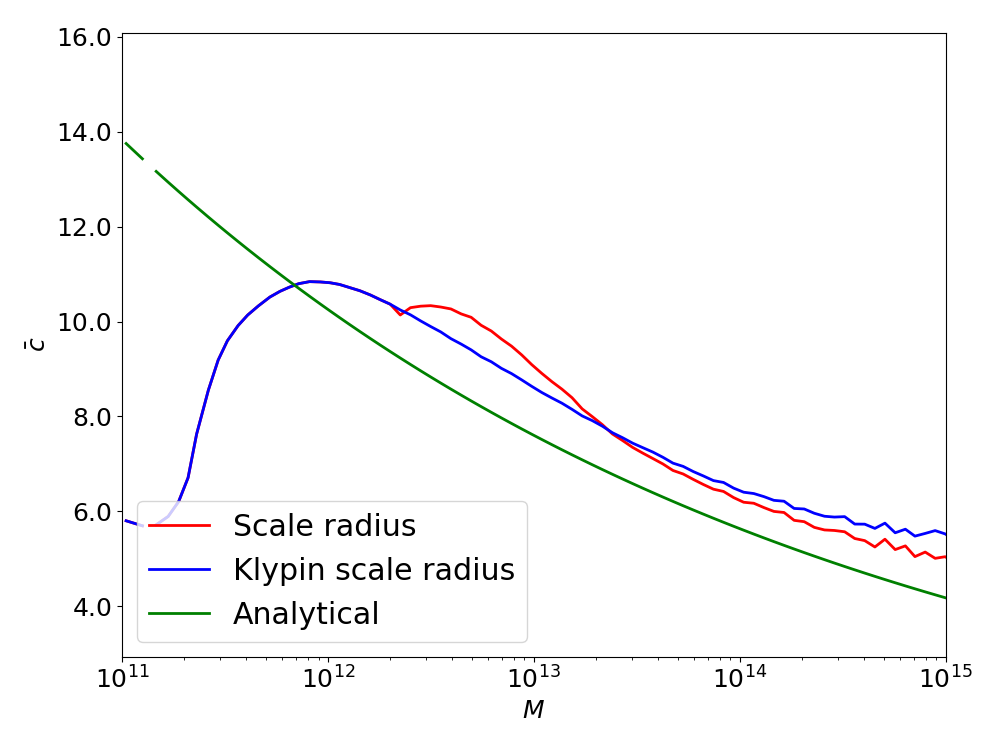}
  \end{center}
  \caption{Mean concentration of the benchmark \texttt{ROCKSTAR}
    halos as a function of their mass, calculated from both the
    scale radius and Klypin scale radius, as well as the analytical
    fit in \citep{concentration} (\Cref{eqn:conc}).
  }
  \label{fig:mean_conc}
\end{figure}

To judge whether the NFW profile is a good choice for our
purposes we first compared the benchmark mean concentration to
the analytical fit in \Cref{eqn:conc}. \texttt{ROCKSTAR} fits an
NFW profile by calculating both the scale radius $r_s$ and the
Klypin scale radius $r_{s,K}$ \citep{Klypin}, which is derived
from $v_{max}$, the maximum circular velocity, and $M_{vir}$.
We have plotted the mean concentration computed from $r_s$
and $r_{s,K}$ against the analytical fit in \Cref{fig:mean_conc}.
While the Klypin concentration demonstrates better numerical
stability overall, it is not clear that it is more robust for
halos with fewer than 100 particles as the authors of
\texttt{ROCKSTAR} claim \citep{rockstar}. We shall be using the Klypin concentration
in all our methods discussed below. We note that while
\Cref{eqn:conc} seems to qualitatively capture the correct power
law behaviour, the magnitude is too low by about 10-20\%.

\begin{figure*}
  \captionsetup[subfigure]{labelformat=empty}
  \begin{center}
    \begin{subfigure}[b]{0.45\textwidth}
      \includegraphics[width=\linewidth]{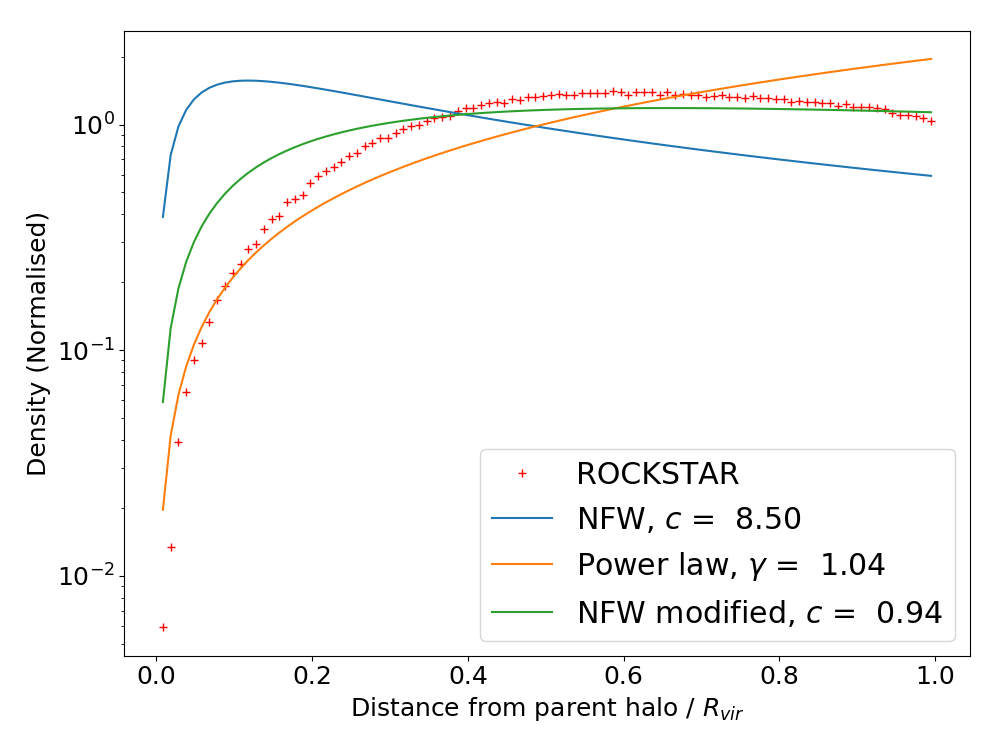}
      \caption{$M=(0.62-2.2)\times10^{13}h^{-1}\,M_{\odot}$}
    \end{subfigure}
    ~
    \begin{subfigure}[b]{0.45\textwidth}
      \includegraphics[width=\linewidth]{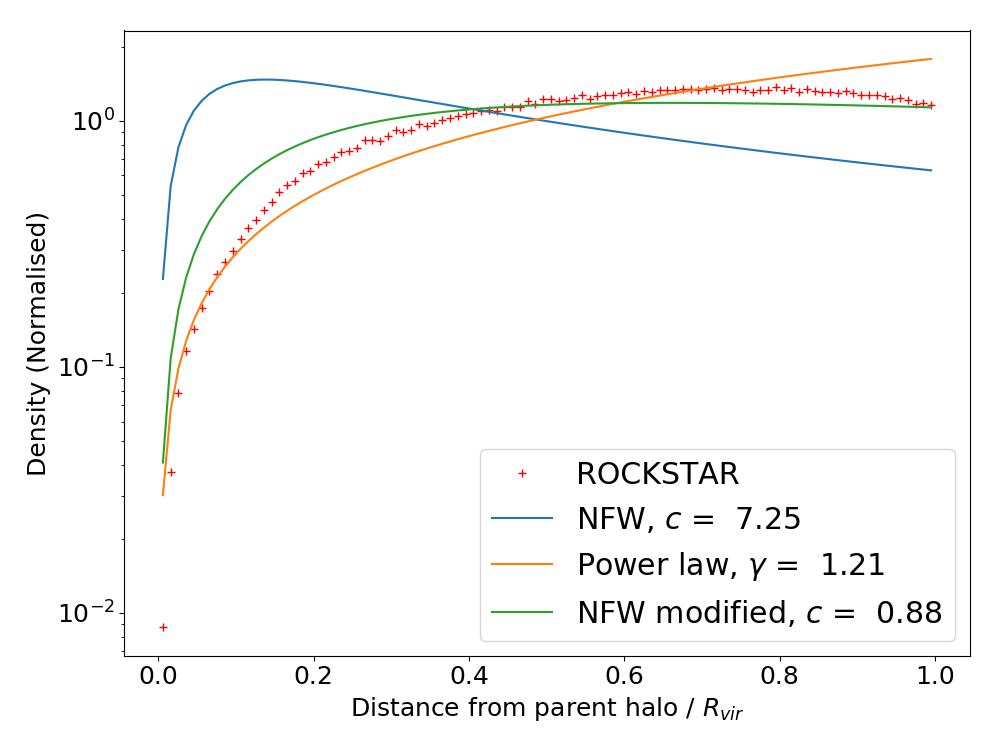}
      \caption{$M=(2.2-7.6)\times10^{13}h^{-1}\,M_{\odot}$}
    \end{subfigure}

    \begin{subfigure}[b]{0.45\textwidth}
      \includegraphics[width=\linewidth]{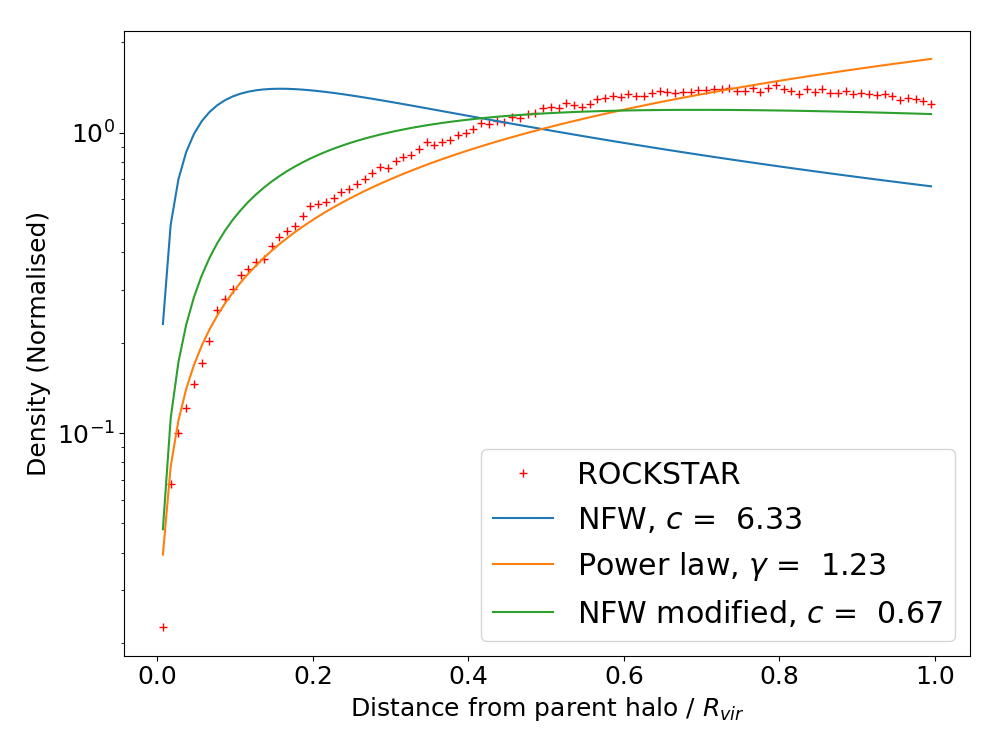}
      \caption{$M=(0.76-2.7)\times10^{14}h^{-1}\,M_{\odot}$}
    \end{subfigure}
    ~
    \begin{subfigure}[b]{0.45\textwidth}
      \includegraphics[width=\linewidth]{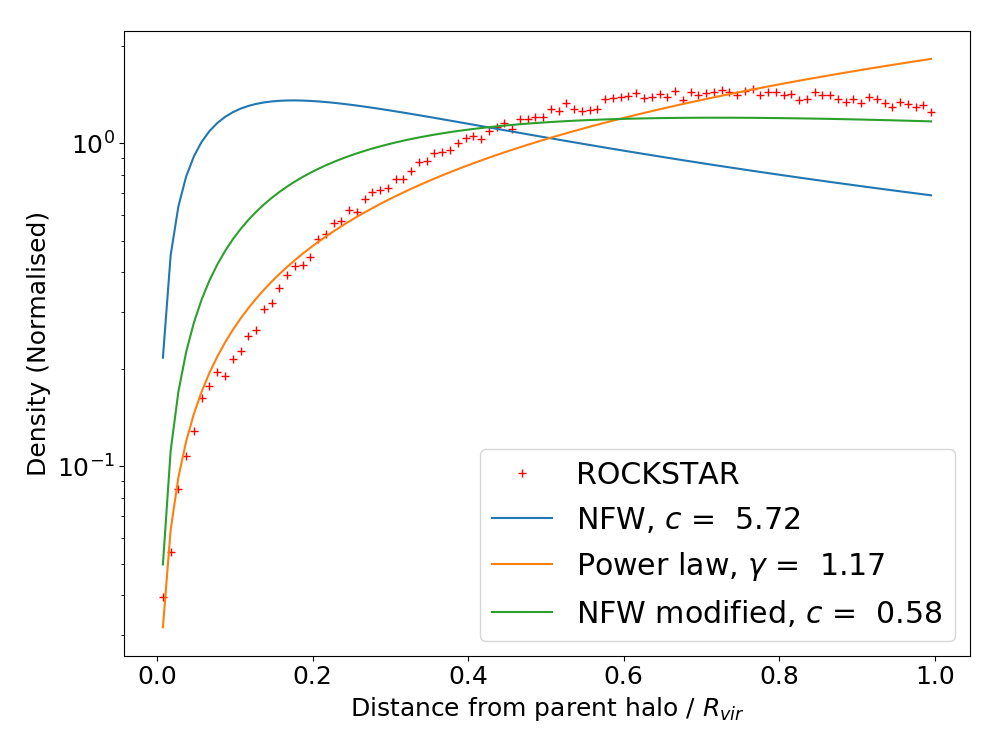}
      \caption{$M=(2.7-9.3)\times10^{14}h^{-1}\,M_{\odot}$}
    \end{subfigure}
    \caption{
      The subhalo number density profile given by \texttt{ROCKSTAR} and NFW for parent halos in
      various mass bins, as well as a power law and modified NFW fits to the
      \texttt{ROCKSTAR} data. Distances are scaled by $R_{vir}$ measured by
      \texttt{ROCKSTAR}. 
    }
    \label{fig:halo_profile1}
  \end{center}
\end{figure*}

More importantly, while the NFW profile is used in the literature to
populate halos with galaxies, it is ultimately a fit to the dark
matter profile and may not reflect the subhalo density profile.
Comparisons between the NFW profile and the number density profile
for the \texttt{ROCKSTAR} benchmark catalogue at different mass bins
is shown in \Cref{fig:halo_profile1}. Throughout the paper we only
populate subhalos to the virial radius $R_{vir}$.
In these plots, the NFW profile is calculated
using the average Klypin concentration given by \texttt{ROCKSTAR} for
the mean halo mass of the bin. Additionally, distances are scaled by
the virial radius $R_{vir}$, since that is the distance \texttt{ROCKSTAR}
uses when fitting the NFW profile.

\begin{figure}
  \begin{center}    
    \includegraphics[width=\linewidth]{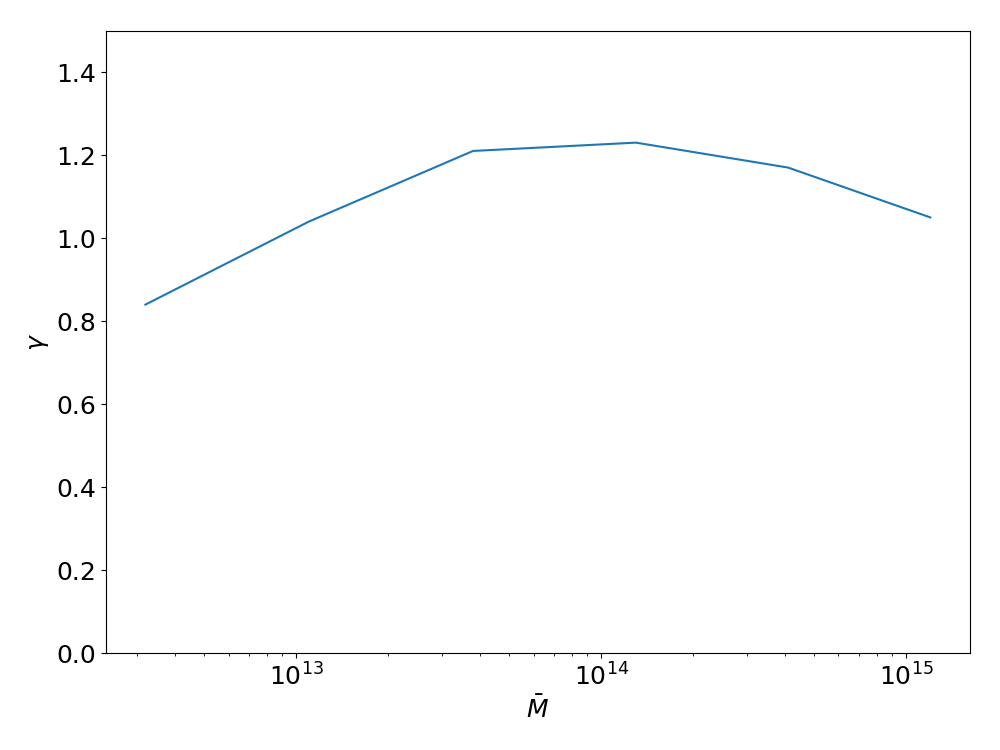}
    \caption{Power law fit to the halo profile at different mass
      bins.
    }
    \label{fig:gamma}
  \end{center}
\end{figure}

We found that the NFW profile is clearly more concentrated near the centre
of the halo than the density profile of the benchmark subhalos
(as observed already in, for example, \citep{subhalo1,subhalo2,subhalo3,subhalo4}).
Consequently, for a NFW profile based galaxy catalogue we expect a stronger
correlation than the benchmark at small scales. We have also modified
the NFW profile by keeping its functional form but changing the
concentration, but this was not a good fit to the \texttt{ROCKSTAR}
profile as shown in \Cref{fig:halo_profile1}.
Following \citep{subgen}, we then adopted a universal
power law $\rho\propto r^{-\gamma}$, where $\gamma\sim1$ is our fiducial
halo profile, such that
\begin{align}
  \label{eq:power_law}
  \text{CDF}_{\text{pow}}(r|M_{vir},c)
  =\left(\frac{r}{R_{vir}}\right)^{3-\gamma}.  
\end{align}
We have found that $\gamma\approx1$ is a satisfactory fit to the subhalo
number distribution, as shown in \Cref{fig:halo_profile1,fig:gamma}.

\subsection{Halo Occupation Distribution (HOD) \label{subsec:HOD}}

Another important consideration in the population of parent halos
is the halo occupation number, i.e. the number of galaxies per halo.
A conventional way to phenomenologically model this is via a Halo
Occupation Distribution (HOD) algorithm \citep{hod1,hod2,hod3} which
gives the mean occupation number as a function of the mass of the halo.
A functional form for this algorithm consisting of 5 parameters is
commonly used in the literature
\citep{zheng,eisenstein1,eisenstein2,mock1}:
\begin{align}
  \label{eqn:5_param_hod}
  \bar{N}_{\text{cent}}(M)
  &=\frac{1}{2}\erfc\left[-\frac{\ln M/M_0}{\sqrt{2}\sigma}\right], \\
  \bar{N}_{\text{sat}}(M)
  &=\left(\frac{M-\kappa M_0}{M_1}\right)^\alpha,
    \label{eq:simpleHODfit}
\end{align}
where $\bar{N}_{\text{cent}}$ is the expected number of central
galaxies and $\bar{N}_{\text{sat}}$ the expected number of
satellite galaxies such that
$\bar{N}_g(M)=\bar{N}_{\text{cent}}(M)+\bar{N}_{\text{sat}}(M)$.
Here $M_0$ denotes the typical minimum mass scale for a halo to have a
central galaxy, and $\sigma$ is the parameter that controls the scatter
around that mass. $\kappa M_0$ sets the cutoff scale for a halo to
host a satellite, $M_1$ is the typical additional mass above $\kappa M_0$
for a halo to have one satellite galaxy, and $\alpha$ is the exponent
that controls the tail of the HOD, and therefore has a strong
influence on the number of high-mass halos. 

Instead of using the error function we employ a Heaviside cut for
$\bar{N}_{\text{cent}}$:
\begin{align}
  \label{eq:N_cent}
  \bar{N}_{\text{cent}}(M)=\theta(M-M_0),
\end{align}
reducing the number of parameters to 4. This is appropriate as we
impose a mass cut on the parent halo when constructing the benchmark
galaxy catalogue. These 4 parameters give us freedom to tweak the
power spectrum and bispectrum of our galaxy mock catalogues to better
reproduce those of the benchmark sample. The total number of galaxies
is 
\begin{align}
  \label{eq:n_g}
  n_g = \int dM\,n(M)\left(\theta(M-M_0)+
  \left(\frac{M-\kappa M_0}{M_1}\right)^{\alpha}\right),
\end{align}
where $n(M)$ is the halo mass function that gives the number density
of halos for a given mass $M$. If the variation in the parameters are
small we obtain the following perturbation to the number of galaxies
to first order:
\begin{align}
  \label{eq:constraint}
  &\Delta n_g \nonumber \\
  ={}& - \int dM\,n(M) \nonumber \\
             &\times\Bigg(\frac{\Delta M_0}{M_0}M_0\left(\delta(M-M_0)+
               \frac{\alpha\kappa}{M_1}
               \left(\frac{M-\kappa M_0}{M_1}\right)^{\alpha-1}\right) \nonumber \\
             &\qquad+\frac{\Delta \kappa}{\kappa}\kappa\frac{\alpha M_0}{M_1}
               \left(\frac{M-\kappa M_0}{M_1}\right)^{\alpha-1} \nonumber \\
             &\qquad+\frac{\Delta M_1}{M_1}M_1\frac{\alpha (M-\kappa M_0)}{M^2_1}
               \left(\frac{M-\kappa M_0}{M_1}\right)^{\alpha-1} \nonumber \\
             &\qquad-\frac{\Delta \alpha}{\alpha}\alpha\log\left(\frac{M-\kappa M_0}{M_1}\right)
               \left(\frac{M-\kappa M_0}{M_1}\right)^\alpha\Bigg),
\end{align}
and we enforce $\Delta n_g=0$ to conserve particle number
when changing the parameters.

\begin{figure*}
  \begin{center}
    \begin{subfigure}[b]{0.49\textwidth}
      \includegraphics[width=\linewidth]{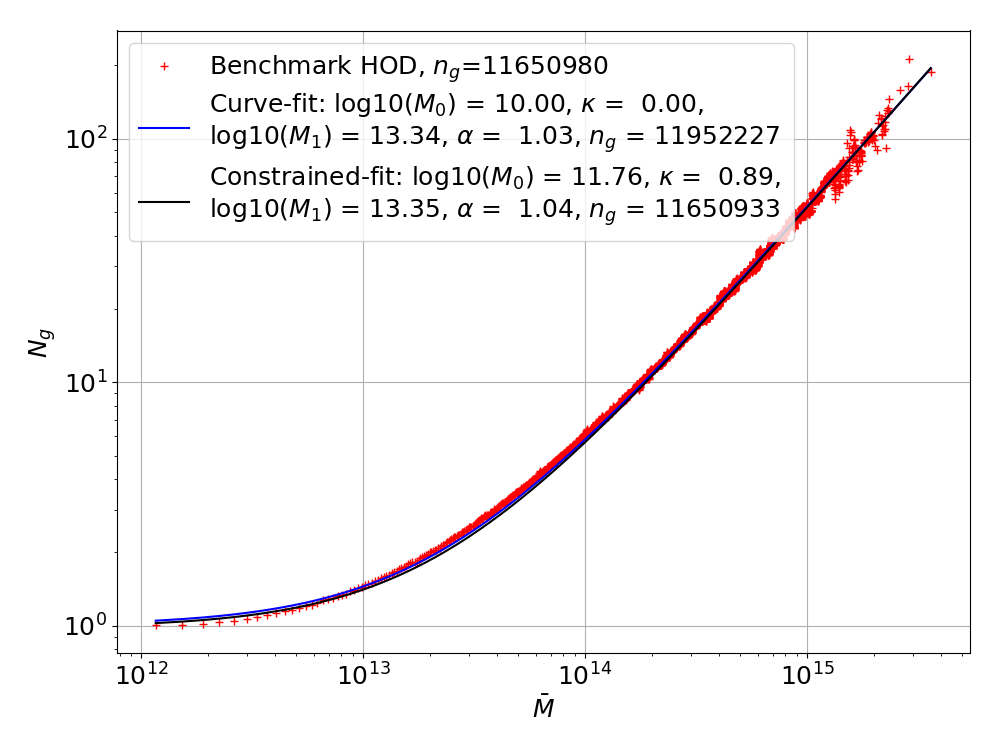}
    \end{subfigure}
    ~
    \begin{subfigure}[b]{0.49\textwidth}
      \includegraphics[width=\linewidth]{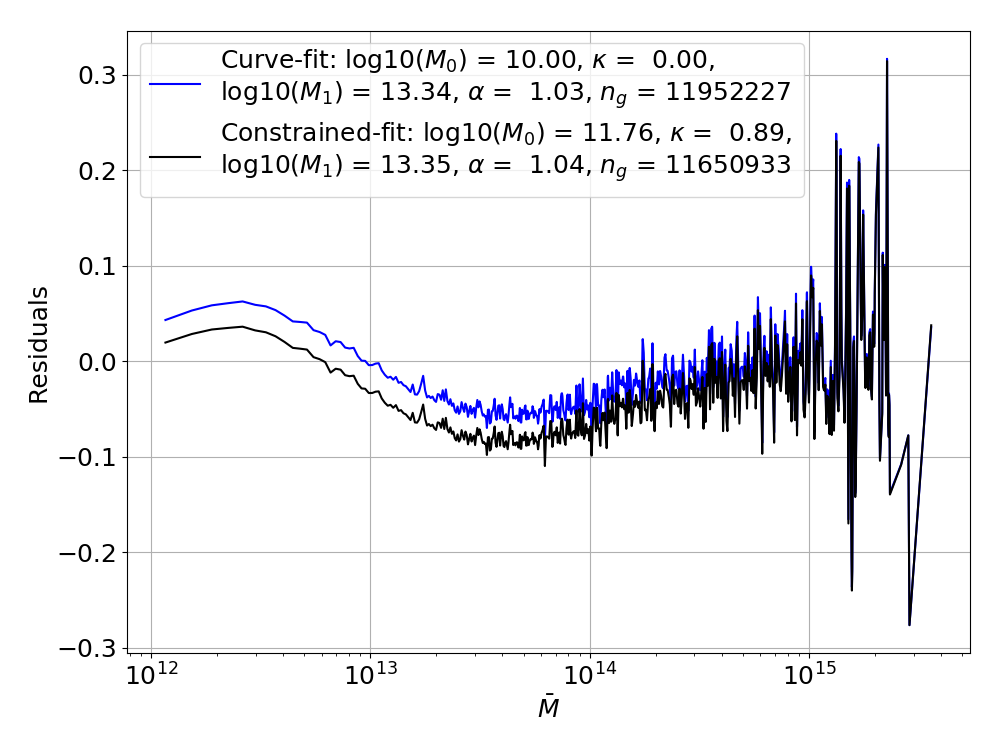}
    \end{subfigure}
    \caption{Left panel: Fits to the benchmark HOD with our 4 parameters,
      both constraining and not constraining the total number of galaxies.
      Right panel: Residuals of these fits.}
    \label{fig:hod_best_fit}
  \end{center}
\end{figure*}

In \Cref{fig:hod_best_fit} we show the HOD $\bar{N}_g(M)$
from our benchmark \texttt{ROCKSTAR} catalogue (which will be referred
to as the benchmark HOD model below), and the best fit for the
4-parameter HOD while keeping the total number of galaxies constant. As
a comparison we also obtain an unconstrained fit to the benchmark HOD.
The best fit parameters for the constrained fit are
$\log(M_0)=11.76$, $\kappa=0.89$,$\log(M_1)=13.35$ and $\alpha=1.04$,
with only a $4\times10^{-4}\%$ deficiency in the number of galaxies.

\section{Halo polyspectra\label{sec:halo-polyspectra}}

\subsection{Power spectrum and Bispectrum}

\begin{figure*}
  \begin{center}
    \begin{subfigure}[b]{0.49\textwidth}
      \includegraphics[width=\textwidth]{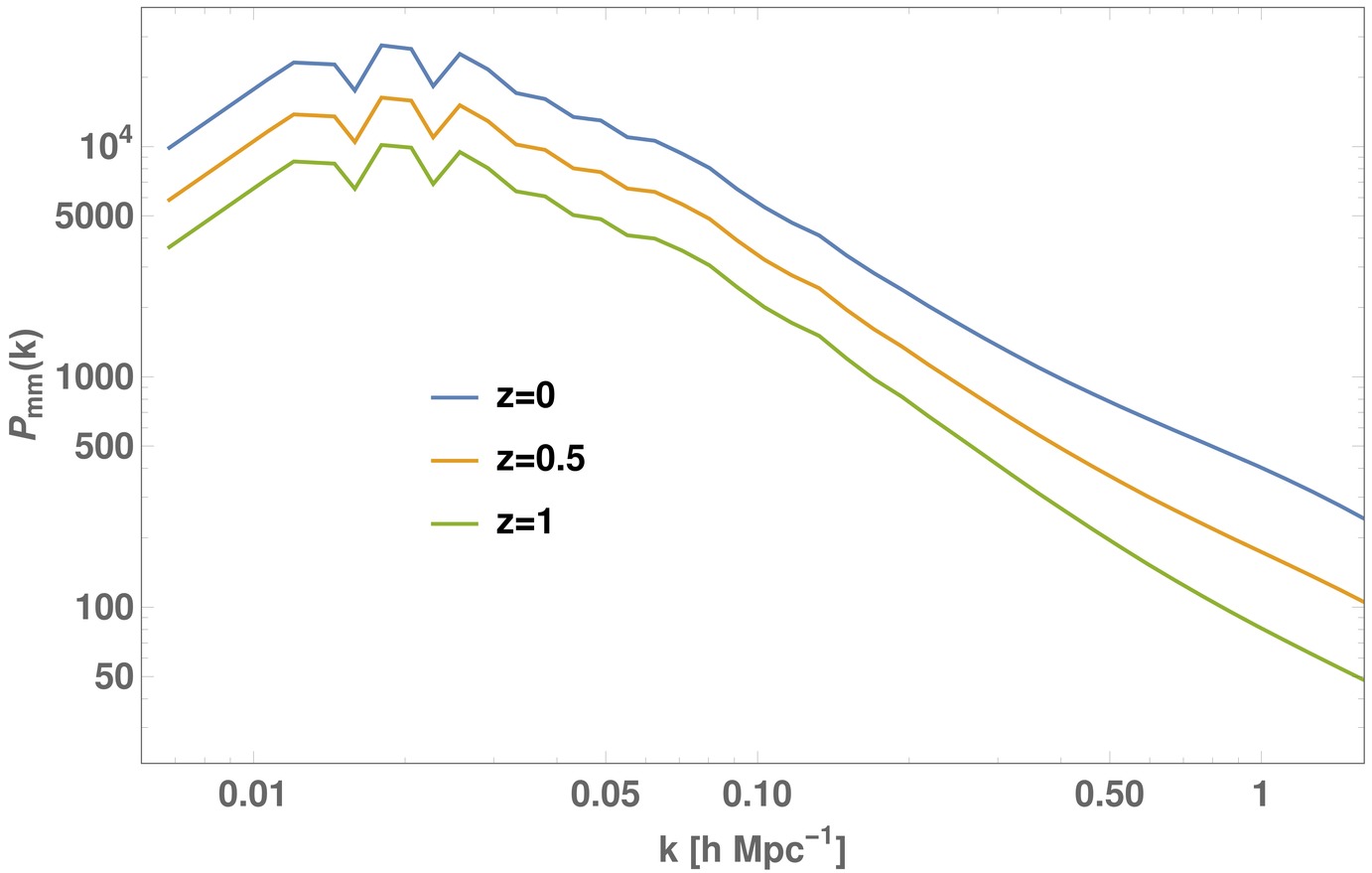} 
    \end{subfigure}
    ~
    \begin{subfigure}[b]{0.49\textwidth}
      \includegraphics[width=\textwidth]{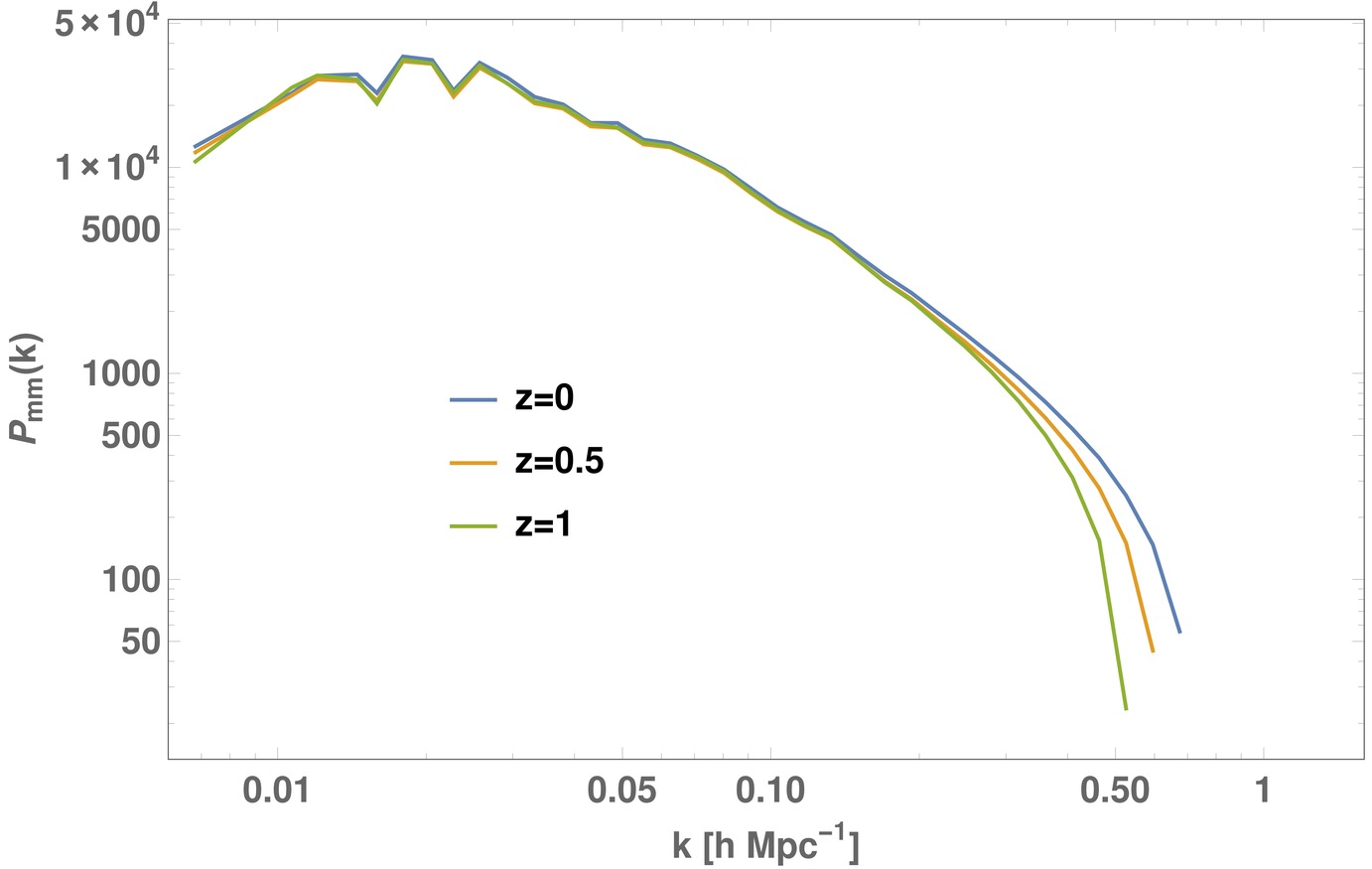}
    \end{subfigure}

    \begin{subfigure}[b]{0.49\textwidth}
      \includegraphics[width=\textwidth]{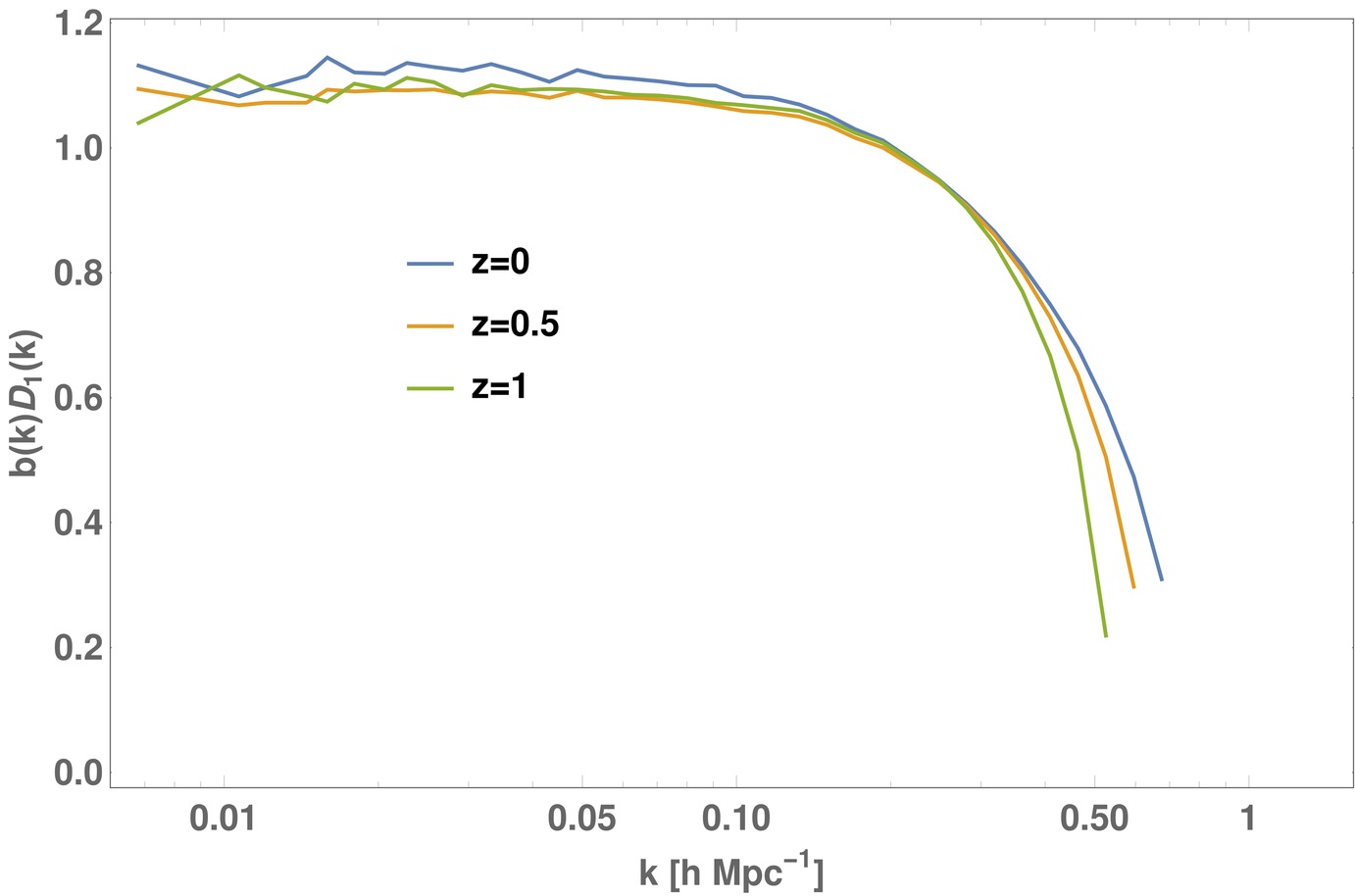}
    \end{subfigure}
    \caption{Redshift evolution of the estimated power spectrum
      of the $1280\,h^{-1}$ Mpc benchmark \GADGET{} dark matter
      simulation (top left), and the benchmark galaxy mock catalogue
      derived from it using \texttt{ROCKSTAR} (after shot noise subtraction,
      top right), plotted
      up to $k_{max}=1.6\,h\,\text{Mpc}^{-1}$. The bottom panel shows
      the product between the bias parameter, obtained from
      $b=\sqrt{P_{hh}/P_{mm}}$, and the linear growth factor $D_1$,
      which tends to a constant at large scales irrespective of
      redshift.
    }
    \label{fig:gadget16_ps}
  \end{center}
\end{figure*}

The leading source of cosmological information, and hence
the principal diagnostic of our methods, is the two-point
correlator, or power spectrum $P(k)$ of an overdensity
field $\delta(\mathbf{x})$:
\begin{align}
  \expval{\delta(\mathbf{k}) \delta(\mathbf{k}')}=(2\pi)^3 
  \delta_D(\mathbf{k}+\mathbf{k}') P(k),
  \label{PS}
\end{align}
where $\delta_D$ is the Dirac delta function. The power
spectra of our benchmark dark matter and galaxy catalogues
at redshifts $z=0,0.5,1$ are plotted in \Cref{fig:gadget16_ps}.
Our galaxy catalogue consists of parent halos with mass
in the range of $1\times10^{12}$ and
$3.2\times10^{15}h^{-1}\,M_{\odot}$ and all their subhalos, and
has a number density of $0.0056\,h^{3}\,\text{Mpc}^{-3}$,
which is similar to the number density of the LOWZ
galaxy sample in BOSS at low redshift \citep{anderson}.
It is well known in the literature that while the dark matter
power spectrum grows with time, the growth of the halo power
spectrum is slow \citep{halo_evo,halo_evo1}. At large scales the
linear bias relationship $b_1=\delta_g/\delta$ between dark
matter and galaxies tends to a constant \citep{linear_bias}, and
since the dark matter power spectrum grows as $D^2_1(z)$ at these
scales, where $D_1(z)$ is the linear growth factor, we expect
$b_1(z)\propto1/D_1(z)$. This is also shown clearly in
\Cref{fig:gadget16_ps}, giving a value of $b_1\approx1.1$.

\begin{figure*}
  \captionsetup[subfigure]{labelformat=empty}
  \begin{center}
    \begin{subfigure}[b]{0.42\textwidth}
      \includegraphics[width=\textwidth]{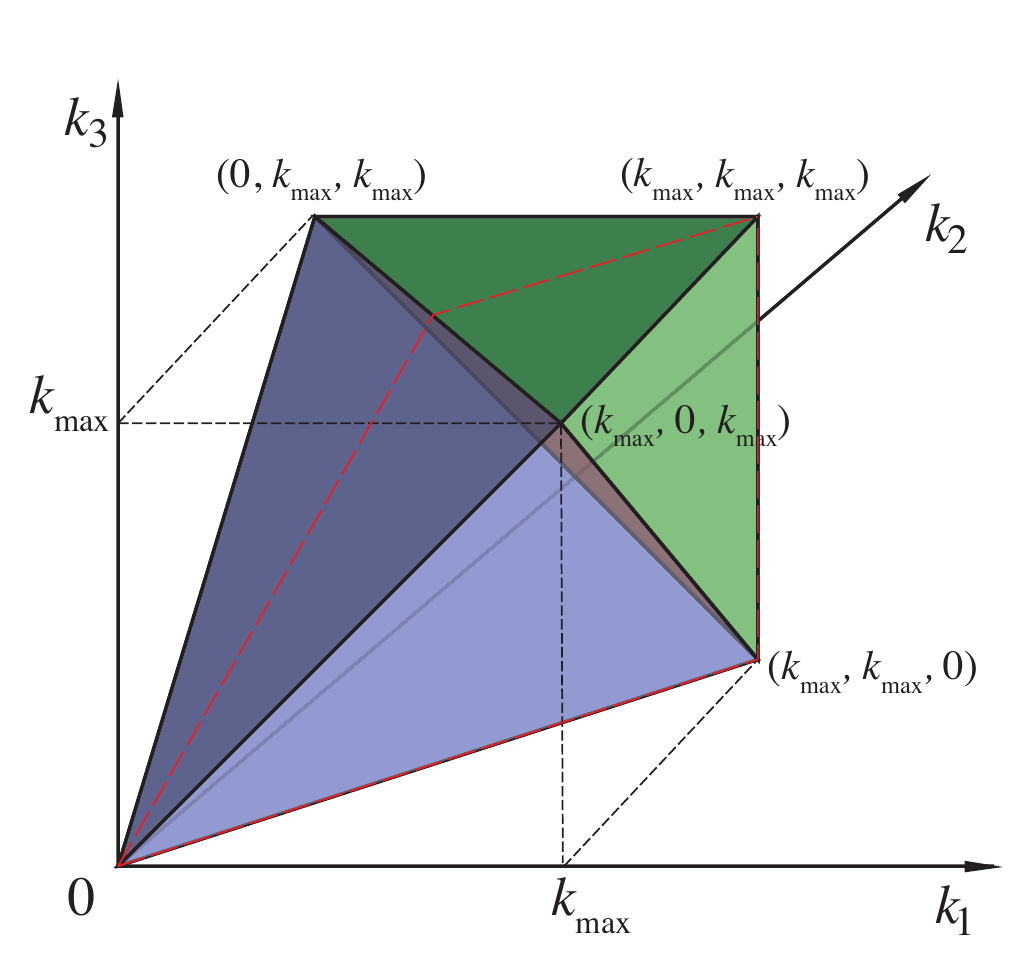} 
    \end{subfigure}
    ~
    \begin{subfigure}[b]{0.42\textwidth}
      \includegraphics[width=\textwidth]{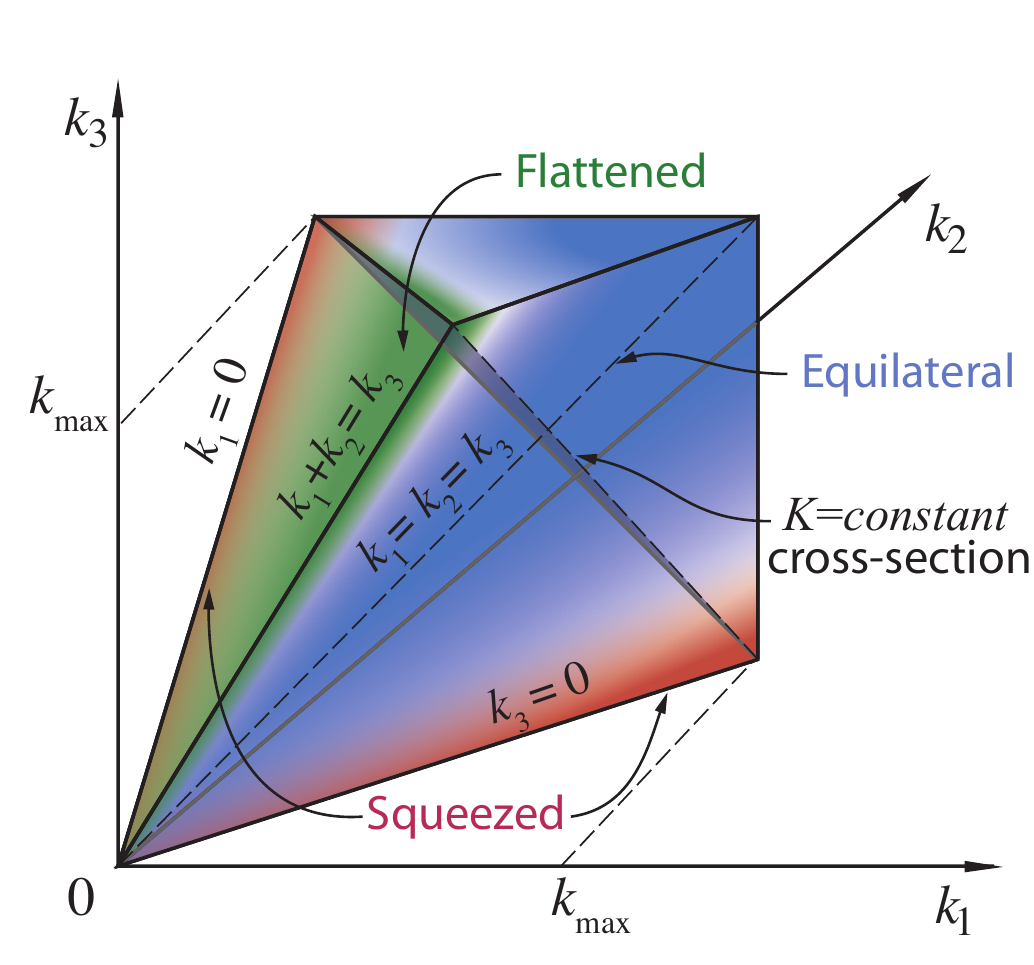}
    \end{subfigure}
    \caption{Left panel: The full tetrapyd bispectrum domain $\mathcal{V}_B$
      consists of a tetrahedral region (blue) defined by the wavevector
      triangle condition $\mathbf{k}_1+\mathbf{k}_2+\mathbf{k}_3=0$, together
      with a pyramidal region (green) bounded by the resolution limit $k_{\text{max}}$.
      Right panel: To show the internal structure of the tetrapyd we split
      it along the red dashed line, showing only the back half with $k_1<k_2$.
      Colour-coded regions show the location of the `squeezed' (red),
      `flattened' (green) and `equilateral' or `constant' (blue) shape signals.
      The scale dependence of the bispectrum is reflected by the
      $K\equiv k_1+k_2+k_3=\text{const.}$ cross sectional planes \citep{Andrei}.
    }
    \label{tetrapyd}
  \end{center}
\end{figure*}

For mildly non-linear scales the primary diagnostic is the
three point correlation function or bispectrum
$B_\delta(k_1, k_2, k_3)$:
\begin{align}
  &\expval{\delta(\mathbf{k}_1) \delta(\mathbf{k}_2) \delta(\mathbf{k}_3)}
    \nonumber \\
  &\qquad=(2\pi)^3 \delta_D (\mathbf{k}_1+\mathbf{k}_2+\mathbf{k}_3)
    B_\delta(k_1,k_2,k_3).
    \label{bispectrum}
\end{align}
Due to statistical isotropy and homogeneity, in configuration space the
bispectrum only depends on the wavenumbers $k_i$ in the absence of redshift
space distortions. Additionally the delta function, arising from momentum
conservation, imposes the triangle condition on the wavevectors so the three
$k_i$ when taken as lengths must be able to form a triangle. Together with
a parameter $k_{max}$ which defines the resolution of the data, the bispectrum
occupies a tetrapydal domain $\mathcal{V}_B$ in $k$-space, as shown in
the left panel of \Cref{tetrapyd}. We have found it useful to split it in half
to make apparent its internal morphology as illustrated in the right panel of
\Cref{tetrapyd}. The bispectra plots in this paper are generated with
\texttt{ParaView} \citep{paraview}, an open source scientific visualisation tool.

\begin{figure*}
  \captionsetup[subfigure]{labelformat=empty}
  \begin{center}
    \begin{subfigure}[b]{0.37\textwidth}
      \includegraphics[width=\linewidth]{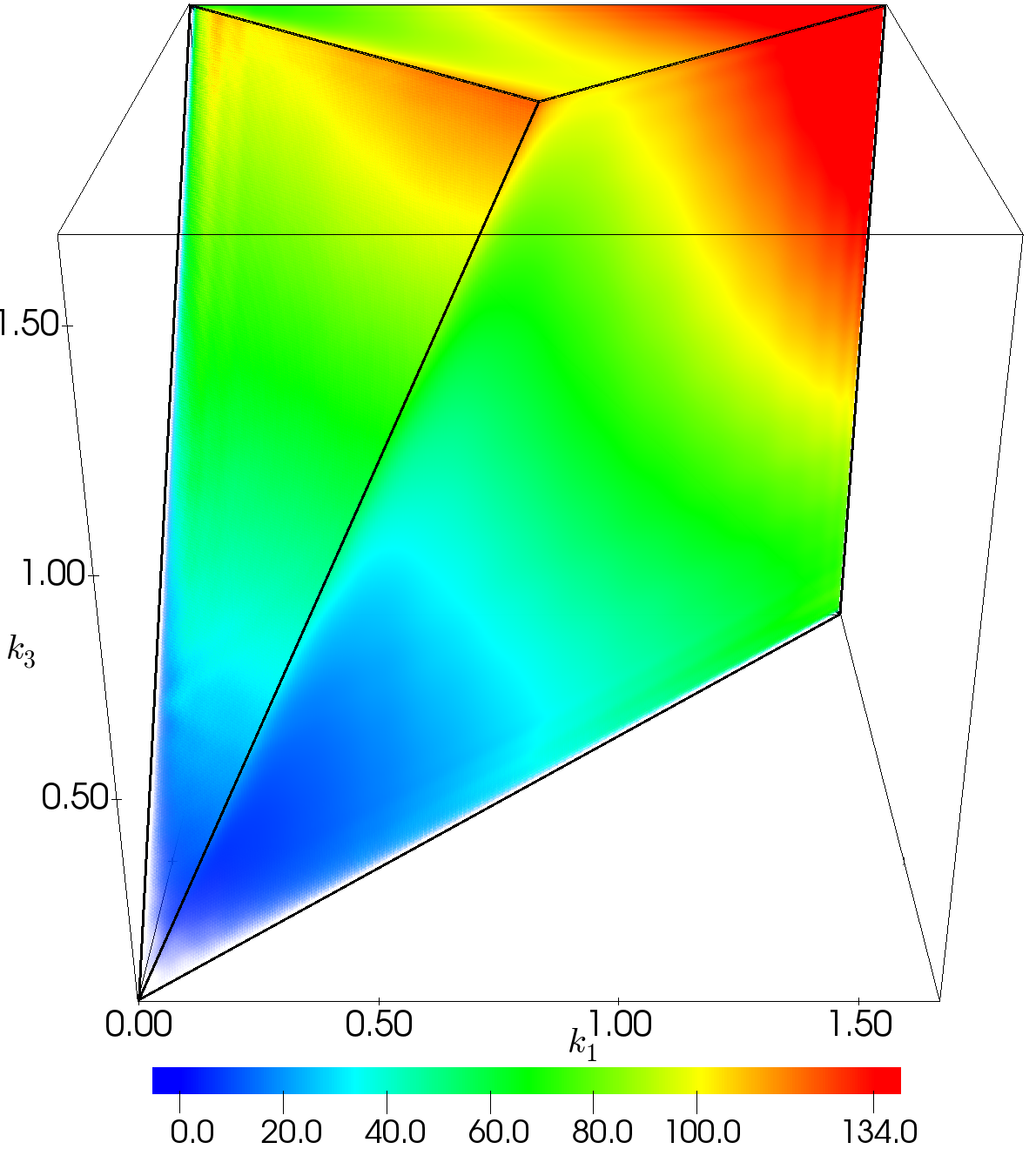}
      \caption{Dark matter, $z=1$}
    \end{subfigure}
    ~
    \begin{subfigure}[b]{0.37\textwidth}
      \includegraphics[width=\linewidth]{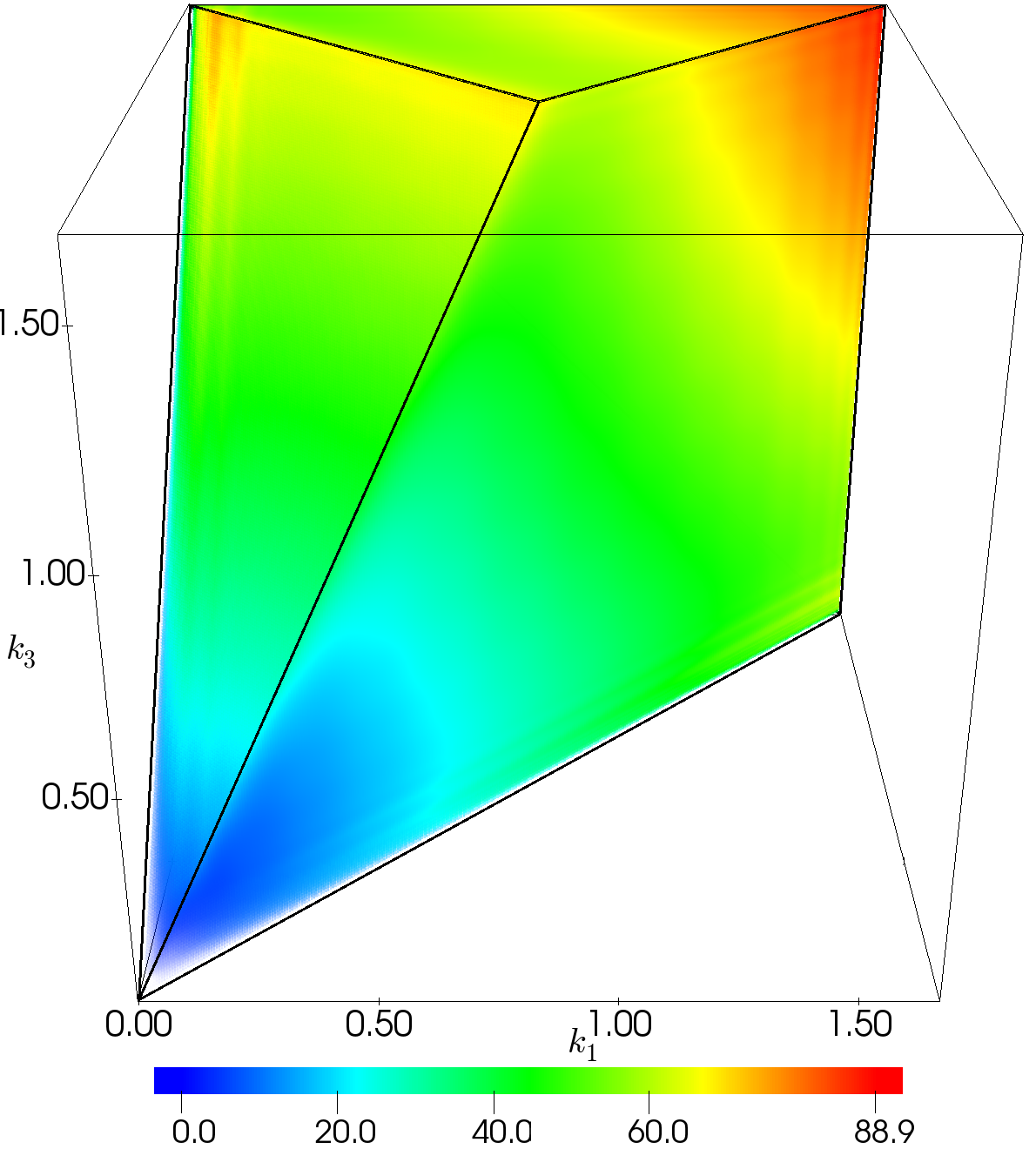}
      \caption{Galaxies, $z=1$}
    \end{subfigure}
    
    \begin{subfigure}[b]{0.37\textwidth}
      \includegraphics[width=\linewidth]{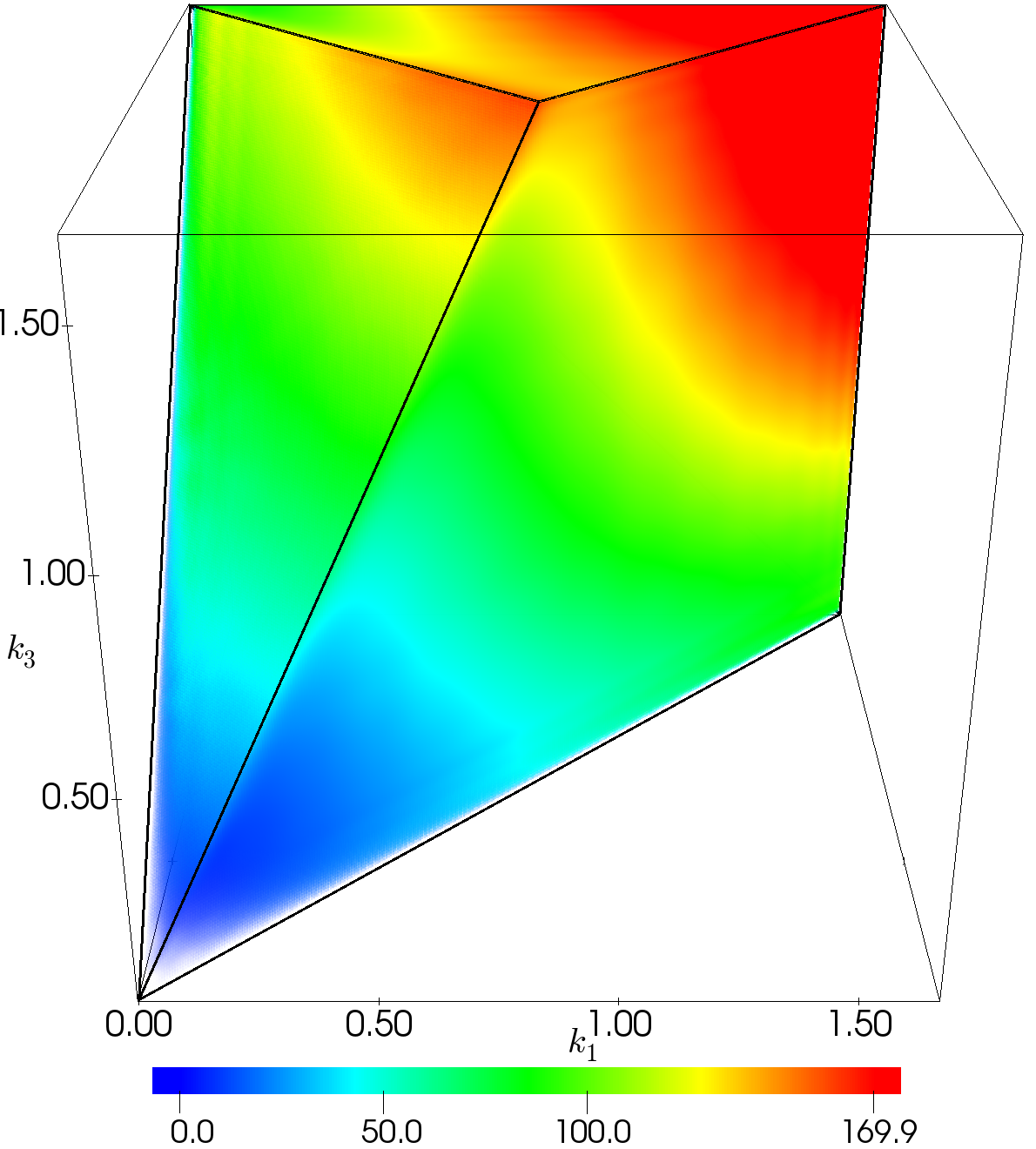}
      \caption{Dark matter, $z=0.5$}
    \end{subfigure}
    ~
    \begin{subfigure}[b]{0.37\textwidth}
      \includegraphics[width=\linewidth]{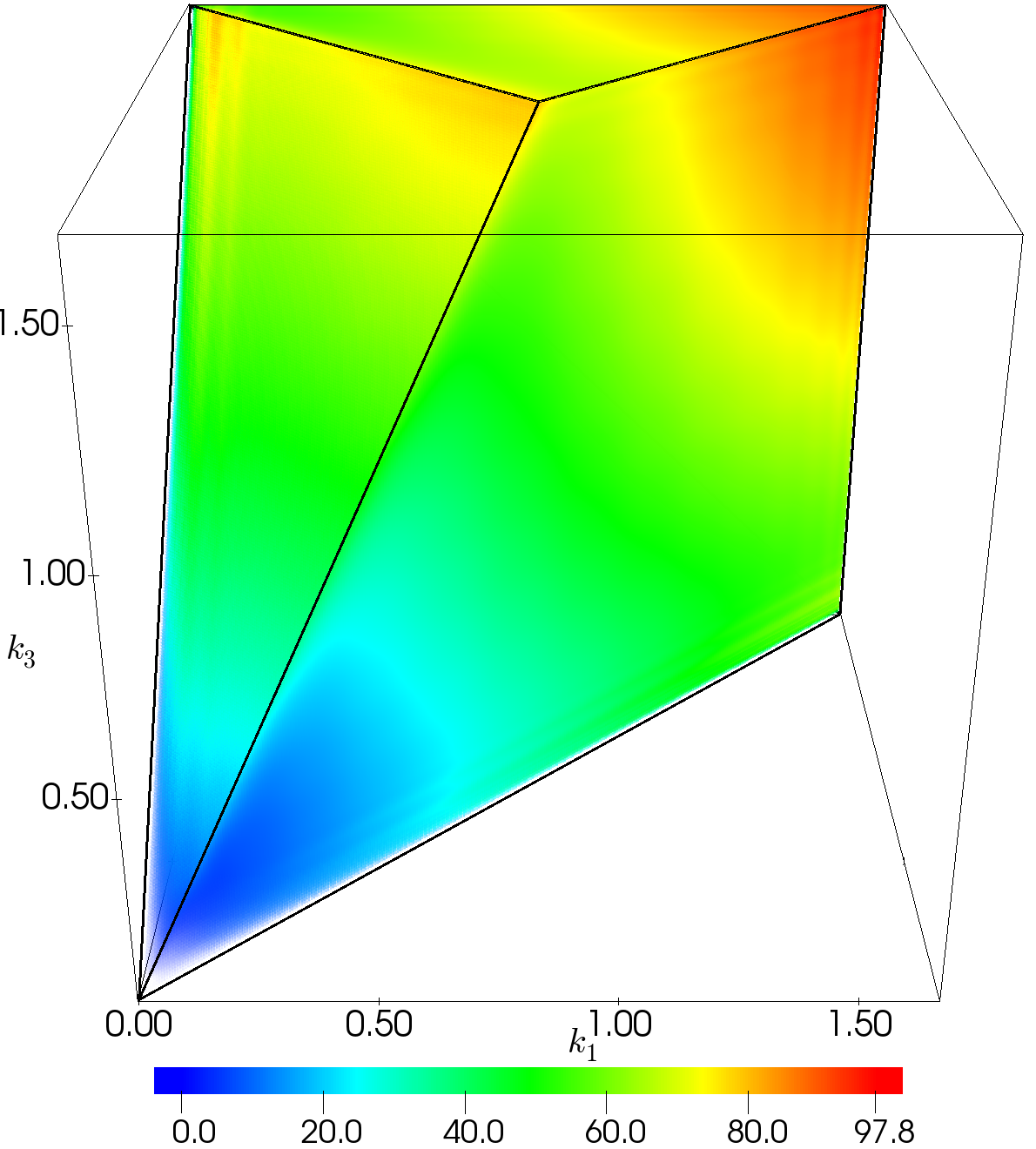}
      \caption{Galaxies, $z=0.5$}
    \end{subfigure}
    
    \begin{subfigure}[b]{0.37\textwidth}
      \includegraphics[width=\linewidth]{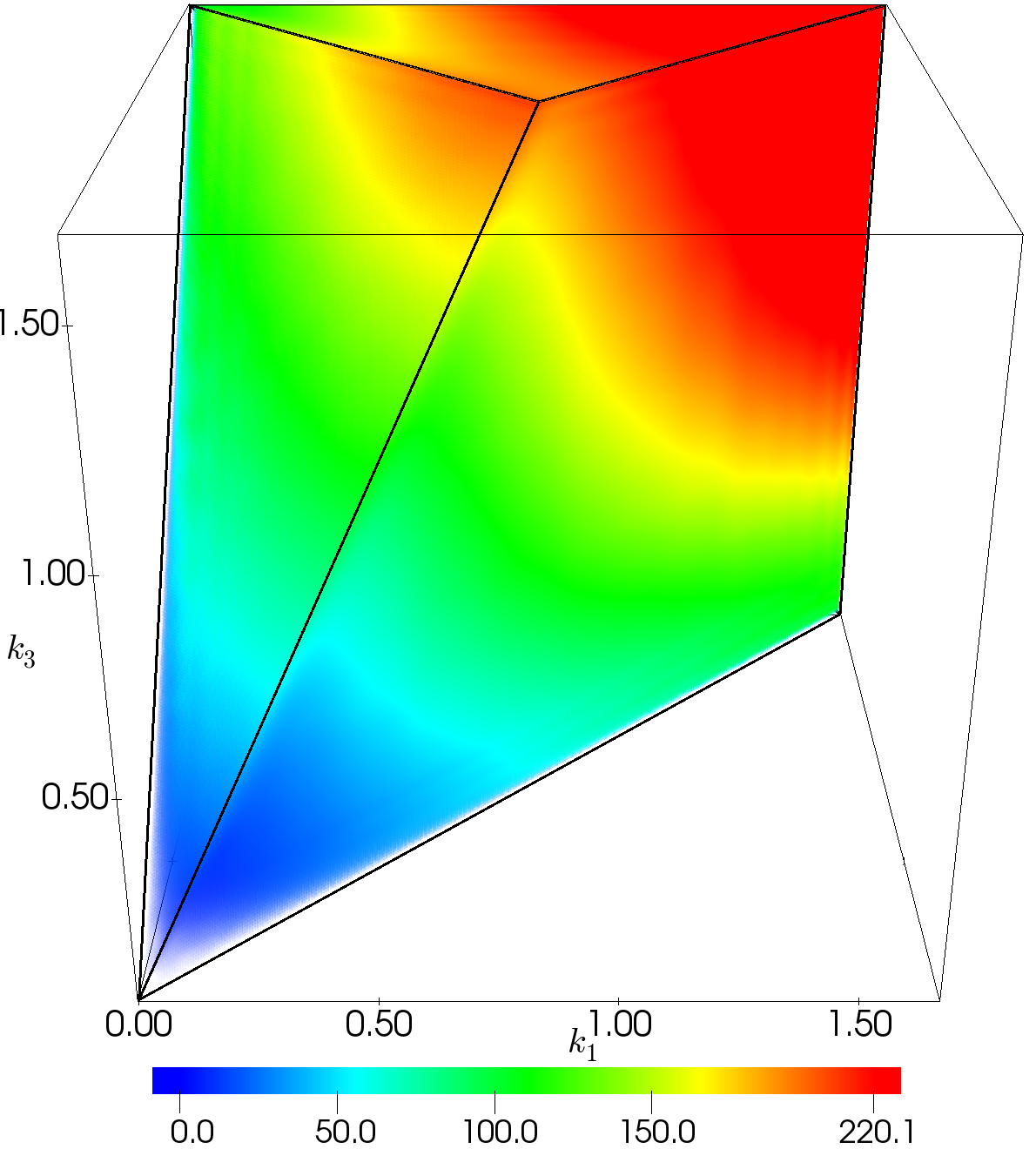}
      \caption{Dark matter, $z=0$}
    \end{subfigure}
    ~
    \begin{subfigure}[b]{0.37\textwidth}
      \includegraphics[width=\linewidth]{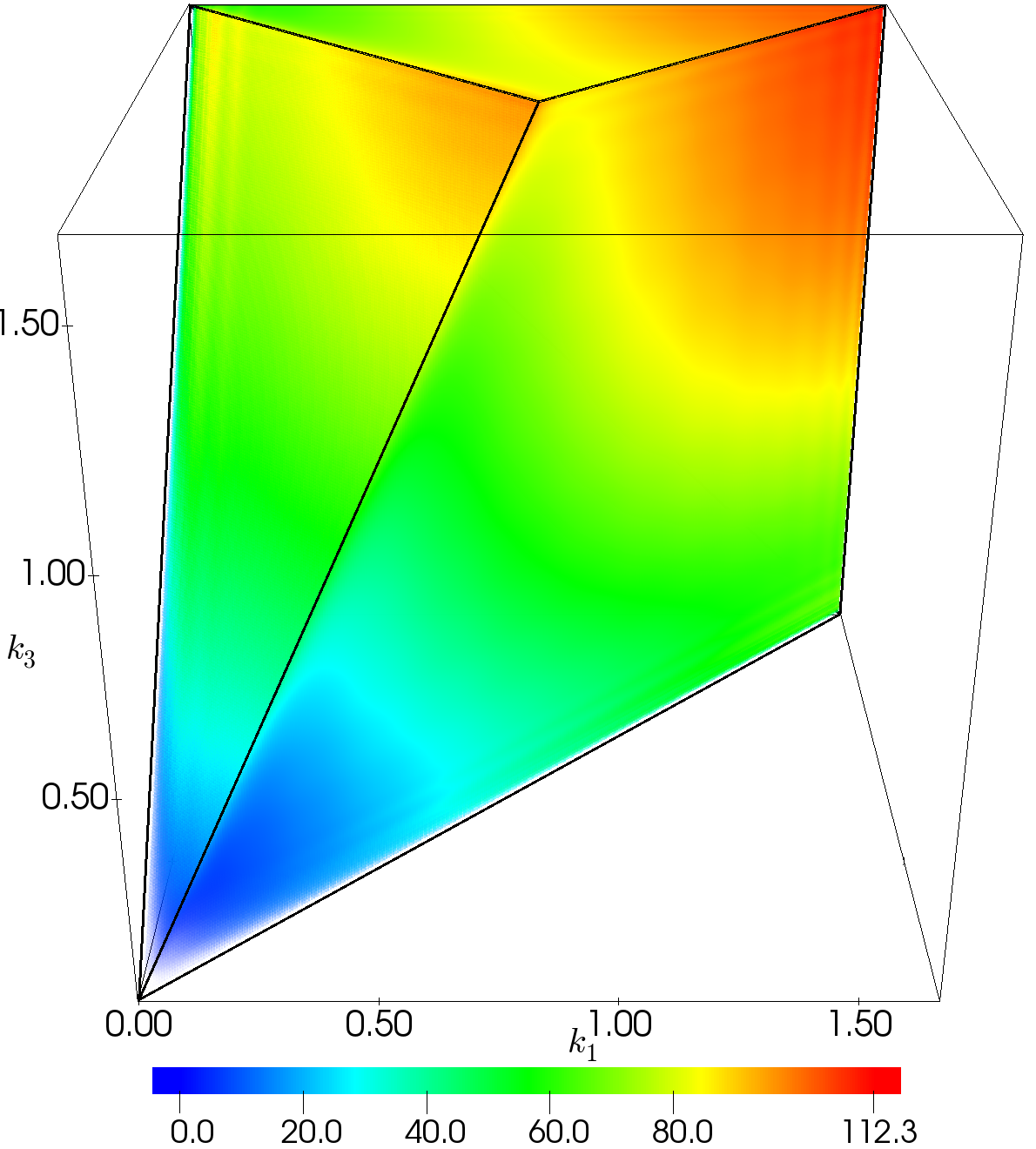}
      \caption{Galaxies, $z=0$}
    \end{subfigure}
    \caption{
      Redshift evolution of the estimated bispectra of the
      $1280\,h^{-1}$ Mpc benchmark \GADGET{} dark matter
      simulation, and the benchmark galaxy mock catalogue
      derived from it using \texttt{ROCKSTAR},
      plotted up to $k_{max}=1.6\,h\,\text{Mpc}^{-1}$.
    }
    \label{fig:gadget16_bis}
  \end{center}
\end{figure*}

Due to the large number of triangle configurations, numerical
estimation of the full bispectrum is computationally expensive.
In this paper we use the newly rewritten \MODALLSS{} method for
the efficient and accurate estimation of the bispectrum for any
overdensity field $\delta$ \citep{DM}. The full bispectra of the
benchmark catalogue at various redshifts thus obtained are shown
in \Cref{fig:gadget16_bis}, along with the corresponding dark
matter bispectra plotted for reference.

\subsection{\MODALLSS{} bispectrum methodology\label{sec:modallss-methodology}}

Here we give a brief summary of the \MODALLSS{} algorithm. We
first approximate the signal-to-noise weighted estimated bispectrum
of a density field,
$\sqrt{\frac{k_1k_2k_3}{P(k_1)P(k_2)P(k_3)}}\hat{B}_\delta(k_1,k_2,k_3)$,
by expanding it in a general separable basis:
\begin{align}
  &\sqrt{\frac{k_1k_2k_3}{P(k_1)P(k_2)P(k_3)}}\hat{B}_\delta(k_1,k_2,k_3) \nonumber \\
  &\quad\approx\sum_{mn}^{n_{max}} \gamma^{-1}_{nm}\beta^Q_m
    Q^{\MODALLSS{}}_n(k_1/k_{max},k_2/k_{max},k_3/k_{max}),
    \label{expand}
\end{align}
where $P(k)$ is the power spectrum of the density field. The
information in the full bispectrum is compressed into these
$\mathcal{O}(1000)$ $\beta^Q_m$ coefficients, and it has been
shown to be superior to other bispectrum estimators in terms
of data compression \citep{sussex}. The basis functions
$Q^{\MODALLSS{}}_n$ are symmetrised products over one dimensional functions
$q_r$:
\begin{align}
  Q^{\MODALLSS{}}_n (x,y,z) \equiv q_{\{r}(x)q_{s}(y)q_{t\}}(z),
\end{align}
with $\{\dots\}$ representing symmetrisation over the indices $r,s,t$, and
each $n$ corresponds to a combination of $r,s,t$. The relationship between
$n$ and $r,s,t$ is `slice ordering' which orders the triples by the sum
$r+s+t$. $k_{max}$ is the resolution of the tetrahedral domain defined above.
$\gamma_{nm}$ is the inner product between $Q^{\MODALLSS{}}_n$ functions over
the tetrapyd domain:
\begin{align}
  \gamma_{nm}\equiv\frac{V}{\pi}\int_{\mathcal{V}_B}dV_kQ_nQ_m,
  \label{gamma}
\end{align} 
where $V=(2\pi)^3\delta_D(\mathbf{0})$ is the volume of the simulation box.
There is a freedom in the choice of $q_r$, provided the $Q^{\MODALLSS{}}_n$
basis is orthogonal, or can be made orthogonal. We employ shifted Legendre
polynomials $\tilde{P}_l(x)=P_l(2x-1)$, such that $\tilde{P}_l(x)$ is
orthogonal over the interval $\left[0,1\right]$ instead of the usual
$\left[-1,1\right]$ for $P_l(x)$.

For $\hat{B}_\delta(k_1,k_2,k_3)=\frac{1}{V}\delta(\mathbf{k}_1)
\delta(\mathbf{k}_2)\delta(\mathbf{k}_3)$ we multiply both sides of
\Cref{expand} by $Q^{\MODALLSS{}}_m(k_1/k_{max},k_2/k_{max},k_3/k_{max})$ and
integrate over $\mathcal{V}_B$ to find 
\begin{align}
  \label{eq:betaQ}
  \beta^Q_n
  &=\frac{1}{\pi}\int_{\mathcal{V}_B}dV_k
    \sqrt{\frac{k_1k_2k_3}{P(k_1)P(k_2)P(k_3)}}
    \delta_{\mathbf{k}_1}\delta_{\mathbf{k}_2}
    \delta_{\mathbf{k}_3}
    \nonumber \\
  &\qquad\qquad\qquad\times q_{\{r}(\frac{k_1}{k_{max}})
    q_s(\frac{k_2}{k_{max}})q_{t\}}(\frac{k_3}{k_{max}})
    \nonumber \\
  &=\int_{\mathbf{k}_1,\mathbf{k}_2,\mathbf{k}_3}(2\pi)^6
    \delta_D(\mathbf{k}_1+\mathbf{k}_2+\mathbf{k}_3)
    \nonumber \\
  &\qquad\qquad\qquad\times\frac{\delta_{\mathbf{k}_1}
    \delta_{\mathbf{k}_2}\delta_{\mathbf{k}_3}}
    {\sqrt{k_1P(k_1)k_2P(k_2)k_3P(k_3)}} \nonumber \\
  &\qquad\qquad\qquad\times q_{\{r}(\frac{k_1}{k_{max}})
    q_s(\frac{k_2}{k_{max}})q_{t\}}(\frac{k_3}{k_{max}})
    \nonumber \\
  &=(2\pi)^3\int d^3 x\int\frac{\prod_i d^3k_i}{(2\pi)^9}
    e^{i(\mathbf{k}_1+\mathbf{k}_2+\mathbf{k}_3)\cdot\mathbf{x}}
    \nonumber \\
  &\qquad\qquad\qquad\times\frac{\delta_{\mathbf{k}_1}
    \delta_{\mathbf{k}_2}\delta_{\mathbf{k}_3}}
    {\sqrt{k_1P(k_1)k_2P(k_2)k_3P(k_3)}} \nonumber \\
  &\qquad\qquad\qquad\times q_{\{r}(\frac{k_1}{k_{max}})
    q_s(\frac{k_2}{k_{max}})q_{t\}}(\frac{k_3}{k_{max}})
    \nonumber \\
  &=(2\pi)^3\int d^3 x\,
    M_r(\mathbf{x})M_s(\mathbf{x})M_t(\mathbf{x}),
\end{align}
where we define
\begin{align}
  M_r(\mathbf{x}) \equiv \int\frac{d^3k}{(2\pi)^3}\frac{\delta_{\mathbf{k}}q_r(k/k_{max})}
  {\sqrt{kP(k)}}e^{i\mathbf{k}\cdot\mathbf{x}},
  \label{Mfunc}
\end{align}
which is an inverse Fourier transform, and
$\int_{\mathbf{k}_1,\mathbf{k}_2,\mathbf{k}_3}=\int\frac{d^3k_1}{(2\pi)^3}
\frac{d^3k_2}{(2\pi)^3}\frac{d^3k_3}{(2\pi)^3}$. In the second line we
used the identity
\begin{align}
  &\int\frac{d^3 k_1}{(2\pi)^3}\frac{d^3 k_2}{(2\pi)^3}
    \frac{d^3 k_3}{(2\pi)^3}(2\pi)^6\delta^2_D
    \left(\mathbf{k}_1+\mathbf{k}_2+\mathbf{k}_3\right)F \nonumber \\
  &\quad=\frac{V}{8\pi^4}\int_{\mathcal{V}_B}dk_1dk_2dk_3\,k_1k_2k_3F.
    \label{integrals}
\end{align} 

In summary, we have reduced the 9-dimensional integrals involved in
bispectrum estimation to a number of (inverse) Fourier transforms
which can be evaluated efficiently with the fast Fourier transform (FFT)
algorithm, together with an integral over the spatial extent of the data
set (\Cref{eq:betaQ}) which can highly parallelised. Additionally we
have compressed the full 3D bispectral information to $\mathcal{O}(1000)$
$\beta^Q_n$ coefficients, which are much easier to manipulate.

To make comparisons between bispectra $B_i$ and $B_j$ we first define
inner products between them as
\begin{align}
  \left[B_i,B_j\right] \equiv \frac{V}{\pi} \int_{\mathcal{V}_B}dV_k\,
  k_1k_2k_3 \frac{B_i(k_1,k_2,k_3)B_j(k_1,k_2,k_3)}{P(k_1)P(k_2)P(k_3)}.
  \label{inner_product}
\end{align}
We define two correlators between bispectra. The first is the total
correlator $\mathcal{T}$:
\begin{align}
  \mathcal{T}(B_i,B_j)
  &\equiv 1- \sqrt{\frac{
    \left[B_j-B_i,B_j-B_i\right]}{\left[B_j,B_j\right]}},
    \label{total}
\end{align}
which is a stringent test of correlation between bispectra, but
is susceptible to degradation by statistical noise. The other one
is the $f_{nl}$ correlator, named as such due to its similarity to the
optimal $\langle\hat{f}_{nl}\rangle$ estimator for the amplitude of a
theoretical shape (see \citep{DM}), as:
\begin{align}
  f_{nl}(B_i,B_j)
  \equiv \frac{\big[B_i,B_j\big]}{\big[B_j,B_j\big]}.
    \label{fnl_corr}
\end{align}
The $f_{nl}$ correlator can be thought of as proportional to the cosine
between the two shapes, weighted by the magnitude of $B_j$. This
correlator is therefore appropriate for the compression of 3D bispectral
information into a one-dimensional function of $k_{max}$.

We further define a `sliced' correlator between bispectra which 
integrates over transverse degrees of freedom
$K\equiv k_1+k_2+k_3=\text{const.}$ on the tetrahedron:
\begin{align}
  \left[B_i,B_j\right]^{S}_{K} \equiv \frac{V}{\pi} \int_{\Delta\mathcal{V}_B}dV_k\,
  k_1k_2k_3 \frac{B_i(k_1,k_2,k_3)B_j(k_1,k_2,k_3)}{P(k_1)P(k_2)P(k_3)}.
  \label{inner_product_sliced}
\end{align}
The new restricted integration region, $\Delta\mathcal{V}_B$,
encompasses a range of these $K$ slices such that:
\begin{align}
  \label{eq:slice}
  K<k_1+k_2+k_3<K+\Delta K.
\end{align}
Similarly we define the sliced $f_{nl}$ correlator as 
\begin{align}
  f^S_{nl}(B_i,B_j,K)
  \equiv \frac{\big[B_i,B_j\big]^{S}_{K}}{\big[B_j,B_j\big]^{S}_{K}}.
    \label{fnl_corr}
\end{align}

\subsection{Halo three-shape model}

\begin{figure}
  \begin{center}    
    \includegraphics[width=0.8\linewidth]{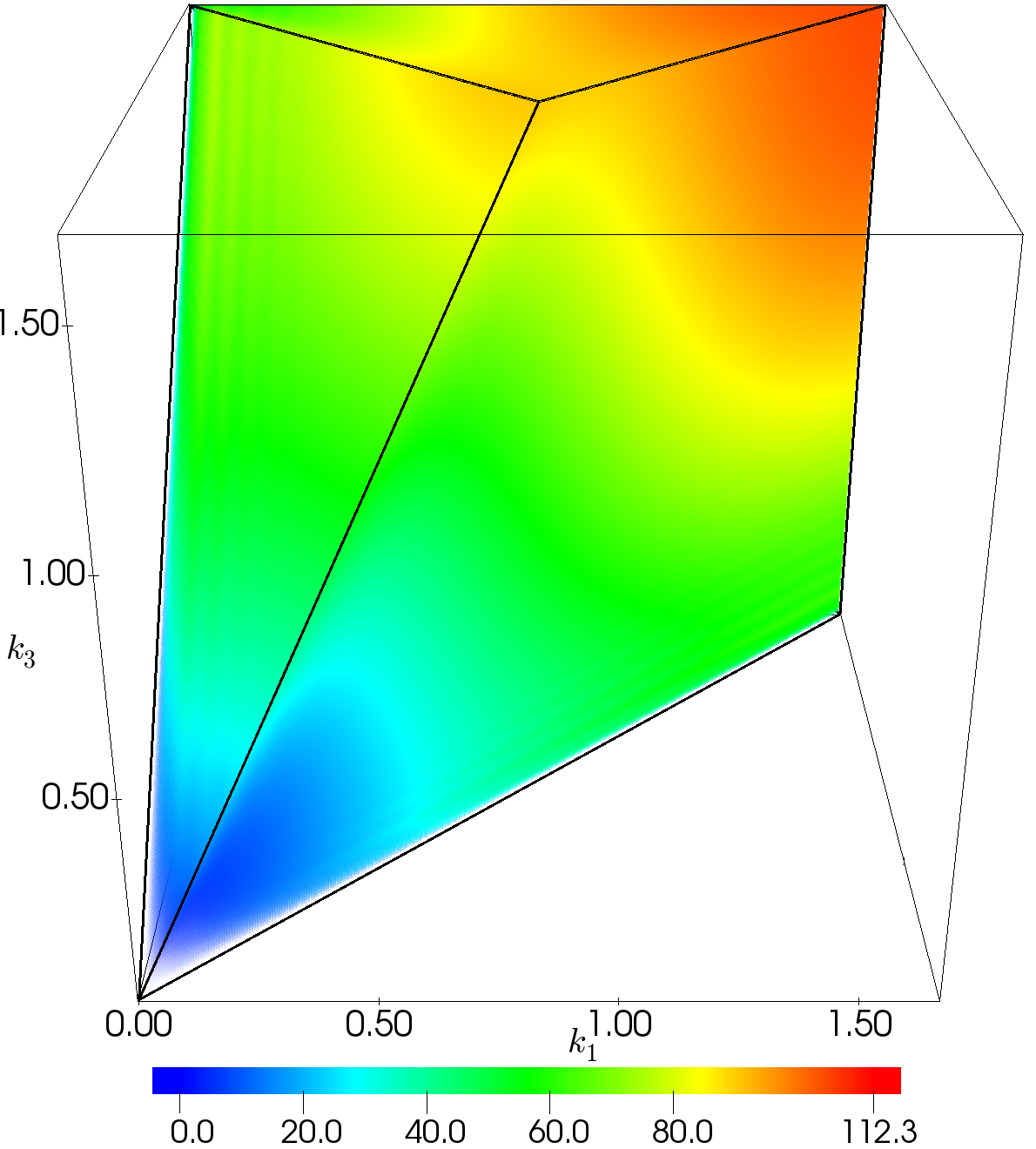}
    \caption{Best fit three-shape model to the bispectrum
      of the benchmark \texttt{ROCKSTAR} catalogue.
    }
    \label{fig:3-shape_bis}
  \end{center}
\end{figure}

\begin{figure}
  \begin{center}    
    \includegraphics[width=\linewidth]{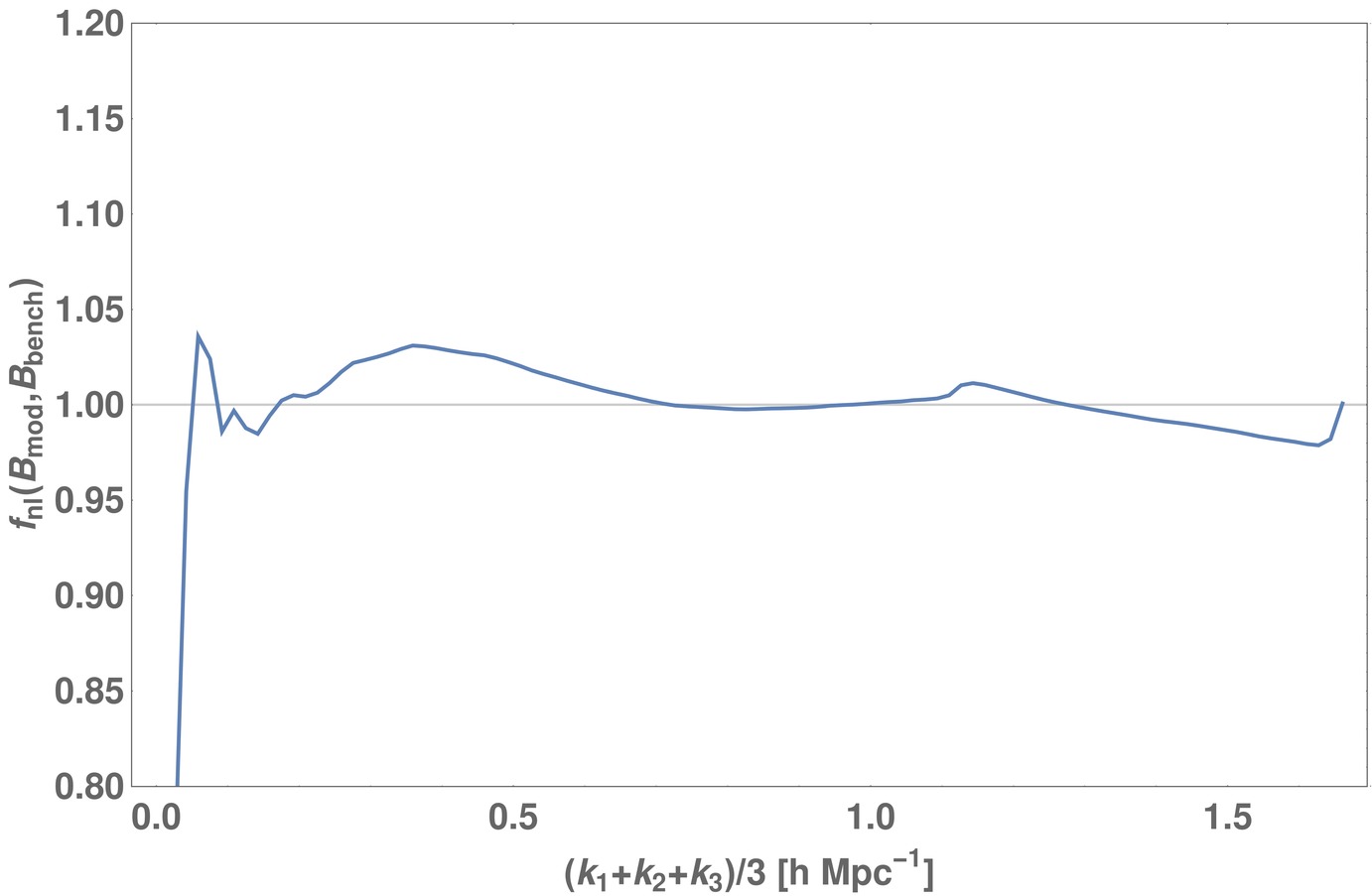}
    \caption{Sliced $f_{nl}$ correlation between the best fit
      three-shape model to the benchmark, and the benchmark.
      The feature observed at $K/3=1.1\,h\,\text{Mpc}^{-1}$ here
      is due to the transition from the tetrahedral region in the
      bottom to the pyramid at the top, causing a kink in the
      sliced correlator, and is not a real physical effect. 
    }
    \label{fig:3-shape_bis_slice}
  \end{center}
\end{figure}

The three-shape model was proposed in \citep{Andrei, Andrei2} 
as a phenomenological model to quantitatively describe the dark matter bispectrum $B_{\rm mmm}(k_1,k_2,k_3)$, consisting
of a linear combination of the `constant' one-halo model on small length scales, the tree-level gravitational bispectrum on the largest, and a local or `squeezed' shape interpolating on intermediate scales.  The combined three-shape model takes the following form:
\begin{align}
  &B_{\text{3-shape}}(k_1,k_2,k_3) \nonumber \\
  &= \sum^3_{i=1}f_i(K)B^i(k_1,k_2,k_3)\nonumber\\
  &= f_{1h}(K)B^{\text{const}}(k_1,k_2,k_3)
    +f_{2h}(K)B^{\text{squeez}}(k_1,k_2,k_3)
    \nonumber \\
  &\quad
    +f_{3h}(K)B^{\text{treeNL}}(k_1,k_2,k_3),
    \label{eqn:3-shape}
\end{align}
where the $f_i(K)$ are scale-dependent amplitudes and the constant, squeezed and tree-level shapes are respectively:
\begin{align}
  \label{eq:shapes}
  &B^{\text{const}}(k_1,k_2,k_3) = 1, \\
  &B^{\text{squeez}}(k_1,k_2,k_3) = 
    \frac{1}{3} [P_{\text{lin}}(k_1)P_{\text{lin}}(k_2)
    \nonumber \\
  &\qquad\qquad +2\,\text{perms.}, \\
  &B^{\text{treeNL}}(k_1,k_2,k_3) =
    2P_{\text{NL}}(k_1) P_{\text{NL}}(k_2)F^{(s),\Lambda}_2
    (\mathbf{k}_1,\mathbf{k}_2)
    \nonumber \\
  &\qquad\qquad +2\,\text{perms.}.
\end{align}
Here, $P_{\text{lin}}$ denotes the linear dark matter power
spectrum, $P_{\text{NL}}$ is the non-linear power
spectrum obtained from simulations, and the gravitational kernel $F^{(s),\Lambda}_2$
is 
\begin{align}
  F^{(s),\Lambda}_2(\mathbf{k}_1,\mathbf{k}_2)
  &=\frac{1}{2}(1+\epsilon)+
    \frac{1}{2}\frac{\mathbf{k}_1\cdot\mathbf{k}_2}{k_1 k_2}
    \left(\frac{k_1}{k_2}+\frac{k_2}{k_1}\right) \nonumber \\
  &\qquad +\frac{1}{2}(1-\epsilon)
    \frac{(\mathbf{k}_1\cdot\mathbf{k}_2)^2}{k_1^2 k_2^2},
    \label{F2}
\end{align}
where $\epsilon\approx-(3/7)\Omega_m^{-1/143}$ to
account for non-zero vacuum energy $\Lambda$ \citep{bouchet}.
A successful fit into highly nonlinear scales was possible using the following physically-motivated functional forms for the amplitudes: 
\begin{align}
  \label{eq:amplitudes}
  f_{1h}(K)&=\frac{A}{(1+BK^2)^2}, \\
  f_{2h}(K)&=\frac{C}{(1+DK^{-1})^3}, \\
  f_{3h}(K)&=F\exp(-K/E). 
\end{align}
The parameters $A-F$ at redshift $z=0$ across the range $0.1\,h\,\text{Mpc}^{-1}< K < 6\,h\,\text{Mpc}^{-1}$ take the values \citep{Andrei}:
\begin{align}
  \label{eq:orig3shape}
  &A=2.45\times10^6,  \quad &&B=0.054,  \nonumber \\
  &C=140, \quad  &&D=1.9, \nonumber \\
  &E=7.5\,k_{\text{NL}}, \quad &&F\equiv 1.0\,
\end{align}
with $k_{\text{NL}}=0.25\,h\,\text{Mpc}^{-1}$.  We note that this approximate fit applies across a much wider set of redshifts $z<10$ (at about 10\% precision) and, here, $F$ has been fixed to unity to match the tree-level gravitational bispectrum as $K\rightarrow 0$ (i.e.\ with unit bias).
Since the dark matter simulation we currently have is of much higher
resolution and precision than previously, we update the best fit parameter values
to the following:
\begin{align}
  \label{eq:dm3shape}
  &A=2.64\times10^6, \quad &&B=0.057,  \nonumber \\
  &C=95, \quad &&D=2.0, \nonumber \\
  &E=10.1\,k_{\text{NL}}, \quad &&F\equiv 1.0,
\end{align}
This yields a high total correlation at
$k_{\rm max} = 1.7\,h\,\text{Mpc}^{-1}$ of 98.4\% with new simulation data,
and 97.1\% with the original three-shape model (\Cref{eq:orig3shape}).  We note that there are some degeneracies between the three shapes, but we leave detailed error estimation of these dark matter parameters for a future publication.
We also note that there are transient grid effects that temporarily increase the tree-level gravitational bispectrum for $N$-body simulations with 2LPT initial conditions (identified in previous papers \citep{glass,Andrei}); even for the high redshift initial conditions used in this paper, this persists at late times leaving an offset in the dark matter bispectrum of a few percent for small $k$.  This small systematic effect can be avoided with `glass' initial conditions for the $N$-body simulations \citep{glass,Andrei} or through quantitative analysis and subtraction (but this is not the focus of the present paper, see the discussion in \citep{DM}). 

We can consider using the same three shapes to fit to our benchmark halo bispectrum  $B_{\rm hhh}(k_1,k_2,k_3)$, but in principle we might require more than three shapes to achieve an adequate correlation.  For example, bias considerations bifurcate the tree-level gravitational bispectrum (\Cref{eq:shapes}) into several apparently different shapes at leading order (LO) \citep{Desjacques}:
\begin{align}
\label{eq:halo3shape}
  &B^{\text{LO}}_\text{hhh}(k_1,k_2,k_3)
  = b_1^3 B^{\text{treeNL}}(k_1,k_2,k_3) \nonumber \\
  +{}& b_1^2 \left[ b_2 + b_{K^2} \left( (\hat{\bf k}_1\cdot\hat{\bf k}_2) ^2 +\textstyle{\frac{1}{3}} \right)\right ]\left(P(k_1) P(k_2) + 2 \,\text{perms}.\right)\nonumber\\
  +{}& B ^{\text{stoch}}_{\cal E}  + b_1^3 \left(P ^{\text{stoch}}_{\cal E}P(k_1) + 2\, \text{perms}.\right)\,,
\end{align}
where $b_1,\, b_2$ are the first- and second-order bias parameters, $b_{K^2}$ is the `tidal' bias parameter, and $P ^{\text{stoch}}_{\cal E}$, $B ^{\text{stoch}}_{\cal E}$ are the stochastic power spectrum and bispectrum respectively.  Closer examination, however, reveals that the second-order bias shape can be incorporated with appropriate scalings in the squeezed two-halo shape $ B^{\text{squeez}}$  and the stochastic bispectrum $B ^{\text{stoch}}_{\cal E} $ in the constant shape $B^{\text{const}}$ (if not subtracted as per usual).   This leaves only the modulated `tidal' bias term, but this can be expected to be relatively small and would be straightforward to include as an additionally modulated version of the squeezed shape $ B^{\text{squeez}}$ (a `four-shape' model).   

For this reason, as a preliminary exercise we endeavour to fit the original three-shape model  \Cref{eq:shapes} to the measured halo bispectrum, finding the best fit parameters as:
\begin{align}
  &A=1.55\times10^6, \quad &&B=0.042,  \nonumber \\
  &C=287, \quad &&D=3.7, \nonumber \\
  &E=8.0\,k_{\text{NL}}, \quad &&F=0.97.
\end{align}
Again we will leave error estimation in these parameters for future work.
The three-shape bispectrum calculated with these values is shown in \Cref{fig:3-shape_bis}.
It gives an overall total correlation of 97.4\% with our benchmark bispectrum,
and a 4\% $f_{nl}$ correlation fit across the entire range of the data apart from the
very tip of the tetrapyd where $K/3<0.2\,h\,\text{Mpc}^{-1}$
(\Cref{fig:3-shape_bis_slice}).   Note again there are degeneracies in the model parameters for the limited wavenumber range we have used; there are significant caveats on large length scales (discussed above), as well as small length scales because we do not probe deep enough into the nonlinear regime on small scales to specify the one-halo parameters.  In principle, we could use this to specify the averaged bias parameter $b_1 \approx 0.99$ (assuming this to be the dominant contribution) or we could estimate $b_1, b_2$ jointly with the power spectrum, but we would have to investigate and calibrate transient grid effects at small $k$ much more carefully \citep{glass} and we leave this for a  future publication.  Nevertheless, this analysis gives an initial indication that an accurate phenomenological fit to the halo (or galaxy) bispectrum is likely to be possible with a few well-motivated bispectrum shapes and a limited number of parameters. 

\section{Phenomenological halo catalogues\label{work}}

Having characterised the halo power spectrum and bispectrum from our benchmark \texttt{ROCKSTAR} catalogue (as a proxy for a galaxy catalogue), we investigate whether these polyspectra can be accurately reproduced using fast statistical prescriptions for populating halos with subhalos, that is, without using costly $N$-body simulations for individual mocks. We first consider minimal approaches by modifying the subhalo distribution using different halo profiles or altering the average occupation number as a function of halo mass.  Next, we develop this further by exploiting halo concentrations, populating individual halos using typical correlations with the occupation number, that is, incorporating statistical information related to the assembly history of halos. 

\subsection{Halo profile}

\begin{figure*}
  \captionsetup[subfigure]{labelformat=empty}
  \begin{center}
    \begin{subfigure}[b]{0.49\textwidth}
      \includegraphics[width=\linewidth]{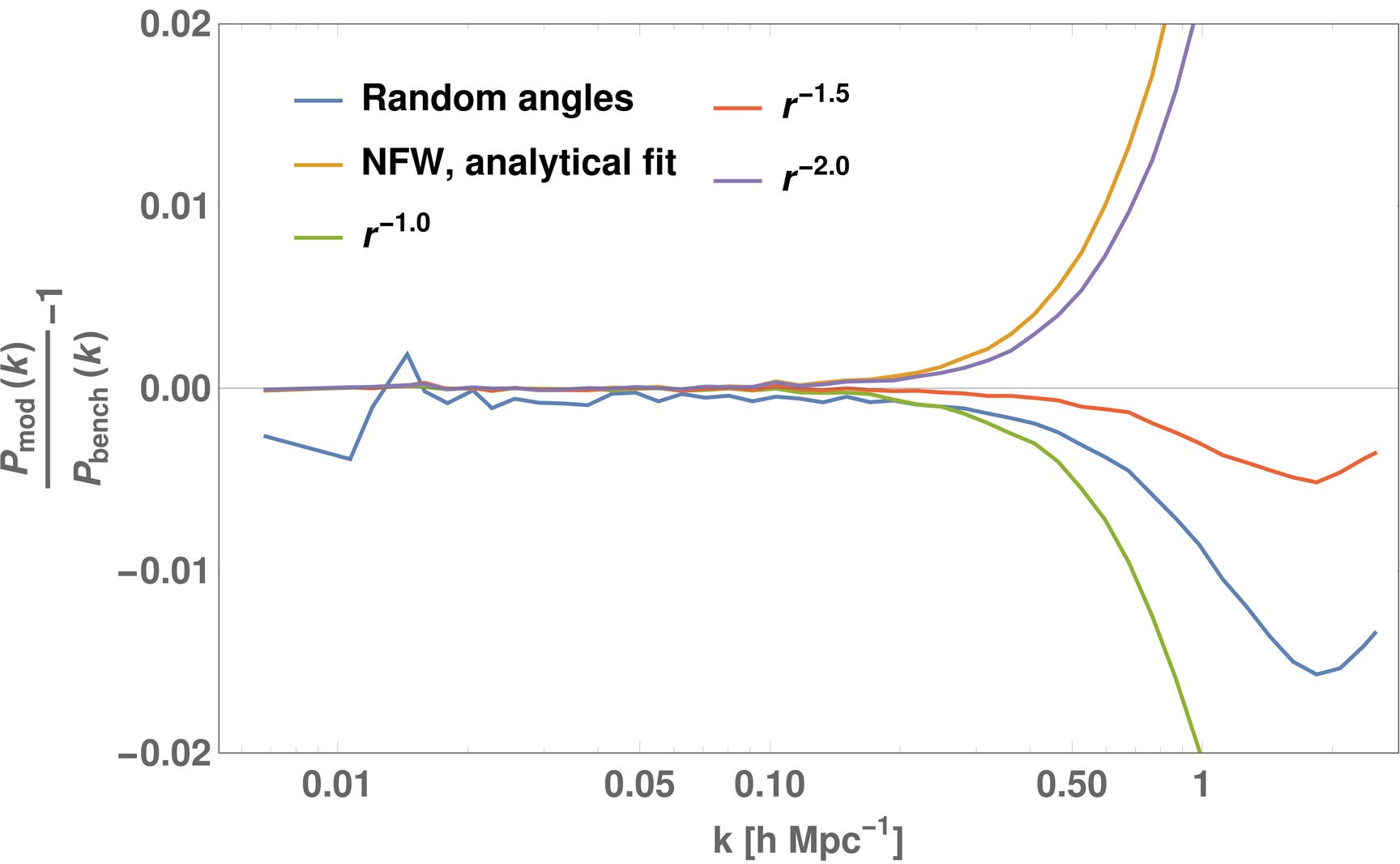}
      \caption{}
    \end{subfigure}
    ~
    \begin{subfigure}[b]{0.455\textwidth}
      \includegraphics[width=\linewidth]{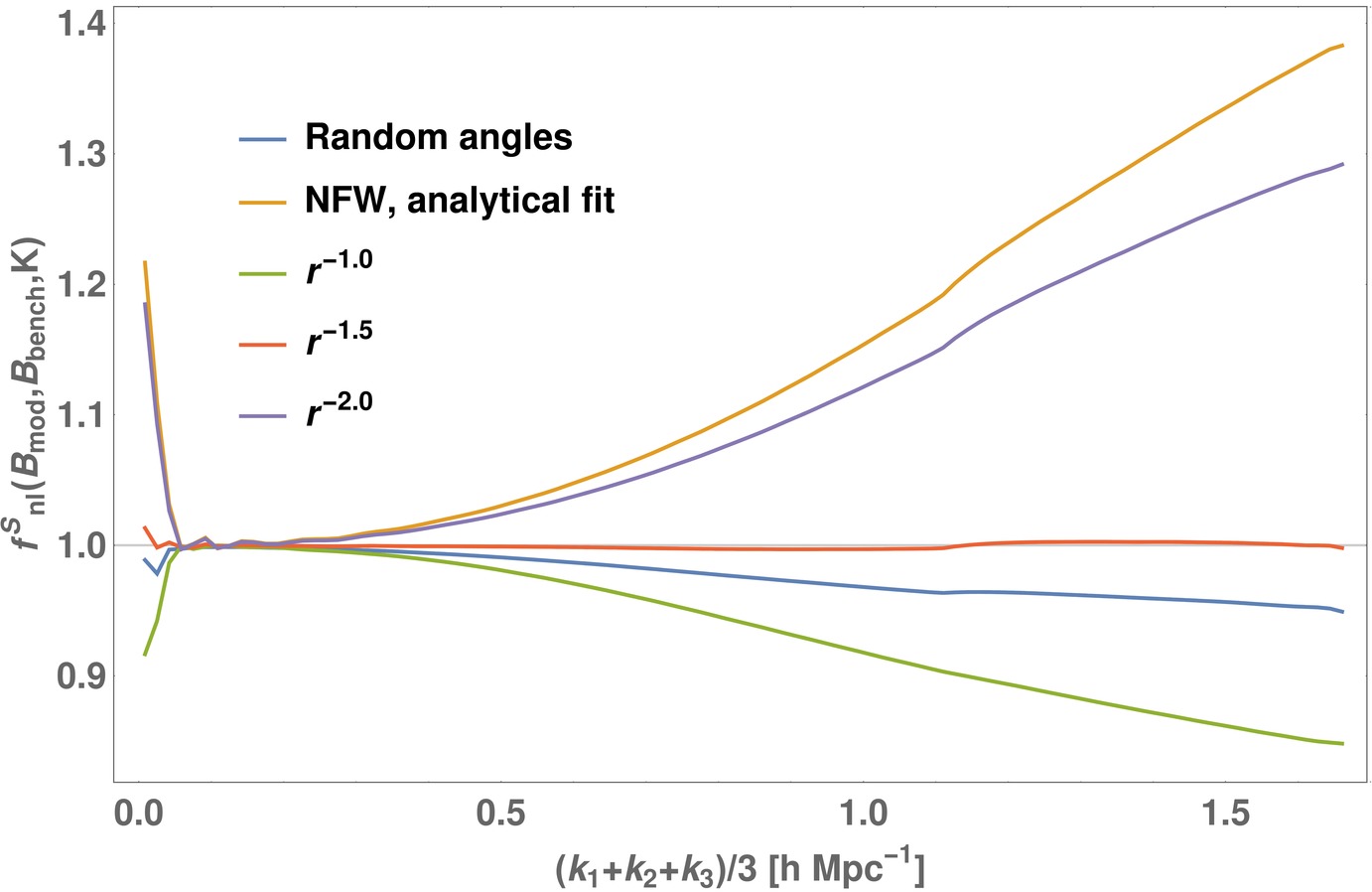}
      \caption{}
    \end{subfigure}
  \end{center}
  \caption{The relative power spectrum (left) and the sliced $f_{nl}$ bispectrum correlator (right) for different radial halo profile prescriptions for populating halos with subhalos.  The power law profile $r^{-\gamma}$ is a much better fit to the actual subhalo distribution than the dark matter NFW profile, although the index $\gamma\approx 1$ suggested by the
    best  fit to the true profile is power deficient. For $\gamma=1.5$ we obtain a  near-perfect fit to both the power spectrum and
    bispectrum to high wavenumbers $k, K/3\le 1.6\,h\,\text{Mpc}^{-1}$. 
  }
  \label{fig:halo_profile2}
\end{figure*}

\begin{figure*}
  \begin{center}    
    \begin{subfigure}[b]{0.49\textwidth}
    \includegraphics[width=\linewidth]{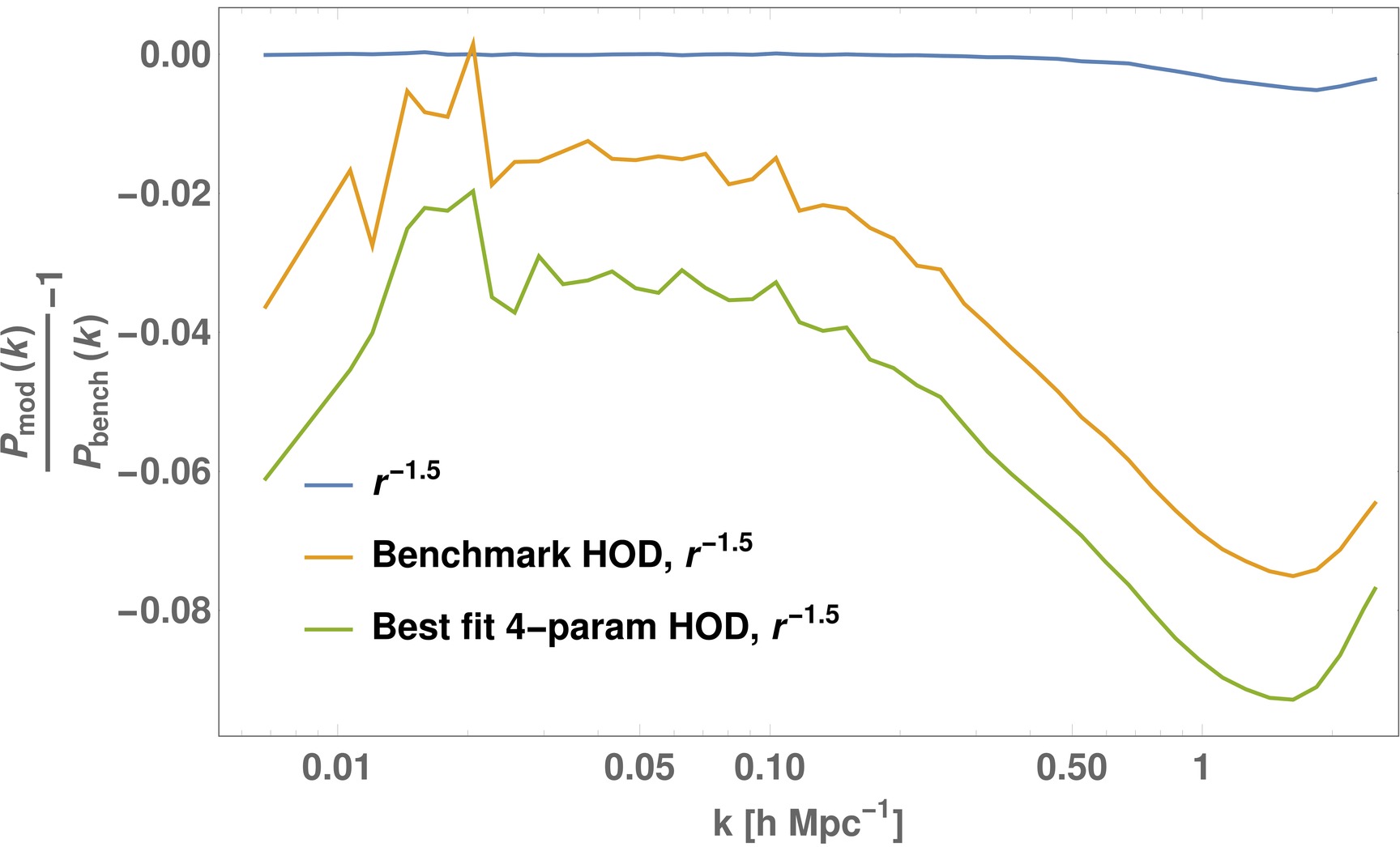}
    \end{subfigure}
    ~
    \begin{subfigure}[b]{0.455\textwidth}
    \includegraphics[width=\linewidth]{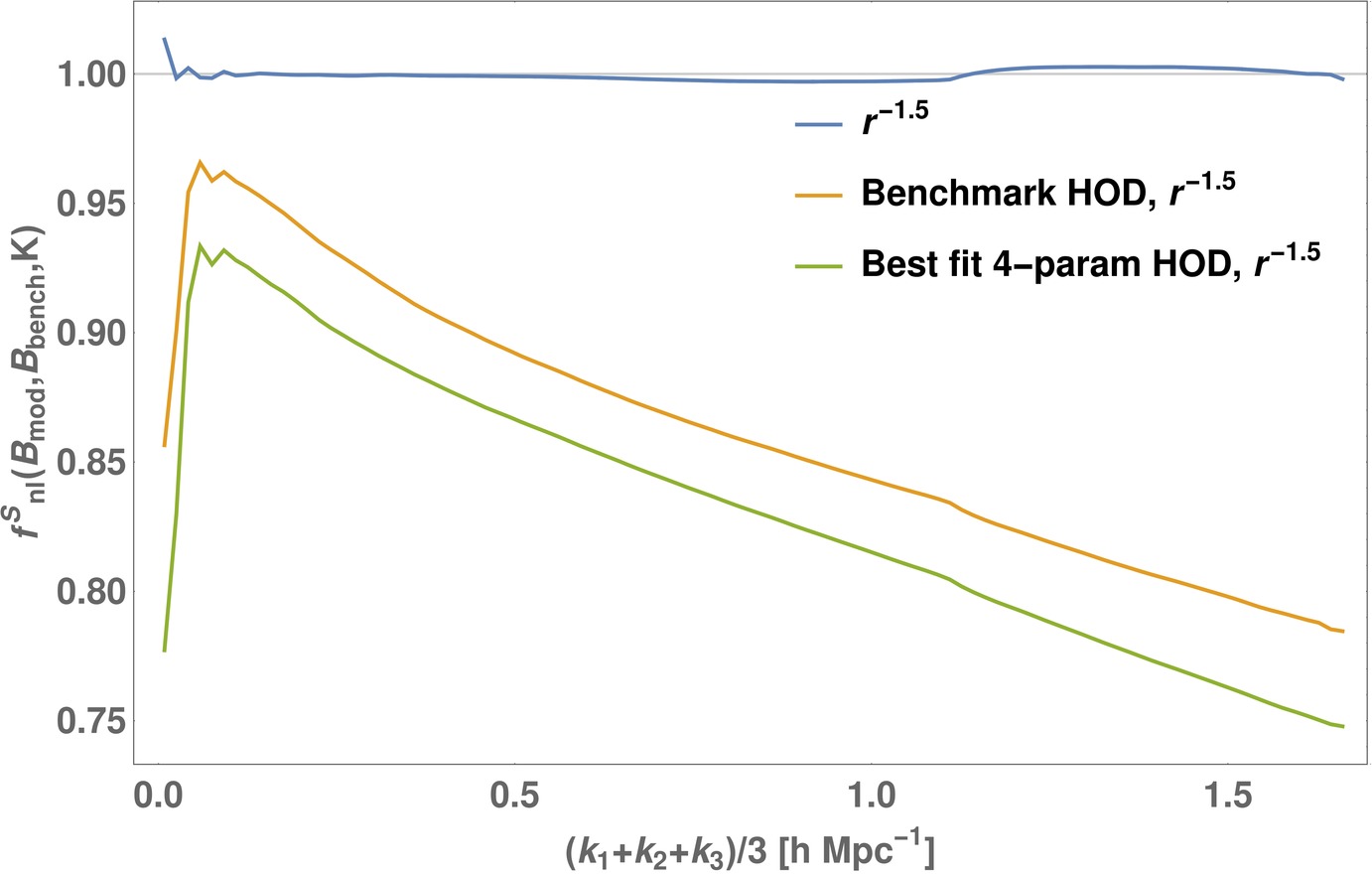}
    \end{subfigure}
    \caption{HOD prescriptions for statistically populating halos with subhalos yield a deficient power spectrum (left) and sliced bispectrum correlation (right).  Neither the benchmark HOD nor the 4-parameter
      HOD model (using best fit parameters) can
      recover the benchmark power spectrum to better than 2\%
      and the bispectrum to better than 4\% at large scales,
      with much larger discrepancies on smaller scales.
    }
    \label{fig:hod}
  \end{center}
\end{figure*}

Modifying the typical halo profile significantly impacts both the power
spectrum and bispectrum, especially on small length scales. We can demonstrate
(see below) this by keeping the number of subhalos fixed in each halo,
while displacing their radial distribution according to a profile of our choosing (such as the popular NFW profile \Cref{eqn:nfw}). 
First, however, we briefly study the importance of halo anisotropy. 
This was motivated by investigations of $N$-body simulations (such as that in \Cref{subsec:profile}), which have
revealed that the dark matter profiles of halos are not spherical, reflecting more complex internal substructure
\citep{triaxial1,triaxial2,triaxial3}. The subhalos
that live within those halos, therefore, also have a non-spherical distribution,
as well as internal structure. We have quantified the importance
of these effects by randomising the solid angular distribution of
the subhalos within a halo, while keeping the radial distance
to the parent halo seed unchanged. This effectively removes halo
triaxiality, destroying the original internal structure of the
halos. For the new `random angle' halo catalogue, we have estimated both the power spectrum and the bispectrum (using the sliced $f_{nl}$ correlator (\Cref{inner_product_sliced}) at a given $K=k_1+k_2+k_3$);  the relative effect is shown by the blue lines in \Cref{fig:halo_profile2}. There is a small diminution of power even at relatively high wavenumbers $k,K/3=1\,h\,\text{Mpc}^{-1}$, with  less than a 1\% and 4\% decrease  for
the power spectrum and bispectrum respectively. Randomisation of the angles tends to reduce subhalo clustering but this remains a subpercent effect on the bispectrum for $K/3\le 0.5\,h\,\text{Mpc}^{-1}$.
The small effect of a randomisation process has on the matter power spectrum
has also been confirmed in \citep{pace}.
This indicates that triaxial effects will predominantly arise from RSDs (see, for instance, \citep{triaxial}).

The radial halo
profile can have a larger effect, notably if we populate subhalos using the NFW profile obtained from the halo dark matter distribution, as shown by the orange line in \Cref{fig:halo_profile2}. In this case, by $k,K/3=1\,h\,\text{Mpc}^{-1}$ there are large deviations of 2\% and 15\% from the halo power spectrum and bispectrum respectively.  This is not unexpected as we have previously seen
that the dark matter NFW profile does not fit the measured subhalo profile from
our benchmark catalogue (given the mass resolution of our $N$-body simulation).   The discrepancies would in fact have been
even larger had we used the measured concentration from \texttt{ROCKSTAR}, instead of the
analytical fit for $\expval{c}$ in \Cref{eqn:conc}.

\begin{figure*}
   \begin{minipage}{0.49\textwidth}
     \centering
     \includegraphics[width=\linewidth]{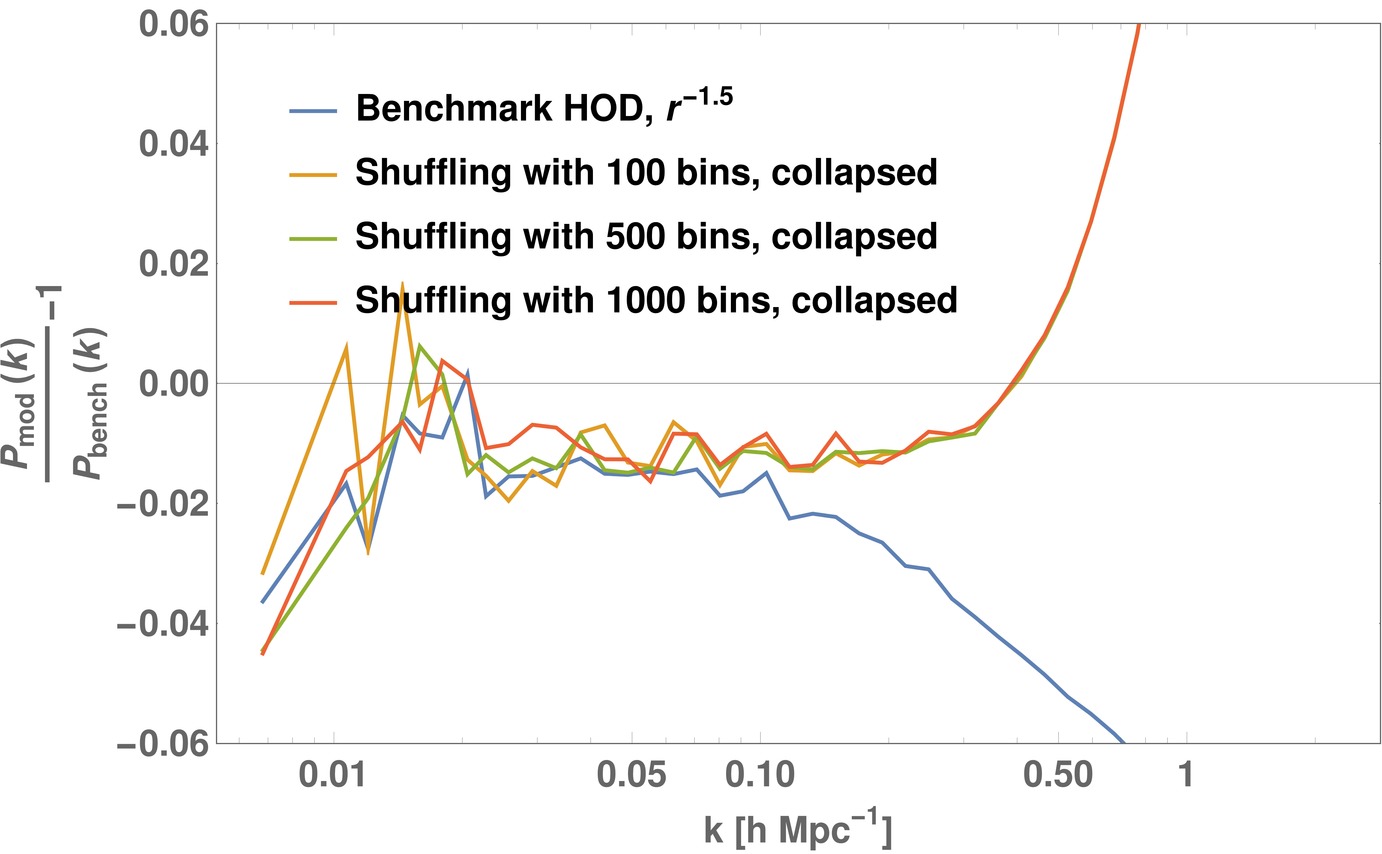}
    \caption{Shuffling the halo occupation number within a mass bin
      has the same effect as using the benchmark HOD.
    }
    \label{fig:shuffled}
   \end{minipage}\hfill
   \begin{minipage}{0.49\textwidth}
     \centering
     \includegraphics[width=\linewidth]{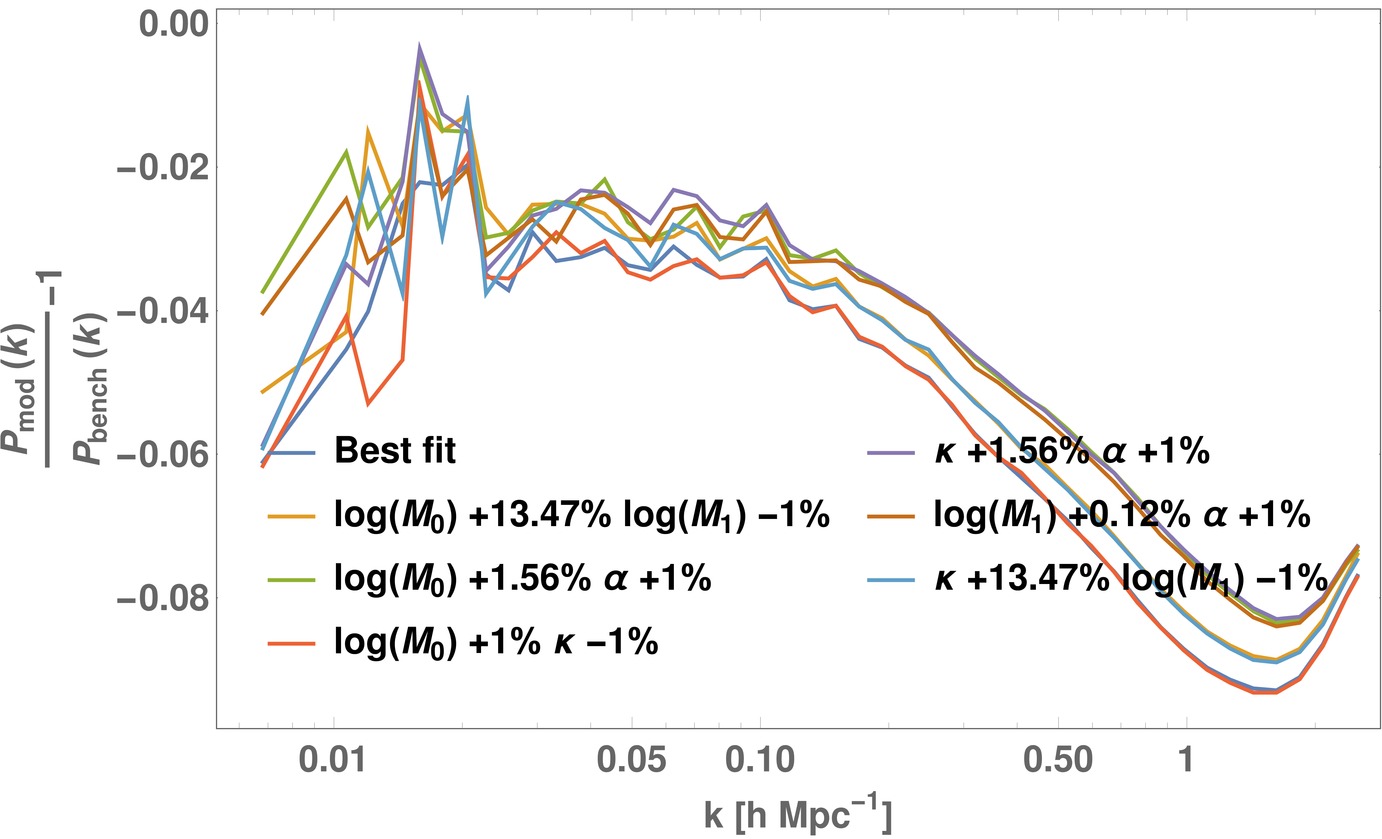}
    \caption{$\alpha$ has a strong influence on the power
      spectrum, unlike the other parameters. A radial
      profile $r^{-1.5}$ is used throughout.
    }
    \label{fig:hod_4-param_mod}
   \end{minipage}
\end{figure*}

\begin{figure*}
  \begin{center}    
    \begin{subfigure}[b]{0.48\textwidth}
      \includegraphics[width=\linewidth]{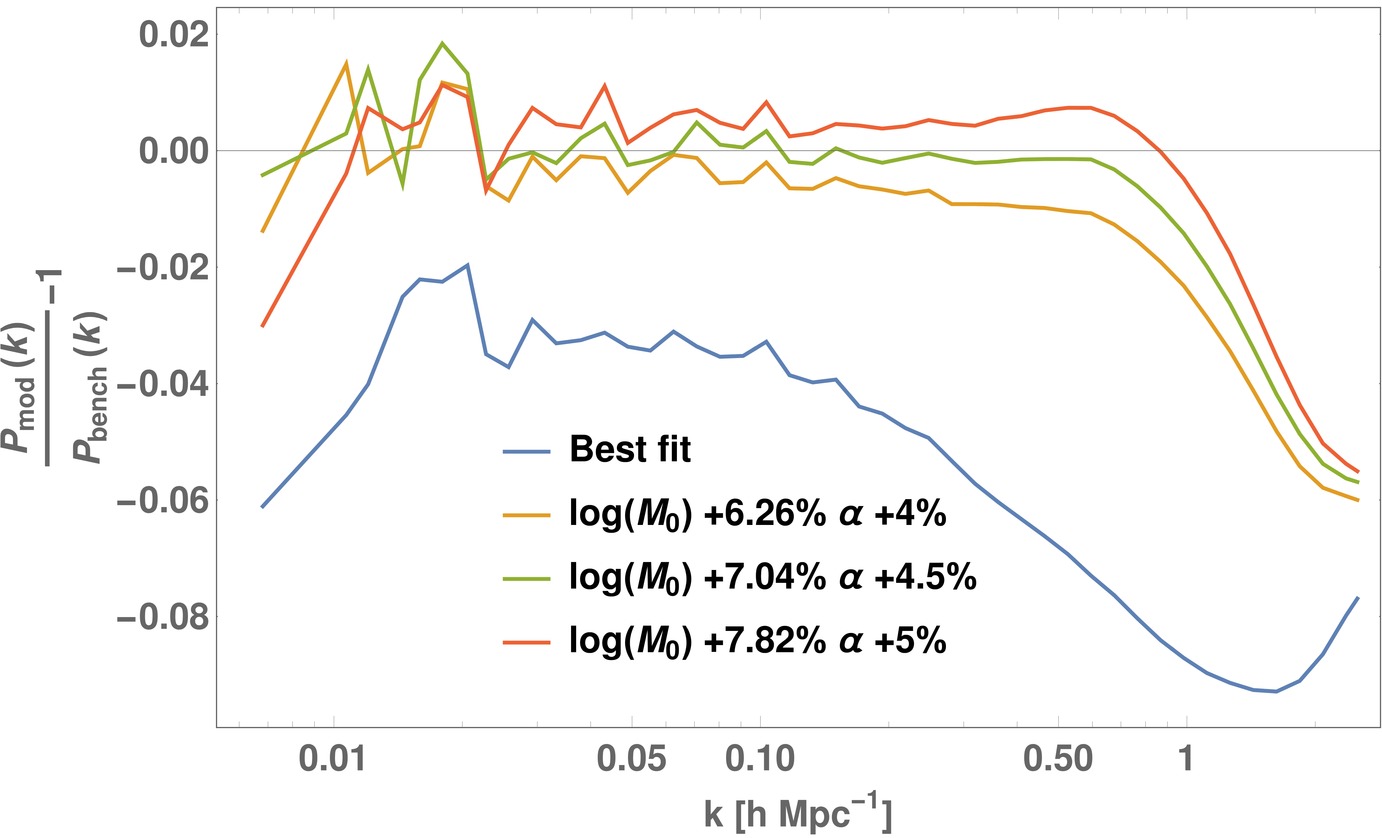}
      \caption{}
    \end{subfigure}
    ~
    \begin{subfigure}[b]{0.48\textwidth}
      \includegraphics[width=\linewidth]{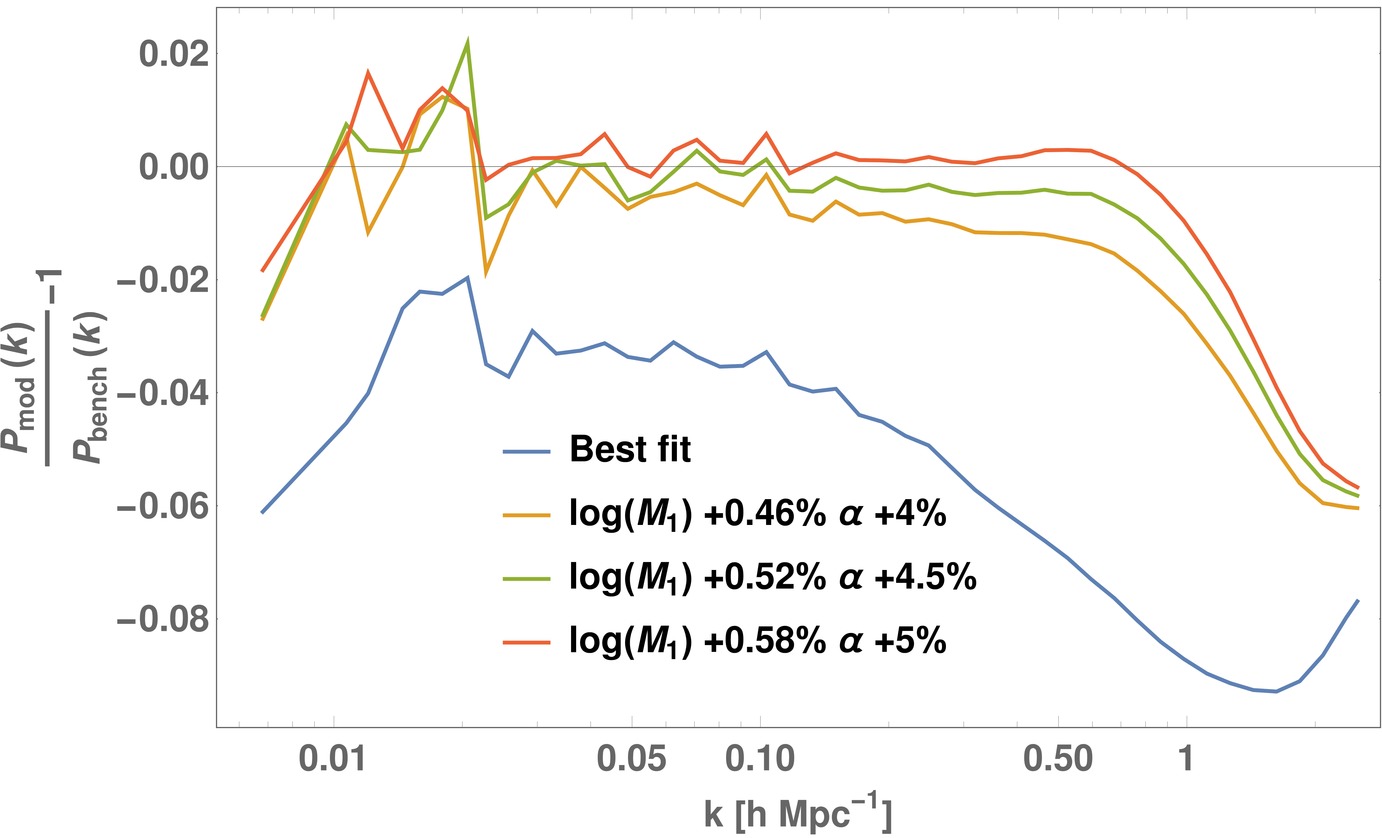}
      \caption{}
    \end{subfigure}
    
    \begin{subfigure}[b]{0.48\textwidth}
      \includegraphics[width=\linewidth]{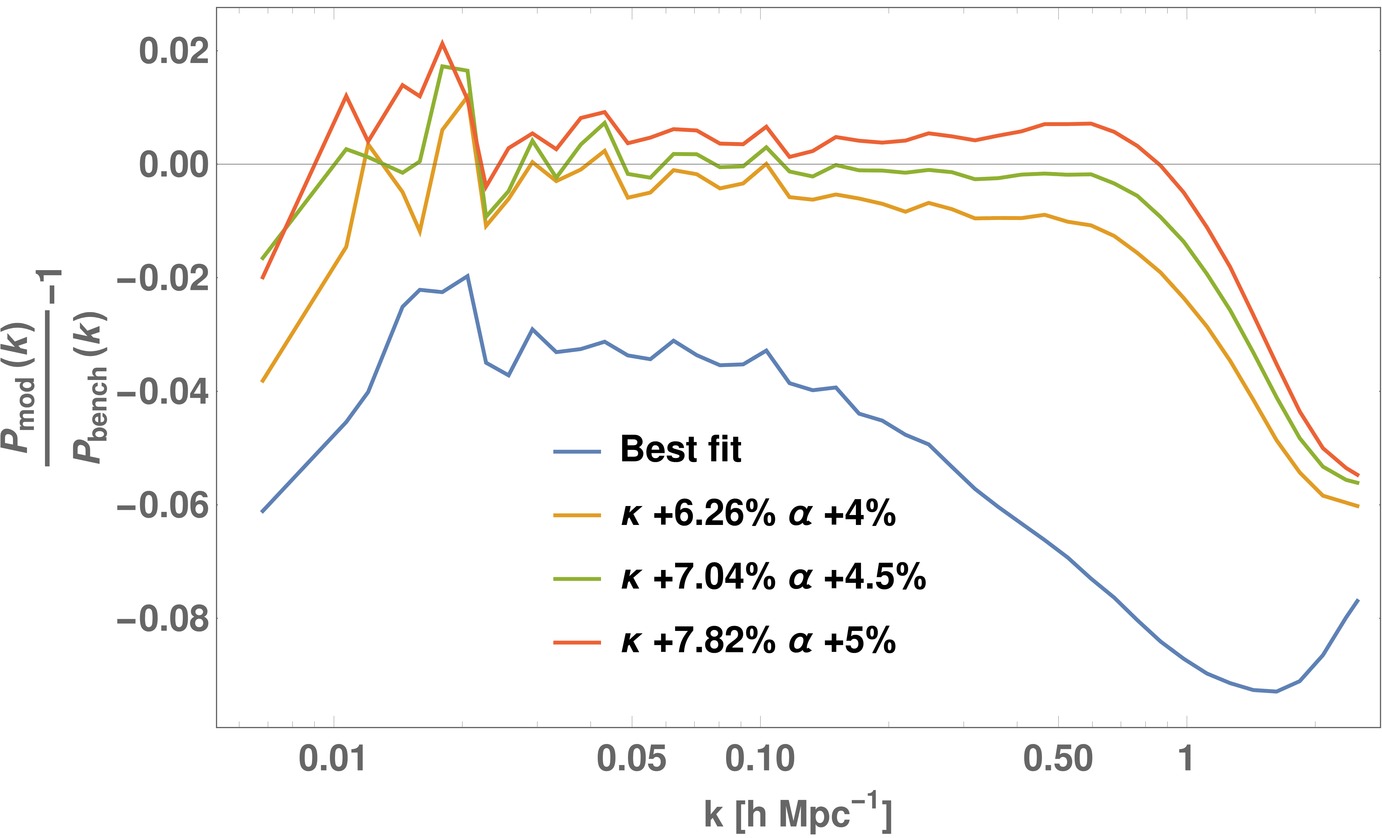}
      \caption{}
    \end{subfigure}
    ~
    \begin{subfigure}[b]{0.445\textwidth}
      \includegraphics[width=\linewidth]{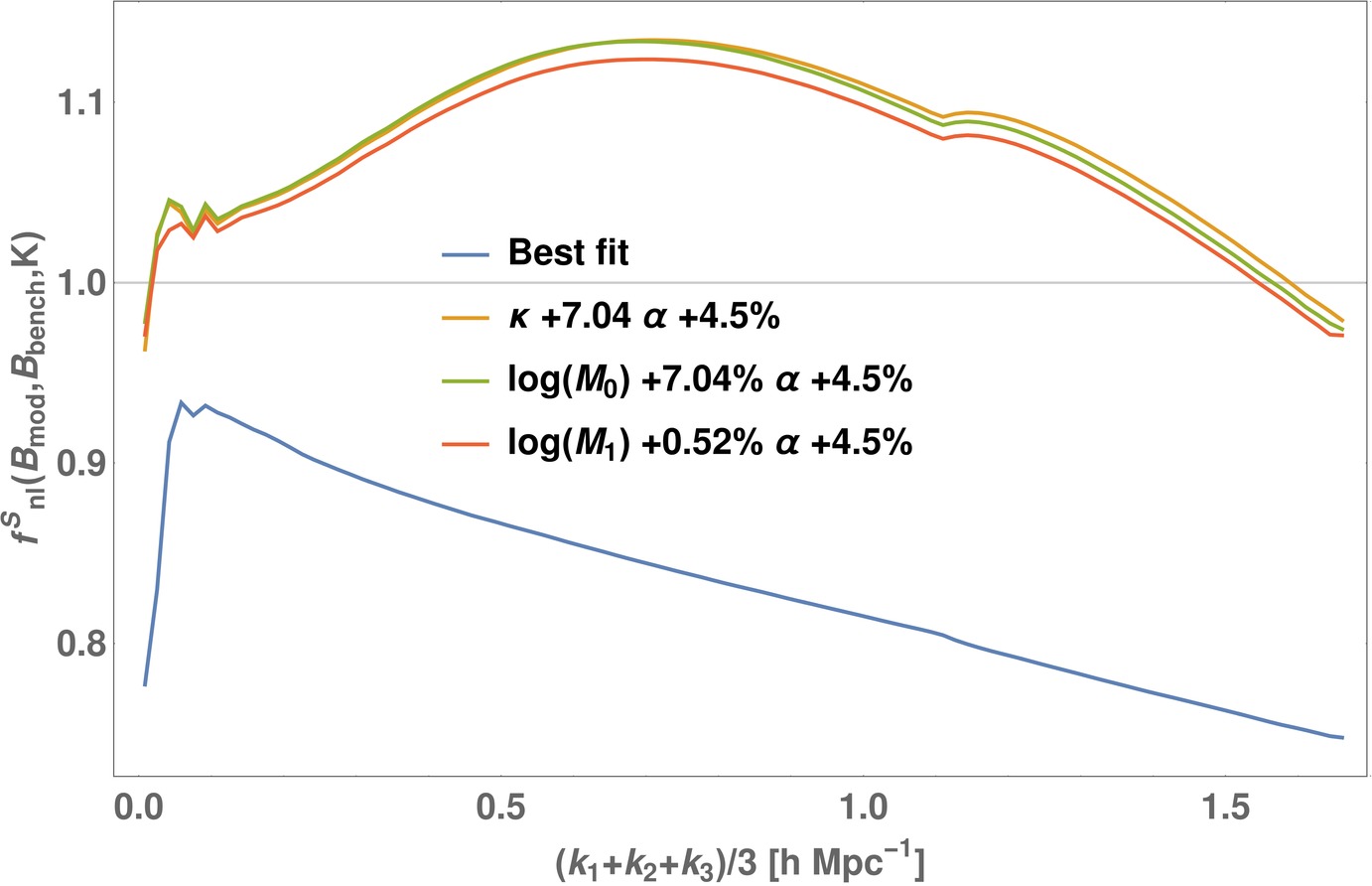}
      \caption{}
    \end{subfigure}
    \caption{Panels (a)-(c): increasing $\alpha$ by 4.5\%
      helps match the power spectrum to the benchmark,
      regardless of choice in the other parameter. Panel (d):
      the boost in power spectrum over-boosts the bispectrum.
    } 
    \label{fig:hod_4-param_mod2}
  \end{center}
\end{figure*}

We turn now to effects of modelling the halo profile with a power law.
As we have seen already in \Cref{fig:gamma}, a power law of $0.8 < \gamma <1.2 $ will
fit most halo profiles for the subhalo distributions found in our benchmark simulation. Modelling the halos
with the best fit power law inevitably removes some signal from the power
spectrum and bispectrum, as the resulting halos have a uniform
solid angular distribution, unlike subhalos in an $N$-body
simulation. The lack of power can be seen in the $\gamma = 1$ profile shown as green line in  \Cref{fig:halo_profile2}. 
We can phenomenologically 
compensate for this effect by considering spherically symmetric halo profiles
with an increased power law exponent. Coincidentally, for $\gamma = 1.5$
both the power spectrum and the bispectrum are very well fitted at all
scales, with a difference of less than 0.5\% up to $k,K/3 \le 1.6\,h\,\text{Mpc}^{-1}$.  We can exploit this dual effect when populating the halos with a statistical halo occupation number rather than that measured from the $N$-body simulation.

\subsection{Halo occupation number}

\begin{figure*}
  \captionsetup[subfigure]{labelformat=empty}
  \begin{center}
    \begin{subfigure}[b]{0.37\textwidth}
      \includegraphics[width=\linewidth]{{Gadget3_2048_1280_run1_z0p000_rockstar_halos_parents_subhalos}.png}
      \caption{Benchmark \texttt{ROCKSTAR} catalogue}
    \end{subfigure}
    ~
    \begin{subfigure}[b]{0.37\textwidth}
      \includegraphics[width=\linewidth]{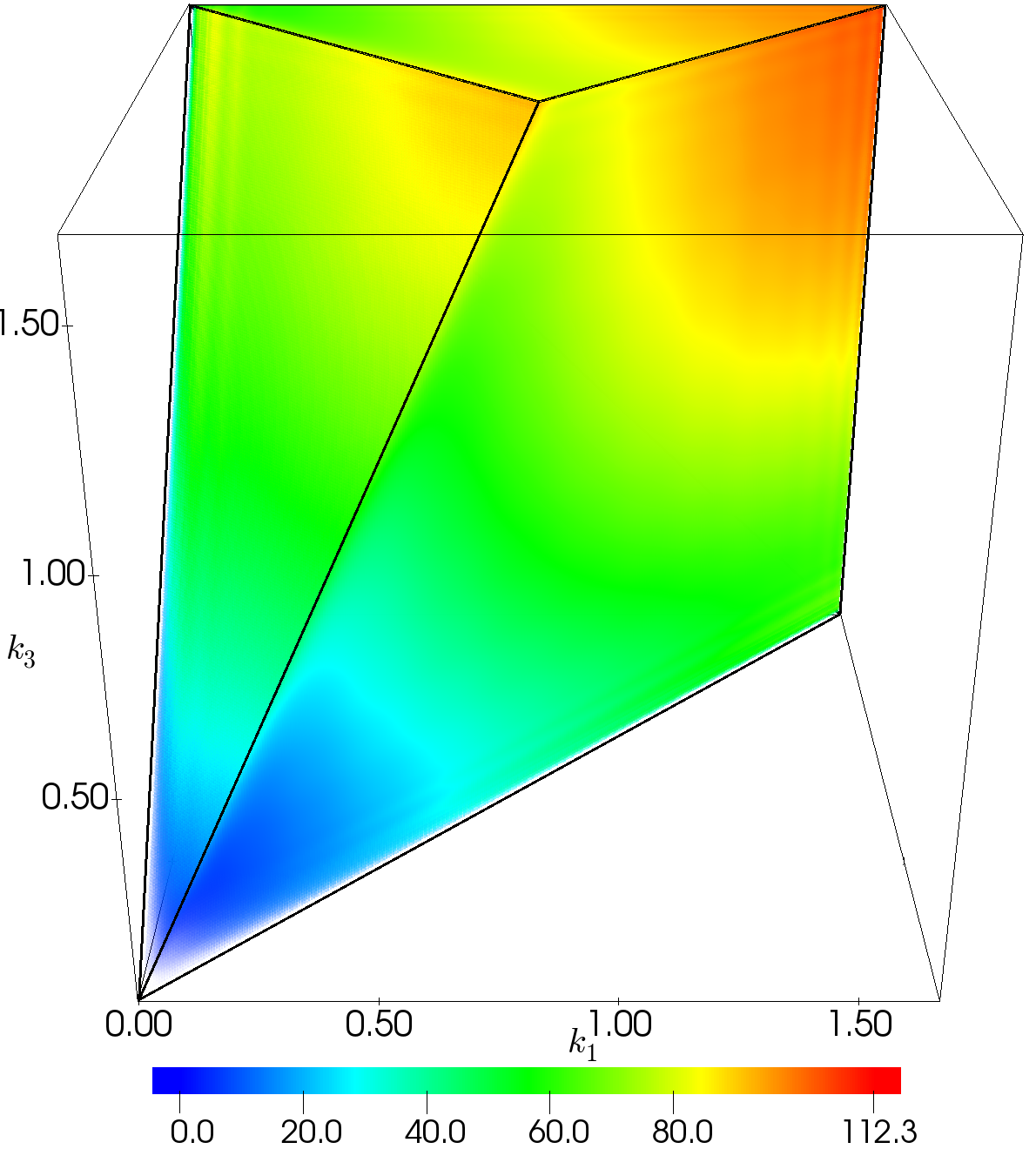}
      \caption{Random solid angle}
    \end{subfigure}
    
    \begin{subfigure}[b]{0.37\textwidth}
      \includegraphics[width=\linewidth]{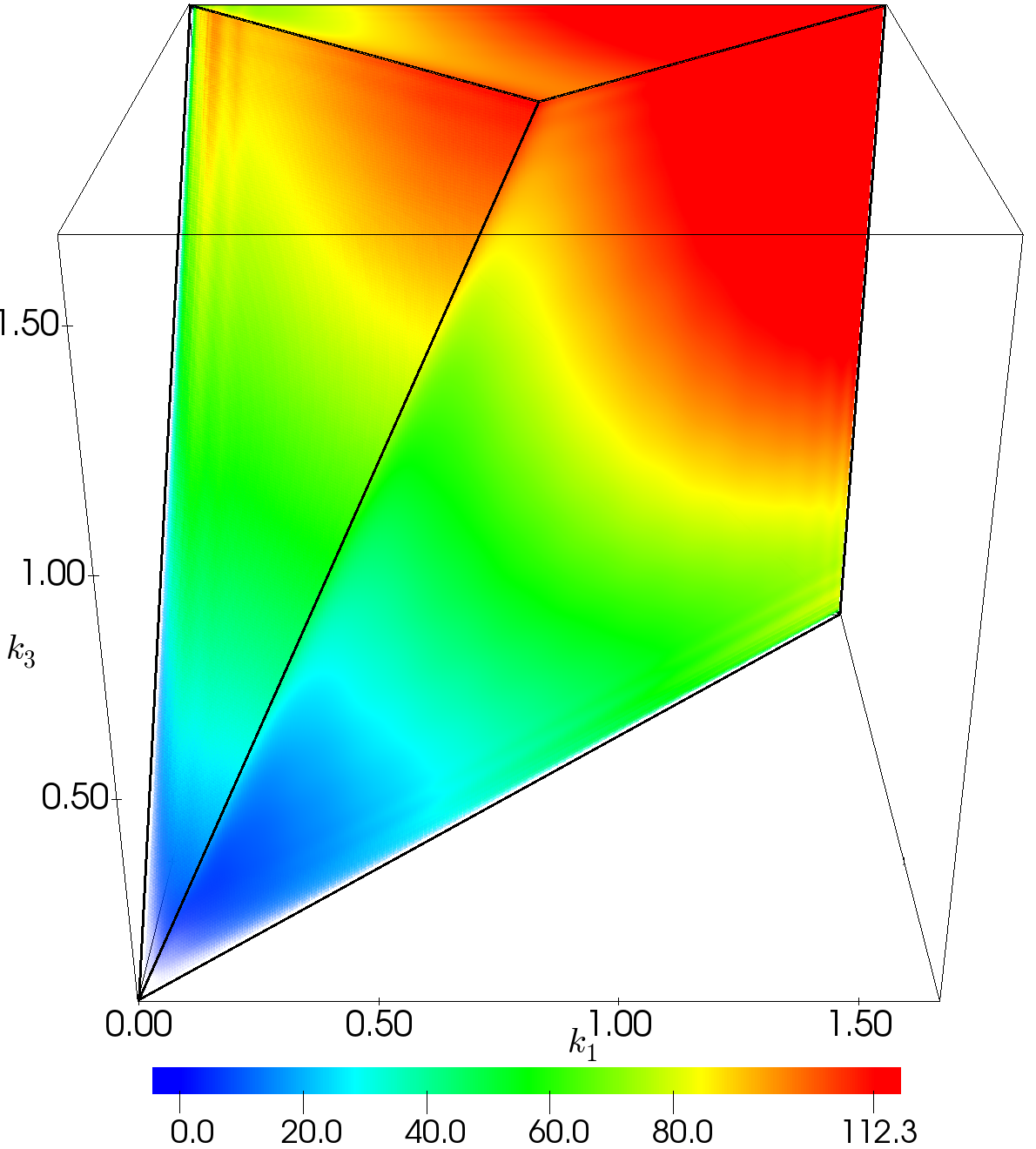}
      \caption{NFW profile}
    \end{subfigure}
    ~
    \begin{subfigure}[b]{0.37\textwidth}
      \includegraphics[width=\linewidth]{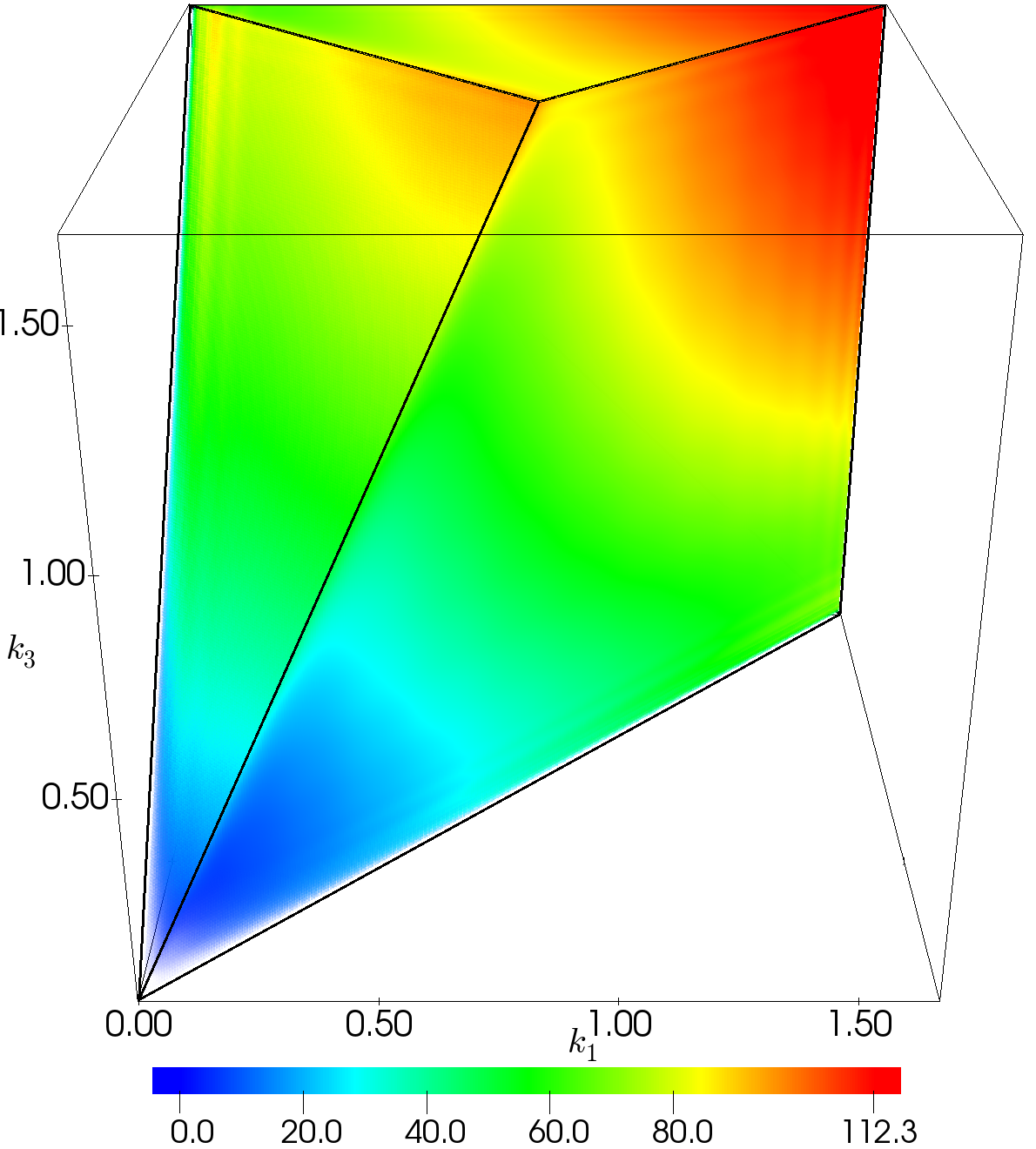}
      \caption{Benchmark HOD and NFW profile}
    \end{subfigure}

    \begin{subfigure}[b]{0.37\textwidth}
      \includegraphics[width=\linewidth]{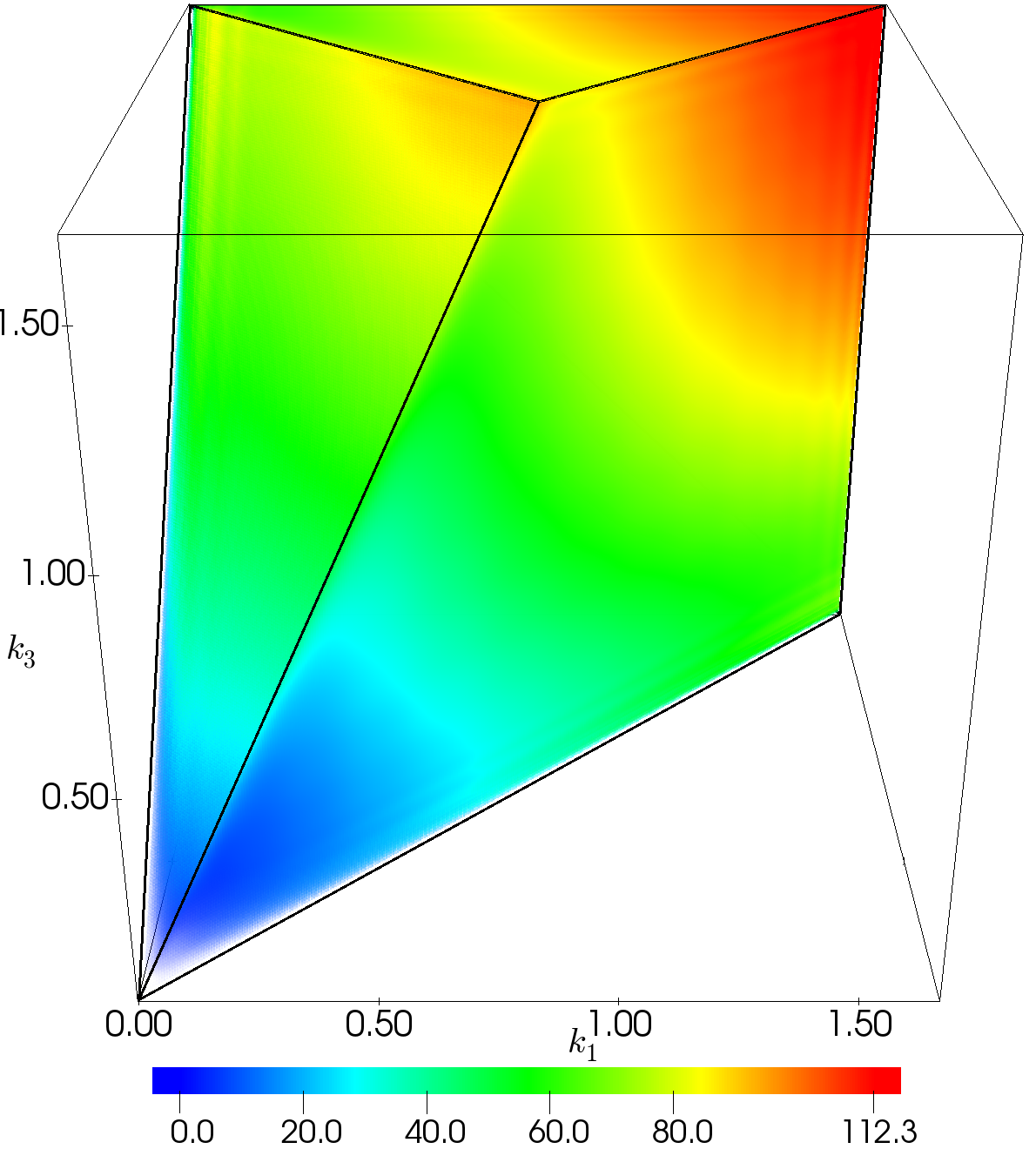}
      \caption{Best fit 4-parameter HOD model}
    \end{subfigure}
    ~
    \begin{subfigure}[b]{0.37\textwidth}
      \includegraphics[width=\linewidth]{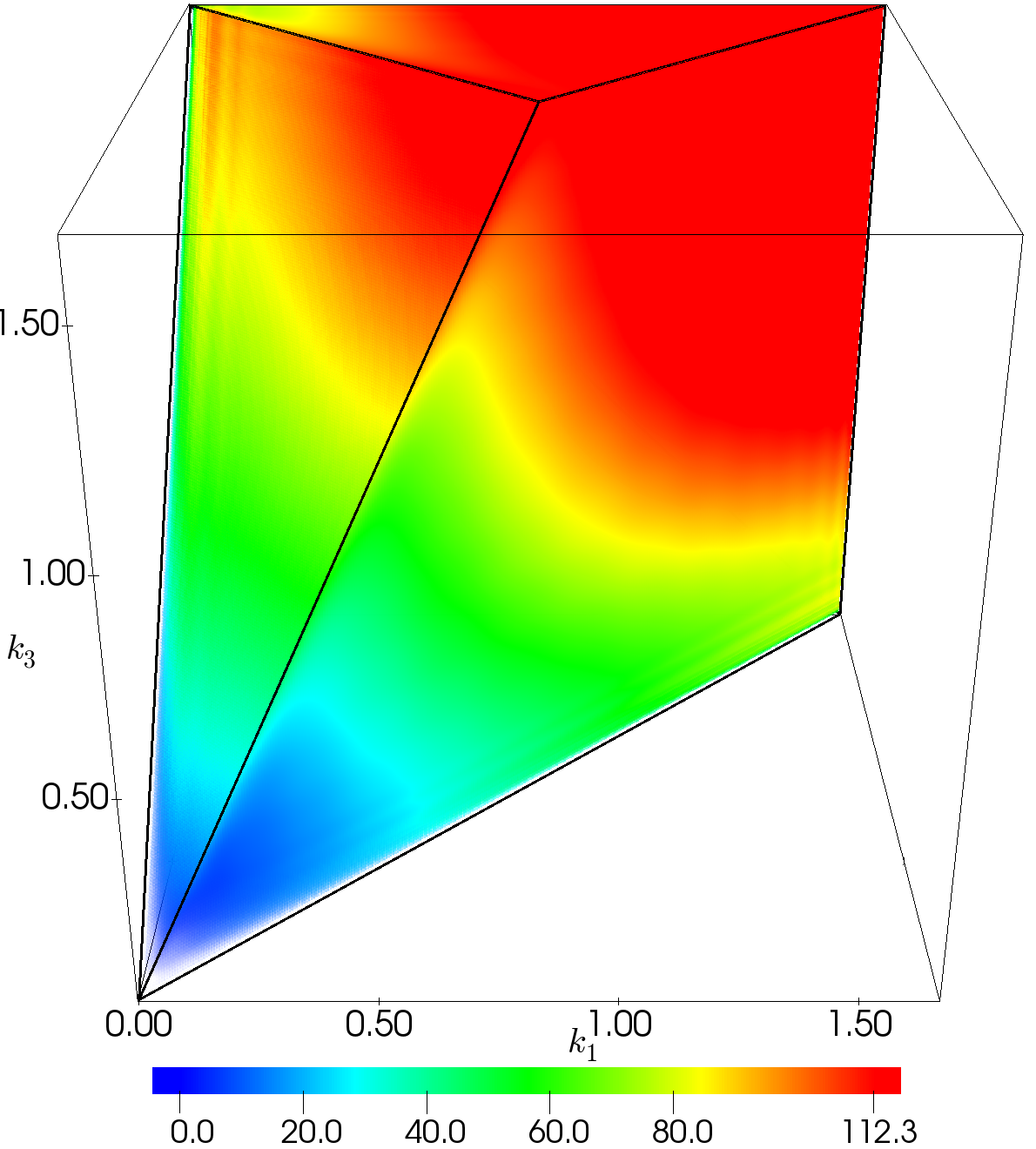}
      \caption{4-parameter HOD model, boost in $\alpha$ and $\kappa$}
    \end{subfigure}
    \caption{
      Bispectra of the simple galaxy mock catalogues.
    }
    \label{fig:halo_bis}
  \end{center}
\end{figure*}

We have also investigated the effect on the power spectrum
and bispectrum of assigning subhalos using the Halo
Occupation Distribution (HOD). 
First, we populated halos using the benchmark HOD model, i.e. we assigned to each halo
the measured mean number of galaxies (subhalos) for a halo of that mass.
This model is shown in \Cref{fig:hod_best_fit} along with
our 4-parameter fit to it. As shown in \Cref{fig:hod}, we
have found that neither the benchmark HOD nor the 4-parameter
HOD fit recovers the power spectrum or the bispectrum to better
than 2\% at large scales $k,K/3 < 0.1\,h\,\text{Mpc}^{-1}$. The 4-parameter 
fit to the benchmark HOD is 4\% below the simulation power
spectrum, and the difference gets rapidly worse at smaller length scales.
The fit to the benchmark HOD is only accurate
to 10\%, indicating a better functional form should be
adopted. The discrepancy in the bispectrum is
considerably higher than the power spectrum, and also demonstrates much
worse scaling in $k$.

To better understand the power deficiency observed in
\Cref{fig:hod} from using the HOD model we first binned the
parent halos by mass, then shuffled around the halo occupation
number within the halos in each mass bin. Since the halo profile plays 
only a marginal role on large length scales, for simplicity we collapsed all objects to the centre of
the parent halo, and the power spectrum of the resulting sample
is shown in \Cref{fig:shuffled}. The fact that this shuffling
method, which preserves the statistical distribution of the halo
occupation number in every mass bin, produces the same effect as
the benchmark HOD strongly implies that number of subhalos in a
halo depends on halo properties other than halo mass.
The shuffling procedure is very similar to populating halos by using a subhalo 
dispersion around the mean HOD; initial experimentation indicated that including such a dispersion 
had no impact resolving the key bispectrum deficit. 

Finally, we explored whether phenomenologically changing the parameters in our
4-parameter HOD could yield a satisfactory fit to both the power spectrum
and bispectrum. As discussed in \Cref{subsec:HOD} we enforce
conservation of galaxy number $\Delta n_g=0$
(\Cref{eq:constraint}) when changing the values of the parameters,
which entails compensating by changing at least 2 parameters simultaneously. By exploring all 6 different ways to
pair up the parameters, it was found that the index $\alpha$ in (\Cref{eq:constraint}), i.e. the
exponent of the power law, appears to make the most dramatic contribution to the
power spectrum relative to the other parameters. As can be seen in panels (a)-(c)
in \Cref{fig:hod_4-param_mod2}, boosting $\alpha$ by 4.5\%
helps match the benchmark power spectrum up to
$k \le 0.5\,h\,\text{Mpc}^{-1}$, regardless of the choice of the
other compensating parameter. However, panel (d) in the same plot reveals that
this boost in $\alpha$ grossly inflates the bispectrum,
resulting in more than 5\% difference between 
$0.2\,h\,\text{Mpc}^{-1} < K/3 < 1.3\,h\,\text{Mpc}^{-1}$. We
conclude that populating halos using an HOD that depends only on mass will not 
simultaneously recover both the benchmark power spectrum and bispectrum (with correlation discrepancies 
in the latter exceeding 4\%).

\subsection{Assembly bias}

Since using the benchmark HOD yields a suppression of power in 
the power spectrum and bispectrum, and tuning the 4-parameter 
HOD model fares no better in matching both the power spectrum and 
bispectrum, we considered alternative methods of modelling 
the halo occupation number that take into account the formation 
history of the halos, known as assembly bias (see, for example, 
\citep{assembly_bias,Gao2005,Sunayama2016,Hearin2016,Wechsler2006}).
Amongst halos with the same mass those formed at higher redshifts in $N$-body simulations are
known to typically have higher concentrations $c$ 
\citep{Zhao2003,Zhao2009,Villarreal2017,Wechsler2002,Wechsler2006} 
(although this relationship should not be over-simplified \citep{Wechsler2017}).
For this reason, we investigate whether incorporating halo concentration
into our HOD model can simultaneously reduce the measured mock catalogue deficit in both the power spectrum and bispectrum.  The probability distribution of the occupation
number $N_g$ becomes $P(N_g|c,M)$, which is a function of both mass and concentration. 

\begin{figure*}
  \begin{center}
    \begin{subfigure}[b]{0.45\textwidth}
      \includegraphics[width=\linewidth]{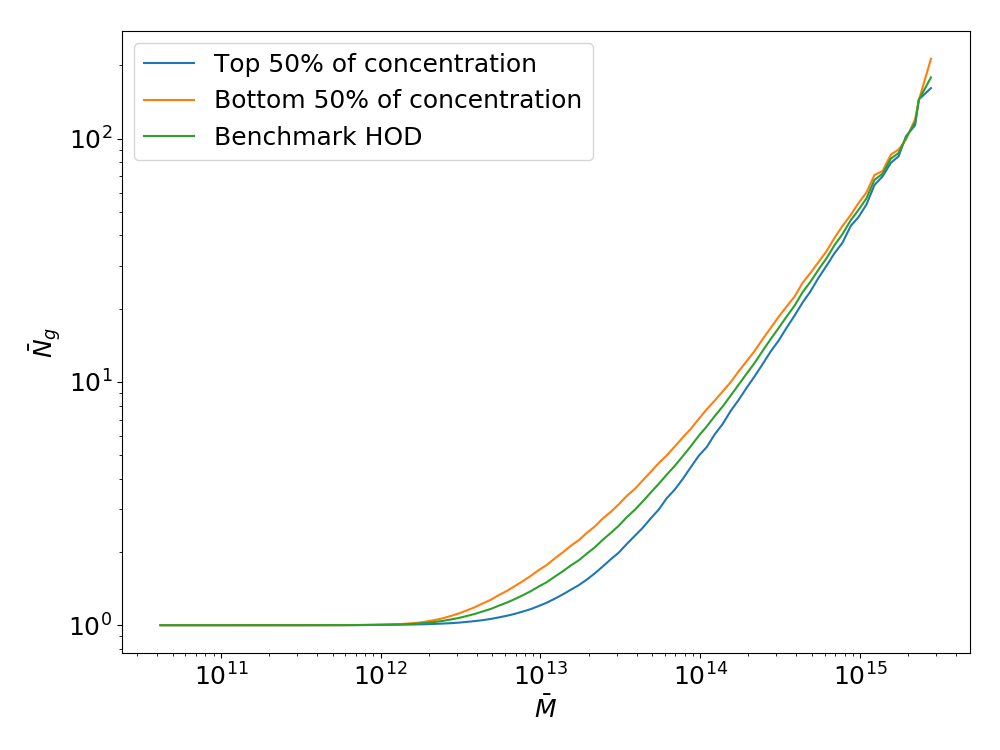}
    \end{subfigure}
    ~
    \begin{subfigure}[b]{0.45\textwidth}
      \includegraphics[width=\linewidth]{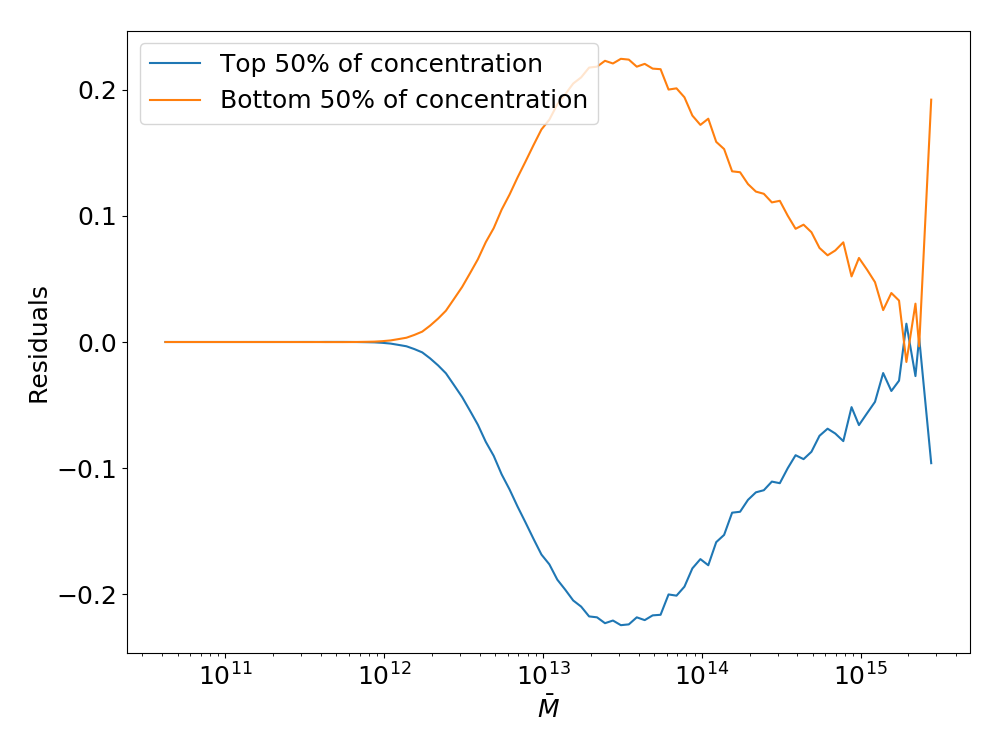}
    \end{subfigure}
    \caption{Left panel: We separate halos within a mass bin into
      2 samples split by the median concentration, and calculate
      their average halo occupation. Right panel: Residuals of those
      2 samples relative to the benchmark HOD.
    }
    \label{fig:hod_conc}
  \end{center}
\end{figure*}

\begin{figure*}
  \captionsetup[subfigure]{labelformat=empty}
  \begin{center}    
    \begin{subfigure}[b]{0.45\textwidth}
      \includegraphics[width=\linewidth]{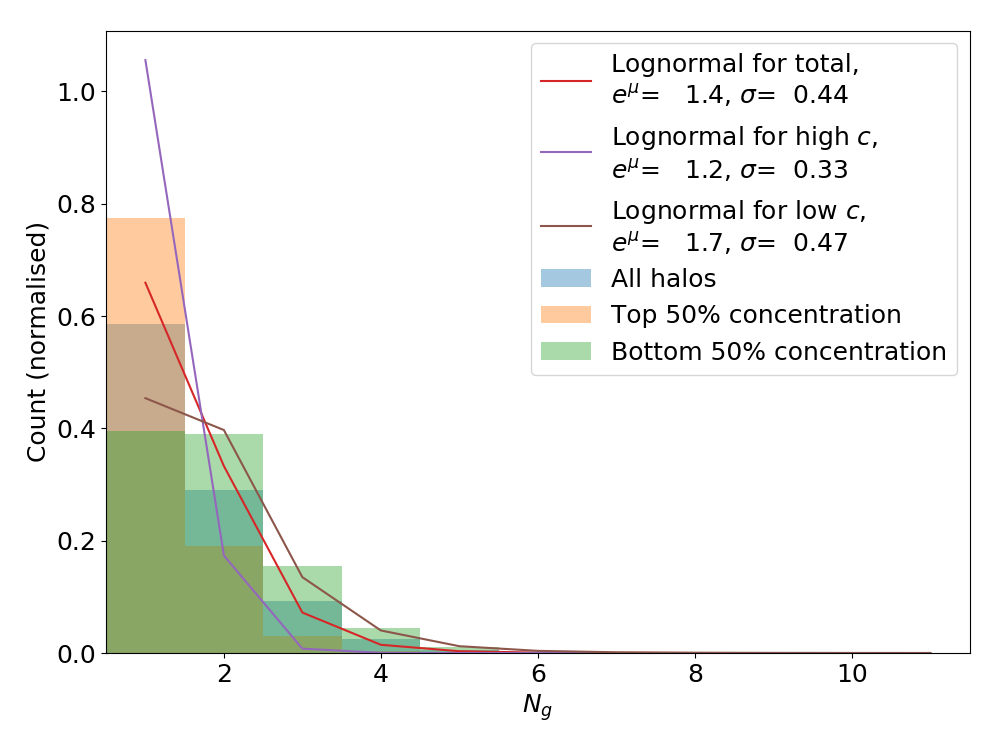}
      \caption{$M=(0.62-2.2)\times10^{13}h^{-1}\,M_{\odot}$}
    \end{subfigure}
    ~
    \begin{subfigure}[b]{0.45\textwidth}
      \includegraphics[width=\linewidth]{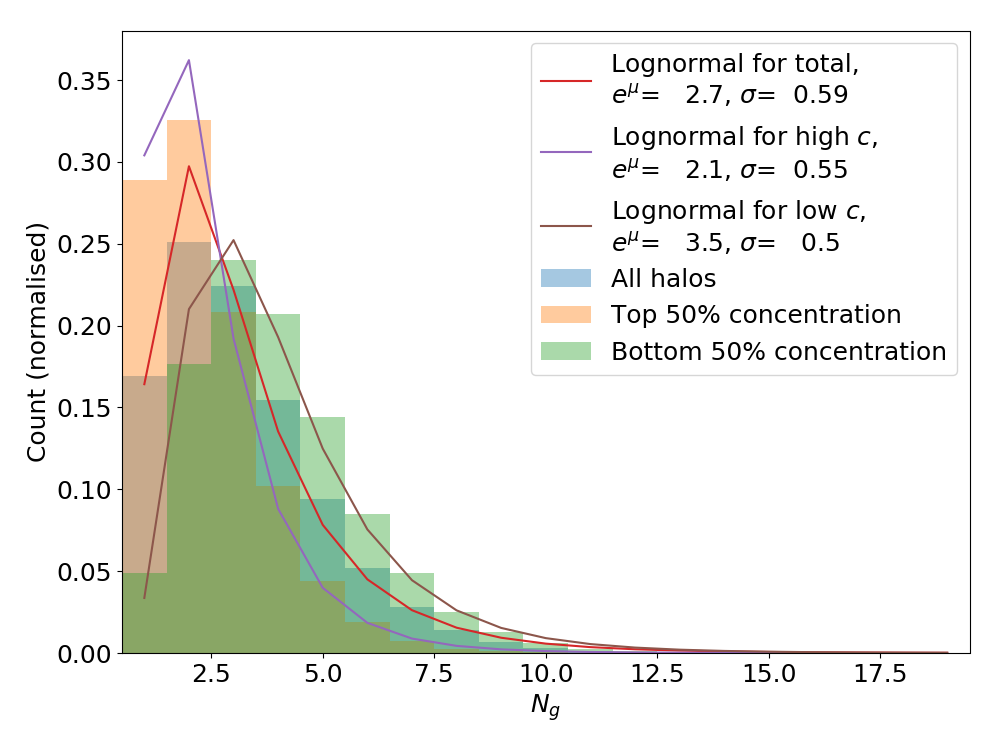}
      \caption{$M=(2.2-7.6)\times10^{13}h^{-1}\,M_{\odot}$}
    \end{subfigure}

    \begin{subfigure}[b]{0.45\textwidth}
      \includegraphics[width=\linewidth]{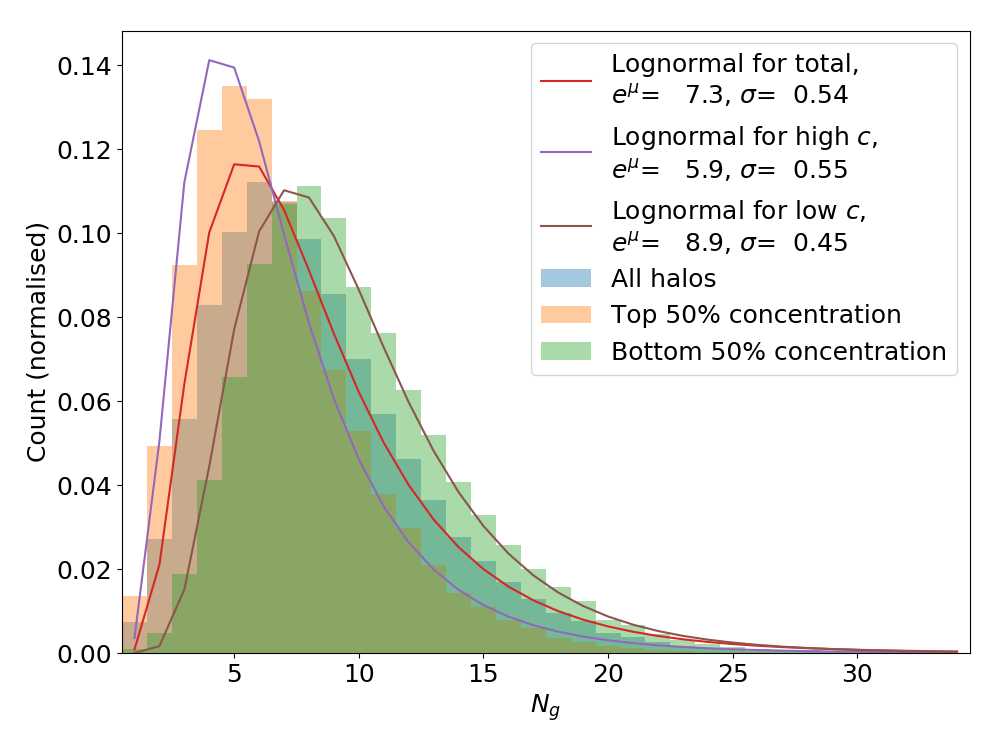}
      \caption{$M=(0.76-2.7)\times10^{14}h^{-1}\,M_{\odot}$}
    \end{subfigure}
    ~
    \begin{subfigure}[b]{0.45\textwidth}
      \includegraphics[width=\linewidth]{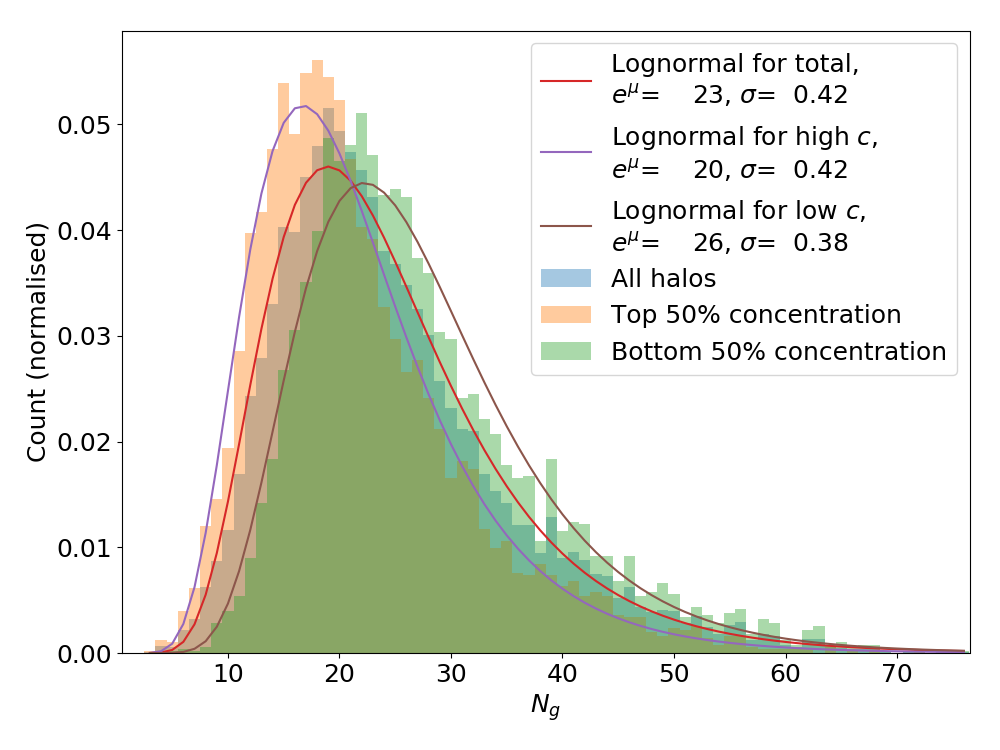}
      \caption{$M=(2.7-9.3)\times10^{14}h^{-1}\,M_{\odot}$}
    \end{subfigure}
    \caption{The standard lognormal distribution
      fitted to the total halo occupation number, as
      well as the occupation number for the high and
      low concentration bins.}
    \label{fig:hod_distribution_lognormal}
  \end{center}
\end{figure*}

To gain insight into how the concentration affects halo
occupation we took inspiration from \citep{eisenstein2} with a simple model that, 
first, bins parent halos by mass and then, secondly, divides these into two bins based on their concentration.  
The threshold for this split into
concentration bins was the median concentration, such that both
the higher and the lower concentration samples at a given 
mass have the same number of subhalos. For each mass bin, we calculated 
the mean occupation number in the high and low concentration
bins (as well as the whole sample). \Cref{fig:hod_conc} shows that halos with lower
concentration clearly have more subhalos than the average,
amounting to a 20\% difference in the mass range between
of $10^{13}h^{-1}\,M_{\odot}$ and $10^{14}h^{-1}\,M_{\odot}$. 
The significant anticorrelation of the concentration with the number of subhalos may or
may not be reflected in actual galaxy distributions because of resolution 
limitations and absent dynamical effects in our DM-only $N$-body simulations. 
If halos with high concentration are indeed typically those
that are formed earlier, then the lower number of subhalos will be affected by 
merging of substructure which is, in turn, influenced by halo resolution (see, for example, \citep{merger}).

\begin{figure*}
  \begin{center}
    \begin{subfigure}[b]{0.49\textwidth}
      \includegraphics[width=\linewidth]{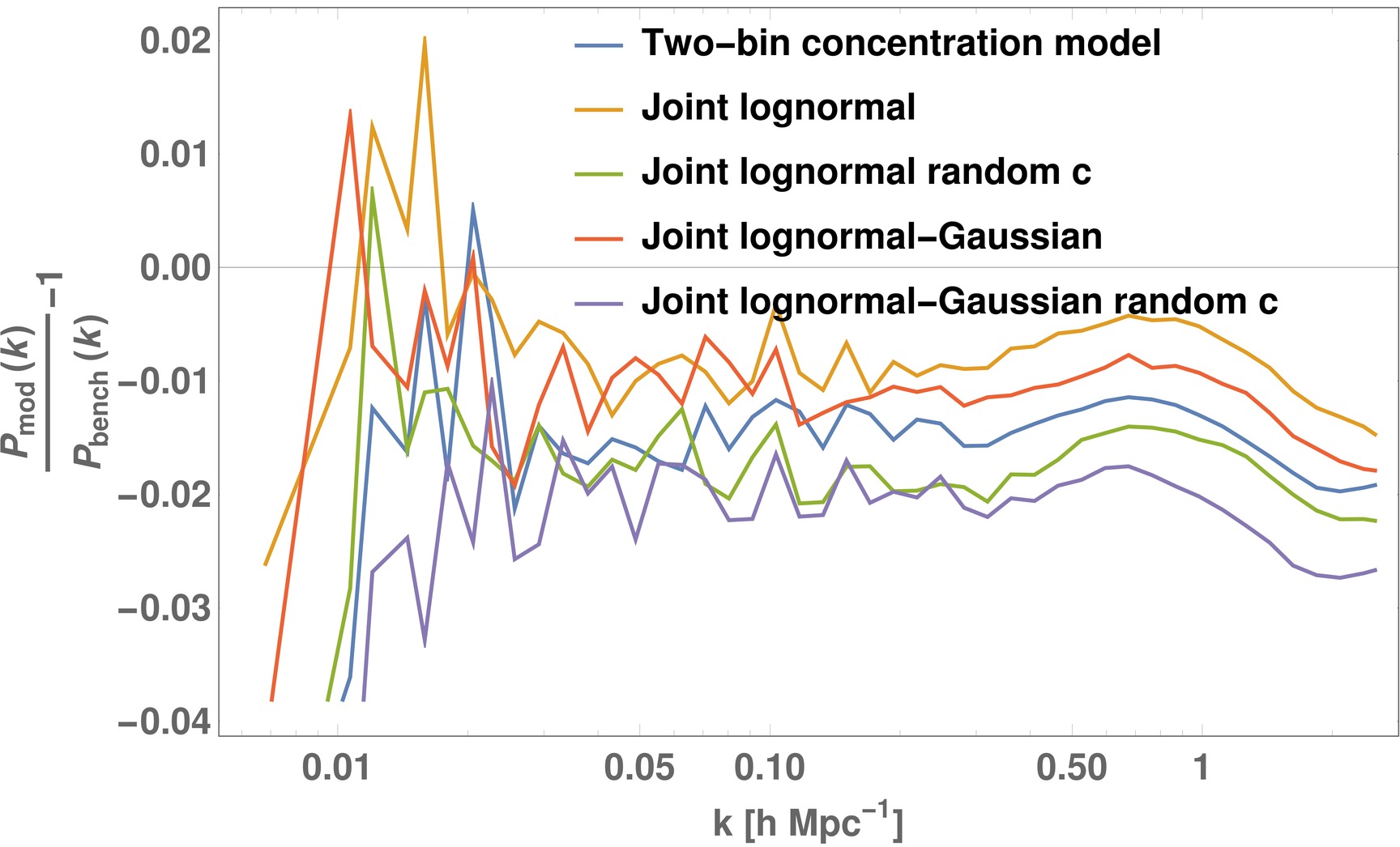}
    \end{subfigure}
    ~
    \begin{subfigure}[b]{0.455\textwidth}
      \includegraphics[width=\linewidth]{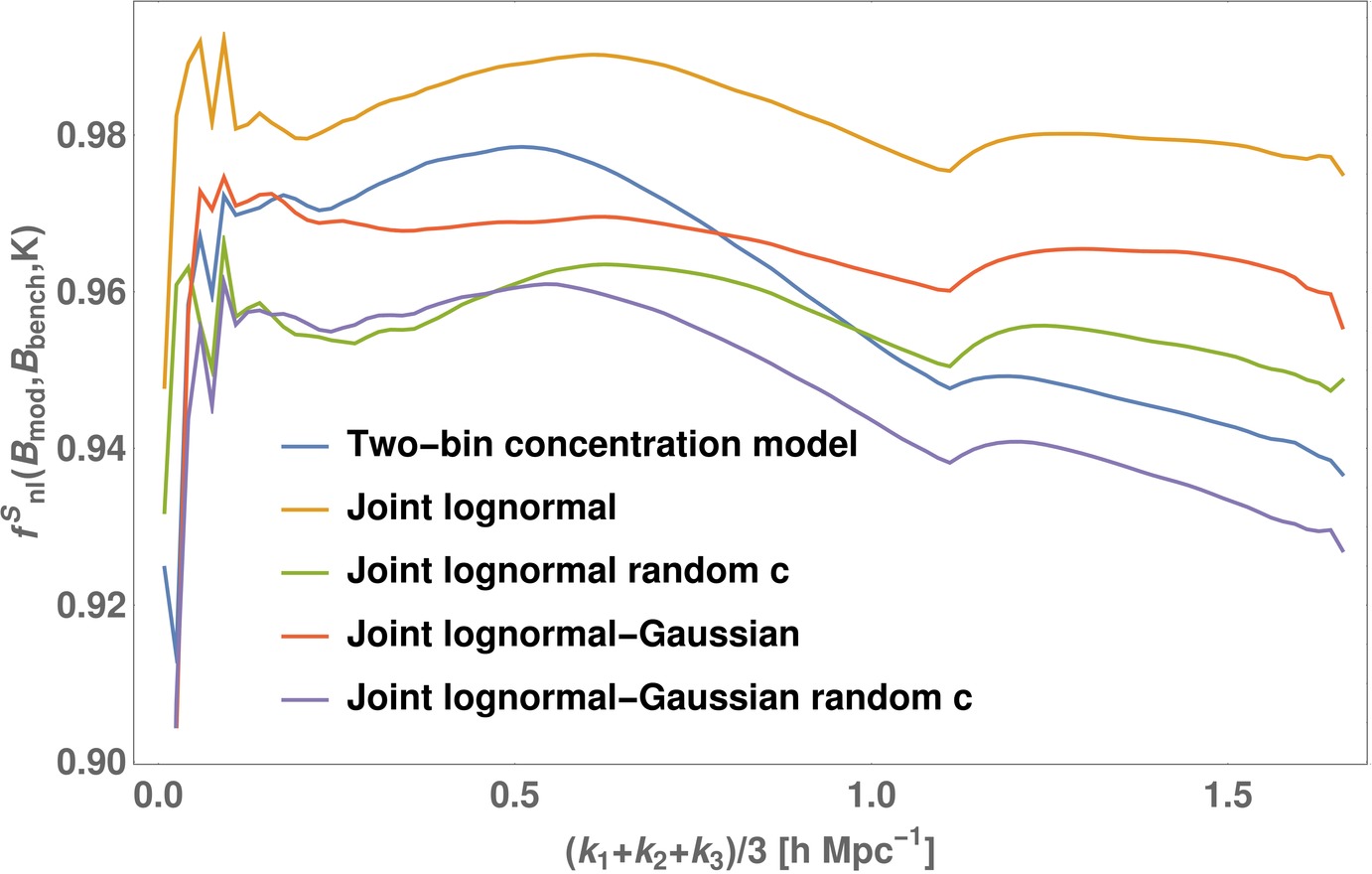}
    \end{subfigure}
    \caption{Power spectra (left) and bispectra (right) comparisons of the two-bin and joint distribution assembly bias
      HOD models relative to the measured benchmark polyspectra.  Prescriptions using the joint probability distribution and information about the individual halo concentrations  improve the fit to better than 2\% for $ k<1.0\,h\,\text{Mpc}^{-1}$.  The halo profile adopted here
      is a power law with $\gamma=1.5$.}
    \label{fig:stats_lognormal}
  \end{center}
\end{figure*}

The positive impact of accounting for concentration with this simple split bin model is illustrated in  \Cref{fig:stats_lognormal} for both the power spectrum and bispectrum. Here, we have populated halos with subhalos drawn from a lognormal distribution to model the total occupation number of the two concentration bins at each mass scale (see below).  These results should be compared with the benchmark HOD model in \Cref{fig:hod_best_fit} where the bispectrum was very discrepant.    In particular, this reduces the deficit in the bispectrum from around 6\% to 3\% at $K/3 = 0.2\,h\,\text{Mpc}^{-1}$, so assembly bias is clearly an important factor which should be taken into account when creating mock catalogues. 

\begin{figure}
  \begin{center}    
    \includegraphics[width=0.9\linewidth]{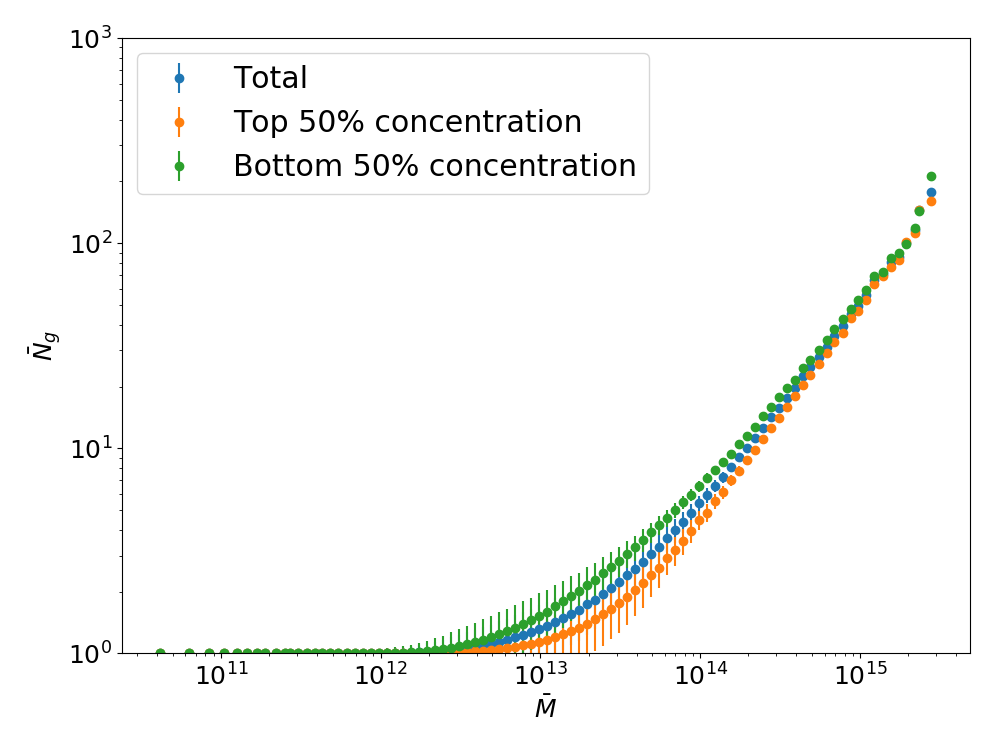}
    \caption{Lognormal fits to the total occupation number
      and the high and low concentration bins.
      The vertical error bars indicate the shape
      parameter $\sigma$ of the fits.}
      \label{fig:hod_distribution_lognormal2}
  \end{center}
\end{figure}

\begin{figure*}
  \begin{center}
    \begin{subfigure}[b]{0.45\textwidth}
      \includegraphics[width=\linewidth]{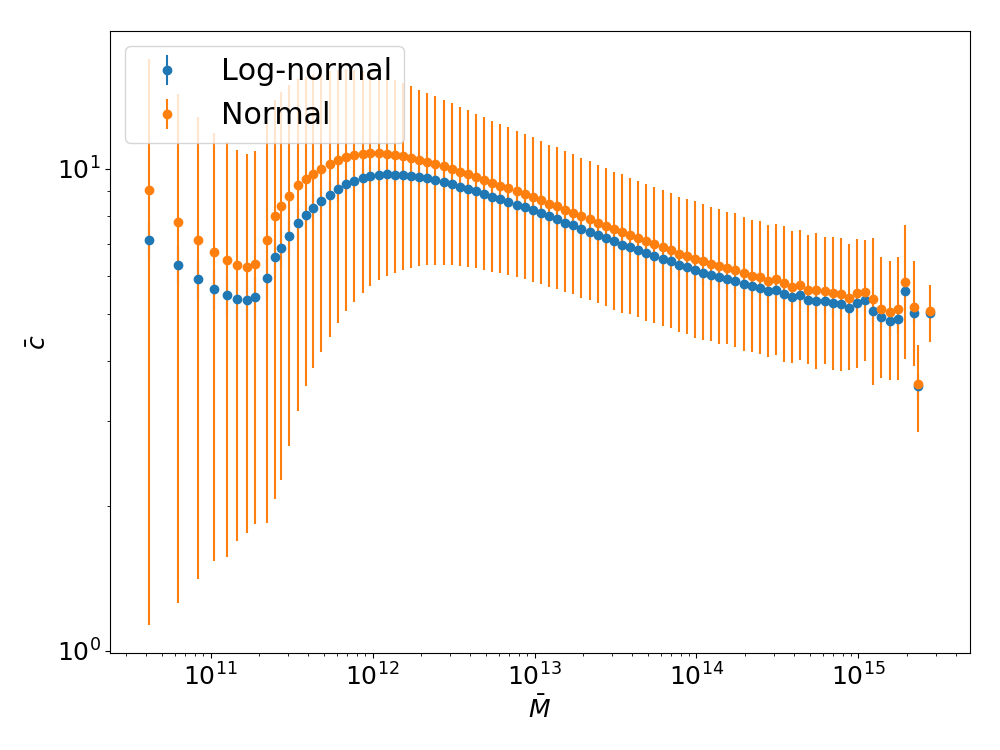}
      \caption{}
    \label{fig:conc_distribution}
    \end{subfigure}
    ~
    \begin{subfigure}[b]{0.45\textwidth}
      \includegraphics[width=\linewidth]{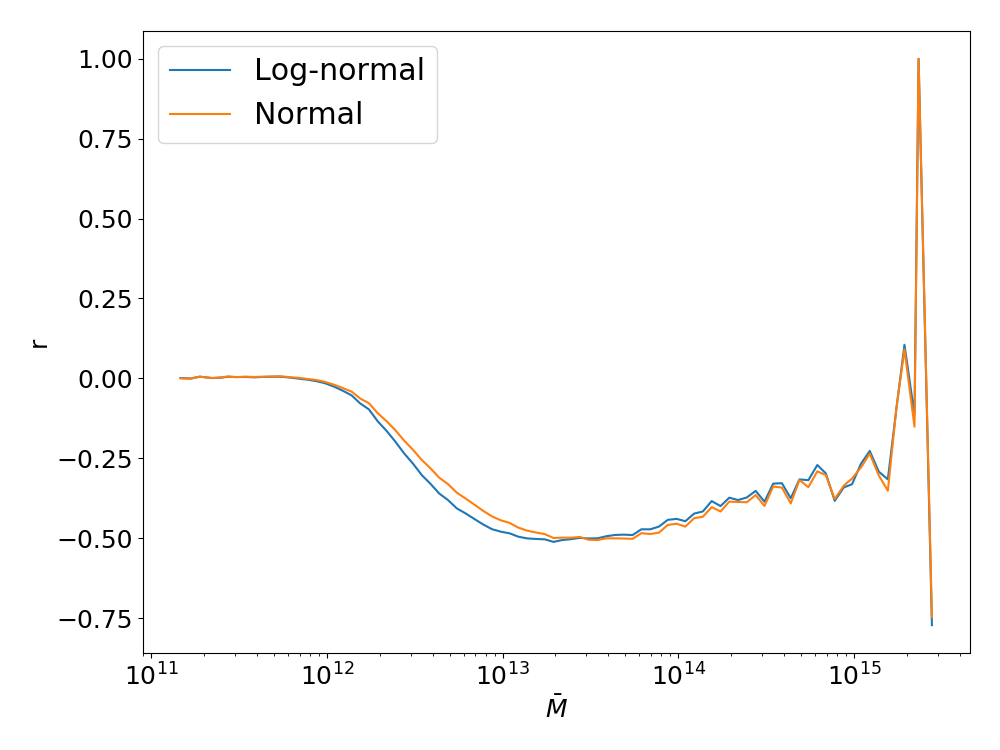}
      \caption{}
      \label{fig:joint_distribution_2}
    \end{subfigure}
    \caption{Left panel: Lognormal and Gaussian fits to $\bar{c}$.
      The vertical error bars indicate the shape parameter $\sigma$ of the
      fits. Right panel: Correlation coefficient $r$
      of the bivariate Gaussian distribution between $\ln(N_g)$ and $X$.}
  \end{center}
\end{figure*}

\begin{figure*}
  \captionsetup[subfigure]{labelformat=empty}
  \begin{center}    
    \begin{subfigure}[b]{0.45\textwidth}
      \includegraphics[width=\linewidth]{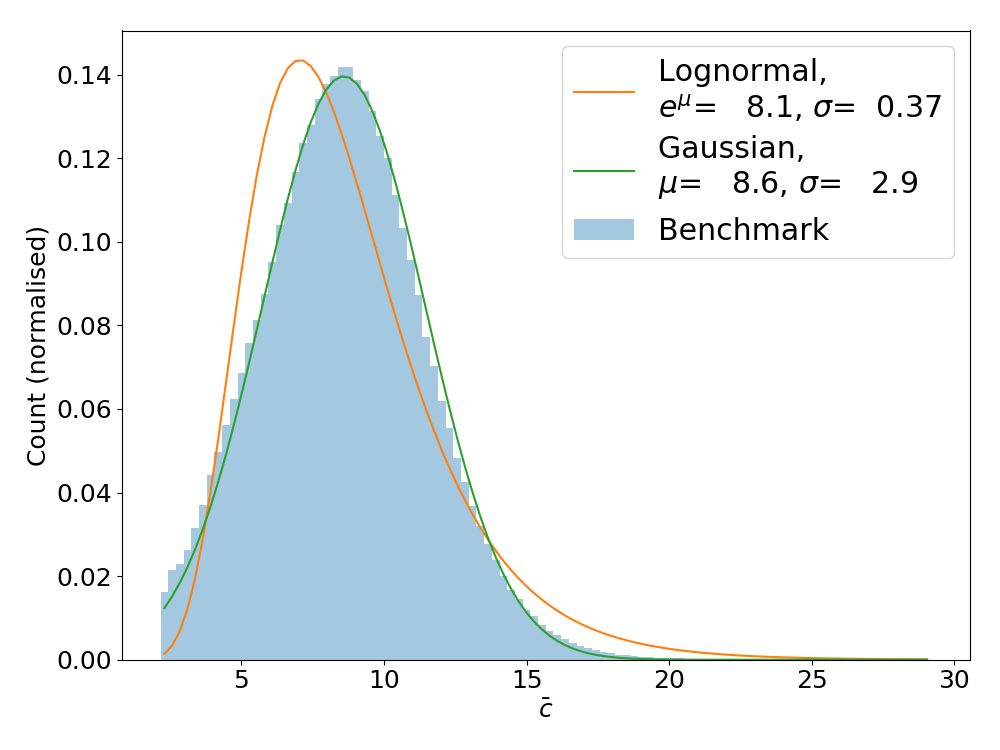}
      \caption{$M=(0.62-2.2)\times10^{13}h^{-1}\,M_{\odot}$}
    \end{subfigure}
    ~
    \begin{subfigure}[b]{0.45\textwidth}
      \includegraphics[width=\linewidth]{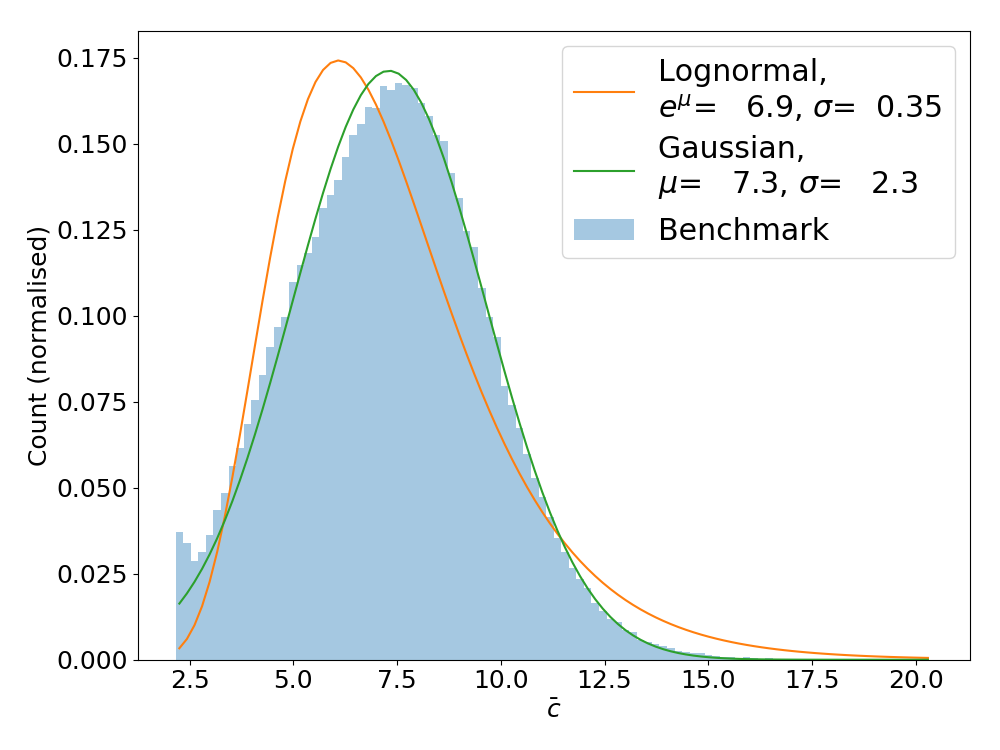}
      \caption{$M=(2.2-7.6)\times10^{13}h^{-1}\,M_{\odot}$}
    \end{subfigure}

    \begin{subfigure}[b]{0.45\textwidth}
      \includegraphics[width=\linewidth]{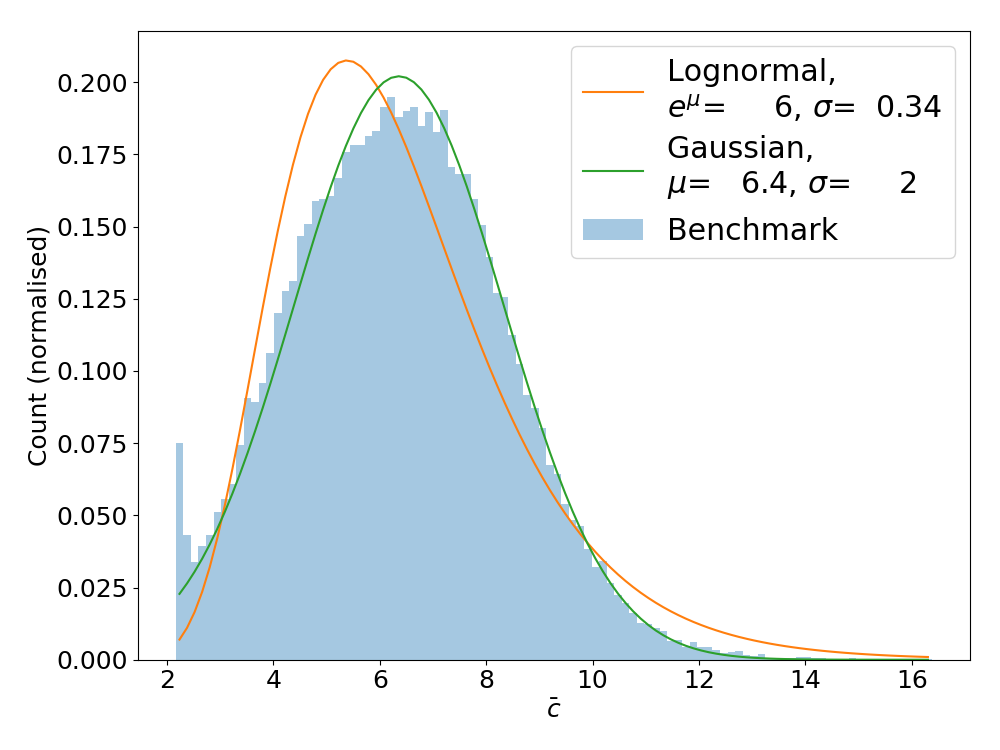}
      \caption{$M=(0.76-2.7)\times10^{14}h^{-1}\,M_{\odot}$}
    \end{subfigure}
    ~
    \begin{subfigure}[b]{0.45\textwidth}
      \includegraphics[width=\linewidth]{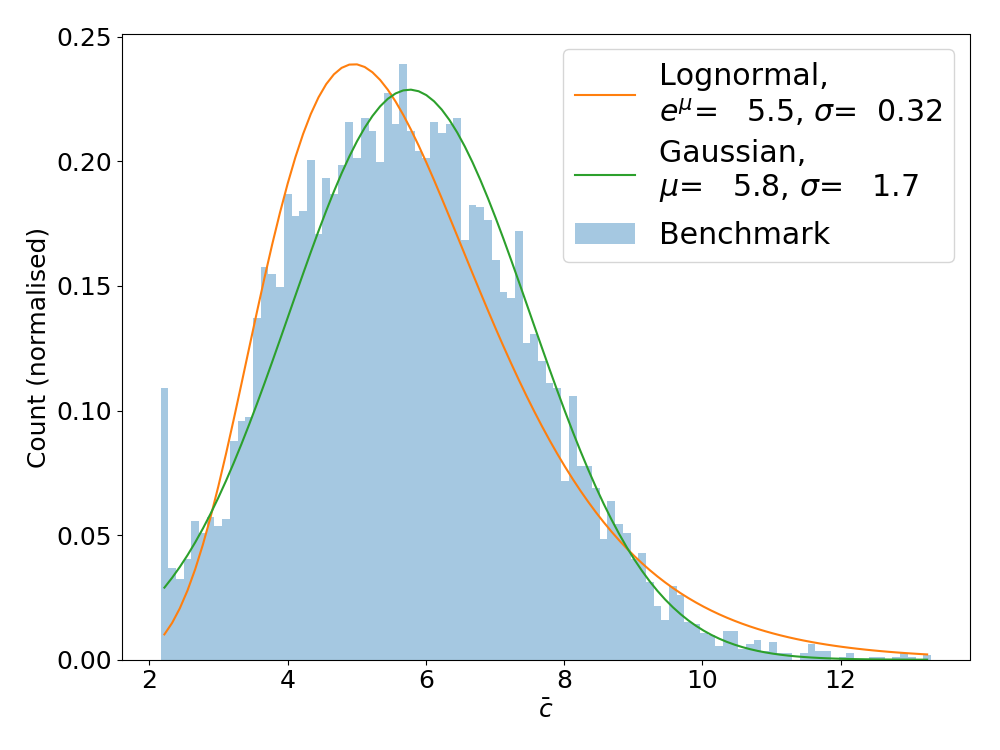}
      \caption{$M=(2.7-9.3)\times10^{14}h^{-1}\,M_{\odot}$}
    \end{subfigure}
    \caption{A lognormal distribution is too skewed to model
      the distribution of halo concentration, but a Gaussian
      fits very well especially at low mass.} 
    \label{fig:conc_distribution_lognormal}
  \end{center}
\end{figure*}

In light of the impact of concentration on subhalo number, our goal is to develop a more sophisticated 
statistical model that allows us to populate individual halos of a given mass, with or without specifying the concentration
from information given by the simulation. To achieve this, we require the joint probability
distribution $P(N_g\cap c\,|M)$ as a function of subhalo number $N_g$ and concentration $c$, so that we can
derive $P(N_g|c,M)$ from Bayes theorem \citep{bayes}:
\begin{align}
  \label{eq:bayes}
  P(N_g|c,M)=\frac{P(N_g\cap c\,|M)}{P(c|M)}.
\end{align}
To find an appropriate joint distribution we first investigate
the marginalised distributions for $N_g$ and $c$. It was found
that the standard lognormal distribution with 2 parameters,
$\text{Lognormal}(\mu,\sigma^2)$ where $e^\mu$ is known as the
scale parameter and $\sigma$ the shape parameter, provides a good fit to the
marginalised halo occupation number.
\Cref{fig:hod_distribution_lognormal} shows the lognormal fits
to the total occupation number, and occupation number in the
high and low concentration bins, for several mass bins. 
In \Cref{fig:hod_distribution_lognormal2} we show the shape
and scale parameters of these fits in 100 mass bins across
the whole range of the benchmark catalogue. Note that we have
adopted the total occupation number, i.e. including the central
galaxy instead of just the satellites, because when
the average number of satellites falls below unity the lognormal fit automatically fails.

For the marginalised concentration distribution, we found that 
it could be more accurately modelled with a Gaussian distribution, particularly
at low masses.  The lognormal distribution provides a significantly worse fit, a comparison which is shown in
\Cref{fig:conc_distribution_lognormal}, where we display the
normalised counts in several mass bins along with the best fit values
for both Gaussian and lognormal fits. 

Either the Gaussian or lognormal distributions for $c$ can be easily
combined with the lognormal distribution for $N_g$ to give a joint
distribution. To do so we simply have to take the natural logarithm
of $N_g$ and calculate the mean $\boldsymbol{\mu}$ and covariance
$\boldsymbol{\Sigma}$ for this joint Gaussian distribution:
\begin{align}
  \label{eq:joint_gaussian}
  \begin{pmatrix}
    \ln(N_g) \\
    X
  \end{pmatrix}
  \sim
  \mathcal{N}(\boldsymbol{\mu},\boldsymbol{\Sigma}),
\end{align}
where $X=c$ or $\ln(c)$ depending on whether a Gaussian
or lognormal distribution for $c$ is desired, and
\begin{align}
  \label{eq:joint_gaussian2}
  \boldsymbol{\mu} =
  \begin{pmatrix}
    \expval{\ln(N_g)} \\
    \expval{X}
  \end{pmatrix}, \qquad
  \boldsymbol{\Sigma} =
  \begin{pmatrix}
    \sigma^2_{\ln(N_g)} & \sigma_{\ln(N_g),X} \\
    \sigma_{\ln(N_g),X} & \sigma^2_{X}
  \end{pmatrix}.
\end{align}
$\sigma^2_{\ln(N_g)}$ and $\sigma^2_{X}$ are the usual
variances for $\ln(N_g)$ and $X$, and
$\sigma_{\ln(N_g),X}=\expval{(\ln(N_g)-
  \expval{\ln(N_g)})(X-\expval{X})}$ is the covariance
between them. 

\begin{figure}
  \captionsetup[subfigure]{labelformat=empty}
  \begin{center}    
    \begin{subfigure}[b]{0.39\textwidth}
      \includegraphics[width=\linewidth]{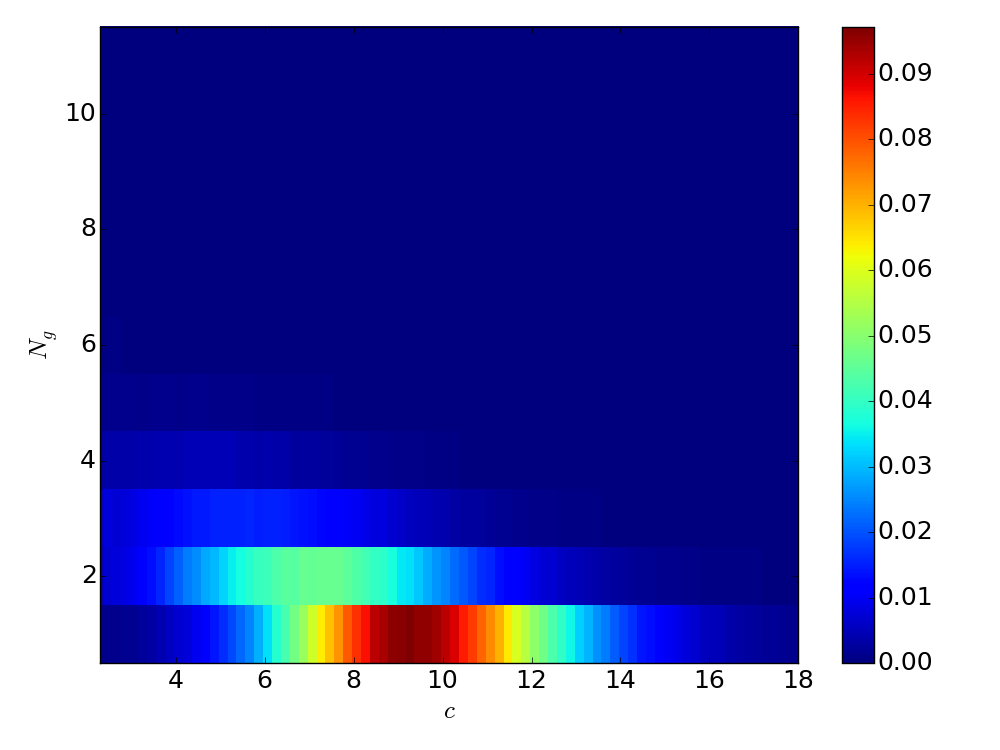}
      \caption{$M=(0.62-2.2)\times10^{13}h^{-1}\,M_{\odot}$}
    \end{subfigure}
    
    \begin{subfigure}[b]{0.39\textwidth}
      \includegraphics[width=\linewidth]{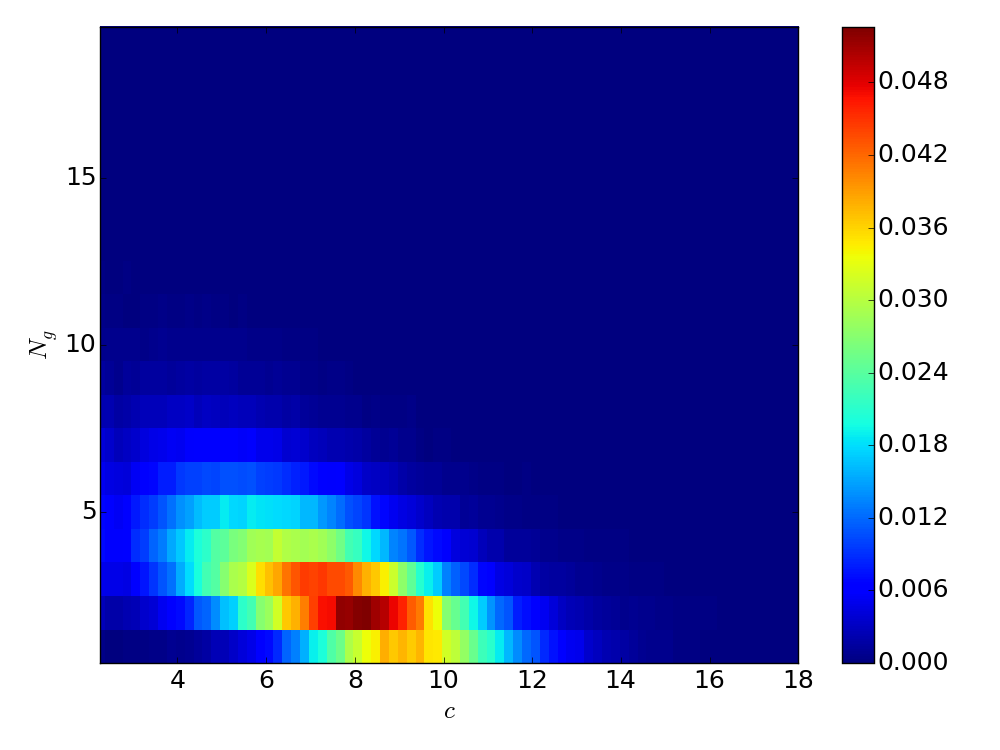}
      \caption{$M=(2.2-7.6)\times10^{13}h^{-1}\,M_{\odot}$}
    \end{subfigure}

    \begin{subfigure}[b]{0.39\textwidth}
      \includegraphics[width=\linewidth]{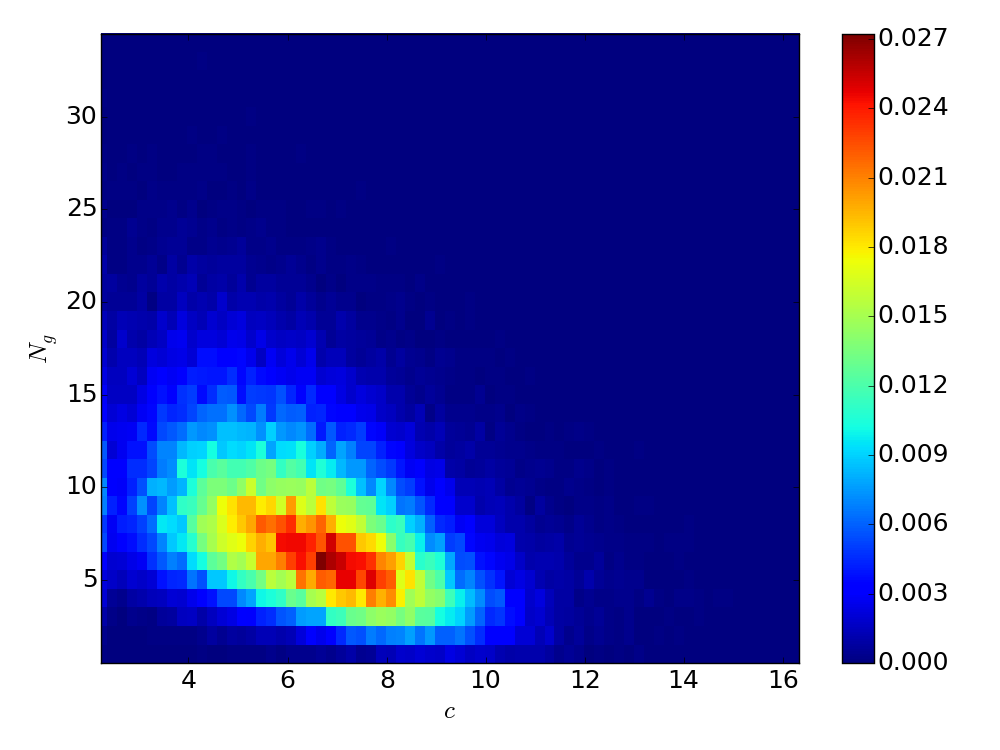}
      \caption{$M=(0.76-2.7)\times10^{14}h^{-1}\,M_{\odot}$}
    \end{subfigure}
    
    \begin{subfigure}[b]{0.39\textwidth}
      \includegraphics[width=\linewidth]{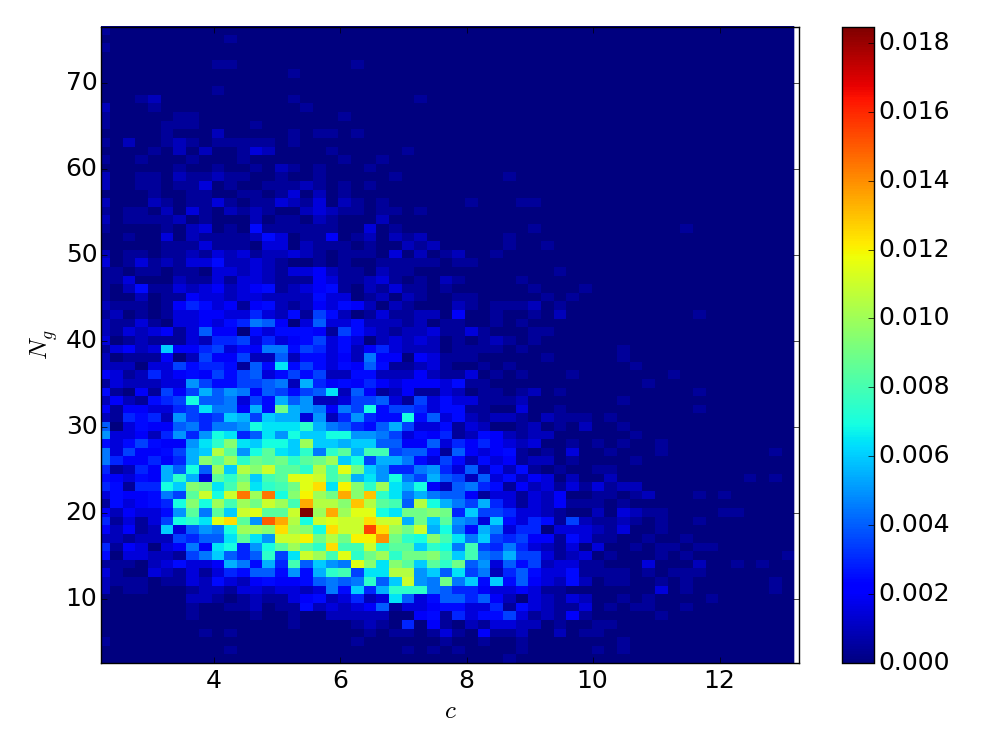}
      \caption{$M=(2.7-9.3)\times10^{14}h^{-1}\,M_{\odot}$}
    \end{subfigure}
    \caption{Joint probability distribution for the subhalo number $N_g$ and concentration $c$ for halos in different mass bins
      of the benchmark \texttt{ROCKSTAR} catalogue.} 
    \label{fig:joint_distribution}
  \end{center}
\end{figure}

\begin{figure}
  \captionsetup[subfigure]{labelformat=empty}
  \begin{center}    
    \begin{subfigure}[b]{0.39\textwidth}
      \includegraphics[width=\linewidth]{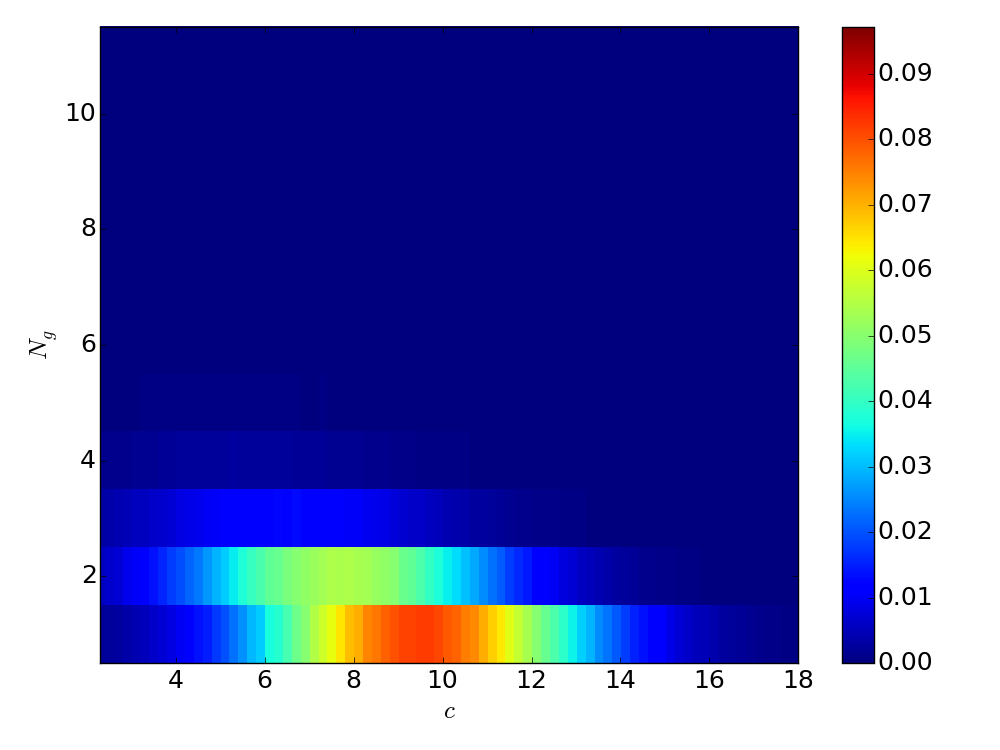}
      \caption{$M=(0.62-2.2)\times10^{13}h^{-1}\,M_{\odot}$}
    \end{subfigure}
    ~
    \begin{subfigure}[b]{0.39\textwidth}
      \includegraphics[width=\linewidth]{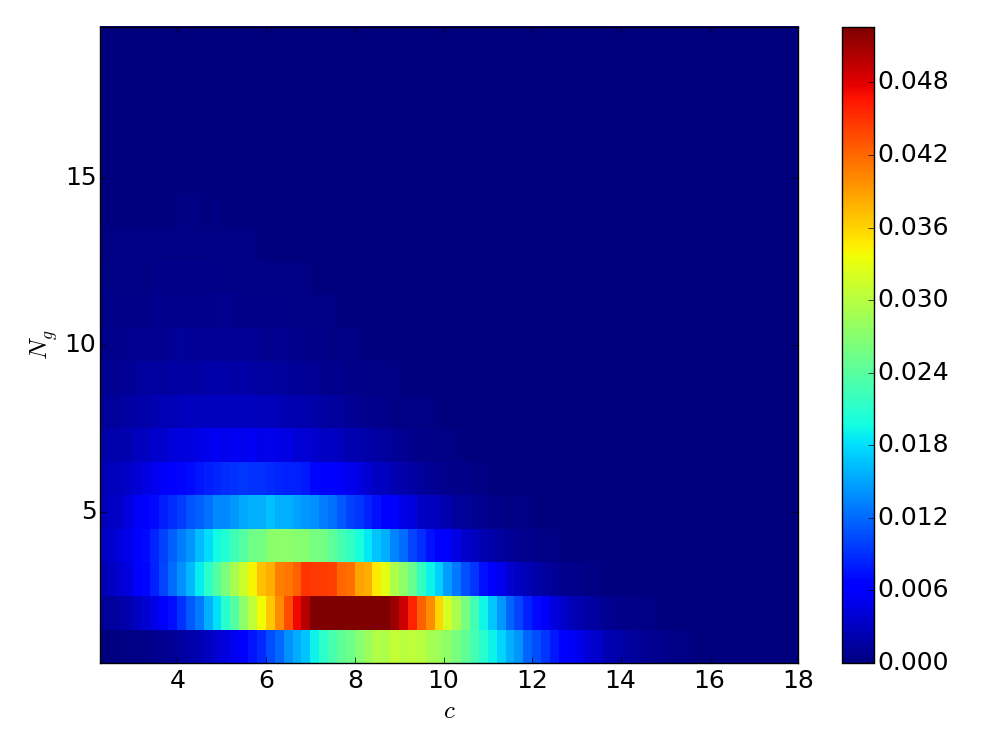}
      \caption{$M=(2.2-7.6)\times10^{13}h^{-1}\,M_{\odot}$}
    \end{subfigure}

    \begin{subfigure}[b]{0.39\textwidth}
      \includegraphics[width=\linewidth]{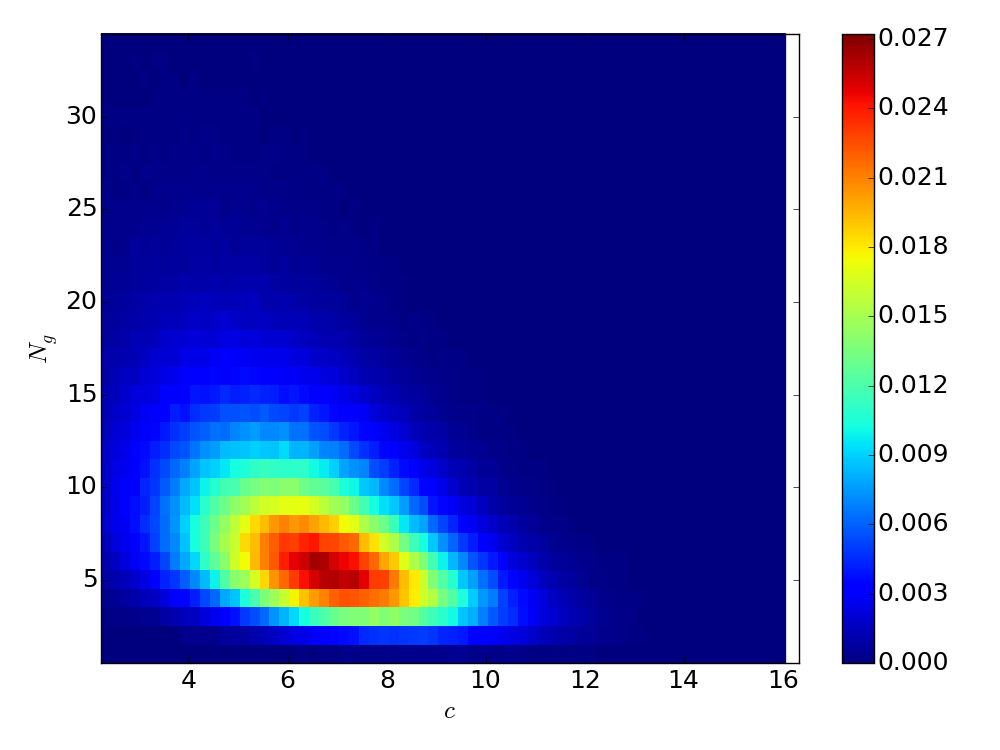}
      \caption{$M=(0.76-2.7)\times10^{14}h^{-1}\,M_{\odot}$}
    \end{subfigure}
    ~
    \begin{subfigure}[b]{0.39\textwidth}
      \includegraphics[width=\linewidth]{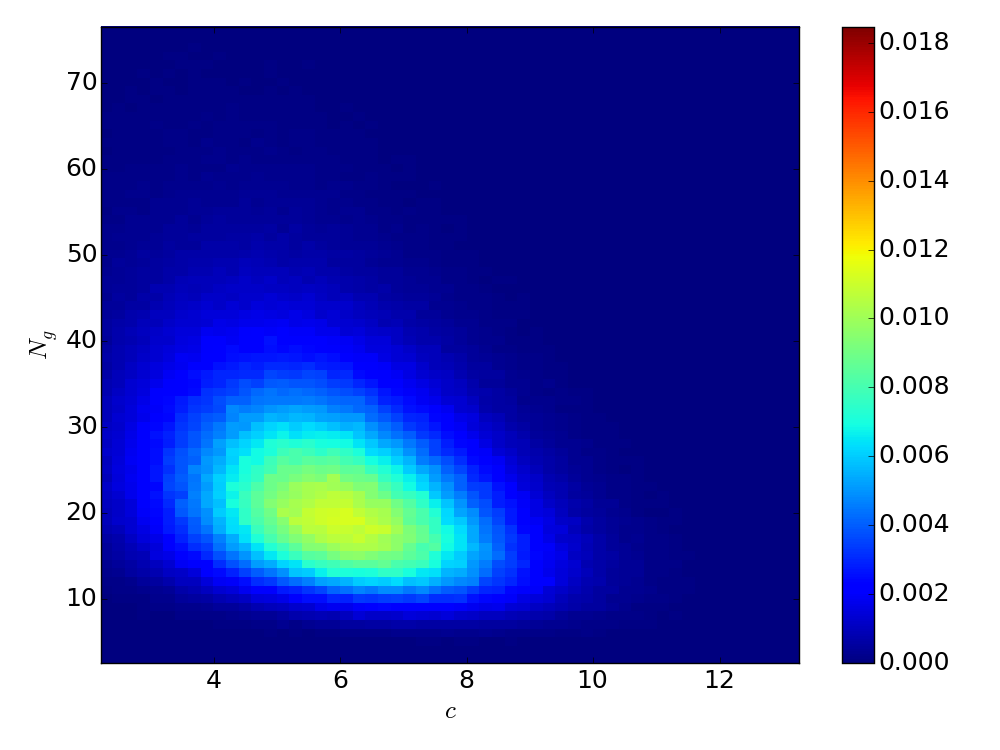}
      \caption{$M=(2.7-9.3)\times10^{14}h^{-1}\,M_{\odot}$}
    \end{subfigure}
    \caption{Joint lognormal-Gaussian fit to the joint distribution in
      \Cref{fig:joint_distribution} which should be compared with benchmark distribution shown in \Cref{fig:joint_distribution}. 
    } 
    \label{fig:joint_distribution_lognormal_normal}
  \end{center}
\end{figure}

\begin{figure}
  \captionsetup[subfigure]{labelformat=empty}
  \begin{center}    
    \begin{subfigure}[b]{0.39\textwidth}
      \includegraphics[width=\linewidth]{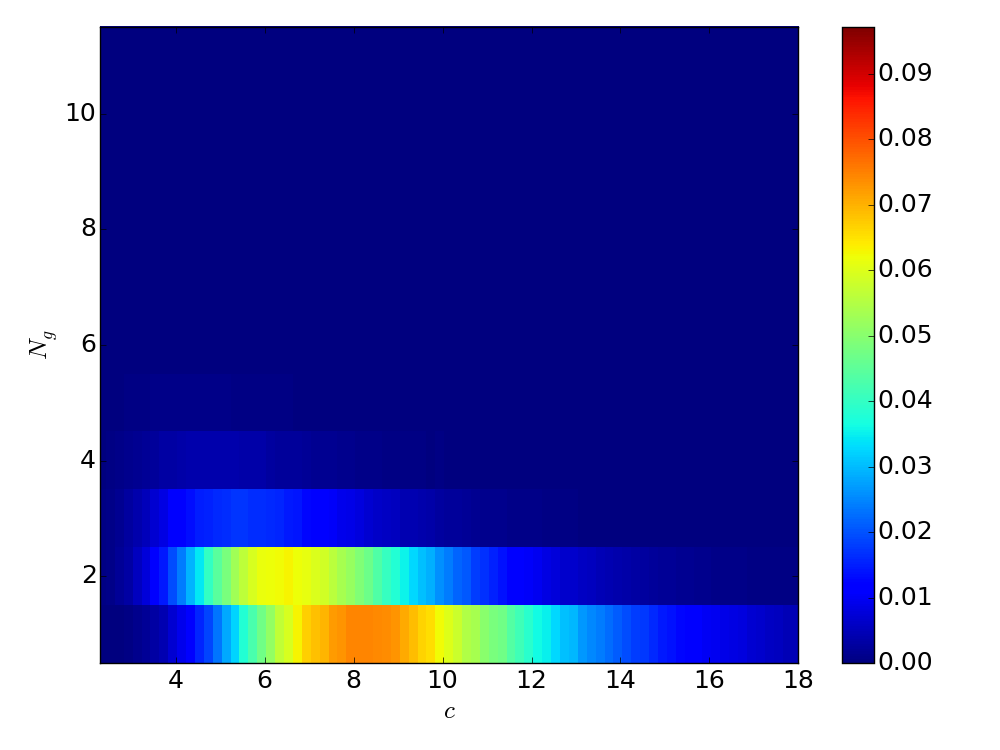}
      \caption{$M=(0.62-2.2)\times10^{13}h^{-1}\,M_{\odot}$}
    \end{subfigure}
    
    \begin{subfigure}[b]{0.39\textwidth}
      \includegraphics[width=\linewidth]{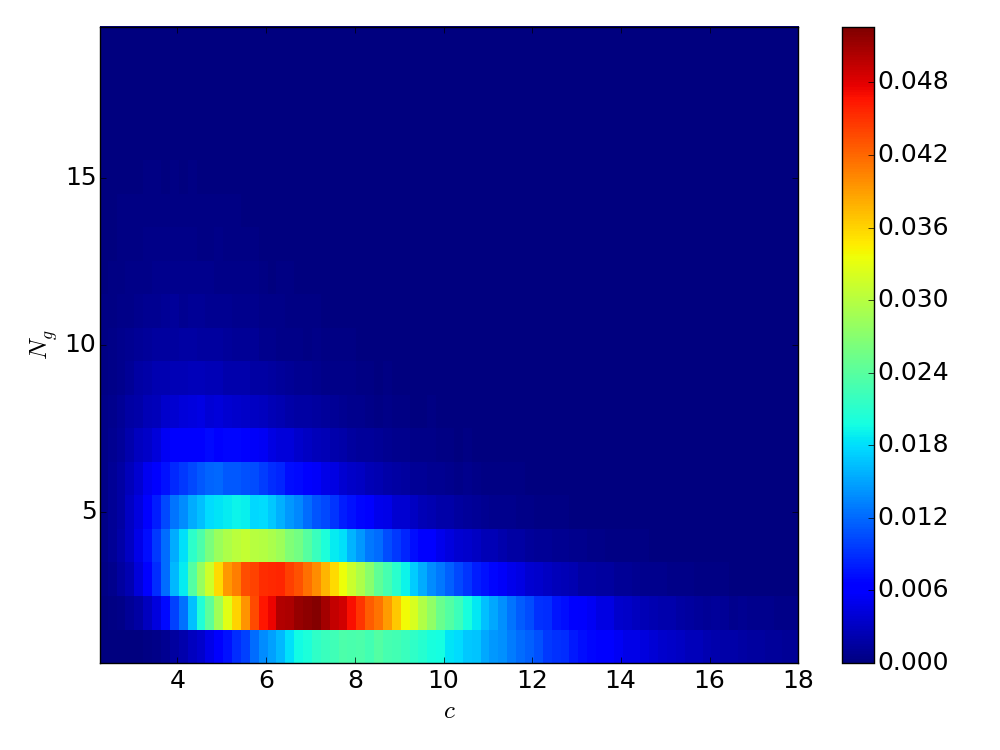}
      \caption{$M=(2.2-7.6)\times10^{13}h^{-1}\,M_{\odot}$}
    \end{subfigure}

    \begin{subfigure}[b]{0.39\textwidth}
      \includegraphics[width=\linewidth]{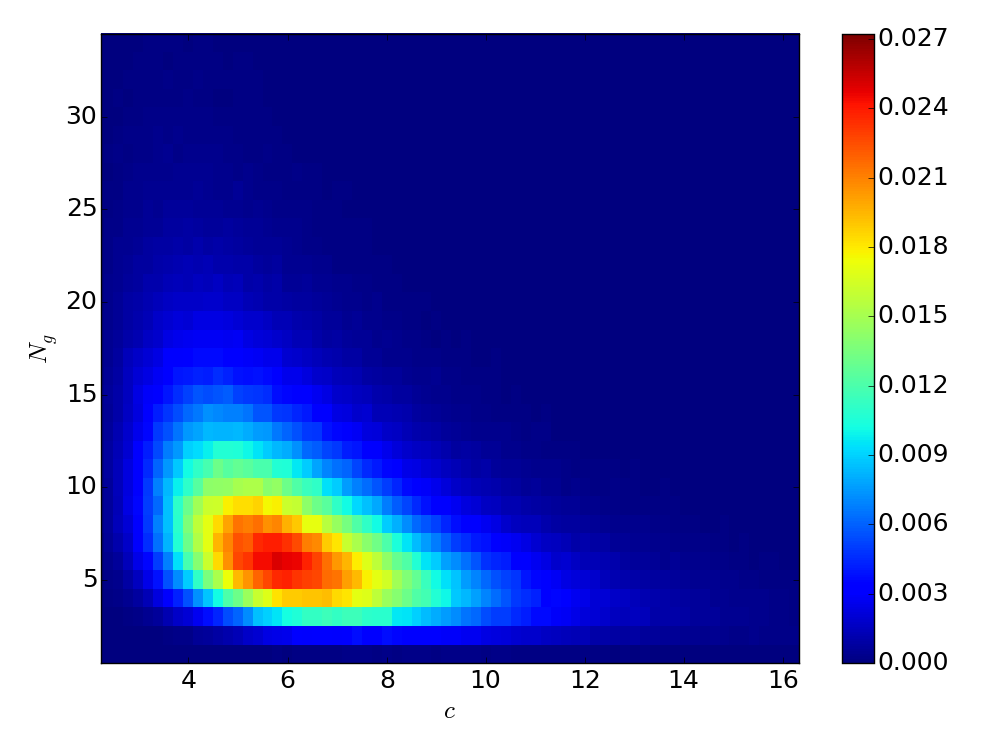}
      \caption{$M=(0.76-2.7)\times10^{14}h^{-1}\,M_{\odot}$}
    \end{subfigure}
    
    \begin{subfigure}[b]{0.39\textwidth}
      \includegraphics[width=\linewidth]{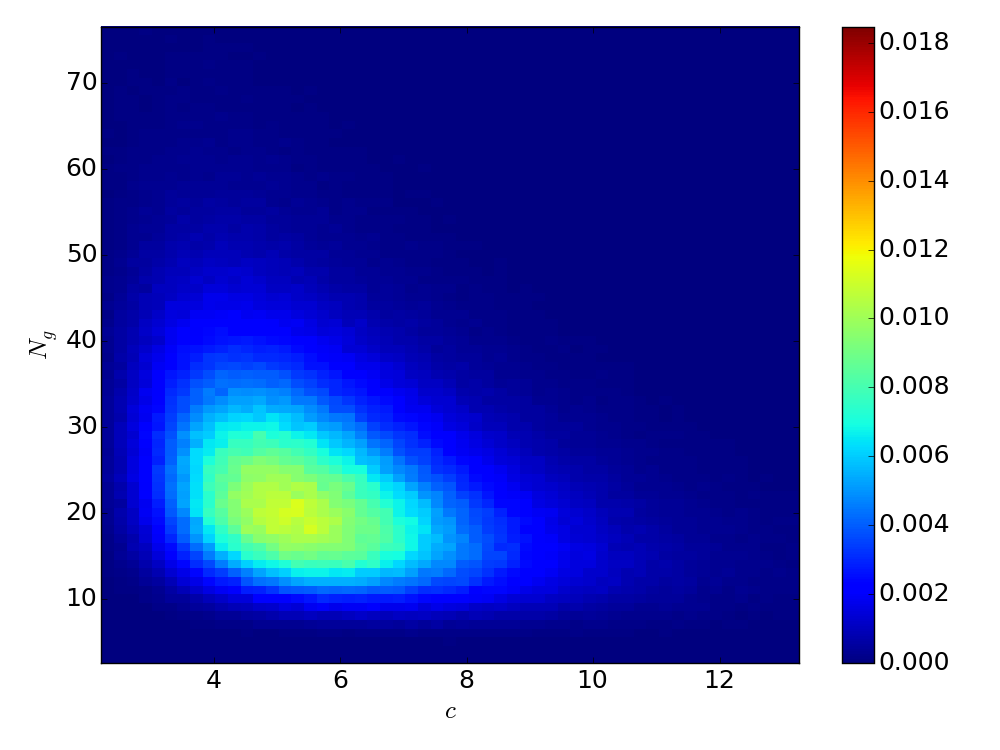}
      \caption{$M=(2.7-9.3)\times10^{14}h^{-1}\,M_{\odot}$}
    \end{subfigure}
    \caption{Joint lognormal fit to the joint distribution in \Cref{fig:joint_distribution}.}
    \label{fig:joint_distribution_lognormal}
  \end{center}
\end{figure}

To draw from the joint distribution one would then sample
from the joint Gaussian distribution and exponentiate the
result as required. The joint distribution obtained from
the \texttt{ROCKSTAR} halo benchmark is shown for various mass bins in
\Cref{fig:joint_distribution}.   For comparison, we show for the same mass bins
calculated both from the joint lognormal distribution 
in \Cref{fig:joint_distribution_lognormal} and from the joint
lognormal-Gaussian distribution in
\Cref{fig:joint_distribution_lognormal_normal}. The joint
lognormal-Gaussian distribution appears to reproduce the
benchmark distribution more accurately, though small discrepancies
remain at high mass. 

In order to obtain $P(N_g|M,c)$ we first shift
the distribution for $\ln(N_g)$ from
$\mathcal{N}(\expval{\ln(N_g)},\sigma^2_{\ln(N_g)})$ to
$\mathcal{N}(\expval{\ln(N_g)}',\sigma^{\prime2}_{\ln(N_g)})$,
where \citep{conditional_gaussian}
\begin{align}
  \label{eq:new_gaussian}
  \expval{\ln(N_g)}'
  &=\expval{\ln(N_g)}+\frac{\sigma_{\ln(N_g),X}}{\sigma^2_{X}}(X-\expval{X}) \\
  \sigma^{\prime2}_{\ln(N_g)}
  &=\sigma^2_{\ln(N_g)}-\frac{\sigma^2_{\ln(N_g),X}}{\sigma^2_{X}},
\end{align}
then exponentiate draws from this shifted Gaussian
distribution. This shift can be derived using the bivariate
Gaussian distribution in \Cref{eq:joint_gaussian}, the
Gaussian distribution for $X$ and Bayes theorem (\Cref{eq:bayes}).
For the benchmark catalogue in \Cref{fig:conc_distribution} we
show the parameters of the lognormal and Gaussian fits to $c$,
and the correlation coefficient
\begin{align} 
r=\frac{\sigma_{\ln(N_g),X}}{\sqrt{\sigma_{\ln(N_g)}\sigma_{X}}}
\end{align}
obtained for the joint Gaussian distribution in
\Cref{fig:joint_distribution_2}. It is worth noting that there
are only minor differences in the correlation coefficient between
the Gaussian and lognormal cases, with a robust value of around $r\approx-0.5$ found
for the mass range $M=10^{13}-10^{14}h^{-1}\,M_{\odot}$.

\begin{figure*}
  \begin{center}
    \begin{subfigure}[b]{0.49\textwidth}
      \includegraphics[width=\linewidth]{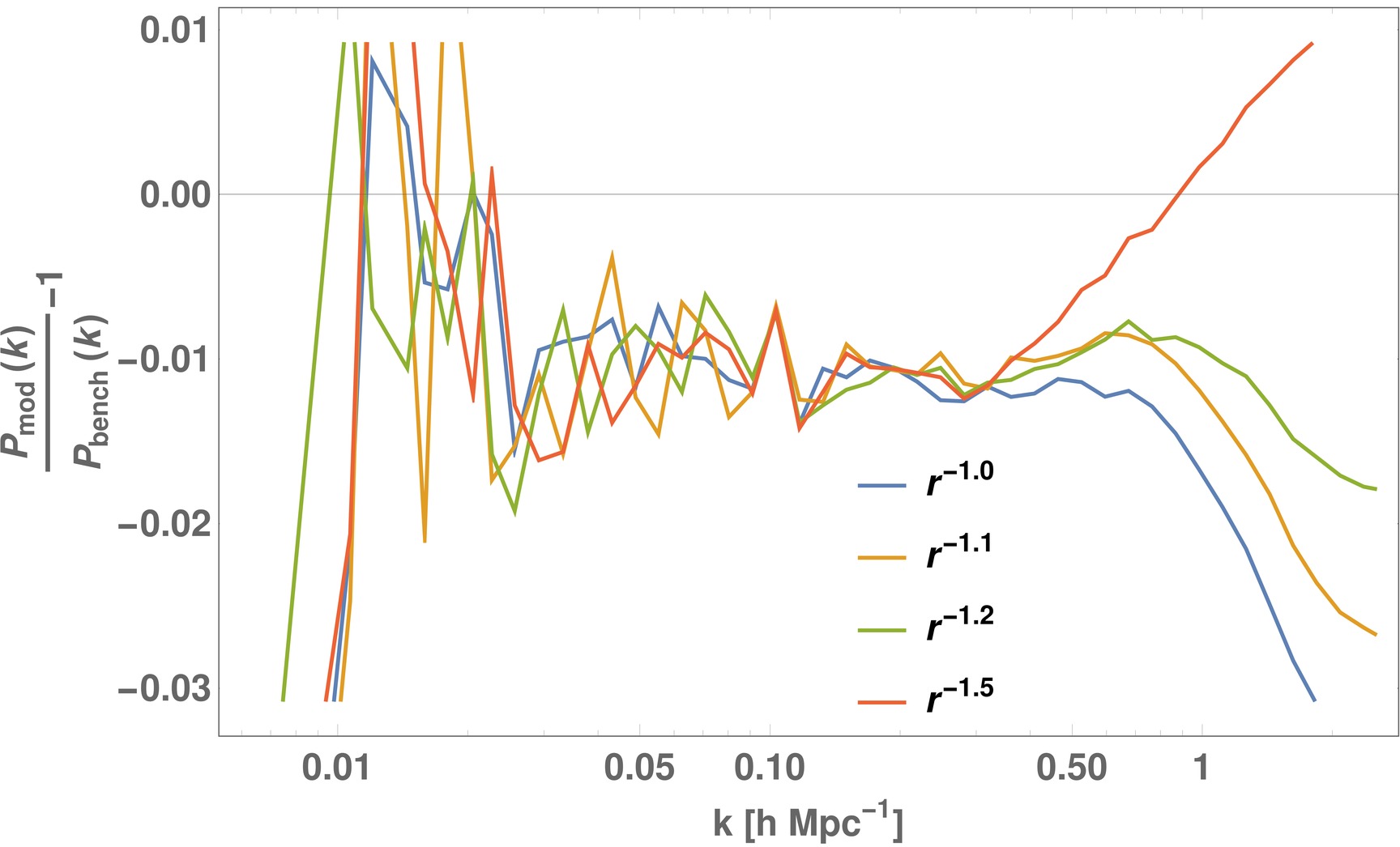}
    \end{subfigure}
    ~
    \begin{subfigure}[b]{0.455\textwidth}
      \includegraphics[width=\linewidth]{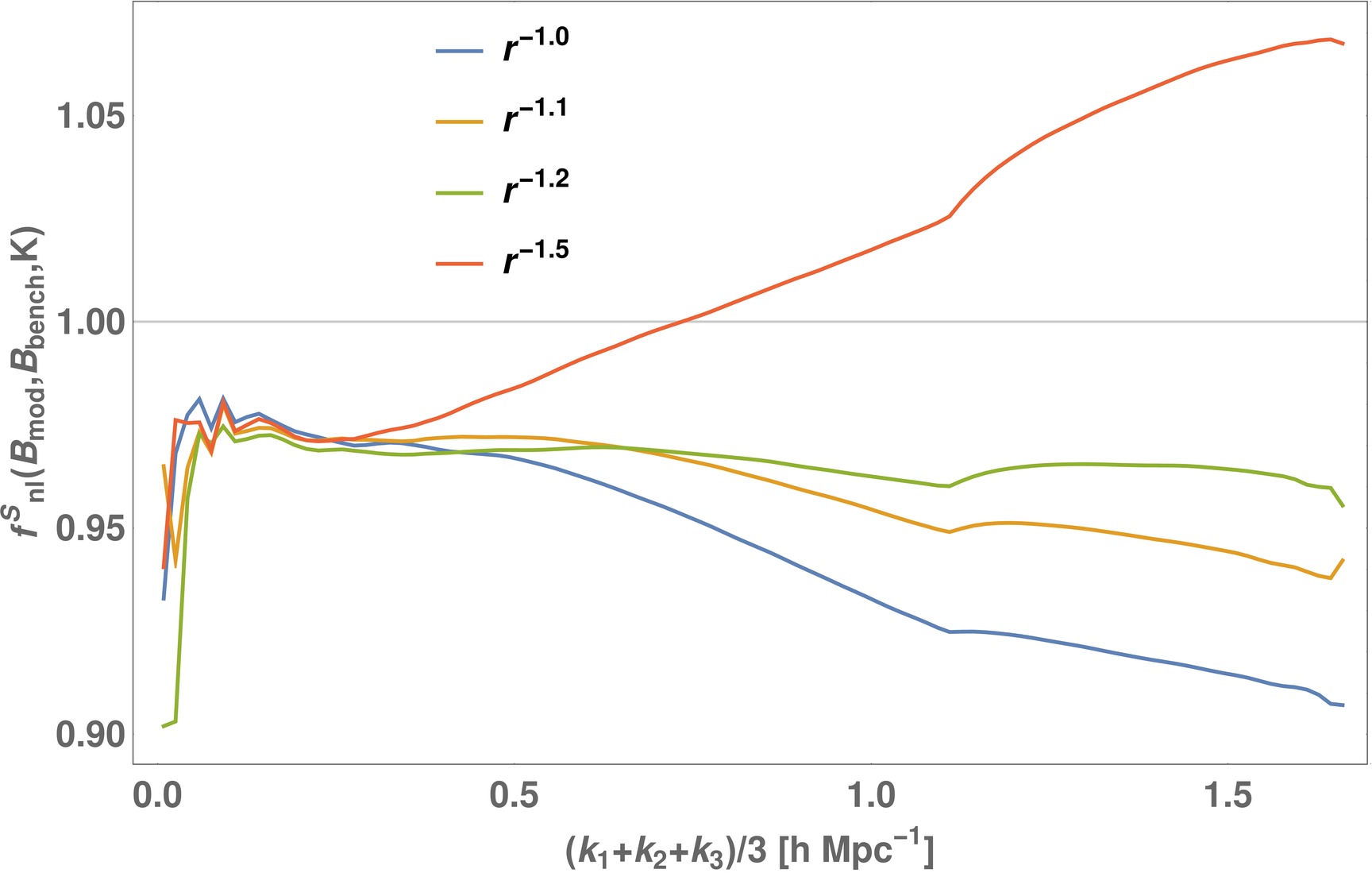}
    \end{subfigure}
    \caption{Fine-tuning the halo profile for the joint lognormal-Gaussian
      model to dampen the high-$k$ tail.
    }
    \label{fig:stats_lognormal_norm}
  \end{center}
\end{figure*}

\begin{figure*}
  \begin{center}
    \begin{subfigure}[b]{0.49\textwidth}
      \includegraphics[width=\linewidth]{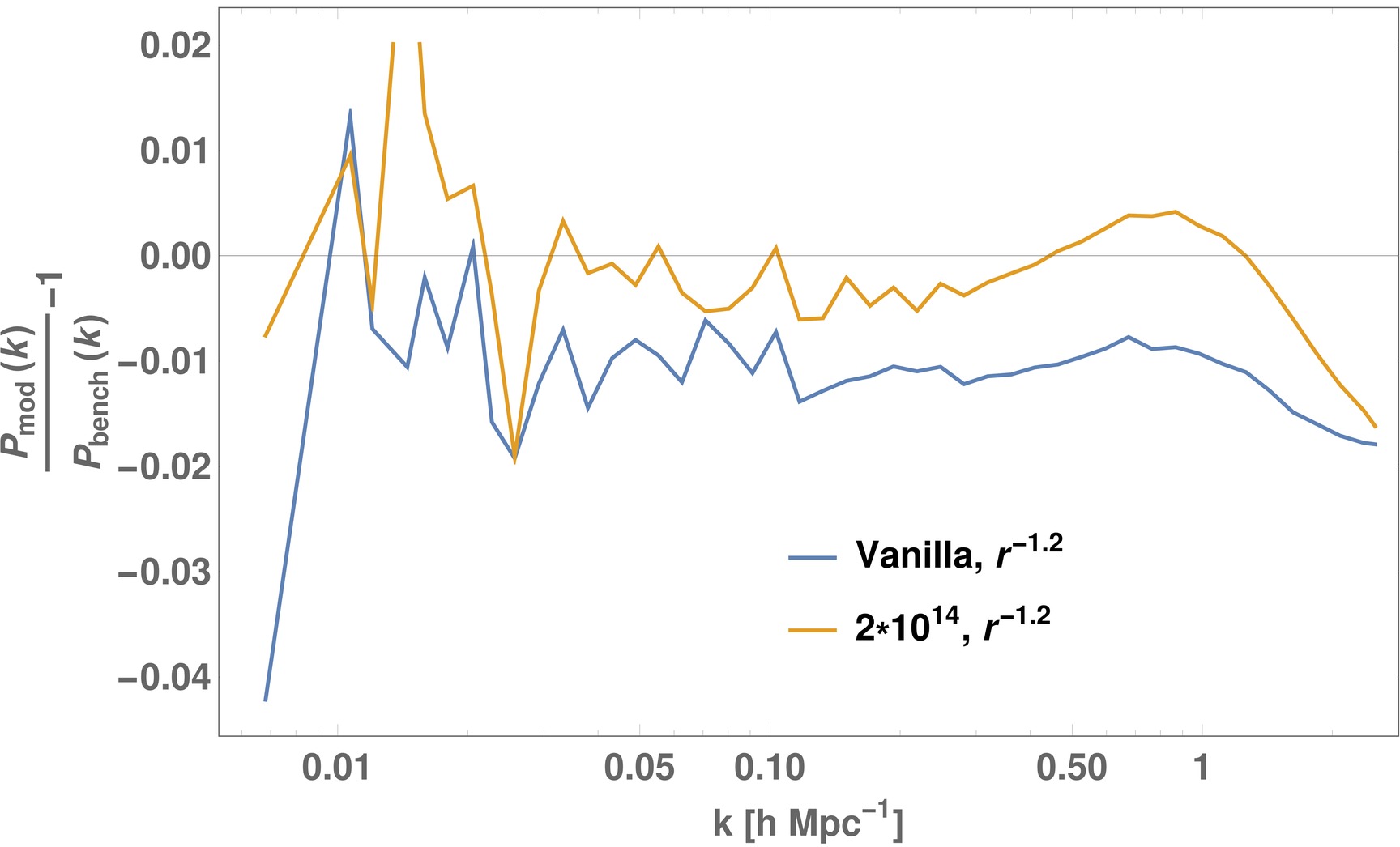}
    \end{subfigure}
    ~
    \begin{subfigure}[b]{0.455\textwidth}
      \includegraphics[width=\linewidth]{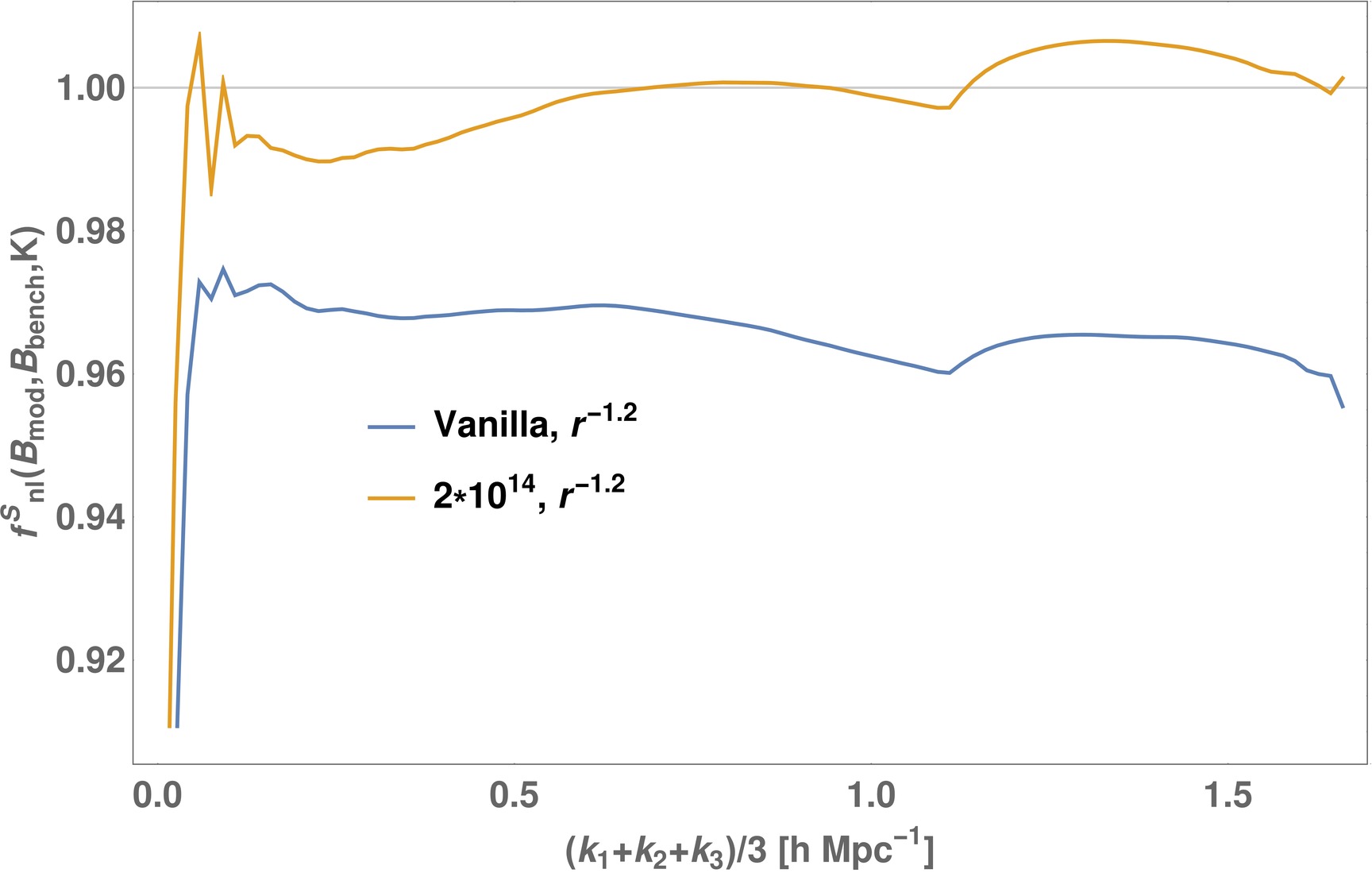}
    \end{subfigure}
    \caption{Improvements to joint lognormal-Gaussian assembly bias model
      by putting an extra galaxy into high mass halos (see text). The number in the
      labels represent the mass threshold, and 'Vanilla' denotes the
      original joint lognormal-Gaussian model without alterations.
    }
    \label{fig:stats_lognormal2}
  \end{center}
\end{figure*}

In summary, we can now implement our assembly bias model using the joint probability distribution $P(N_g|M,c)$ using one of four possible methods:
\begin{itemize}
\item[1.] For an individual halo, use the joint lognormal distribution to draw a suitable value for $N_g$ by 
  shifting the Gaussian distribution for $\ln(N_g)$ using the 
  the concentration $c$ given for that halo by \texttt{ROCKSTAR};
\item[2.] Follow the same procedure as in 1 but with the joint lognormal-Gaussian, 
      shifting the Gaussian distribution for $\ln(N_g)$ using the individual halo concentration given by \texttt{ROCKSTAR};
 \item[3.] Use the joint lognormal distribution for $N_g$ and $c$, 
      but draw values at random for  $c$ from the Gaussian
      distribution for $\ln(c)$, thus eliminating the need for
      the simulation to provide this information.
\item[4.] Follow the same procedure as in 3 but with the joint lognormal-Gaussian 
      distribution,  drawing both $c$ and $N_g$ randomly, so the simulation again does not provide information about concentration.  (For methods 3 and 4 we impose a lower bound of 2 for random
draws of $c$, which is lowest value of $c$ calculated by
\texttt{ROCKSTAR}.)
\end{itemize}

\begin{figure*}
  \begin{center}
    \begin{subfigure}[b]{0.49\textwidth}
      \includegraphics[width=\linewidth]{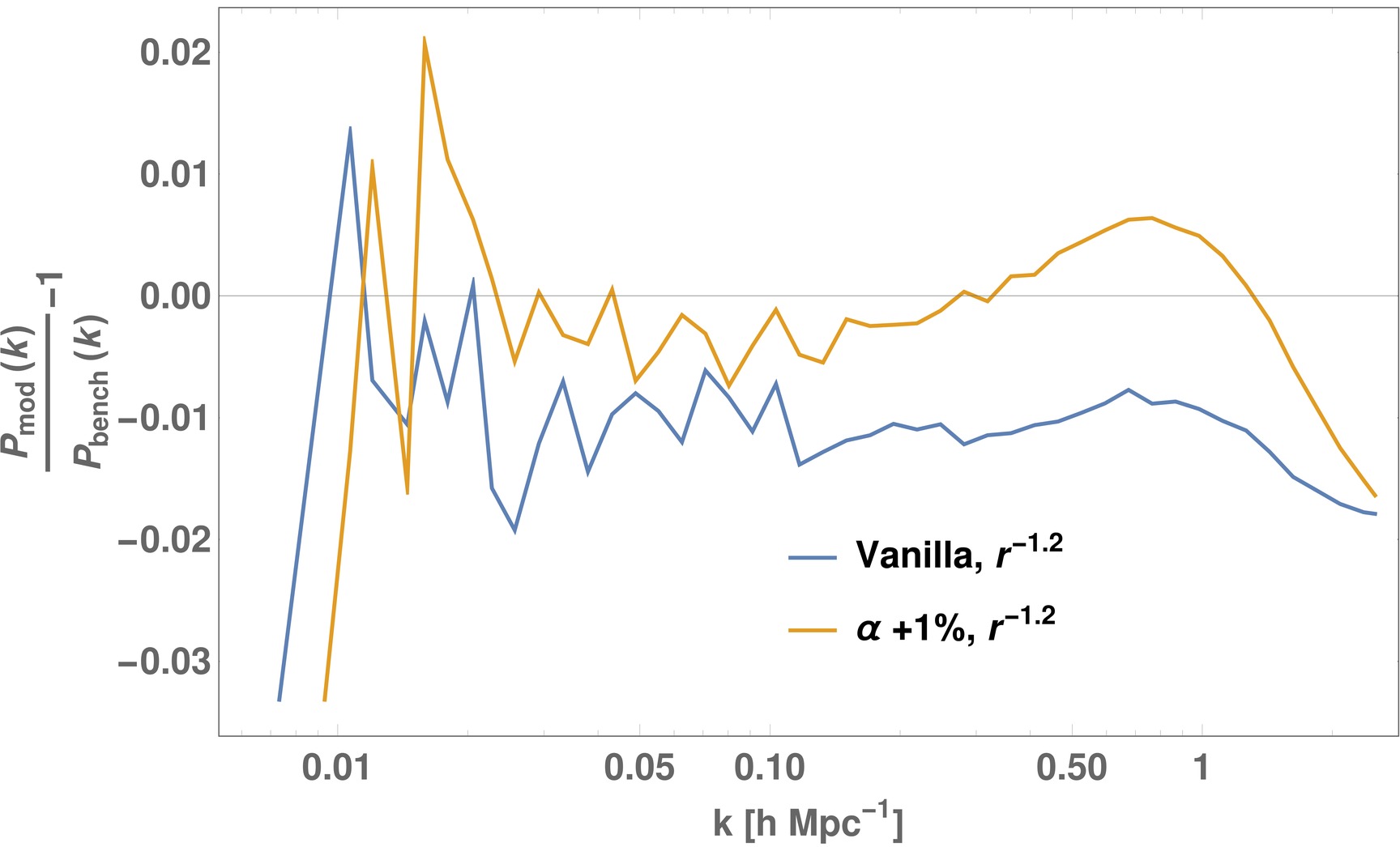}
    \end{subfigure}
    ~
    \begin{subfigure}[b]{0.455\textwidth}
      \includegraphics[width=\linewidth]{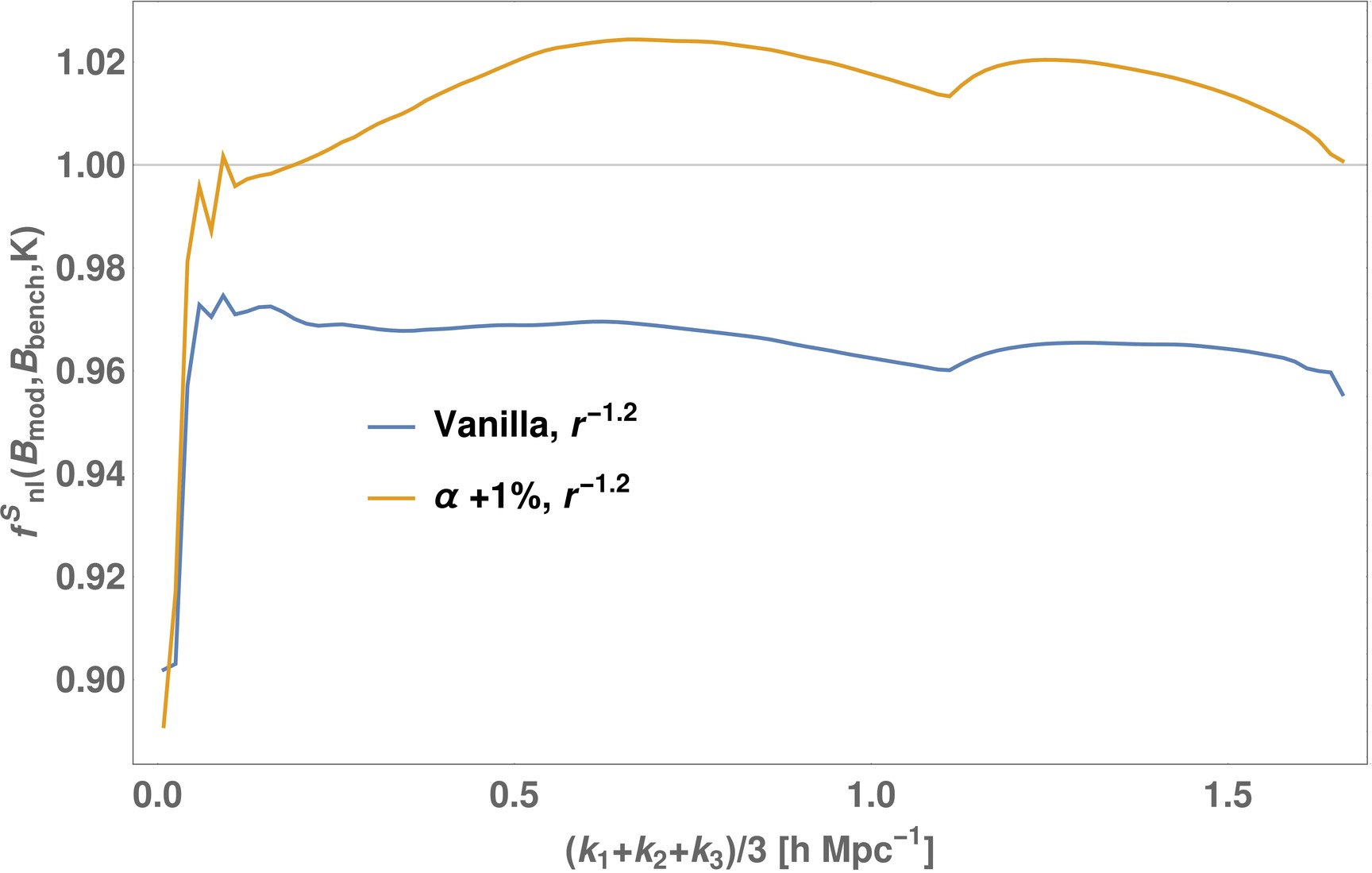}
    \end{subfigure}
    \caption{Improvements to joint lognormal-Gaussian assembly bias model
      by boosting $\alpha$ (see text).
    }
    \label{fig:stats_lognormal2}
  \end{center}
\end{figure*}

\begin{figure*}
  \captionsetup[subfigure]{labelformat=empty}
  \begin{center}
    \begin{subfigure}[b]{0.37\textwidth}
      \includegraphics[width=\linewidth]{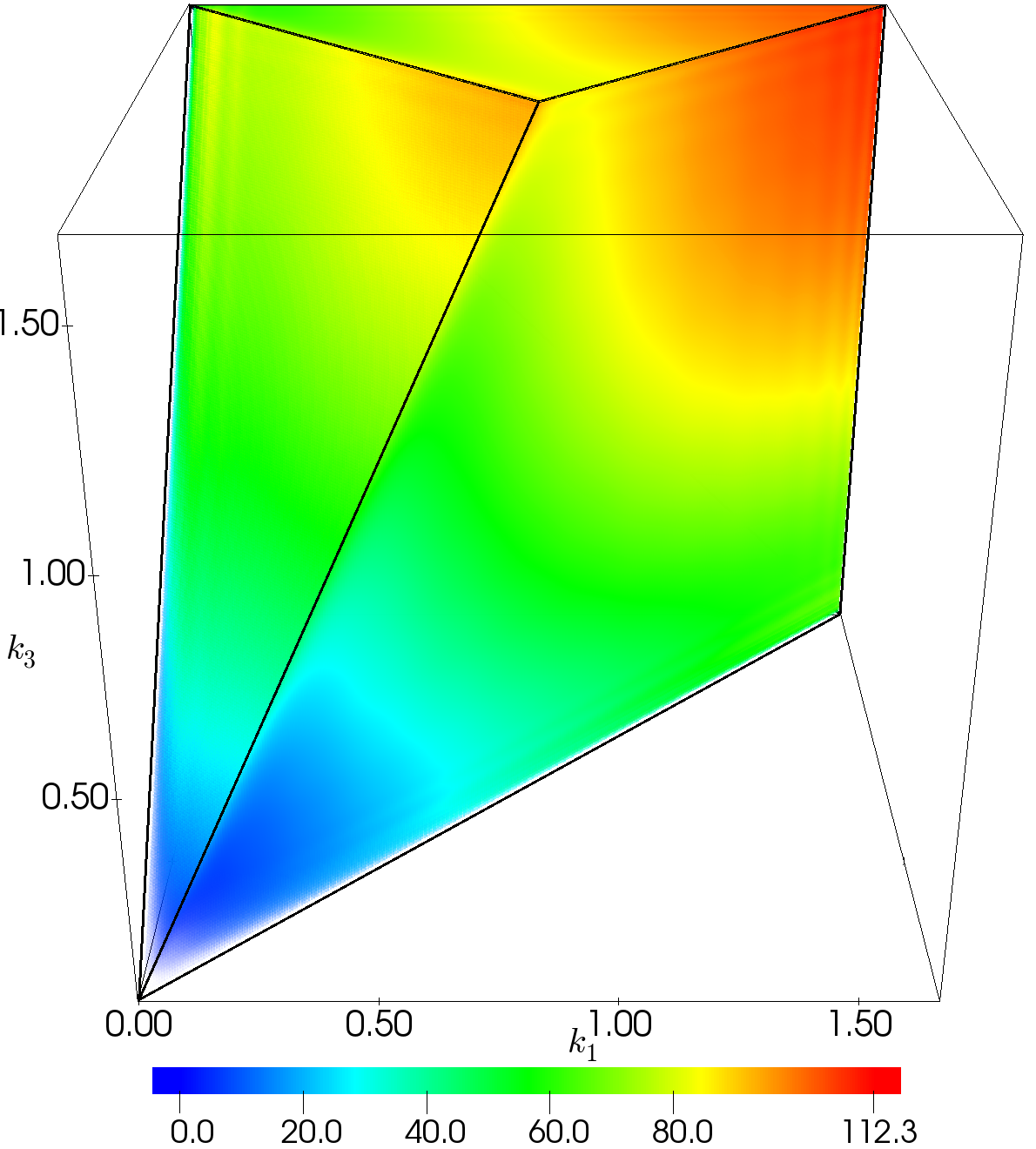}
    \end{subfigure}
    ~
    \begin{subfigure}[b]{0.37\textwidth}
      \includegraphics[width=\linewidth]{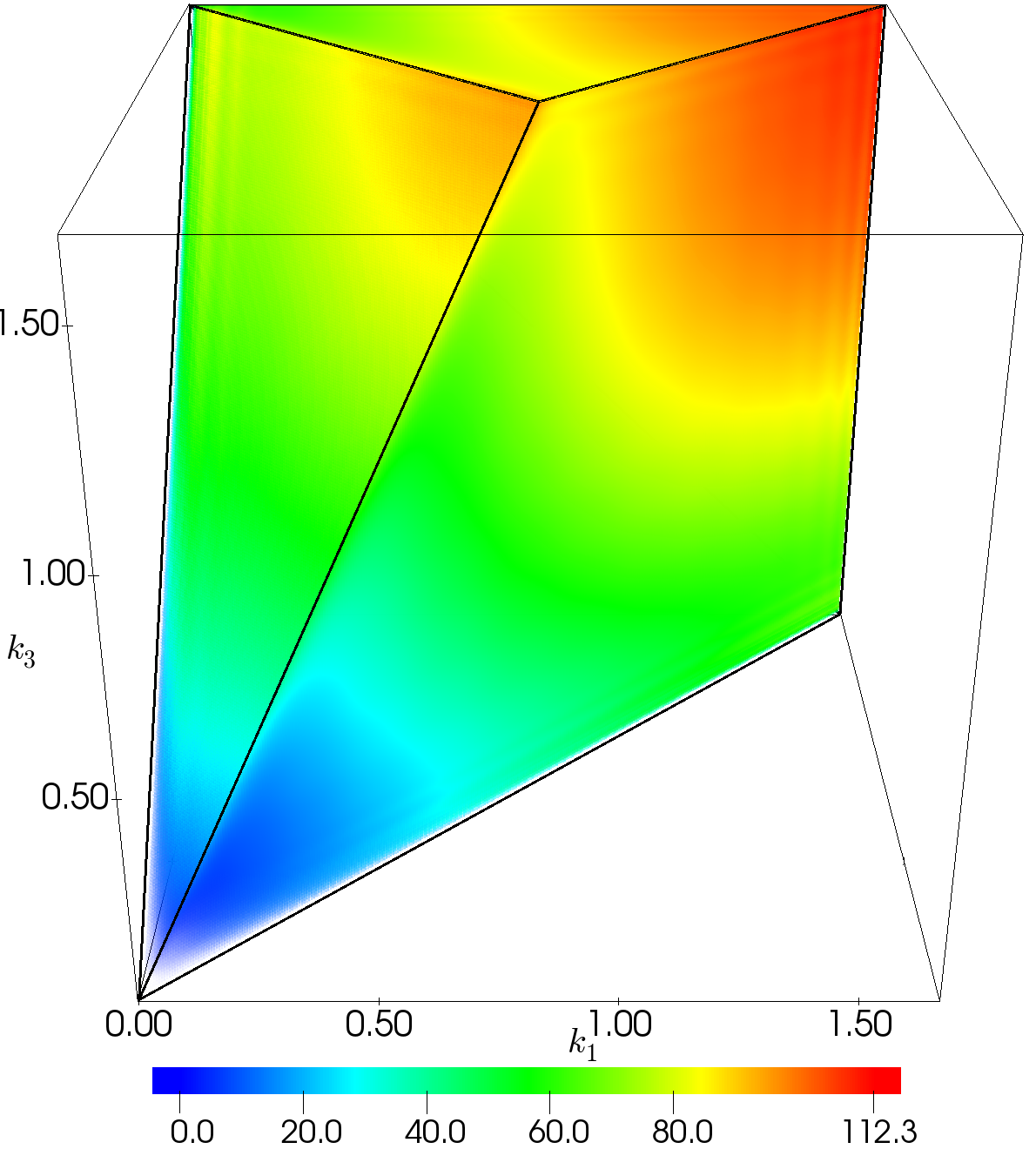}
    \end{subfigure}
    \caption{
      Improvements to joint lognormal-Gaussian assembly bias model.
      The halo profile is radial with $\gamma=1.2$. Left panel:
      Adding an extra galaxy to all parent halos with mass greater
      than $2\times10^{13}\,h^{-1} M_\odot$. Right panel: Boosting
      $\alpha$ by 1\%.
    }
    \label{fig:halo_bis_assembly}
  \end{center}
\end{figure*}

The resulting power spectra and bispectra from these prescriptions for creating mock catalogues are shown in \Cref{fig:stats_lognormal}, with a comparison also to the two-bin concentration model described above. As explained in \Cref{fig:3-shape_bis_slice} the kink at $K/3=1.1\,h\,\text{Mpc}^{-1}$ is due to the geometry of the tetrapyd rather than a physical discontinuation. All these methods, endeavouring to incorporate assembly bias in some form, offer a very substantial improvement over the simplest HOD case shown in \Cref{fig:hod_best_fit}.   Out of these four possibilities,  the superior methods also exploit knowledge of individual halo concentrations given by the \texttt{ROCKSTAR} simulation (which to some extent also includes the simpler two-bin method described earlier). 
The well-motivated joint lognormal-Gaussian modelling of the occupation number
and concentration, with a power law halo profile of $\gamma=1.2$, yields a better than 1\% accuracy in the power spectrum
and 4\% accuracy in the bispectrum for $ k,K/3<1.0\,h\,\text{Mpc}^{-1}$, which is significantly better than methods previously investigated in this paper. Moreover, both its power spectrum and bispectrum are flatter than the joint lognormal-lognormal case which makes it the more suitable model. It is clear that 
some information about the assembly history of halos is certainly helpful when creating mock catalogues targeting an accurate halo bispectrum, as it can be used a proxy for concentration.   Information about the merger history of halos can be 
obtained by fast simulation methods without resorting to $N$-body simulations (see, for example, PINOCCHIO \citep{pinocchio}). A number of methods have been developed to correlate halo concentration with halo mass and redshift \citep{merger_history_1,merger_history_2,merger_history_3}, and furthermore the authors of \citep{merger_history_4} have shown that these models, combined with an empirical model of environmental effects on halo formation times, gives the correct mean concentration and scatter as a function of halo mass.

As can be seen from \Cref{fig:stats_lognormal}, there is still some room for improvement to obtain high precision mock 
power spectra and bispectra to match the benchmark results.   In
\Cref{fig:stats_lognormal_norm} we explore changes to the halo
profile for the joint lognormal-Gaussian model to curtail the
excess power at small scales. It is clear that a value of
$\gamma=1.2$, which is in the range of best fit values shown in
\Cref{fig:gamma}, gives both a flat relative power spectrum and bispectrum.   We also studied methods by which we might be able to generically boost the  power spectrum and
bispectrum across all scales, notable large length scales. From our investigations of different mass halos, we found that the high mass halos dominate the power at large scales, due to their high occupation number. One way to boost the power is
therefore to add an extra galaxy to every parent halo above a
certain mass threshold. We tested this tweak using the joint
lognormal-Gaussian model, the results of which are shown in
\Cref{fig:stats_lognormal2}. We found that
$M=2\times10^{14}\,h^{-1} M_\odot$ seems to be the appropriate mass
threshold, which coupled with a radial profile of $r^{-1.2}$ allows us to 
obtain a fit to both the power spectrum and bispectrum to 1\%
accuracy between $0.04\,h\,\text{Mpc}^{-1}<k<1.1\,h\,\text{Mpc}^{-1}$.
The average occupation number at this mass threshold is about 11,
therefore this boost is at the 10\% level in magnitude. A more natural, continuous transition, such as the
$\erfc$ function used in the 5-parameter HOD model, can be adopted
instead of a step function to obtain smoother behaviour.   This may seem a rather contrived way to boost
power, but it is presumably compensating some missing physical correlation (such as triaxiality etc.).

Another means by which to achieve a power boost is to raise the 
the power law exponent index $\alpha$ in the marginalised HOD (\Cref{eq:simpleHODfit}), as we did
with the 4-parameter model. Instead of using the analytical
form as we did previously, we change the occupation number
drawn from the joint distribution by scaling the number of
satellites by this factor:
\begin{align}
  \label{eq:log_alpha_boost}
  \left(\frac{M-\kappa' M_0}{M_1}\right)^{\alpha'}/\left(\frac{M-\kappa M_0}{M_1}\right)^\alpha,
\end{align}
as we boost both $\alpha$ and $\kappa$ to conserve particle
number. We use the best fit parameters for $\alpha$, $\kappa$,
$M_0$ and $M_1$, and the results are shown in
\Cref{fig:stats_lognormal2}. The power spectrum results are
comparably to the extra galaxy method above, but this has the additional property of over-boosting the bispectrum,  as we have observed in the
4-parameter HOD case. Finally, we show the 3D bispectrum tetrapyd of these
improved models in \Cref{fig:halo_bis_assembly} which are qualitatively indistinguishable from the bispectrum obtained directly from the benchmark halo distribution. 

\section{Conclusions \label{sec:conclusions}}

In this paper we have applied the fast bispectrum estimator \MODALLSS{} \citep{DM} to accurately measure the bispectrum from a large mock galaxy catalogue.  This catalogue was generated from a \GADGET{} $N$-body simulation using the \texttt{ROCKSTAR} halo-finder.   We have provided a quantitative three-shape fit to the resulting halo bispectrum, comparing it with the corresponding bispectrum of the underlying dark matter, studied previously \citep{Andrei}.    A key goal has been to determine phenomenological
methods to create fast mock catalogues that can reproduce the benchmark halo bispectrum from 
\texttt{ROCKSTAR}. In doing so we have restricted ourselves to 
using only the mass, position and concentration information for parent halos,  relying on
statistical modelling of the halo profile and
occupation number to recover the benchmark power
spectrum and bispectrum. We modelled these effects in configuration space to obtain
accurate mock power spectra and bispectra, and we aim to incorporate further observational
effects such as RSDs in a future paper.

\subsection{Halo profile}

An important ingredient in a phenomenological galaxy
catalogue is the spatial distribution of the galaxies within
a parent halo. The subhalo radial number density 
found for parent halos (separated into a number of mass bins) was not 
well matched by the average NFW dark matter profile found in the same halo mass range. 
On the other hand, as suggested, e.g., by \citep{subgen} a
power law profile of the form $\rho\propto r^{-\gamma}$, with
$\gamma\sim1$, works well as a universal profile across a
wide range of halo masses spanning three orders of magnitude. 

By randomising the solid angular distribution of the benchmark
halos we have also quantified the power loss in the power
spectrum and bispectrum if halo substructure and triaxiality
are not preserved, that is, by fixing the original subhalo number and 
then retaining radial distances while randomising angular positions.  
The effect of this internal redistribution was modest with deviations less than 1\% and 4\% at
$k,K/3=1\,h\,\text{Mpc}^{-1}$ for the power spectrum and
bispectrum correlator respectively. These lost correlations mean that the
best fit power law profile near $\gamma\approx1$ is necessarily power deficient
at small scales.  However, we have found that phenomenological
values around $\gamma \approx1.5$ apparently help to recover this power loss to less
than 0.5\% up to $k,K/3=2\,h\,\text{Mpc}^{-1}$ in both power spectrum and bispectrum.
Note that these profile modifications are constrained by using the original occupation numbers for individual halos, which is information generally only available from costly nonlinear simulations. 

\subsection{Halo occupation distribution}

To statistically model the number of galaxies within a parent halo,
we have investigated the popular practice of using an HOD that only depends on halo mass, $\bar{N}_g(M)$.
We observed that using the measured mean number of galaxies
for a halo of a given mass $M$ to repopulate the parent halos leads
to a power deficit of about 2\% in the power spectrum at large
scales where $k < 0.1\,h\,\text{Mpc}^{-1}$, and greater
differences at smaller length scales. The loss of power in the
bispectrum is more pronounced with much poorer scaling, yielding deviations 
exceeding 10\% by $k = 0.5\,h\,\text{Mpc}^{-1}$.  We found
that the same effect can be reproduced if one shuffles the given
halo occupation numbers within the mass bins (or by using a dispersion around 
the mean HOD value).  Clearly this simple HOD prescription for populating halos destroys important correlations, so it suggests that other 
physical mechanisms are contributing to the number of galaxies per
halo, rather than just the halo mass.  

Nevertheless, we have attempted to recover this power loss by tuning the
four-parameter HOD model given by (\Cref{eq:constraint}). The best fit parameters actually lead
to further power loss at all scales, perhaps because the HOD fit is only accurate up to 10\%,
which suggests that a better functional form should be adopted to match HODs
from simulations. After tweaking the parameters while keeping
galaxy number constant we found that boosting the power law exponent $\alpha$ by
4.5\% raised the power spectrum to the correct level to 
$k = 0.5\,h\,\text{Mpc}^{-1}$ irrespective of choice in the other
parameter, but unfortunately this results in substantially over-boosting
the bispectrum (overcompensating at around the 10\% level). 
We infer that an HOD model which
only depends on halo mass cannot accurately reproduce both the power
spectrum and bispectrum of a benchmark mock catalogue.

\subsection{Assembly bias}

These investigations led us to incorporate
further information in the HOD that takes into account the formation
history of the halos to determine the halo occupation number.
Motivated by other assembly bias studies in the literature such as
\citep{assembly_bias,Gao2005,Sunayama2016,Hearin2016,Wechsler2006}, we have 
developed a new prescription using a joint probability distribution to model 
correlations between the halo occupation number $N_g$ and concentration $c$ found in the 
benchmark catalogue.   Even an extension which just separates halos of a given mass into two concentration bins \citep{eisenstein2} - representing above and below median values for $c$ - yields more accurate power spectra and bispectra with improved scaling.  

We have found that the marginalised distribution for halo
concentration is well described by a Gaussian distribution
across the entire mass range of the benchmark, while
taking care to impose an appropriate lower bound when drawing
from the distribution. On the other hand the
marginalised halo occupation number is well fitted with a
lognormal distribution. Our assembly bias model is therefore
a joint lognormal-Gaussian bivariate distribution which
depends on halo mass, $P(N_g\cap c\,|M)$. A non-zero covariance
between the two variables imply that halo concentration is
correlated with halo occupation number, and we find the
correlation coefficient is $r\approx-0.5$ for a mass range of
$M=10^{13}-10^{14}h^{-1}\,M_{\odot}$.   In terms of the assembly history within an $N$-body simulation, 
we can interpret higher halo concentration causing fewer subhalos because of 
earlier halo formation, that is, in this case there is more time for the merger of substructure
(a factor which depends to some extent on our benchmark resolution).  
We were also able to obtain very similar results using a 
joint lognormal-lognormal distribution for the  halo number and concentration. 

\subsection{Prescriptions for fast mock catalogue polyspectra}

One of the key results of our paper is that our assembly bias model 
for populating halos can recover the benchmark
power spectrum to within 1\% and the bispectrum to within 4\% across the entire
range of scales of the simulation.  In its most accurate form this involves using a joint lognormal-Gaussian probability distribution for $N_g$ and $c$, coupled with a radial power law halo profile with $\gamma=1.2$, together with the concentrations found for individual halos.  Without use of individual halo concentrations, we could assign both concentration and halo number statistically, obtaining good bispectrum scaling though with a 2\% and 5\% deficit emerging for the power spectrum and bispectrum respectively.  These assembly bias prescriptions represent a considerable improvement
over all the other methods we investigated in this paper and can be deployed with fast mock catalogue generators. 

We also explored ways to phenomenologically reduce this small remaining power
deficit. Modifying the index in the four-parameter HOD model, as before, encountered the problem of
over-boosting the bispectrum.   However, motivated by the dominant contributions of high mass halos, we considered enhancing this by adding an extra galaxy to all parent halos above a certain mass threshold $M>2\times10^{14}\,h^{-1} M_\odot$.  We were able to obtain a 1\% fit to both the benchmark power spectrum and bispectrum in the range
$0.04\,h\,\text{Mpc}^{-1}<k<1.1\,h\,\text{Mpc}^{-1}$.     

Finally we note a few caveats about the mock catalogue population methods we have proposed. Our assumption that galaxies can be identified with subhalos will have an important impact on both the spatial
distribution and occupation number of the parent halos; clearly this approach can be developed further and made more realistic by increasing resolution and incorporating more physical mechanisms in the simulations.   For example, our present mass resolution with a particle mass of $M_p=2.093\times10^{10}$ may be insufficient to ensure finer substructures are resolved and preserved during halo mergers; it would be prudent in future to
expand these investigations by exploring the dependence on simulation resolution.  We also note that our most accurate assembly bias model relies on concentration information for individual halos obtained from the mock catalogue simulation. This is not necessarily available from all fast simulation generators and halo finder codes, but algorithms such as PINOCCHIO can provide the merger history of dark matter halos, which in turn could be converted into halo concentrations. Nevertheless, by statistically sampling the Gaussian distribution for concentration we were still able to obtain a good power spectrum and bispectrum fit, and this model can be further fine-tuned with the galaxy boost.

In summary, motivated by assembly bias, we have developed a statistical prescription for populating parent halos with subhalos which can simultaneously reproduce both the halo power spectrum and bispectrum obtained from nonlinear $N$-body simulations.   We anticipate that this robust approach can be adapted to match polyspectra obtained from more sophisticated $N$-body and hydrodynamic simulations.  Combining this relatively simple methodology with fast estimators like \MODALLSS{} \citep{DM} should enable the bispectrum to become a key diagnostic tool, both for breaking degeneracies in cosmological parameter estimation and for quantitatively analysing gravitational collapse and other physical effects on highly nonlinear length scales. 

\section{Acknowledgements}
\label{sec:acknowledgements}

Many thanks to Tobias Baldauf and James Fergusson for many
enlightening conversations, and to Oliver Friedrich and Cora
Uhlemann for comments on the manuscript. Kacper Kornet provided invaluable
technical support for which we are very grateful. 

This work was undertaken on the COSMOS Shared Memory system at
DAMTP, University of Cambridge operated on behalf of the STFC
DiRAC HPC Facility. This equipment is funded by BIS National
E-infrastructure capital grant ST/J005673/1 and STFC grants
ST/H008586/1, ST/K00333X/1.

This work used the COSMA Data Centric system at Durham University, 
operated by the Institute for Computational Cosmology on behalf of 
the STFC DiRAC HPC Facility (www.dirac.ac.uk. This equipment was 
funded by a BIS National E-infrastructure capital grant ST/K00042X/1, 
DiRAC Operations grant ST/K003267/1 and Durham University. DiRAC is 
part of the National E-Infrastructure.

MM acknowledges support from the European Union’s Horizon 2020
research and innovation program under Marie Sklodowska-Curie grant
agreement No 6655919.

\bibliographystyle{apsrev4-1}
\bibliography{main}{}

\begin{thebibliography}{111}%
\makeatletter
\providecommand \@ifxundefined [1]{%
 \@ifx{#1\undefined}
}%
\providecommand \@ifnum [1]{%
 \ifnum #1\expandafter \@firstoftwo
 \else \expandafter \@secondoftwo
 \fi
}%
\providecommand \@ifx [1]{%
 \ifx #1\expandafter \@firstoftwo
 \else \expandafter \@secondoftwo
 \fi
}%
\providecommand \natexlab [1]{#1}%
\providecommand \enquote  [1]{``#1''}%
\providecommand \bibnamefont  [1]{#1}%
\providecommand \bibfnamefont [1]{#1}%
\providecommand \citenamefont [1]{#1}%
\providecommand \href@noop [0]{\@secondoftwo}%
\providecommand \href [0]{\begingroup \@sanitize@url \@href}%
\providecommand \@href[1]{\@@startlink{#1}\@@href}%
\providecommand \@@href[1]{\endgroup#1\@@endlink}%
\providecommand \@sanitize@url [0]{\catcode `\\12\catcode `\$12\catcode
  `\&12\catcode `\#12\catcode `\^12\catcode `\_12\catcode `\%12\relax}%
\providecommand \@@startlink[1]{}%
\providecommand \@@endlink[0]{}%
\providecommand \url  [0]{\begingroup\@sanitize@url \@url }%
\providecommand \@url [1]{\endgroup\@href {#1}{\urlprefix }}%
\providecommand \urlprefix  [0]{URL }%
\providecommand \Eprint [0]{\href }%
\providecommand \doibase [0]{http://dx.doi.org/}%
\providecommand \selectlanguage [0]{\@gobble}%
\providecommand \bibinfo  [0]{\@secondoftwo}%
\providecommand \bibfield  [0]{\@secondoftwo}%
\providecommand \translation [1]{[#1]}%
\providecommand \BibitemOpen [0]{}%
\providecommand \bibitemStop [0]{}%
\providecommand \bibitemNoStop [0]{.\EOS\space}%
\providecommand \EOS [0]{\spacefactor3000\relax}%
\providecommand \BibitemShut  [1]{\csname bibitem#1\endcsname}%
\let\auto@bib@innerbib\@empty
\bibitem [{\citenamefont {{The Dark Energy Survey
  Collaboration}}(2005)}]{DES1}%
  \BibitemOpen
  \bibfield  {author} {\bibinfo {author} {\bibnamefont {{The Dark Energy Survey
  Collaboration}}},\ }\href@noop {} {\enquote {\bibinfo {title} {{The Dark
  Energy Survey}},}\ } (\bibinfo {year} {2005}),\ \Eprint
  {http://arxiv.org/abs/arXiv:astro-ph/0510346} {arXiv:astro-ph/0510346}
  \BibitemShut {NoStop}%
\bibitem [{\citenamefont {Diehl}\ \emph {et~al.}(2014)\citenamefont {Diehl}
  \emph {et~al.}}]{DES2}%
  \BibitemOpen
  \bibfield  {author} {\bibinfo {author} {\bibfnamefont {H.~T.}\ \bibnamefont
  {Diehl}} \emph {et~al.},\ }\href {\doibase 10.1117/12.2056982} {\bibfield
  {journal} {\bibinfo  {journal} {Proc.SPIE}\ }\textbf {\bibinfo {volume}
  {9149}} (\bibinfo {year} {2014}),\ 10.1117/12.2056982}\BibitemShut {NoStop}%
\bibitem [{\citenamefont {Ivezic}\ \emph {et~al.}(2008)\citenamefont {Ivezic}
  \emph {et~al.}}]{LSST}%
  \BibitemOpen
  \bibfield  {author} {\bibinfo {author} {\bibfnamefont {Z.}~\bibnamefont
  {Ivezic}} \emph {et~al.},\ }\href@noop {} {\enquote {\bibinfo {title} {{LSST:
  from Science Drivers to Reference Design and Anticipated Data Products}},}\ }
  (\bibinfo {year} {2008}),\ \Eprint {http://arxiv.org/abs/arXiv:0805.2366}
  {arXiv:0805.2366} \BibitemShut {NoStop}%
\bibitem [{\citenamefont {Laureijs}\ \emph {et~al.}(2011)\citenamefont
  {Laureijs} \emph {et~al.}}]{Euclid}%
  \BibitemOpen
  \bibfield  {author} {\bibinfo {author} {\bibfnamefont {R.}~\bibnamefont
  {Laureijs}} \emph {et~al.},\ }\href@noop {} {\enquote {\bibinfo {title}
  {{Euclid Definition Study Report}},}\ } (\bibinfo {year} {2011}),\ \Eprint
  {http://arxiv.org/abs/arXiv:1110.3193} {arXiv:1110.3193} \BibitemShut
  {NoStop}%
\bibitem [{\citenamefont {Brenna~Flaugher}(2014)}]{DESI}%
  \BibitemOpen
  \bibfield  {author} {\bibinfo {author} {\bibfnamefont {C.~B.}\ \bibnamefont
  {Brenna~Flaugher}},\ }\href {\doibase 10.1117/12.2057105} {\bibfield
  {journal} {\bibinfo  {journal} {Proc.SPIE}\ }\textbf {\bibinfo {volume}
  {9147}} (\bibinfo {year} {2014}),\ 10.1117/12.2057105}\BibitemShut {NoStop}%
\bibitem [{\citenamefont {Karagiannis}\ \emph {et~al.}(2018)\citenamefont
  {Karagiannis}, \citenamefont {Lazanu}, \citenamefont {Liguori}, \citenamefont
  {Raccanelli}, \citenamefont {Bartolo},\ and\ \citenamefont
  {Verde}}]{forecast}%
  \BibitemOpen
  \bibfield  {author} {\bibinfo {author} {\bibfnamefont {D.}~\bibnamefont
  {Karagiannis}}, \bibinfo {author} {\bibfnamefont {A.}~\bibnamefont {Lazanu}},
  \bibinfo {author} {\bibfnamefont {M.}~\bibnamefont {Liguori}}, \bibinfo
  {author} {\bibfnamefont {A.}~\bibnamefont {Raccanelli}}, \bibinfo {author}
  {\bibfnamefont {N.}~\bibnamefont {Bartolo}}, \ and\ \bibinfo {author}
  {\bibfnamefont {L.}~\bibnamefont {Verde}},\ }\href@noop {} {\enquote
  {\bibinfo {title} {Constraining primordial non-gaussianity with bispectrum
  and power spectum from upcoming optical and radio surveys},}\ } (\bibinfo
  {year} {2018}),\ \Eprint {http://arxiv.org/abs/arXiv:1801.09280}
  {arXiv:1801.09280} \BibitemShut {NoStop}%
\bibitem [{\citenamefont {{Planck Collaboration}}(2014)}]{bias}%
  \BibitemOpen
  \bibfield  {author} {\bibinfo {author} {\bibnamefont {{Planck
  Collaboration}}},\ }\href {\doibase 10.1051/0004-6361/201321521} {\bibfield
  {journal} {\bibinfo  {journal} {A\&A}\ }\textbf {\bibinfo {volume} {571}},\
  \bibinfo {pages} {A20} (\bibinfo {year} {2014})}\BibitemShut {NoStop}%
\bibitem [{\citenamefont {Hung}\ \emph {et~al.}(2019)\citenamefont {Hung},
  \citenamefont {Fergusson},\ and\ \citenamefont {Shellard}}]{DM}%
  \BibitemOpen
  \bibfield  {author} {\bibinfo {author} {\bibfnamefont {J.}~\bibnamefont
  {Hung}}, \bibinfo {author} {\bibfnamefont {J.~R.}\ \bibnamefont {Fergusson}},
  \ and\ \bibinfo {author} {\bibfnamefont {E.~P.~S.}\ \bibnamefont
  {Shellard}},\ }\href@noop {} {\enquote {\bibinfo {title} {Advancing the
  matter bispectrum estimation of large-scale structure: a comparison of dark
  matter codes},}\ } (\bibinfo {year} {2019}),\ \Eprint
  {http://arxiv.org/abs/arXiv:1902.01830} {arXiv:1902.01830} \BibitemShut
  {NoStop}%
\bibitem [{\citenamefont {Wagner}\ \emph {et~al.}(2010)\citenamefont {Wagner},
  \citenamefont {Verde},\ and\ \citenamefont {Boubekeur}}]{wagner}%
  \BibitemOpen
  \bibfield  {author} {\bibinfo {author} {\bibfnamefont {C.}~\bibnamefont
  {Wagner}}, \bibinfo {author} {\bibfnamefont {L.}~\bibnamefont {Verde}}, \
  and\ \bibinfo {author} {\bibfnamefont {L.}~\bibnamefont {Boubekeur}},\ }\href
  {\doibase 10.1088/1475-7516/2010/10/022} {\bibfield  {journal} {\bibinfo
  {journal} {Journal of Cosmology and Astroparticle Physics}\ }\textbf
  {\bibinfo {volume} {2010}},\ \bibinfo {pages} {022} (\bibinfo {year}
  {2010})}\BibitemShut {NoStop}%
\bibitem [{\citenamefont {Regan}\ \emph {et~al.}(2012)\citenamefont {Regan},
  \citenamefont {Schmittfull}, \citenamefont {Shellard},\ and\ \citenamefont
  {Fergusson}}]{regan}%
  \BibitemOpen
  \bibfield  {author} {\bibinfo {author} {\bibfnamefont {D.~M.}\ \bibnamefont
  {Regan}}, \bibinfo {author} {\bibfnamefont {M.~M.}\ \bibnamefont
  {Schmittfull}}, \bibinfo {author} {\bibfnamefont {E.~P.~S.}\ \bibnamefont
  {Shellard}}, \ and\ \bibinfo {author} {\bibfnamefont {J.~R.}\ \bibnamefont
  {Fergusson}},\ }\href {\doibase 10.1103/PhysRevD.86.123524} {\bibfield
  {journal} {\bibinfo  {journal} {Phys. Rev. D}\ }\textbf {\bibinfo {volume}
  {86}},\ \bibinfo {pages} {123524} (\bibinfo {year} {2012})}\BibitemShut
  {NoStop}%
\bibitem [{\citenamefont {Gil-Marín}\ \emph {et~al.}(2012)\citenamefont
  {Gil-Marín}, \citenamefont {Wagner}, \citenamefont {Fragkoudi},
  \citenamefont {Jimenez},\ and\ \citenamefont {Verde}}]{9param}%
  \BibitemOpen
  \bibfield  {author} {\bibinfo {author} {\bibfnamefont {H.}~\bibnamefont
  {Gil-Marín}}, \bibinfo {author} {\bibfnamefont {C.}~\bibnamefont {Wagner}},
  \bibinfo {author} {\bibfnamefont {F.}~\bibnamefont {Fragkoudi}}, \bibinfo
  {author} {\bibfnamefont {R.}~\bibnamefont {Jimenez}}, \ and\ \bibinfo
  {author} {\bibfnamefont {L.}~\bibnamefont {Verde}},\ }\href@noop {}
  {\bibfield  {journal} {\bibinfo  {journal} {Journal of Cosmology and
  Astroparticle Physics}\ }\textbf {\bibinfo {volume} {2012}},\ \bibinfo
  {pages} {047} (\bibinfo {year} {2012})}\BibitemShut {NoStop}%
\bibitem [{\citenamefont {Schmittfull}\ \emph {et~al.}(2013)\citenamefont
  {Schmittfull}, \citenamefont {Regan},\ and\ \citenamefont
  {Shellard}}]{glass}%
  \BibitemOpen
  \bibfield  {author} {\bibinfo {author} {\bibfnamefont {M.~M.}\ \bibnamefont
  {Schmittfull}}, \bibinfo {author} {\bibfnamefont {D.~M.}\ \bibnamefont
  {Regan}}, \ and\ \bibinfo {author} {\bibfnamefont {E.~P.~S.}\ \bibnamefont
  {Shellard}},\ }\href {\doibase 10.1103/PhysRevD.88.063512} {\bibfield
  {journal} {\bibinfo  {journal} {Phys. Rev. D}\ }\textbf {\bibinfo {volume}
  {88}},\ \bibinfo {pages} {063512} (\bibinfo {year} {2013})}\BibitemShut
  {NoStop}%
\bibitem [{\citenamefont {Lazanu}\ \emph {et~al.}(2016)\citenamefont {Lazanu},
  \citenamefont {Giannantonio}, \citenamefont {Schmittfull},\ and\
  \citenamefont {Shellard}}]{Andrei}%
  \BibitemOpen
  \bibfield  {author} {\bibinfo {author} {\bibfnamefont {A.}~\bibnamefont
  {Lazanu}}, \bibinfo {author} {\bibfnamefont {T.}~\bibnamefont
  {Giannantonio}}, \bibinfo {author} {\bibfnamefont {M.}~\bibnamefont
  {Schmittfull}}, \ and\ \bibinfo {author} {\bibfnamefont {E.~P.~S.}\
  \bibnamefont {Shellard}},\ }\href {\doibase 10.1103/PhysRevD.93.083517}
  {\bibfield  {journal} {\bibinfo  {journal} {Phys. Rev. D}\ }\textbf {\bibinfo
  {volume} {93}},\ \bibinfo {pages} {083517} (\bibinfo {year}
  {2016})}\BibitemShut {NoStop}%
\bibitem [{\citenamefont {Eisenstein}\ \emph {et~al.}(2011)\citenamefont
  {Eisenstein} \emph {et~al.}}]{BOSS1}%
  \BibitemOpen
  \bibfield  {author} {\bibinfo {author} {\bibfnamefont {D.~J.}\ \bibnamefont
  {Eisenstein}} \emph {et~al.},\ }\href@noop {} {\bibfield  {journal} {\bibinfo
   {journal} {The Astronomical Journal}\ }\textbf {\bibinfo {volume} {142}},\
  \bibinfo {pages} {72} (\bibinfo {year} {2011})}\BibitemShut {NoStop}%
\bibitem [{\citenamefont {Dawson}\ \emph {et~al.}(2013)\citenamefont {Dawson}
  \emph {et~al.}}]{BOSS2}%
  \BibitemOpen
  \bibfield  {author} {\bibinfo {author} {\bibfnamefont {K.~S.}\ \bibnamefont
  {Dawson}} \emph {et~al.},\ }\href@noop {} {\bibfield  {journal} {\bibinfo
  {journal} {The Astronomical Journal}\ }\textbf {\bibinfo {volume} {145}},\
  \bibinfo {pages} {10} (\bibinfo {year} {2013})}\BibitemShut {NoStop}%
\bibitem [{\citenamefont {Gil-Marín}\ \emph
  {et~al.}(2015{\natexlab{a}})\citenamefont {Gil-Marín}, \citenamefont
  {Noreña}, \citenamefont {Verde}, \citenamefont {Percival}, \citenamefont
  {Wagner}, \citenamefont {Manera},\ and\ \citenamefont {Schneider}}]{sdssi}%
  \BibitemOpen
  \bibfield  {author} {\bibinfo {author} {\bibfnamefont {H.}~\bibnamefont
  {Gil-Marín}}, \bibinfo {author} {\bibfnamefont {J.}~\bibnamefont {Noreña}},
  \bibinfo {author} {\bibfnamefont {L.}~\bibnamefont {Verde}}, \bibinfo
  {author} {\bibfnamefont {W.~J.}\ \bibnamefont {Percival}}, \bibinfo {author}
  {\bibfnamefont {C.}~\bibnamefont {Wagner}}, \bibinfo {author} {\bibfnamefont
  {M.}~\bibnamefont {Manera}}, \ and\ \bibinfo {author} {\bibfnamefont {D.~P.}\
  \bibnamefont {Schneider}},\ }\href {\doibase 10.1093/mnras/stv961} {\bibfield
   {journal} {\bibinfo  {journal} {Monthly Notices of the Royal Astronomical
  Society}\ }\textbf {\bibinfo {volume} {451}},\ \bibinfo {pages} {539}
  (\bibinfo {year} {2015}{\natexlab{a}})}\BibitemShut {NoStop}%
\bibitem [{\citenamefont {Gil-Marín}\ \emph
  {et~al.}(2015{\natexlab{b}})\citenamefont {Gil-Marín}, \citenamefont
  {Verde}, \citenamefont {Noreña}, \citenamefont {Cuesta}, \citenamefont
  {Samushia}, \citenamefont {Percival}, \citenamefont {Wagner}, \citenamefont
  {Manera},\ and\ \citenamefont {Schneider}}]{sdssii}%
  \BibitemOpen
  \bibfield  {author} {\bibinfo {author} {\bibfnamefont {H.}~\bibnamefont
  {Gil-Marín}}, \bibinfo {author} {\bibfnamefont {L.}~\bibnamefont {Verde}},
  \bibinfo {author} {\bibfnamefont {J.}~\bibnamefont {Noreña}}, \bibinfo
  {author} {\bibfnamefont {A.~J.}\ \bibnamefont {Cuesta}}, \bibinfo {author}
  {\bibfnamefont {L.}~\bibnamefont {Samushia}}, \bibinfo {author}
  {\bibfnamefont {W.~J.}\ \bibnamefont {Percival}}, \bibinfo {author}
  {\bibfnamefont {C.}~\bibnamefont {Wagner}}, \bibinfo {author} {\bibfnamefont
  {M.}~\bibnamefont {Manera}}, \ and\ \bibinfo {author} {\bibfnamefont {D.~P.}\
  \bibnamefont {Schneider}},\ }\href {\doibase 10.1093/mnras/stv1359}
  {\bibfield  {journal} {\bibinfo  {journal} {Monthly Notices of the Royal
  Astronomical Society}\ }\textbf {\bibinfo {volume} {452}},\ \bibinfo {pages}
  {1914} (\bibinfo {year} {2015}{\natexlab{b}})}\BibitemShut {NoStop}%
\bibitem [{\citenamefont {Gil-Marín}\ \emph {et~al.}(2017)\citenamefont
  {Gil-Marín} \emph {et~al.}}]{sdssiii}%
  \BibitemOpen
  \bibfield  {author} {\bibinfo {author} {\bibfnamefont {H.}~\bibnamefont
  {Gil-Marín}} \emph {et~al.},\ }\href {\doibase 10.1093/mnras/stw2679}
  {\bibfield  {journal} {\bibinfo  {journal} {Monthly Notices of the Royal
  Astronomical Society}\ }\textbf {\bibinfo {volume} {465}},\ \bibinfo {pages}
  {1757} (\bibinfo {year} {2017})}\BibitemShut {NoStop}%
\bibitem [{\citenamefont {Howlett}\ \emph {et~al.}(2015)\citenamefont
  {Howlett}, \citenamefont {Manera},\ and\ \citenamefont
  {Percival}}]{l-picola}%
  \BibitemOpen
  \bibfield  {author} {\bibinfo {author} {\bibfnamefont {C.}~\bibnamefont
  {Howlett}}, \bibinfo {author} {\bibfnamefont {M.}~\bibnamefont {Manera}}, \
  and\ \bibinfo {author} {\bibfnamefont {W.~J.}\ \bibnamefont {Percival}},\
  }\href@noop {} {\enquote {\bibinfo {title} {{L-PICOLA: A parallel code for
  fast dark matter simulation}},}\ } (\bibinfo {year} {2015}),\ \Eprint
  {http://arxiv.org/abs/arXiv:1506.03737} {arXiv:1506.03737} \BibitemShut
  {NoStop}%
\bibitem [{\citenamefont {Gualdi}\ \emph
  {et~al.}(2019{\natexlab{a}})\citenamefont {Gualdi}, \citenamefont
  {Gil-Marín}, \citenamefont {Manera}, \citenamefont {Joachimi},\ and\
  \citenamefont {Lahav}}]{gualdi1}%
  \BibitemOpen
  \bibfield  {author} {\bibinfo {author} {\bibfnamefont {D.}~\bibnamefont
  {Gualdi}}, \bibinfo {author} {\bibfnamefont {H.}~\bibnamefont {Gil-Marín}},
  \bibinfo {author} {\bibfnamefont {M.}~\bibnamefont {Manera}}, \bibinfo
  {author} {\bibfnamefont {B.}~\bibnamefont {Joachimi}}, \ and\ \bibinfo
  {author} {\bibfnamefont {O.}~\bibnamefont {Lahav}},\ }\href {\doibase
  10.1093/mnrasl/sly242} {\bibfield  {journal} {\bibinfo  {journal} {Monthly
  Notices of the Royal Astronomical Society: Letters}\ }\textbf {\bibinfo
  {volume} {484}},\ \bibinfo {pages} {L29} (\bibinfo {year}
  {2019}{\natexlab{a}})},\ \Eprint
  {http://arxiv.org/abs/http://oup.prod.sis.lan/mnrasl/article-pdf/484/1/L29/27496796/sly242.pdf}
  {http://oup.prod.sis.lan/mnrasl/article-pdf/484/1/L29/27496796/sly242.pdf}
  \BibitemShut {NoStop}%
\bibitem [{\citenamefont {Gualdi}\ \emph {et~al.}(2018)\citenamefont {Gualdi},
  \citenamefont {Manera}, \citenamefont {Joachimi},\ and\ \citenamefont
  {Lahav}}]{gualdi2}%
  \BibitemOpen
  \bibfield  {author} {\bibinfo {author} {\bibfnamefont {D.}~\bibnamefont
  {Gualdi}}, \bibinfo {author} {\bibfnamefont {M.}~\bibnamefont {Manera}},
  \bibinfo {author} {\bibfnamefont {B.}~\bibnamefont {Joachimi}}, \ and\
  \bibinfo {author} {\bibfnamefont {O.}~\bibnamefont {Lahav}},\ }\href
  {\doibase 10.1093/mnras/sty261} {\bibfield  {journal} {\bibinfo  {journal}
  {Monthly Notices of the Royal Astronomical Society}\ }\textbf {\bibinfo
  {volume} {476}},\ \bibinfo {pages} {4045} (\bibinfo {year} {2018})},\ \Eprint
  {http://arxiv.org/abs/http://oup.prod.sis.lan/mnras/article-pdf/476/3/4045/24541764/sty261.pdf}
  {http://oup.prod.sis.lan/mnras/article-pdf/476/3/4045/24541764/sty261.pdf}
  \BibitemShut {NoStop}%
\bibitem [{\citenamefont {Gualdi}\ \emph
  {et~al.}(2019{\natexlab{b}})\citenamefont {Gualdi}, \citenamefont
  {Gil-Marín}, \citenamefont {Schuhmann}, \citenamefont {Manera},
  \citenamefont {Joachimi},\ and\ \citenamefont {Lahav}}]{gualdi3}%
  \BibitemOpen
  \bibfield  {author} {\bibinfo {author} {\bibfnamefont {D.}~\bibnamefont
  {Gualdi}}, \bibinfo {author} {\bibfnamefont {H.}~\bibnamefont {Gil-Marín}},
  \bibinfo {author} {\bibfnamefont {R.~L.}\ \bibnamefont {Schuhmann}}, \bibinfo
  {author} {\bibfnamefont {M.}~\bibnamefont {Manera}}, \bibinfo {author}
  {\bibfnamefont {B.}~\bibnamefont {Joachimi}}, \ and\ \bibinfo {author}
  {\bibfnamefont {O.}~\bibnamefont {Lahav}},\ }\href {\doibase
  10.1093/mnras/stz051} {\bibfield  {journal} {\bibinfo  {journal} {Monthly
  Notices of the Royal Astronomical Society}\ }\textbf {\bibinfo {volume}
  {484}},\ \bibinfo {pages} {3713} (\bibinfo {year} {2019}{\natexlab{b}})},\
  \Eprint
  {http://arxiv.org/abs/http://oup.prod.sis.lan/mnras/article-pdf/484/3/3713/27723669/stz051.pdf}
  {http://oup.prod.sis.lan/mnras/article-pdf/484/3/3713/27723669/stz051.pdf}
  \BibitemShut {NoStop}%
\bibitem [{\citenamefont {{Heavens}}\ \emph {et~al.}(2017)\citenamefont
  {{Heavens}}, \citenamefont {{Sellentin}}, \citenamefont {{de Mijolla}},\ and\
  \citenamefont {{Vianello}}}]{heavens}%
  \BibitemOpen
  \bibfield  {author} {\bibinfo {author} {\bibfnamefont {A.~F.}\ \bibnamefont
  {{Heavens}}}, \bibinfo {author} {\bibfnamefont {E.}~\bibnamefont
  {{Sellentin}}}, \bibinfo {author} {\bibfnamefont {D.}~\bibnamefont {{de
  Mijolla}}}, \ and\ \bibinfo {author} {\bibfnamefont {A.}~\bibnamefont
  {{Vianello}}},\ }\href {\doibase 10.1093/mnras/stx2326} {\bibfield  {journal}
  {\bibinfo  {journal} {Monthly Notices of the Royal Astronomical Society}\
  }\textbf {\bibinfo {volume} {472}},\ \bibinfo {pages} {4244} (\bibinfo {year}
  {2017})},\ \Eprint {http://arxiv.org/abs/1707.06529} {arXiv:1707.06529
  [astro-ph.CO]} \BibitemShut {NoStop}%
\bibitem [{\citenamefont {Alsing}\ and\ \citenamefont
  {Wandelt}(2018)}]{alsing}%
  \BibitemOpen
  \bibfield  {author} {\bibinfo {author} {\bibfnamefont {J.}~\bibnamefont
  {Alsing}}\ and\ \bibinfo {author} {\bibfnamefont {B.}~\bibnamefont
  {Wandelt}},\ }\href {\doibase 10.1093/mnrasl/sly029} {\bibfield  {journal}
  {\bibinfo  {journal} {Monthly Notices of the Royal Astronomical Society:
  Letters}\ }\textbf {\bibinfo {volume} {476}},\ \bibinfo {pages} {L60}
  (\bibinfo {year} {2018})},\ \Eprint
  {http://arxiv.org/abs/http://oup.prod.sis.lan/mnrasl/article-pdf/476/1/L60/24566044/sly029.pdf}
  {http://oup.prod.sis.lan/mnrasl/article-pdf/476/1/L60/24566044/sly029.pdf}
  \BibitemShut {NoStop}%
\bibitem [{\citenamefont {{Kaiser}}(1984)}]{kaiser}%
  \BibitemOpen
  \bibfield  {author} {\bibinfo {author} {\bibfnamefont {N.}~\bibnamefont
  {{Kaiser}}},\ }\href {\doibase 10.1086/184341} {\bibfield  {journal}
  {\bibinfo  {journal} {The Astrophysical Journal}\ }\textbf {\bibinfo {volume}
  {284}},\ \bibinfo {pages} {L9} (\bibinfo {year} {1984})}\BibitemShut
  {NoStop}%
\bibitem [{\citenamefont {{Gelb}}\ and\ \citenamefont
  {{Bertschinger}}(1994)}]{halo1}%
  \BibitemOpen
  \bibfield  {author} {\bibinfo {author} {\bibfnamefont {J.~M.}\ \bibnamefont
  {{Gelb}}}\ and\ \bibinfo {author} {\bibfnamefont {E.}~\bibnamefont
  {{Bertschinger}}},\ }\href {\doibase 10.1086/174922} {\bibfield  {journal}
  {\bibinfo  {journal} {The Astrophysical Journal}\ }\textbf {\bibinfo {volume}
  {436}},\ \bibinfo {pages} {467} (\bibinfo {year} {1994})},\ \Eprint
  {http://arxiv.org/abs/astro-ph/9408028} {astro-ph/9408028} \BibitemShut
  {NoStop}%
\bibitem [{\citenamefont {Klypin}\ and\ \citenamefont
  {Holtzman}(1997)}]{halo2}%
  \BibitemOpen
  \bibfield  {author} {\bibinfo {author} {\bibfnamefont {A.}~\bibnamefont
  {Klypin}}\ and\ \bibinfo {author} {\bibfnamefont {J.}~\bibnamefont
  {Holtzman}},\ }\href@noop {} {\enquote {\bibinfo {title} {Particle-mesh code
  for cosmological simulations},}\ } (\bibinfo {year} {1997}),\ \Eprint
  {http://arxiv.org/abs/arXiv:astro-ph/9712217} {arXiv:astro-ph/9712217}
  \BibitemShut {NoStop}%
\bibitem [{\citenamefont {{Eisenstein}}\ and\ \citenamefont
  {{Hut}}(1998)}]{halo3}%
  \BibitemOpen
  \bibfield  {author} {\bibinfo {author} {\bibfnamefont {D.~J.}\ \bibnamefont
  {{Eisenstein}}}\ and\ \bibinfo {author} {\bibfnamefont {P.}~\bibnamefont
  {{Hut}}},\ }\href {\doibase 10.1086/305535} {\bibfield  {journal} {\bibinfo
  {journal} {The Astrophysical Journal}\ }\textbf {\bibinfo {volume} {498}},\
  \bibinfo {pages} {137} (\bibinfo {year} {1998})},\ \Eprint
  {http://arxiv.org/abs/astro-ph/9712200} {astro-ph/9712200} \BibitemShut
  {NoStop}%
\bibitem [{\citenamefont {{Bullock}}\ \emph {et~al.}(2001)\citenamefont
  {{Bullock}}, \citenamefont {{Kolatt}}, \citenamefont {{Sigad}}, \citenamefont
  {{Somerville}}, \citenamefont {{Kravtsov}}, \citenamefont {{Klypin}},
  \citenamefont {{Primack}},\ and\ \citenamefont {{Dekel}}}]{halo4}%
  \BibitemOpen
  \bibfield  {author} {\bibinfo {author} {\bibfnamefont {J.~S.}\ \bibnamefont
  {{Bullock}}}, \bibinfo {author} {\bibfnamefont {T.~S.}\ \bibnamefont
  {{Kolatt}}}, \bibinfo {author} {\bibfnamefont {Y.}~\bibnamefont {{Sigad}}},
  \bibinfo {author} {\bibfnamefont {R.~S.}\ \bibnamefont {{Somerville}}},
  \bibinfo {author} {\bibfnamefont {A.~V.}\ \bibnamefont {{Kravtsov}}},
  \bibinfo {author} {\bibfnamefont {A.~A.}\ \bibnamefont {{Klypin}}}, \bibinfo
  {author} {\bibfnamefont {J.~R.}\ \bibnamefont {{Primack}}}, \ and\ \bibinfo
  {author} {\bibfnamefont {A.}~\bibnamefont {{Dekel}}},\ }\href {\doibase
  10.1046/j.1365-8711.2001.04068.x} {\bibfield  {journal} {\bibinfo  {journal}
  {Monthly Notices of the Royal Astronomical Society}\ }\textbf {\bibinfo
  {volume} {321}},\ \bibinfo {pages} {559} (\bibinfo {year} {2001})},\ \Eprint
  {http://arxiv.org/abs/astro-ph/9908159} {astro-ph/9908159} \BibitemShut
  {NoStop}%
\bibitem [{\citenamefont {Springel}\ \emph {et~al.}(2001)\citenamefont
  {Springel}, \citenamefont {White}, \citenamefont {Tormen},\ and\
  \citenamefont {Kauffmann}}]{halo5}%
  \BibitemOpen
  \bibfield  {author} {\bibinfo {author} {\bibfnamefont {V.}~\bibnamefont
  {Springel}}, \bibinfo {author} {\bibfnamefont {S.~D.~M.}\ \bibnamefont
  {White}}, \bibinfo {author} {\bibfnamefont {G.}~\bibnamefont {Tormen}}, \
  and\ \bibinfo {author} {\bibfnamefont {G.}~\bibnamefont {Kauffmann}},\ }\href
  {\doibase 10.1046/j.1365-8711.2001.04912.x} {\bibfield  {journal} {\bibinfo
  {journal} {Monthly Notices of the Royal Astronomical Society}\ }\textbf
  {\bibinfo {volume} {328}},\ \bibinfo {pages} {726} (\bibinfo {year}
  {2001})}\BibitemShut {NoStop}%
\bibitem [{\citenamefont {{Stadel}}(2001)}]{halo6}%
  \BibitemOpen
  \bibfield  {author} {\bibinfo {author} {\bibfnamefont {J.~G.}\ \bibnamefont
  {{Stadel}}},\ }\emph {\bibinfo {title} {{Cosmological N-body simulations and
  their analysis}}},\ \href@noop {} {Ph.D. thesis},\ \bibinfo  {school}
  {UNIVERSITY OF WASHINGTON} (\bibinfo {year} {2001})\BibitemShut {NoStop}%
\bibitem [{\citenamefont {{Aubert}}\ \emph {et~al.}(2004)\citenamefont
  {{Aubert}}, \citenamefont {{Pichon}},\ and\ \citenamefont
  {{Colombi}}}]{halo7}%
  \BibitemOpen
  \bibfield  {author} {\bibinfo {author} {\bibfnamefont {D.}~\bibnamefont
  {{Aubert}}}, \bibinfo {author} {\bibfnamefont {C.}~\bibnamefont {{Pichon}}},
  \ and\ \bibinfo {author} {\bibfnamefont {S.}~\bibnamefont {{Colombi}}},\
  }\href {\doibase 10.1111/j.1365-2966.2004.07883.x} {\bibfield  {journal}
  {\bibinfo  {journal} {Monthly Notices of the Royal Astronomical Society}\
  }\textbf {\bibinfo {volume} {352}},\ \bibinfo {pages} {376} (\bibinfo {year}
  {2004})},\ \Eprint {http://arxiv.org/abs/astro-ph/0402405} {astro-ph/0402405}
  \BibitemShut {NoStop}%
\bibitem [{\citenamefont {{Gill}}\ \emph {et~al.}(2004)\citenamefont {{Gill}},
  \citenamefont {{Knebe}},\ and\ \citenamefont {{Gibson}}}]{ahf2}%
  \BibitemOpen
  \bibfield  {author} {\bibinfo {author} {\bibfnamefont {S.~P.~D.}\
  \bibnamefont {{Gill}}}, \bibinfo {author} {\bibfnamefont {A.}~\bibnamefont
  {{Knebe}}}, \ and\ \bibinfo {author} {\bibfnamefont {B.~K.}\ \bibnamefont
  {{Gibson}}},\ }\href {\doibase 10.1111/j.1365-2966.2004.07786.x} {\bibfield
  {journal} {\bibinfo  {journal} {Monthly Notices of the Royal Astronomical
  Society}\ }\textbf {\bibinfo {volume} {351}},\ \bibinfo {pages} {399}
  (\bibinfo {year} {2004})},\ \Eprint {http://arxiv.org/abs/astro-ph/0404258}
  {astro-ph/0404258} \BibitemShut {NoStop}%
\bibitem [{\citenamefont {{Neyrinck}}\ \emph {et~al.}(2005)\citenamefont
  {{Neyrinck}}, \citenamefont {{Gnedin}},\ and\ \citenamefont
  {{Hamilton}}}]{halo9}%
  \BibitemOpen
  \bibfield  {author} {\bibinfo {author} {\bibfnamefont {M.~C.}\ \bibnamefont
  {{Neyrinck}}}, \bibinfo {author} {\bibfnamefont {N.~Y.}\ \bibnamefont
  {{Gnedin}}}, \ and\ \bibinfo {author} {\bibfnamefont {A.~J.~S.}\ \bibnamefont
  {{Hamilton}}},\ }\href {\doibase 10.1111/j.1365-2966.2004.08505.x} {\bibfield
   {journal} {\bibinfo  {journal} {Monthly Notices of the Royal Astronomical
  Society}\ }\textbf {\bibinfo {volume} {356}},\ \bibinfo {pages} {1222}
  (\bibinfo {year} {2005})},\ \Eprint {http://arxiv.org/abs/astro-ph/0402346}
  {astro-ph/0402346} \BibitemShut {NoStop}%
\bibitem [{\citenamefont {{Weller}}\ \emph {et~al.}(2005)\citenamefont
  {{Weller}}, \citenamefont {{Ostriker}}, \citenamefont {{Bode}},\ and\
  \citenamefont {{Shaw}}}]{halo10}%
  \BibitemOpen
  \bibfield  {author} {\bibinfo {author} {\bibfnamefont {J.}~\bibnamefont
  {{Weller}}}, \bibinfo {author} {\bibfnamefont {J.~P.}\ \bibnamefont
  {{Ostriker}}}, \bibinfo {author} {\bibfnamefont {P.}~\bibnamefont {{Bode}}},
  \ and\ \bibinfo {author} {\bibfnamefont {L.}~\bibnamefont {{Shaw}}},\ }\href
  {\doibase 10.1111/j.1365-2966.2005.09602.x} {\bibfield  {journal} {\bibinfo
  {journal} {Monthly Notices of the Royal Astronomical Society}\ }\textbf
  {\bibinfo {volume} {364}},\ \bibinfo {pages} {823} (\bibinfo {year}
  {2005})},\ \Eprint {http://arxiv.org/abs/astro-ph/0405445} {astro-ph/0405445}
  \BibitemShut {NoStop}%
\bibitem [{\citenamefont {{Diemand}}\ \emph {et~al.}(2006)\citenamefont
  {{Diemand}}, \citenamefont {{Kuhlen}},\ and\ \citenamefont
  {{Madau}}}]{halo11}%
  \BibitemOpen
  \bibfield  {author} {\bibinfo {author} {\bibfnamefont {J.}~\bibnamefont
  {{Diemand}}}, \bibinfo {author} {\bibfnamefont {M.}~\bibnamefont {{Kuhlen}}},
  \ and\ \bibinfo {author} {\bibfnamefont {P.}~\bibnamefont {{Madau}}},\ }\href
  {\doibase 10.1086/506377} {\bibfield  {journal} {\bibinfo  {journal} {The
  Astrophysical Journal}\ }\textbf {\bibinfo {volume} {649}},\ \bibinfo {pages}
  {1} (\bibinfo {year} {2006})},\ \Eprint
  {http://arxiv.org/abs/astro-ph/0603250} {astro-ph/0603250} \BibitemShut
  {NoStop}%
\bibitem [{\citenamefont {{Kim}}\ and\ \citenamefont {{Park}}(2006)}]{halo12}%
  \BibitemOpen
  \bibfield  {author} {\bibinfo {author} {\bibfnamefont {J.}~\bibnamefont
  {{Kim}}}\ and\ \bibinfo {author} {\bibfnamefont {C.}~\bibnamefont {{Park}}},\
  }\href {\doibase 10.1086/499761} {\bibfield  {journal} {\bibinfo  {journal}
  {The Astrophysical Journal}\ }\textbf {\bibinfo {volume} {639}},\ \bibinfo
  {pages} {600} (\bibinfo {year} {2006})},\ \Eprint
  {http://arxiv.org/abs/astro-ph/0401386} {astro-ph/0401386} \BibitemShut
  {NoStop}%
\bibitem [{\citenamefont {Gardner}\ \emph {et~al.}(2007)\citenamefont
  {Gardner}, \citenamefont {Connolly},\ and\ \citenamefont {McBride}}]{halo13}%
  \BibitemOpen
  \bibfield  {author} {\bibinfo {author} {\bibfnamefont {J.~P.}\ \bibnamefont
  {Gardner}}, \bibinfo {author} {\bibfnamefont {A.}~\bibnamefont {Connolly}}, \
  and\ \bibinfo {author} {\bibfnamefont {C.}~\bibnamefont {McBride}},\ }in\
  \href {\doibase 10.1145/1273404.1273410} {\emph {\bibinfo {booktitle}
  {Proceedings of the 5th IEEE Workshop on Challenges of Large Applications in
  Distributed Environments}}},\ \bibinfo {series and number} {CLADE '07}\
  (\bibinfo  {publisher} {ACM},\ \bibinfo {address} {New York, NY, USA},\
  \bibinfo {year} {2007})\ pp.\ \bibinfo {pages} {1--10}\BibitemShut {NoStop}%
\bibitem [{\citenamefont {{Shaw}}\ \emph {et~al.}(2007)\citenamefont {{Shaw}},
  \citenamefont {{Weller}}, \citenamefont {{Ostriker}},\ and\ \citenamefont
  {{Bode}}}]{halo14}%
  \BibitemOpen
  \bibfield  {author} {\bibinfo {author} {\bibfnamefont {L.~D.}\ \bibnamefont
  {{Shaw}}}, \bibinfo {author} {\bibfnamefont {J.}~\bibnamefont {{Weller}}},
  \bibinfo {author} {\bibfnamefont {J.~P.}\ \bibnamefont {{Ostriker}}}, \ and\
  \bibinfo {author} {\bibfnamefont {P.}~\bibnamefont {{Bode}}},\ }\href
  {\doibase 10.1086/511849} {\bibfield  {journal} {\bibinfo  {journal} {The
  Astrophysical Journal}\ }\textbf {\bibinfo {volume} {659}},\ \bibinfo {pages}
  {1082} (\bibinfo {year} {2007})},\ \Eprint
  {http://arxiv.org/abs/astro-ph/0603150} {astro-ph/0603150} \BibitemShut
  {NoStop}%
\bibitem [{\citenamefont {{Habib}}\ \emph {et~al.}(2009)\citenamefont
  {{Habib}}, \citenamefont {{Pope}}, \citenamefont {{Luki{\'c}}}, \citenamefont
  {{Daniel}}, \citenamefont {{Fasel}}, \citenamefont {{Desai}}, \citenamefont
  {{Heitmann}}, \citenamefont {{Hsu}}, \citenamefont {{Ankeny}}, \citenamefont
  {{Mark}}, \citenamefont {{Bhattacharya}},\ and\ \citenamefont
  {{Ahrens}}}]{halo15}%
  \BibitemOpen
  \bibfield  {author} {\bibinfo {author} {\bibfnamefont {S.}~\bibnamefont
  {{Habib}}}, \bibinfo {author} {\bibfnamefont {A.}~\bibnamefont {{Pope}}},
  \bibinfo {author} {\bibfnamefont {Z.}~\bibnamefont {{Luki{\'c}}}}, \bibinfo
  {author} {\bibfnamefont {D.}~\bibnamefont {{Daniel}}}, \bibinfo {author}
  {\bibfnamefont {P.}~\bibnamefont {{Fasel}}}, \bibinfo {author} {\bibfnamefont
  {N.}~\bibnamefont {{Desai}}}, \bibinfo {author} {\bibfnamefont
  {K.}~\bibnamefont {{Heitmann}}}, \bibinfo {author} {\bibfnamefont {C.-H.}\
  \bibnamefont {{Hsu}}}, \bibinfo {author} {\bibfnamefont {L.}~\bibnamefont
  {{Ankeny}}}, \bibinfo {author} {\bibfnamefont {G.}~\bibnamefont {{Mark}}},
  \bibinfo {author} {\bibfnamefont {S.}~\bibnamefont {{Bhattacharya}}}, \ and\
  \bibinfo {author} {\bibfnamefont {J.}~\bibnamefont {{Ahrens}}},\ }in\ \href
  {\doibase 10.1088/1742-6596/180/1/012019} {\emph {\bibinfo {booktitle}
  {Journal of Physics Conference Series}}},\ \bibinfo {series} {Journal of
  Physics Conference Series}, Vol.\ \bibinfo {volume} {180}\ (\bibinfo {year}
  {2009})\ p.\ \bibinfo {pages} {012019}\BibitemShut {NoStop}%
\bibitem [{\citenamefont {{Knollmann}}\ and\ \citenamefont
  {{Knebe}}(2009)}]{ahf}%
  \BibitemOpen
  \bibfield  {author} {\bibinfo {author} {\bibfnamefont {S.~R.}\ \bibnamefont
  {{Knollmann}}}\ and\ \bibinfo {author} {\bibfnamefont {A.}~\bibnamefont
  {{Knebe}}},\ }\href {\doibase 10.1088/0067-0049/182/2/608} {\bibfield
  {journal} {\bibinfo  {journal} {The Astrophysical Journal Supplement}\
  }\textbf {\bibinfo {volume} {182}},\ \bibinfo {pages} {608} (\bibinfo {year}
  {2009})},\ \Eprint {http://arxiv.org/abs/0904.3662} {arXiv:0904.3662}
  \BibitemShut {NoStop}%
\bibitem [{\citenamefont {{Maciejewski}}\ \emph {et~al.}(2009)\citenamefont
  {{Maciejewski}}, \citenamefont {{Colombi}}, \citenamefont {{Springel}},
  \citenamefont {{Alard}},\ and\ \citenamefont {{Bouchet}}}]{halo17}%
  \BibitemOpen
  \bibfield  {author} {\bibinfo {author} {\bibfnamefont {M.}~\bibnamefont
  {{Maciejewski}}}, \bibinfo {author} {\bibfnamefont {S.}~\bibnamefont
  {{Colombi}}}, \bibinfo {author} {\bibfnamefont {V.}~\bibnamefont
  {{Springel}}}, \bibinfo {author} {\bibfnamefont {C.}~\bibnamefont {{Alard}}},
  \ and\ \bibinfo {author} {\bibfnamefont {F.~R.}\ \bibnamefont {{Bouchet}}},\
  }\href {\doibase 10.1111/j.1365-2966.2009.14825.x} {\bibfield  {journal}
  {\bibinfo  {journal} {Monthly Notices of the Royal Astronomical Society}\
  }\textbf {\bibinfo {volume} {396}},\ \bibinfo {pages} {1329} (\bibinfo {year}
  {2009})},\ \Eprint {http://arxiv.org/abs/0812.0288} {arXiv:0812.0288}
  \BibitemShut {NoStop}%
\bibitem [{\citenamefont {Ascasibar}(2010)}]{halo18}%
  \BibitemOpen
  \bibfield  {author} {\bibinfo {author} {\bibfnamefont {Y.}~\bibnamefont
  {Ascasibar}},\ }\href {\doibase https://doi.org/10.1016/j.cpc.2010.04.011}
  {\bibfield  {journal} {\bibinfo  {journal} {Computer Physics Communications}\
  }\textbf {\bibinfo {volume} {181}},\ \bibinfo {pages} {1438 } (\bibinfo
  {year} {2010})}\BibitemShut {NoStop}%
\bibitem [{\citenamefont {Behroozi}\ \emph {et~al.}(2013)\citenamefont
  {Behroozi}, \citenamefont {Wechsler},\ and\ \citenamefont {Wu}}]{rockstar}%
  \BibitemOpen
  \bibfield  {author} {\bibinfo {author} {\bibfnamefont {P.~S.}\ \bibnamefont
  {Behroozi}}, \bibinfo {author} {\bibfnamefont {R.~H.}\ \bibnamefont
  {Wechsler}}, \ and\ \bibinfo {author} {\bibfnamefont {H.-Y.}\ \bibnamefont
  {Wu}},\ }\href {http://stacks.iop.org/0004-637X/762/i=2/a=109} {\bibfield
  {journal} {\bibinfo  {journal} {The Astrophysical Journal}\ }\textbf
  {\bibinfo {volume} {762}},\ \bibinfo {pages} {109} (\bibinfo {year}
  {2013})}\BibitemShut {NoStop}%
\bibitem [{\citenamefont {{Planelles}}\ and\ \citenamefont
  {{Quilis}}(2010)}]{halo20}%
  \BibitemOpen
  \bibfield  {author} {\bibinfo {author} {\bibfnamefont {S.}~\bibnamefont
  {{Planelles}}}\ and\ \bibinfo {author} {\bibfnamefont {V.}~\bibnamefont
  {{Quilis}}},\ }\href {\doibase 10.1051/0004-6361/201014214} {\bibfield
  {journal} {\bibinfo  {journal} {Astronomy and Astrophysics}\ }\textbf
  {\bibinfo {volume} {519}},\ \bibinfo {eid} {A94} (\bibinfo {year} {2010})},\
  \Eprint {http://arxiv.org/abs/1006.3205} {arXiv:1006.3205} \BibitemShut
  {NoStop}%
\bibitem [{\citenamefont {{Rasera}}\ \emph {et~al.}(2010)\citenamefont
  {{Rasera}}, \citenamefont {{Alimi}}, \citenamefont {{Courtin}}, \citenamefont
  {{Roy}}, \citenamefont {{Corasaniti}}, \citenamefont {{F{\"u}zfa}},\ and\
  \citenamefont {{Boucher}}}]{halo21}%
  \BibitemOpen
  \bibfield  {author} {\bibinfo {author} {\bibfnamefont {Y.}~\bibnamefont
  {{Rasera}}}, \bibinfo {author} {\bibfnamefont {J.-M.}\ \bibnamefont
  {{Alimi}}}, \bibinfo {author} {\bibfnamefont {J.}~\bibnamefont {{Courtin}}},
  \bibinfo {author} {\bibfnamefont {F.}~\bibnamefont {{Roy}}}, \bibinfo
  {author} {\bibfnamefont {P.-S.}\ \bibnamefont {{Corasaniti}}}, \bibinfo
  {author} {\bibfnamefont {A.}~\bibnamefont {{F{\"u}zfa}}}, \ and\ \bibinfo
  {author} {\bibfnamefont {V.}~\bibnamefont {{Boucher}}},\ }in\ \href {\doibase
  10.1063/1.3462610} {\emph {\bibinfo {booktitle} {American Institute of
  Physics Conference Series}}},\ \bibinfo {series} {American Institute of
  Physics Conference Series}, Vol.\ \bibinfo {volume} {1241},\ \bibinfo
  {editor} {edited by\ \bibinfo {editor} {\bibfnamefont {J.-M.}\ \bibnamefont
  {{Alimi}}}\ and\ \bibinfo {editor} {\bibfnamefont {A.}~\bibnamefont
  {{Fu{\"o}zfa}}}}\ (\bibinfo {year} {2010})\ pp.\ \bibinfo {pages}
  {1134--1139},\ \Eprint {http://arxiv.org/abs/1002.4950} {arXiv:1002.4950}
  \BibitemShut {NoStop}%
\bibitem [{\citenamefont {{Skory}}\ \emph {et~al.}(2010)\citenamefont
  {{Skory}}, \citenamefont {{Turk}}, \citenamefont {{Norman}},\ and\
  \citenamefont {{Coil}}}]{halo22}%
  \BibitemOpen
  \bibfield  {author} {\bibinfo {author} {\bibfnamefont {S.}~\bibnamefont
  {{Skory}}}, \bibinfo {author} {\bibfnamefont {M.~J.}\ \bibnamefont {{Turk}}},
  \bibinfo {author} {\bibfnamefont {M.~L.}\ \bibnamefont {{Norman}}}, \ and\
  \bibinfo {author} {\bibfnamefont {A.~L.}\ \bibnamefont {{Coil}}},\ }\href
  {\doibase 10.1088/0067-0049/191/1/43} {\bibfield  {journal} {\bibinfo
  {journal} {The Astrophysical Journal Supplement}\ }\textbf {\bibinfo {volume}
  {191}},\ \bibinfo {pages} {43} (\bibinfo {year} {2010})},\ \Eprint
  {http://arxiv.org/abs/1001.3411} {arXiv:1001.3411} \BibitemShut {NoStop}%
\bibitem [{\citenamefont {{Sutter}}\ and\ \citenamefont
  {{Ricker}}(2010)}]{halo23}%
  \BibitemOpen
  \bibfield  {author} {\bibinfo {author} {\bibfnamefont {P.~M.}\ \bibnamefont
  {{Sutter}}}\ and\ \bibinfo {author} {\bibfnamefont {P.~M.}\ \bibnamefont
  {{Ricker}}},\ }\href {\doibase 10.1088/0004-637X/723/2/1308} {\bibfield
  {journal} {\bibinfo  {journal} {The Astrophysical Journal}\ }\textbf
  {\bibinfo {volume} {723}},\ \bibinfo {pages} {1308} (\bibinfo {year}
  {2010})},\ \Eprint {http://arxiv.org/abs/1006.2879} {arXiv:1006.2879
  [astro-ph.CO]} \BibitemShut {NoStop}%
\bibitem [{\citenamefont {Falck}\ \emph {et~al.}(2012)\citenamefont {Falck},
  \citenamefont {Neyrinck},\ and\ \citenamefont {Szalay}}]{halo24}%
  \BibitemOpen
  \bibfield  {author} {\bibinfo {author} {\bibfnamefont {B.~L.}\ \bibnamefont
  {Falck}}, \bibinfo {author} {\bibfnamefont {M.~C.}\ \bibnamefont {Neyrinck}},
  \ and\ \bibinfo {author} {\bibfnamefont {A.~S.}\ \bibnamefont {Szalay}},\
  }\href {http://stacks.iop.org/0004-637X/754/i=2/a=126} {\bibfield  {journal}
  {\bibinfo  {journal} {The Astrophysical Journal}\ }\textbf {\bibinfo {volume}
  {754}},\ \bibinfo {pages} {126} (\bibinfo {year} {2012})}\BibitemShut
  {NoStop}%
\bibitem [{\citenamefont {{Wu}}\ \emph {et~al.}(2010)\citenamefont {{Wu}},
  \citenamefont {{Zentner}},\ and\ \citenamefont {{Wechsler}}}]{percent1}%
  \BibitemOpen
  \bibfield  {author} {\bibinfo {author} {\bibfnamefont {H.-Y.}\ \bibnamefont
  {{Wu}}}, \bibinfo {author} {\bibfnamefont {A.~R.}\ \bibnamefont {{Zentner}}},
  \ and\ \bibinfo {author} {\bibfnamefont {R.~H.}\ \bibnamefont {{Wechsler}}},\
  }\href {\doibase 10.1088/0004-637X/713/2/856} {\bibfield  {journal} {\bibinfo
   {journal} {The Astrophysical Journal}\ }\textbf {\bibinfo {volume} {713}},\
  \bibinfo {pages} {856} (\bibinfo {year} {2010})},\ \Eprint
  {http://arxiv.org/abs/0910.3668} {arXiv:0910.3668} \BibitemShut {NoStop}%
\bibitem [{\citenamefont {{Cunha}}\ and\ \citenamefont
  {{Evrard}}(2010)}]{percent2}%
  \BibitemOpen
  \bibfield  {author} {\bibinfo {author} {\bibfnamefont {C.~E.}\ \bibnamefont
  {{Cunha}}}\ and\ \bibinfo {author} {\bibfnamefont {A.~E.}\ \bibnamefont
  {{Evrard}}},\ }\href {\doibase 10.1103/PhysRevD.81.083509} {\bibfield
  {journal} {\bibinfo  {journal} {Physical Review D}\ }\textbf {\bibinfo
  {volume} {81}},\ \bibinfo {eid} {083509} (\bibinfo {year} {2010})},\ \Eprint
  {http://arxiv.org/abs/0908.0526} {arXiv:0908.0526 [astro-ph.CO]} \BibitemShut
  {NoStop}%
\bibitem [{\citenamefont {{Davis}}\ \emph {et~al.}(1985)\citenamefont
  {{Davis}}, \citenamefont {{Efstathiou}}, \citenamefont {{Frenk}},\ and\
  \citenamefont {{White}}}]{FoF_Davis}%
  \BibitemOpen
  \bibfield  {author} {\bibinfo {author} {\bibfnamefont {M.}~\bibnamefont
  {{Davis}}}, \bibinfo {author} {\bibfnamefont {G.}~\bibnamefont
  {{Efstathiou}}}, \bibinfo {author} {\bibfnamefont {C.~S.}\ \bibnamefont
  {{Frenk}}}, \ and\ \bibinfo {author} {\bibfnamefont {S.~D.~M.}\ \bibnamefont
  {{White}}},\ }\href@noop {} {\bibfield  {journal} {\bibinfo  {journal} {The
  Astrophysical Journal}\ }\textbf {\bibinfo {volume} {292}},\ \bibinfo {pages}
  {371} (\bibinfo {year} {1985})}\BibitemShut {NoStop}%
\bibitem [{\citenamefont {{Press}}\ and\ \citenamefont
  {{Schechter}}(1974)}]{SO_PS}%
  \BibitemOpen
  \bibfield  {author} {\bibinfo {author} {\bibfnamefont {W.~H.}\ \bibnamefont
  {{Press}}}\ and\ \bibinfo {author} {\bibfnamefont {P.}~\bibnamefont
  {{Schechter}}},\ }\href {\doibase 10.1086/152650} {\bibfield  {journal}
  {\bibinfo  {journal} {The Astrophysical Journal}\ }\textbf {\bibinfo {volume}
  {187}},\ \bibinfo {pages} {425} (\bibinfo {year} {1974})}\BibitemShut
  {NoStop}%
\bibitem [{\citenamefont {Knebe}\ \emph {et~al.}(2011)\citenamefont {Knebe},
  \citenamefont {Knollmann}, \citenamefont {Muldrew}, \citenamefont {Pearce},
  \citenamefont {Aragon-Calvo}, \citenamefont {Ascasibar}, \citenamefont
  {Behroozi}, \citenamefont {Ceverino}, \citenamefont {Colombi}, \citenamefont
  {Diemand}, \citenamefont {Dolag}, \citenamefont {Falck}, \citenamefont
  {Fasel}, \citenamefont {Gardner}, \citenamefont {Gottlöber}, \citenamefont
  {Hsu}, \citenamefont {Iannuzzi}, \citenamefont {Klypin}, \citenamefont
  {Lukić}, \citenamefont {Maciejewski}, \citenamefont {McBride}, \citenamefont
  {Neyrinck}, \citenamefont {Planelles}, \citenamefont {Potter}, \citenamefont
  {Quilis}, \citenamefont {Rasera}, \citenamefont {Read}, \citenamefont
  {Ricker}, \citenamefont {Roy}, \citenamefont {Springel}, \citenamefont
  {Stadel}, \citenamefont {Stinson}, \citenamefont {Sutter}, \citenamefont
  {Turchaninov}, \citenamefont {Tweed}, \citenamefont {Yepes},\ and\
  \citenamefont {Zemp}}]{mad}%
  \BibitemOpen
  \bibfield  {author} {\bibinfo {author} {\bibfnamefont {A.}~\bibnamefont
  {Knebe}}, \bibinfo {author} {\bibfnamefont {S.~R.}\ \bibnamefont
  {Knollmann}}, \bibinfo {author} {\bibfnamefont {S.~I.}\ \bibnamefont
  {Muldrew}}, \bibinfo {author} {\bibfnamefont {F.~R.}\ \bibnamefont {Pearce}},
  \bibinfo {author} {\bibfnamefont {M.~A.}\ \bibnamefont {Aragon-Calvo}},
  \bibinfo {author} {\bibfnamefont {Y.}~\bibnamefont {Ascasibar}}, \bibinfo
  {author} {\bibfnamefont {P.~S.}\ \bibnamefont {Behroozi}}, \bibinfo {author}
  {\bibfnamefont {D.}~\bibnamefont {Ceverino}}, \bibinfo {author}
  {\bibfnamefont {S.}~\bibnamefont {Colombi}}, \bibinfo {author} {\bibfnamefont
  {J.}~\bibnamefont {Diemand}}, \bibinfo {author} {\bibfnamefont
  {K.}~\bibnamefont {Dolag}}, \bibinfo {author} {\bibfnamefont {B.~L.}\
  \bibnamefont {Falck}}, \bibinfo {author} {\bibfnamefont {P.}~\bibnamefont
  {Fasel}}, \bibinfo {author} {\bibfnamefont {J.}~\bibnamefont {Gardner}},
  \bibinfo {author} {\bibfnamefont {S.}~\bibnamefont {Gottlöber}}, \bibinfo
  {author} {\bibfnamefont {C.-H.}\ \bibnamefont {Hsu}}, \bibinfo {author}
  {\bibfnamefont {F.}~\bibnamefont {Iannuzzi}}, \bibinfo {author}
  {\bibfnamefont {A.}~\bibnamefont {Klypin}}, \bibinfo {author} {\bibfnamefont
  {Z.}~\bibnamefont {Lukić}}, \bibinfo {author} {\bibfnamefont
  {M.}~\bibnamefont {Maciejewski}}, \bibinfo {author} {\bibfnamefont
  {C.}~\bibnamefont {McBride}}, \bibinfo {author} {\bibfnamefont {M.~C.}\
  \bibnamefont {Neyrinck}}, \bibinfo {author} {\bibfnamefont {S.}~\bibnamefont
  {Planelles}}, \bibinfo {author} {\bibfnamefont {D.}~\bibnamefont {Potter}},
  \bibinfo {author} {\bibfnamefont {V.}~\bibnamefont {Quilis}}, \bibinfo
  {author} {\bibfnamefont {Y.}~\bibnamefont {Rasera}}, \bibinfo {author}
  {\bibfnamefont {J.~I.}\ \bibnamefont {Read}}, \bibinfo {author}
  {\bibfnamefont {P.~M.}\ \bibnamefont {Ricker}}, \bibinfo {author}
  {\bibfnamefont {F.}~\bibnamefont {Roy}}, \bibinfo {author} {\bibfnamefont
  {V.}~\bibnamefont {Springel}}, \bibinfo {author} {\bibfnamefont
  {J.}~\bibnamefont {Stadel}}, \bibinfo {author} {\bibfnamefont
  {G.}~\bibnamefont {Stinson}}, \bibinfo {author} {\bibfnamefont {P.~M.}\
  \bibnamefont {Sutter}}, \bibinfo {author} {\bibfnamefont {V.}~\bibnamefont
  {Turchaninov}}, \bibinfo {author} {\bibfnamefont {D.}~\bibnamefont {Tweed}},
  \bibinfo {author} {\bibfnamefont {G.}~\bibnamefont {Yepes}}, \ and\ \bibinfo
  {author} {\bibfnamefont {M.}~\bibnamefont {Zemp}},\ }\href {\doibase
  10.1111/j.1365-2966.2011.18858.x} {\bibfield  {journal} {\bibinfo  {journal}
  {Monthly Notices of the Royal Astronomical Society}\ }\textbf {\bibinfo
  {volume} {415}},\ \bibinfo {pages} {2293} (\bibinfo {year} {2011})},\ \Eprint
  {http://arxiv.org/abs/http://oup.prod.sis.lan/mnras/article-pdf/415/3/2293/5972749/mnras0415-2293.pdf}
  {http://oup.prod.sis.lan/mnras/article-pdf/415/3/2293/5972749/mnras0415-2293.pdf}
  \BibitemShut {NoStop}%
\bibitem [{\citenamefont {{Planck Collaboration}}(2016)}]{planck2015}%
  \BibitemOpen
  \bibfield  {author} {\bibinfo {author} {\bibnamefont {{Planck
  Collaboration}}},\ }\href {\doibase 10.1051/0004-6361/201525830} {\bibfield
  {journal} {\bibinfo  {journal} {Astronomy \& Astrophysics}\ }\textbf
  {\bibinfo {volume} {594}},\ \bibinfo {pages} {A13} (\bibinfo {year}
  {2016})}\BibitemShut {NoStop}%
\bibitem [{\citenamefont {McCullagh}\ \emph {et~al.}(2016)\citenamefont
  {McCullagh}, \citenamefont {Jeong},\ and\ \citenamefont
  {Szalay}}]{transients}%
  \BibitemOpen
  \bibfield  {author} {\bibinfo {author} {\bibfnamefont {N.}~\bibnamefont
  {McCullagh}}, \bibinfo {author} {\bibfnamefont {D.}~\bibnamefont {Jeong}}, \
  and\ \bibinfo {author} {\bibfnamefont {A.~S.}\ \bibnamefont {Szalay}},\
  }\href {\doibase 10.1093/mnras/stv2525} {\bibfield  {journal} {\bibinfo
  {journal} {Monthly Notices of the Royal Astronomical Society}\ }\textbf
  {\bibinfo {volume} {455}},\ \bibinfo {pages} {2945} (\bibinfo {year}
  {2016})}\BibitemShut {NoStop}%
\bibitem [{\citenamefont {Crocce}\ \emph {et~al.}(2006)\citenamefont {Crocce},
  \citenamefont {Pueblas},\ and\ \citenamefont {Scoccimarro}}]{crocce2006}%
  \BibitemOpen
  \bibfield  {author} {\bibinfo {author} {\bibfnamefont {M.}~\bibnamefont
  {Crocce}}, \bibinfo {author} {\bibfnamefont {S.}~\bibnamefont {Pueblas}}, \
  and\ \bibinfo {author} {\bibfnamefont {R.}~\bibnamefont {Scoccimarro}},\
  }\href {\doibase 10.1111/j.1365-2966.2006.11040.x} {\bibfield  {journal}
  {\bibinfo  {journal} {Monthly Notices of the Royal Astronomical Society}\
  }\textbf {\bibinfo {volume} {373}},\ \bibinfo {pages} {369} (\bibinfo {year}
  {2006})}\BibitemShut {NoStop}%
\bibitem [{\citenamefont {Scoccimarro}\ \emph {et~al.}(2012)\citenamefont
  {Scoccimarro}, \citenamefont {Hui}, \citenamefont {Manera},\ and\
  \citenamefont {Chan}}]{scoccimarro}%
  \BibitemOpen
  \bibfield  {author} {\bibinfo {author} {\bibfnamefont {R.}~\bibnamefont
  {Scoccimarro}}, \bibinfo {author} {\bibfnamefont {L.}~\bibnamefont {Hui}},
  \bibinfo {author} {\bibfnamefont {M.}~\bibnamefont {Manera}}, \ and\ \bibinfo
  {author} {\bibfnamefont {K.~C.}\ \bibnamefont {Chan}},\ }\href {\doibase
  10.1103/PhysRevD.85.083002} {\bibfield  {journal} {\bibinfo  {journal} {Phys.
  Rev. D}\ }\textbf {\bibinfo {volume} {85}},\ \bibinfo {pages} {083002}
  (\bibinfo {year} {2012})}\BibitemShut {NoStop}%
\bibitem [{\citenamefont {Lewis}\ \emph {et~al.}(2000)\citenamefont {Lewis},
  \citenamefont {Challinor},\ and\ \citenamefont {Lasenby}}]{CAMB}%
  \BibitemOpen
  \bibfield  {author} {\bibinfo {author} {\bibfnamefont {A.}~\bibnamefont
  {Lewis}}, \bibinfo {author} {\bibfnamefont {A.}~\bibnamefont {Challinor}}, \
  and\ \bibinfo {author} {\bibfnamefont {A.}~\bibnamefont {Lasenby}},\
  }\href@noop {} {\bibfield  {journal} {\bibinfo  {journal} {The Astrophysical
  Journal}\ }\textbf {\bibinfo {volume} {538}},\ \bibinfo {pages} {473}
  (\bibinfo {year} {2000})}\BibitemShut {NoStop}%
\bibitem [{\citenamefont {Eisenstein}\ \emph {et~al.}(2017)\citenamefont
  {Eisenstein}, \citenamefont {Garrison},\ and\ \citenamefont
  {Yuan}}]{eisenstein1}%
  \BibitemOpen
  \bibfield  {author} {\bibinfo {author} {\bibfnamefont {D.~J.}\ \bibnamefont
  {Eisenstein}}, \bibinfo {author} {\bibfnamefont {L.~H.}\ \bibnamefont
  {Garrison}}, \ and\ \bibinfo {author} {\bibfnamefont {S.}~\bibnamefont
  {Yuan}},\ }\href {\doibase 10.1093/mnras/stx2032} {\bibfield  {journal}
  {\bibinfo  {journal} {Monthly Notices of the Royal Astronomical Society}\
  }\textbf {\bibinfo {volume} {472}},\ \bibinfo {pages} {577} (\bibinfo {year}
  {2017})}\BibitemShut {NoStop}%
\bibitem [{\citenamefont {Eisenstein}\ \emph {et~al.}(2018)\citenamefont
  {Eisenstein}, \citenamefont {Garrison},\ and\ \citenamefont
  {Yuan}}]{eisenstein2}%
  \BibitemOpen
  \bibfield  {author} {\bibinfo {author} {\bibfnamefont {D.~J.}\ \bibnamefont
  {Eisenstein}}, \bibinfo {author} {\bibfnamefont {L.~H.}\ \bibnamefont
  {Garrison}}, \ and\ \bibinfo {author} {\bibfnamefont {S.}~\bibnamefont
  {Yuan}},\ }\href {\doibase 10.1093/mnras/sty1089} {\bibfield  {journal}
  {\bibinfo  {journal} {Monthly Notices of the Royal Astronomical Society}\
  }\textbf {\bibinfo {volume} {478}},\ \bibinfo {pages} {2019} (\bibinfo {year}
  {2018})}\BibitemShut {NoStop}%
\bibitem [{\citenamefont {{Padmanabhan}}\ \emph {et~al.}(2007)\citenamefont
  {{Padmanabhan}}, \citenamefont {{Schlegel}}, \citenamefont {{Seljak}},
  \citenamefont {{Makarov}}, \citenamefont {{Bahcall}}, \citenamefont
  {{Blanton}}, \citenamefont {{Brinkmann}}, \citenamefont {{Eisenstein}},
  \citenamefont {{Finkbeiner}}, \citenamefont {{Gunn}}, \citenamefont {{Hogg}},
  \citenamefont {{Ivezi{\'c}}}, \citenamefont {{Knapp}}, \citenamefont
  {{Loveday}}, \citenamefont {{Lupton}}, \citenamefont {{Nichol}},
  \citenamefont {{Schneider}}, \citenamefont {{Strauss}}, \citenamefont
  {{Tegmark}},\ and\ \citenamefont {{York}}}]{projection}%
  \BibitemOpen
  \bibfield  {author} {\bibinfo {author} {\bibfnamefont {N.}~\bibnamefont
  {{Padmanabhan}}}, \bibinfo {author} {\bibfnamefont {D.~J.}\ \bibnamefont
  {{Schlegel}}}, \bibinfo {author} {\bibfnamefont {U.}~\bibnamefont
  {{Seljak}}}, \bibinfo {author} {\bibfnamefont {A.}~\bibnamefont {{Makarov}}},
  \bibinfo {author} {\bibfnamefont {N.~A.}\ \bibnamefont {{Bahcall}}}, \bibinfo
  {author} {\bibfnamefont {M.~R.}\ \bibnamefont {{Blanton}}}, \bibinfo {author}
  {\bibfnamefont {J.}~\bibnamefont {{Brinkmann}}}, \bibinfo {author}
  {\bibfnamefont {D.~J.}\ \bibnamefont {{Eisenstein}}}, \bibinfo {author}
  {\bibfnamefont {D.~P.}\ \bibnamefont {{Finkbeiner}}}, \bibinfo {author}
  {\bibfnamefont {J.~E.}\ \bibnamefont {{Gunn}}}, \bibinfo {author}
  {\bibfnamefont {D.~W.}\ \bibnamefont {{Hogg}}}, \bibinfo {author}
  {\bibfnamefont {{\v Z}.}~\bibnamefont {{Ivezi{\'c}}}}, \bibinfo {author}
  {\bibfnamefont {G.~R.}\ \bibnamefont {{Knapp}}}, \bibinfo {author}
  {\bibfnamefont {J.}~\bibnamefont {{Loveday}}}, \bibinfo {author}
  {\bibfnamefont {R.~H.}\ \bibnamefont {{Lupton}}}, \bibinfo {author}
  {\bibfnamefont {R.~C.}\ \bibnamefont {{Nichol}}}, \bibinfo {author}
  {\bibfnamefont {D.~P.}\ \bibnamefont {{Schneider}}}, \bibinfo {author}
  {\bibfnamefont {M.~A.}\ \bibnamefont {{Strauss}}}, \bibinfo {author}
  {\bibfnamefont {M.}~\bibnamefont {{Tegmark}}}, \ and\ \bibinfo {author}
  {\bibfnamefont {D.~G.}\ \bibnamefont {{York}}},\ }\href {\doibase
  10.1111/j.1365-2966.2007.11593.x} {\bibfield  {journal} {\bibinfo  {journal}
  {Monthly Notices of the Royal Astronomical Society}\ }\textbf {\bibinfo
  {volume} {378}},\ \bibinfo {pages} {852} (\bibinfo {year} {2007})},\ \Eprint
  {http://arxiv.org/abs/astro-ph/0605302} {astro-ph/0605302} \BibitemShut
  {NoStop}%
\bibitem [{\citenamefont {Scoccimarro}(2000)}]{monopole}%
  \BibitemOpen
  \bibfield  {author} {\bibinfo {author} {\bibfnamefont {R.}~\bibnamefont
  {Scoccimarro}},\ }\href {\doibase 10.1086/317248} {\bibfield  {journal}
  {\bibinfo  {journal} {The Astrophysical Journal}\ }\textbf {\bibinfo {volume}
  {544}},\ \bibinfo {pages} {597} (\bibinfo {year} {2000})}\BibitemShut
  {NoStop}%
\bibitem [{\citenamefont {{Manera}}\ \emph {et~al.}(2013)\citenamefont
  {{Manera}}, \citenamefont {{Scoccimarro}}, \citenamefont {{Percival}},
  \citenamefont {{Samushia}}, \citenamefont {{McBride}}, \citenamefont
  {{Ross}}, \citenamefont {{Sheth}}, \citenamefont {{White}}, \citenamefont
  {{Reid}}, \citenamefont {{S{\'a}nchez}}, \citenamefont {{de Putter}},
  \citenamefont {{Xu}}, \citenamefont {{Berlind}}, \citenamefont {{Brinkmann}},
  \citenamefont {{Maraston}}, \citenamefont {{Nichol}}, \citenamefont
  {{Montesano}}, \citenamefont {{Padmanabhan}}, \citenamefont {{Skibba}},
  \citenamefont {{Tojeiro}},\ and\ \citenamefont {{Weaver}}}]{mock1}%
  \BibitemOpen
  \bibfield  {author} {\bibinfo {author} {\bibfnamefont {M.}~\bibnamefont
  {{Manera}}}, \bibinfo {author} {\bibfnamefont {R.}~\bibnamefont
  {{Scoccimarro}}}, \bibinfo {author} {\bibfnamefont {W.~J.}\ \bibnamefont
  {{Percival}}}, \bibinfo {author} {\bibfnamefont {L.}~\bibnamefont
  {{Samushia}}}, \bibinfo {author} {\bibfnamefont {C.~K.}\ \bibnamefont
  {{McBride}}}, \bibinfo {author} {\bibfnamefont {A.~J.}\ \bibnamefont
  {{Ross}}}, \bibinfo {author} {\bibfnamefont {R.~K.}\ \bibnamefont {{Sheth}}},
  \bibinfo {author} {\bibfnamefont {M.}~\bibnamefont {{White}}}, \bibinfo
  {author} {\bibfnamefont {B.~A.}\ \bibnamefont {{Reid}}}, \bibinfo {author}
  {\bibfnamefont {A.~G.}\ \bibnamefont {{S{\'a}nchez}}}, \bibinfo {author}
  {\bibfnamefont {R.}~\bibnamefont {{de Putter}}}, \bibinfo {author}
  {\bibfnamefont {X.}~\bibnamefont {{Xu}}}, \bibinfo {author} {\bibfnamefont
  {A.~A.}\ \bibnamefont {{Berlind}}}, \bibinfo {author} {\bibfnamefont
  {J.}~\bibnamefont {{Brinkmann}}}, \bibinfo {author} {\bibfnamefont
  {C.}~\bibnamefont {{Maraston}}}, \bibinfo {author} {\bibfnamefont
  {B.}~\bibnamefont {{Nichol}}}, \bibinfo {author} {\bibfnamefont
  {F.}~\bibnamefont {{Montesano}}}, \bibinfo {author} {\bibfnamefont
  {N.}~\bibnamefont {{Padmanabhan}}}, \bibinfo {author} {\bibfnamefont {R.~A.}\
  \bibnamefont {{Skibba}}}, \bibinfo {author} {\bibfnamefont {R.}~\bibnamefont
  {{Tojeiro}}}, \ and\ \bibinfo {author} {\bibfnamefont {B.~A.}\ \bibnamefont
  {{Weaver}}},\ }\href {\doibase 10.1093/mnras/sts084} {\bibfield  {journal}
  {\bibinfo  {journal} {Monthly Notices of the Royal Astronomical Society}\
  }\textbf {\bibinfo {volume} {428}},\ \bibinfo {pages} {1036} (\bibinfo {year}
  {2013})},\ \Eprint {http://arxiv.org/abs/1203.6609} {arXiv:1203.6609}
  \BibitemShut {NoStop}%
\bibitem [{\citenamefont {{Kitaura}}\ \emph {et~al.}(2016)\citenamefont
  {{Kitaura}}, \citenamefont {{Rodr{\'{\i}}guez-Torres}}, \citenamefont
  {{Chuang}}, \citenamefont {{Zhao}}, \citenamefont {{Prada}}, \citenamefont
  {{Gil-Mar{\'{\i}}n}}, \citenamefont {{Guo}}, \citenamefont {{Yepes}},
  \citenamefont {{Klypin}}, \citenamefont {{Sc{\'o}ccola}}, \citenamefont
  {{Tinker}}, \citenamefont {{McBride}}, \citenamefont {{Reid}}, \citenamefont
  {{S{\'a}nchez}}, \citenamefont {{Salazar-Albornoz}}, \citenamefont {{Grieb}},
  \citenamefont {{Vargas-Magana}}, \citenamefont {{Cuesta}}, \citenamefont
  {{Neyrinck}}, \citenamefont {{Beutler}}, \citenamefont {{Comparat}},
  \citenamefont {{Percival}},\ and\ \citenamefont {{Ross}}}]{mock2}%
  \BibitemOpen
  \bibfield  {author} {\bibinfo {author} {\bibfnamefont {F.-S.}\ \bibnamefont
  {{Kitaura}}}, \bibinfo {author} {\bibfnamefont {S.}~\bibnamefont
  {{Rodr{\'{\i}}guez-Torres}}}, \bibinfo {author} {\bibfnamefont {C.-H.}\
  \bibnamefont {{Chuang}}}, \bibinfo {author} {\bibfnamefont {C.}~\bibnamefont
  {{Zhao}}}, \bibinfo {author} {\bibfnamefont {F.}~\bibnamefont {{Prada}}},
  \bibinfo {author} {\bibfnamefont {H.}~\bibnamefont {{Gil-Mar{\'{\i}}n}}},
  \bibinfo {author} {\bibfnamefont {H.}~\bibnamefont {{Guo}}}, \bibinfo
  {author} {\bibfnamefont {G.}~\bibnamefont {{Yepes}}}, \bibinfo {author}
  {\bibfnamefont {A.}~\bibnamefont {{Klypin}}}, \bibinfo {author}
  {\bibfnamefont {C.~G.}\ \bibnamefont {{Sc{\'o}ccola}}}, \bibinfo {author}
  {\bibfnamefont {J.}~\bibnamefont {{Tinker}}}, \bibinfo {author}
  {\bibfnamefont {C.}~\bibnamefont {{McBride}}}, \bibinfo {author}
  {\bibfnamefont {B.}~\bibnamefont {{Reid}}}, \bibinfo {author} {\bibfnamefont
  {A.~G.}\ \bibnamefont {{S{\'a}nchez}}}, \bibinfo {author} {\bibfnamefont
  {S.}~\bibnamefont {{Salazar-Albornoz}}}, \bibinfo {author} {\bibfnamefont
  {J.~N.}\ \bibnamefont {{Grieb}}}, \bibinfo {author} {\bibfnamefont
  {M.}~\bibnamefont {{Vargas-Magana}}}, \bibinfo {author} {\bibfnamefont
  {A.~J.}\ \bibnamefont {{Cuesta}}}, \bibinfo {author} {\bibfnamefont
  {M.}~\bibnamefont {{Neyrinck}}}, \bibinfo {author} {\bibfnamefont
  {F.}~\bibnamefont {{Beutler}}}, \bibinfo {author} {\bibfnamefont
  {J.}~\bibnamefont {{Comparat}}}, \bibinfo {author} {\bibfnamefont {W.~J.}\
  \bibnamefont {{Percival}}}, \ and\ \bibinfo {author} {\bibfnamefont
  {A.}~\bibnamefont {{Ross}}},\ }\href {\doibase 10.1093/mnras/stv2826}
  {\bibfield  {journal} {\bibinfo  {journal} {Monthly Notices of the Royal
  Astronomical Society}\ }\textbf {\bibinfo {volume} {456}},\ \bibinfo {pages}
  {4156} (\bibinfo {year} {2016})},\ \Eprint {http://arxiv.org/abs/1509.06400}
  {arXiv:1509.06400} \BibitemShut {NoStop}%
\bibitem [{\citenamefont {Vega-Ferrero}\ \emph {et~al.}(2017)\citenamefont
  {Vega-Ferrero}, \citenamefont {Yepes},\ and\ \citenamefont
  {Gottlöber}}]{triaxial1}%
  \BibitemOpen
  \bibfield  {author} {\bibinfo {author} {\bibfnamefont {J.}~\bibnamefont
  {Vega-Ferrero}}, \bibinfo {author} {\bibfnamefont {G.}~\bibnamefont {Yepes}},
  \ and\ \bibinfo {author} {\bibfnamefont {S.}~\bibnamefont {Gottlöber}},\
  }\href {\doibase 10.1093/mnras/stx282} {\bibfield  {journal} {\bibinfo
  {journal} {Monthly Notices of the Royal Astronomical Society}\ }\textbf
  {\bibinfo {volume} {467}},\ \bibinfo {pages} {3226} (\bibinfo {year}
  {2017})},\ \Eprint
  {http://arxiv.org/abs/http://oup.prod.sis.lan/mnras/article-pdf/467/3/3226/10875743/stx282.pdf}
  {http://oup.prod.sis.lan/mnras/article-pdf/467/3/3226/10875743/stx282.pdf}
  \BibitemShut {NoStop}%
\bibitem [{\citenamefont {Schneider}\ \emph {et~al.}(2012)\citenamefont
  {Schneider}, \citenamefont {Frenk},\ and\ \citenamefont {Cole}}]{triaxial2}%
  \BibitemOpen
  \bibfield  {author} {\bibinfo {author} {\bibfnamefont {M.~D.}\ \bibnamefont
  {Schneider}}, \bibinfo {author} {\bibfnamefont {C.~S.}\ \bibnamefont
  {Frenk}}, \ and\ \bibinfo {author} {\bibfnamefont {S.}~\bibnamefont {Cole}},\
  }\href {\doibase 10.1088/1475-7516/2012/05/030} {\bibfield  {journal}
  {\bibinfo  {journal} {Journal of Cosmology and Astroparticle Physics}\
  }\textbf {\bibinfo {volume} {2012}},\ \bibinfo {pages} {030} (\bibinfo {year}
  {2012})}\BibitemShut {NoStop}%
\bibitem [{\citenamefont {Vera-Ciro}\ \emph {et~al.}(2011)\citenamefont
  {Vera-Ciro}, \citenamefont {Sales}, \citenamefont {Helmi}, \citenamefont
  {Frenk}, \citenamefont {Navarro}, \citenamefont {Springel}, \citenamefont
  {Vogelsberger},\ and\ \citenamefont {White}}]{triaxial3}%
  \BibitemOpen
  \bibfield  {author} {\bibinfo {author} {\bibfnamefont {C.~A.}\ \bibnamefont
  {Vera-Ciro}}, \bibinfo {author} {\bibfnamefont {L.~V.}\ \bibnamefont
  {Sales}}, \bibinfo {author} {\bibfnamefont {A.}~\bibnamefont {Helmi}},
  \bibinfo {author} {\bibfnamefont {C.~S.}\ \bibnamefont {Frenk}}, \bibinfo
  {author} {\bibfnamefont {J.~F.}\ \bibnamefont {Navarro}}, \bibinfo {author}
  {\bibfnamefont {V.}~\bibnamefont {Springel}}, \bibinfo {author}
  {\bibfnamefont {M.}~\bibnamefont {Vogelsberger}}, \ and\ \bibinfo {author}
  {\bibfnamefont {S.~D.~M.}\ \bibnamefont {White}},\ }\href {\doibase
  10.1111/j.1365-2966.2011.19134.x} {\bibfield  {journal} {\bibinfo  {journal}
  {Monthly Notices of the Royal Astronomical Society}\ }\textbf {\bibinfo
  {volume} {416}},\ \bibinfo {pages} {1377} (\bibinfo {year} {2011})},\ \Eprint
  {http://arxiv.org/abs/http://oup.prod.sis.lan/mnras/article-pdf/416/2/1377/18594489/mnras0416-1377.pdf}
  {http://oup.prod.sis.lan/mnras/article-pdf/416/2/1377/18594489/mnras0416-1377.pdf}
  \BibitemShut {NoStop}%
\bibitem [{\citenamefont {{Angrick, C.}}\ and\ \citenamefont {{Bartelmann,
  M.}}(2010)}]{triaxial4}%
  \BibitemOpen
  \bibfield  {author} {\bibinfo {author} {\bibnamefont {{Angrick, C.}}}\ and\
  \bibinfo {author} {\bibnamefont {{Bartelmann, M.}}},\ }\href {\doibase
  10.1051/0004-6361/201014147} {\bibfield  {journal} {\bibinfo  {journal}
  {Astronomy \& Astrophysics}\ }\textbf {\bibinfo {volume} {518}},\ \bibinfo
  {pages} {A38} (\bibinfo {year} {2010})}\BibitemShut {NoStop}%
\bibitem [{\citenamefont {{Navarro}}\ \emph {et~al.}(1996)\citenamefont
  {{Navarro}}, \citenamefont {{Frenk}},\ and\ \citenamefont {{White}}}]{nfw}%
  \BibitemOpen
  \bibfield  {author} {\bibinfo {author} {\bibfnamefont {J.~F.}\ \bibnamefont
  {{Navarro}}}, \bibinfo {author} {\bibfnamefont {C.~S.}\ \bibnamefont
  {{Frenk}}}, \ and\ \bibinfo {author} {\bibfnamefont {S.~D.~M.}\ \bibnamefont
  {{White}}},\ }\href {\doibase 10.1086/177173} {\bibfield  {journal} {\bibinfo
   {journal} {The Astrophysical Journal}\ }\textbf {\bibinfo {volume} {462}},\
  \bibinfo {pages} {563} (\bibinfo {year} {1996})},\ \Eprint
  {http://arxiv.org/abs/astro-ph/9508025} {astro-ph/9508025} \BibitemShut
  {NoStop}%
\bibitem [{\citenamefont {{Bryan}}\ and\ \citenamefont
  {{Norman}}(1998)}]{virial}%
  \BibitemOpen
  \bibfield  {author} {\bibinfo {author} {\bibfnamefont {G.~L.}\ \bibnamefont
  {{Bryan}}}\ and\ \bibinfo {author} {\bibfnamefont {M.~L.}\ \bibnamefont
  {{Norman}}},\ }\href {\doibase 10.1086/305262} {\bibfield  {journal}
  {\bibinfo  {journal} {The Astrophysical Journal}\ }\textbf {\bibinfo {volume}
  {495}},\ \bibinfo {pages} {80} (\bibinfo {year} {1998})},\ \Eprint
  {http://arxiv.org/abs/astro-ph/9710107} {astro-ph/9710107} \BibitemShut
  {NoStop}%
\bibitem [{\citenamefont {Cooray}\ and\ \citenamefont
  {Sheth}(2002)}]{concentration}%
  \BibitemOpen
  \bibfield  {author} {\bibinfo {author} {\bibfnamefont {A.}~\bibnamefont
  {Cooray}}\ and\ \bibinfo {author} {\bibfnamefont {R.}~\bibnamefont {Sheth}},\
  }\href {\doibase https://doi.org/10.1016/S0370-1573(02)00276-4} {\bibfield
  {journal} {\bibinfo  {journal} {Physics Reports}\ }\textbf {\bibinfo {volume}
  {372}},\ \bibinfo {pages} {1 } (\bibinfo {year} {2002})}\BibitemShut
  {NoStop}%
\bibitem [{\citenamefont {Klypin}\ \emph {et~al.}(2011)\citenamefont {Klypin},
  \citenamefont {Trujillo-Gomez},\ and\ \citenamefont {Primack}}]{Klypin}%
  \BibitemOpen
  \bibfield  {author} {\bibinfo {author} {\bibfnamefont {A.~A.}\ \bibnamefont
  {Klypin}}, \bibinfo {author} {\bibfnamefont {S.}~\bibnamefont
  {Trujillo-Gomez}}, \ and\ \bibinfo {author} {\bibfnamefont {J.}~\bibnamefont
  {Primack}},\ }\href {\doibase 10.1088/0004-637x/740/2/102} {\bibfield
  {journal} {\bibinfo  {journal} {The Astrophysical Journal}\ }\textbf
  {\bibinfo {volume} {740}},\ \bibinfo {pages} {102} (\bibinfo {year}
  {2011})}\BibitemShut {NoStop}%
\bibitem [{\citenamefont {Jenkins}\ \emph {et~al.}(2004)\citenamefont
  {Jenkins}, \citenamefont {De~Lucia}, \citenamefont {White},\ and\
  \citenamefont {Gao}}]{subhalo1}%
  \BibitemOpen
  \bibfield  {author} {\bibinfo {author} {\bibfnamefont {A.}~\bibnamefont
  {Jenkins}}, \bibinfo {author} {\bibfnamefont {G.}~\bibnamefont {De~Lucia}},
  \bibinfo {author} {\bibfnamefont {S.~D.~M.}\ \bibnamefont {White}}, \ and\
  \bibinfo {author} {\bibfnamefont {L.}~\bibnamefont {Gao}},\ }\href {\doibase
  10.1111/j.1365-2966.2004.08098.x} {\bibfield  {journal} {\bibinfo  {journal}
  {Monthly Notices of the Royal Astronomical Society}\ }\textbf {\bibinfo
  {volume} {352}},\ \bibinfo {pages} {L1} (\bibinfo {year} {2004})}\BibitemShut
  {NoStop}%
\bibitem [{\citenamefont {Libeskind}\ \emph {et~al.}(2005)\citenamefont
  {Libeskind}, \citenamefont {Frenk}, \citenamefont {Cole}, \citenamefont
  {Helly}, \citenamefont {Jenkins}, \citenamefont {Navarro},\ and\
  \citenamefont {Power}}]{subhalo2}%
  \BibitemOpen
  \bibfield  {author} {\bibinfo {author} {\bibfnamefont {N.~I.}\ \bibnamefont
  {Libeskind}}, \bibinfo {author} {\bibfnamefont {C.~S.}\ \bibnamefont
  {Frenk}}, \bibinfo {author} {\bibfnamefont {S.}~\bibnamefont {Cole}},
  \bibinfo {author} {\bibfnamefont {J.~C.}\ \bibnamefont {Helly}}, \bibinfo
  {author} {\bibfnamefont {A.}~\bibnamefont {Jenkins}}, \bibinfo {author}
  {\bibfnamefont {J.~F.}\ \bibnamefont {Navarro}}, \ and\ \bibinfo {author}
  {\bibfnamefont {C.}~\bibnamefont {Power}},\ }\href {\doibase
  10.1111/j.1365-2966.2005.09425.x} {\bibfield  {journal} {\bibinfo  {journal}
  {Monthly Notices of the Royal Astronomical Society}\ }\textbf {\bibinfo
  {volume} {363}},\ \bibinfo {pages} {146} (\bibinfo {year}
  {2005})}\BibitemShut {NoStop}%
\bibitem [{\citenamefont {Moore}\ \emph {et~al.}(2004)\citenamefont {Moore},
  \citenamefont {Stadel},\ and\ \citenamefont {Diemand}}]{subhalo3}%
  \BibitemOpen
  \bibfield  {author} {\bibinfo {author} {\bibfnamefont {B.}~\bibnamefont
  {Moore}}, \bibinfo {author} {\bibfnamefont {J.}~\bibnamefont {Stadel}}, \
  and\ \bibinfo {author} {\bibfnamefont {J.}~\bibnamefont {Diemand}},\ }\href
  {\doibase 10.1111/j.1365-2966.2004.07940.x} {\bibfield  {journal} {\bibinfo
  {journal} {Monthly Notices of the Royal Astronomical Society}\ }\textbf
  {\bibinfo {volume} {352}},\ \bibinfo {pages} {535} (\bibinfo {year}
  {2004})}\BibitemShut {NoStop}%
\bibitem [{\citenamefont {Gao}\ \emph {et~al.}(2004)\citenamefont {Gao},
  \citenamefont {White}, \citenamefont {Jenkins}, \citenamefont {Stoehr},\ and\
  \citenamefont {Springel}}]{subhalo4}%
  \BibitemOpen
  \bibfield  {author} {\bibinfo {author} {\bibfnamefont {L.}~\bibnamefont
  {Gao}}, \bibinfo {author} {\bibfnamefont {S.~D.~M.}\ \bibnamefont {White}},
  \bibinfo {author} {\bibfnamefont {A.}~\bibnamefont {Jenkins}}, \bibinfo
  {author} {\bibfnamefont {F.}~\bibnamefont {Stoehr}}, \ and\ \bibinfo {author}
  {\bibfnamefont {V.}~\bibnamefont {Springel}},\ }\href {\doibase
  10.1111/j.1365-2966.2004.08360.x} {\bibfield  {journal} {\bibinfo  {journal}
  {Monthly Notices of the Royal Astronomical Society}\ }\textbf {\bibinfo
  {volume} {355}},\ \bibinfo {pages} {819} (\bibinfo {year}
  {2004})}\BibitemShut {NoStop}%
\bibitem [{\citenamefont {Frenk}\ \emph {et~al.}(2016)\citenamefont {Frenk},
  \citenamefont {Han}, \citenamefont {Cole},\ and\ \citenamefont
  {Jing}}]{subgen}%
  \BibitemOpen
  \bibfield  {author} {\bibinfo {author} {\bibfnamefont {C.~S.}\ \bibnamefont
  {Frenk}}, \bibinfo {author} {\bibfnamefont {J.}~\bibnamefont {Han}}, \bibinfo
  {author} {\bibfnamefont {S.}~\bibnamefont {Cole}}, \ and\ \bibinfo {author}
  {\bibfnamefont {Y.}~\bibnamefont {Jing}},\ }\href {\doibase
  10.1093/mnras/stv2900} {\bibfield  {journal} {\bibinfo  {journal} {Monthly
  Notices of the Royal Astronomical Society}\ }\textbf {\bibinfo {volume}
  {457}},\ \bibinfo {pages} {1208} (\bibinfo {year} {2016})},\ \Eprint
  {http://arxiv.org/abs/http://oup.prod.sis.lan/mnras/article-pdf/457/2/1208/2882862/stv2900.pdf}
  {http://oup.prod.sis.lan/mnras/article-pdf/457/2/1208/2882862/stv2900.pdf}
  \BibitemShut {NoStop}%
\bibitem [{\citenamefont {{Peacock}}\ and\ \citenamefont
  {{Smith}}(2000)}]{hod1}%
  \BibitemOpen
  \bibfield  {author} {\bibinfo {author} {\bibfnamefont {J.~A.}\ \bibnamefont
  {{Peacock}}}\ and\ \bibinfo {author} {\bibfnamefont {R.~E.}\ \bibnamefont
  {{Smith}}},\ }\href {\doibase 10.1046/j.1365-8711.2000.03779.x} {\bibfield
  {journal} {\bibinfo  {journal} {Monthly Notices of the Royal Astronomical
  Society}\ }\textbf {\bibinfo {volume} {318}},\ \bibinfo {pages} {1144}
  (\bibinfo {year} {2000})},\ \Eprint {http://arxiv.org/abs/astro-ph/0005010}
  {astro-ph/0005010} \BibitemShut {NoStop}%
\bibitem [{\citenamefont {{Scoccimarro}}\ \emph {et~al.}(2001)\citenamefont
  {{Scoccimarro}}, \citenamefont {{Sheth}}, \citenamefont {{Hui}},\ and\
  \citenamefont {{Jain}}}]{hod2}%
  \BibitemOpen
  \bibfield  {author} {\bibinfo {author} {\bibfnamefont {R.}~\bibnamefont
  {{Scoccimarro}}}, \bibinfo {author} {\bibfnamefont {R.~K.}\ \bibnamefont
  {{Sheth}}}, \bibinfo {author} {\bibfnamefont {L.}~\bibnamefont {{Hui}}}, \
  and\ \bibinfo {author} {\bibfnamefont {B.}~\bibnamefont {{Jain}}},\ }\href
  {\doibase 10.1086/318261} {\bibfield  {journal} {\bibinfo  {journal} {The
  Astrophysical Journal}\ }\textbf {\bibinfo {volume} {546}},\ \bibinfo {pages}
  {20} (\bibinfo {year} {2001})},\ \Eprint
  {http://arxiv.org/abs/astro-ph/0006319} {astro-ph/0006319} \BibitemShut
  {NoStop}%
\bibitem [{\citenamefont {{Berlind}}\ and\ \citenamefont
  {{Weinberg}}(2002)}]{hod3}%
  \BibitemOpen
  \bibfield  {author} {\bibinfo {author} {\bibfnamefont {A.~A.}\ \bibnamefont
  {{Berlind}}}\ and\ \bibinfo {author} {\bibfnamefont {D.~H.}\ \bibnamefont
  {{Weinberg}}},\ }\href {\doibase 10.1086/341469} {\bibfield  {journal}
  {\bibinfo  {journal} {The Astrophysical Journal}\ }\textbf {\bibinfo {volume}
  {575}},\ \bibinfo {pages} {587} (\bibinfo {year} {2002})},\ \Eprint
  {http://arxiv.org/abs/astro-ph/0109001} {astro-ph/0109001} \BibitemShut
  {NoStop}%
\bibitem [{\citenamefont {Zheng}\ \emph {et~al.}(2009)\citenamefont {Zheng},
  \citenamefont {Zehavi}, \citenamefont {Eisenstein}, \citenamefont
  {Weinberg},\ and\ \citenamefont {Jing}}]{zheng}%
  \BibitemOpen
  \bibfield  {author} {\bibinfo {author} {\bibfnamefont {Z.}~\bibnamefont
  {Zheng}}, \bibinfo {author} {\bibfnamefont {I.}~\bibnamefont {Zehavi}},
  \bibinfo {author} {\bibfnamefont {D.~J.}\ \bibnamefont {Eisenstein}},
  \bibinfo {author} {\bibfnamefont {D.~H.}\ \bibnamefont {Weinberg}}, \ and\
  \bibinfo {author} {\bibfnamefont {Y.~P.}\ \bibnamefont {Jing}},\ }\href
  {\doibase 10.1088/0004-637x/707/1/554} {\bibfield  {journal} {\bibinfo
  {journal} {The Astrophysical Journal}\ }\textbf {\bibinfo {volume} {707}},\
  \bibinfo {pages} {554} (\bibinfo {year} {2009})}\BibitemShut {NoStop}%
\bibitem [{\citenamefont {{Anderson}}\ \emph {et~al.}(2014)\citenamefont
  {{Anderson}}, \citenamefont {{Aubourg}}, \citenamefont {{Bailey}},
  \citenamefont {{Beutler}}, \citenamefont {{Bhardwaj}}, \citenamefont
  {{Blanton}}, \citenamefont {{Bolton}}, \citenamefont {{Brinkmann}},
  \citenamefont {{Brownstein}}, \citenamefont {{Burden}}, \citenamefont
  {{Chuang}}, \citenamefont {{Cuesta}}, \citenamefont {{Dawson}}, \citenamefont
  {{Eisenstein}}, \citenamefont {{Escoffier}}, \citenamefont {{Gunn}},
  \citenamefont {{Guo}}, \citenamefont {{Ho}}, \citenamefont {{Honscheid}},
  \citenamefont {{Howlett}}, \citenamefont {{Kirkby}}, \citenamefont
  {{Lupton}}, \citenamefont {{Manera}}, \citenamefont {{Maraston}},
  \citenamefont {{McBride}}, \citenamefont {{Mena}}, \citenamefont
  {{Montesano}}, \citenamefont {{Nichol}}, \citenamefont {{Nuza}},
  \citenamefont {{Olmstead}}, \citenamefont {{Padmanabhan}}, \citenamefont
  {{Palanque-Delabrouille}}, \citenamefont {{Parejko}}, \citenamefont
  {{Percival}}, \citenamefont {{Petitjean}}, \citenamefont {{Prada}},
  \citenamefont {{Price-Whelan}}, \citenamefont {{Reid}}, \citenamefont
  {{Roe}}, \citenamefont {{Ross}}, \citenamefont {{Ross}}, \citenamefont
  {{Sabiu}}, \citenamefont {{Saito}}, \citenamefont {{Samushia}}, \citenamefont
  {{S{\'a}nchez}}, \citenamefont {{Schlegel}}, \citenamefont {{Schneider}},
  \citenamefont {{Scoccola}}, \citenamefont {{Seo}}, \citenamefont {{Skibba}},
  \citenamefont {{Strauss}}, \citenamefont {{Swanson}}, \citenamefont
  {{Thomas}}, \citenamefont {{Tinker}}, \citenamefont {{Tojeiro}},
  \citenamefont {{Maga{\~n}a}}, \citenamefont {{Verde}}, \citenamefont
  {{Wake}}, \citenamefont {{Weaver}}, \citenamefont {{Weinberg}}, \citenamefont
  {{White}}, \citenamefont {{Xu}}, \citenamefont {{Y{\`e}che}}, \citenamefont
  {{Zehavi}},\ and\ \citenamefont {{Zhao}}}]{anderson}%
  \BibitemOpen
  \bibfield  {author} {\bibinfo {author} {\bibfnamefont {L.}~\bibnamefont
  {{Anderson}}}, \bibinfo {author} {\bibfnamefont {{\'E}.}~\bibnamefont
  {{Aubourg}}}, \bibinfo {author} {\bibfnamefont {S.}~\bibnamefont {{Bailey}}},
  \bibinfo {author} {\bibfnamefont {F.}~\bibnamefont {{Beutler}}}, \bibinfo
  {author} {\bibfnamefont {V.}~\bibnamefont {{Bhardwaj}}}, \bibinfo {author}
  {\bibfnamefont {M.}~\bibnamefont {{Blanton}}}, \bibinfo {author}
  {\bibfnamefont {A.~S.}\ \bibnamefont {{Bolton}}}, \bibinfo {author}
  {\bibfnamefont {J.}~\bibnamefont {{Brinkmann}}}, \bibinfo {author}
  {\bibfnamefont {J.~R.}\ \bibnamefont {{Brownstein}}}, \bibinfo {author}
  {\bibfnamefont {A.}~\bibnamefont {{Burden}}}, \bibinfo {author}
  {\bibfnamefont {C.-H.}\ \bibnamefont {{Chuang}}}, \bibinfo {author}
  {\bibfnamefont {A.~J.}\ \bibnamefont {{Cuesta}}}, \bibinfo {author}
  {\bibfnamefont {K.~S.}\ \bibnamefont {{Dawson}}}, \bibinfo {author}
  {\bibfnamefont {D.~J.}\ \bibnamefont {{Eisenstein}}}, \bibinfo {author}
  {\bibfnamefont {S.}~\bibnamefont {{Escoffier}}}, \bibinfo {author}
  {\bibfnamefont {J.~E.}\ \bibnamefont {{Gunn}}}, \bibinfo {author}
  {\bibfnamefont {H.}~\bibnamefont {{Guo}}}, \bibinfo {author} {\bibfnamefont
  {S.}~\bibnamefont {{Ho}}}, \bibinfo {author} {\bibfnamefont {K.}~\bibnamefont
  {{Honscheid}}}, \bibinfo {author} {\bibfnamefont {C.}~\bibnamefont
  {{Howlett}}}, \bibinfo {author} {\bibfnamefont {D.}~\bibnamefont {{Kirkby}}},
  \bibinfo {author} {\bibfnamefont {R.~H.}\ \bibnamefont {{Lupton}}}, \bibinfo
  {author} {\bibfnamefont {M.}~\bibnamefont {{Manera}}}, \bibinfo {author}
  {\bibfnamefont {C.}~\bibnamefont {{Maraston}}}, \bibinfo {author}
  {\bibfnamefont {C.~K.}\ \bibnamefont {{McBride}}}, \bibinfo {author}
  {\bibfnamefont {O.}~\bibnamefont {{Mena}}}, \bibinfo {author} {\bibfnamefont
  {F.}~\bibnamefont {{Montesano}}}, \bibinfo {author} {\bibfnamefont {R.~C.}\
  \bibnamefont {{Nichol}}}, \bibinfo {author} {\bibfnamefont {S.~E.}\
  \bibnamefont {{Nuza}}}, \bibinfo {author} {\bibfnamefont {M.~D.}\
  \bibnamefont {{Olmstead}}}, \bibinfo {author} {\bibfnamefont
  {N.}~\bibnamefont {{Padmanabhan}}}, \bibinfo {author} {\bibfnamefont
  {N.}~\bibnamefont {{Palanque-Delabrouille}}}, \bibinfo {author}
  {\bibfnamefont {J.}~\bibnamefont {{Parejko}}}, \bibinfo {author}
  {\bibfnamefont {W.~J.}\ \bibnamefont {{Percival}}}, \bibinfo {author}
  {\bibfnamefont {P.}~\bibnamefont {{Petitjean}}}, \bibinfo {author}
  {\bibfnamefont {F.}~\bibnamefont {{Prada}}}, \bibinfo {author} {\bibfnamefont
  {A.~M.}\ \bibnamefont {{Price-Whelan}}}, \bibinfo {author} {\bibfnamefont
  {B.}~\bibnamefont {{Reid}}}, \bibinfo {author} {\bibfnamefont {N.~A.}\
  \bibnamefont {{Roe}}}, \bibinfo {author} {\bibfnamefont {A.~J.}\ \bibnamefont
  {{Ross}}}, \bibinfo {author} {\bibfnamefont {N.~P.}\ \bibnamefont {{Ross}}},
  \bibinfo {author} {\bibfnamefont {C.~G.}\ \bibnamefont {{Sabiu}}}, \bibinfo
  {author} {\bibfnamefont {S.}~\bibnamefont {{Saito}}}, \bibinfo {author}
  {\bibfnamefont {L.}~\bibnamefont {{Samushia}}}, \bibinfo {author}
  {\bibfnamefont {A.~G.}\ \bibnamefont {{S{\'a}nchez}}}, \bibinfo {author}
  {\bibfnamefont {D.~J.}\ \bibnamefont {{Schlegel}}}, \bibinfo {author}
  {\bibfnamefont {D.~P.}\ \bibnamefont {{Schneider}}}, \bibinfo {author}
  {\bibfnamefont {C.~G.}\ \bibnamefont {{Scoccola}}}, \bibinfo {author}
  {\bibfnamefont {H.-J.}\ \bibnamefont {{Seo}}}, \bibinfo {author}
  {\bibfnamefont {R.~A.}\ \bibnamefont {{Skibba}}}, \bibinfo {author}
  {\bibfnamefont {M.~A.}\ \bibnamefont {{Strauss}}}, \bibinfo {author}
  {\bibfnamefont {M.~E.~C.}\ \bibnamefont {{Swanson}}}, \bibinfo {author}
  {\bibfnamefont {D.}~\bibnamefont {{Thomas}}}, \bibinfo {author}
  {\bibfnamefont {J.~L.}\ \bibnamefont {{Tinker}}}, \bibinfo {author}
  {\bibfnamefont {R.}~\bibnamefont {{Tojeiro}}}, \bibinfo {author}
  {\bibfnamefont {M.~V.}\ \bibnamefont {{Maga{\~n}a}}}, \bibinfo {author}
  {\bibfnamefont {L.}~\bibnamefont {{Verde}}}, \bibinfo {author} {\bibfnamefont
  {D.~A.}\ \bibnamefont {{Wake}}}, \bibinfo {author} {\bibfnamefont {B.~A.}\
  \bibnamefont {{Weaver}}}, \bibinfo {author} {\bibfnamefont {D.~H.}\
  \bibnamefont {{Weinberg}}}, \bibinfo {author} {\bibfnamefont
  {M.}~\bibnamefont {{White}}}, \bibinfo {author} {\bibfnamefont
  {X.}~\bibnamefont {{Xu}}}, \bibinfo {author} {\bibfnamefont {C.}~\bibnamefont
  {{Y{\`e}che}}}, \bibinfo {author} {\bibfnamefont {I.}~\bibnamefont
  {{Zehavi}}}, \ and\ \bibinfo {author} {\bibfnamefont {G.-B.}\ \bibnamefont
  {{Zhao}}},\ }\href {\doibase 10.1093/mnras/stu523} {\bibfield  {journal}
  {\bibinfo  {journal} {Monthly Notices of the Royal Astronomical Society}\
  }\textbf {\bibinfo {volume} {441}},\ \bibinfo {pages} {24} (\bibinfo {year}
  {2014})},\ \Eprint {http://arxiv.org/abs/1312.4877} {arXiv:1312.4877}
  \BibitemShut {NoStop}%
\bibitem [{\citenamefont {{Gottl{\"o}ber}}\ \emph {et~al.}(1999)\citenamefont
  {{Gottl{\"o}ber}}, \citenamefont {{Klypin}},\ and\ \citenamefont
  {{Kravtsov}}}]{halo_evo}%
  \BibitemOpen
  \bibfield  {author} {\bibinfo {author} {\bibfnamefont {S.}~\bibnamefont
  {{Gottl{\"o}ber}}}, \bibinfo {author} {\bibfnamefont {A.~A.}\ \bibnamefont
  {{Klypin}}}, \ and\ \bibinfo {author} {\bibfnamefont {A.~V.}\ \bibnamefont
  {{Kravtsov}}},\ }in\ \href@noop {} {\emph {\bibinfo {booktitle}
  {Observational Cosmology: The Development of Galaxy Systems}}},\ \bibinfo
  {series} {Astronomical Society of the Pacific Conference Series}, Vol.\
  \bibinfo {volume} {176},\ \bibinfo {editor} {edited by\ \bibinfo {editor}
  {\bibfnamefont {G.}~\bibnamefont {{Giuricin}}}, \bibinfo {editor}
  {\bibfnamefont {M.}~\bibnamefont {{Mezzetti}}}, \ and\ \bibinfo {editor}
  {\bibfnamefont {P.}~\bibnamefont {{Salucci}}}}\ (\bibinfo {year} {1999})\ p.\
  \bibinfo {pages} {418},\ \Eprint {http://arxiv.org/abs/astro-ph/9810445}
  {arXiv:astro-ph/9810445 [astro-ph]} \BibitemShut {NoStop}%
\bibitem [{\citenamefont {Kravtsov}\ and\ \citenamefont
  {Klypin}(1999)}]{halo_evo1}%
  \BibitemOpen
  \bibfield  {author} {\bibinfo {author} {\bibfnamefont {A.~V.}\ \bibnamefont
  {Kravtsov}}\ and\ \bibinfo {author} {\bibfnamefont {A.~A.}\ \bibnamefont
  {Klypin}},\ }\href {\doibase 10.1086/307495} {\bibfield  {journal} {\bibinfo
  {journal} {The Astrophysical Journal}\ }\textbf {\bibinfo {volume} {520}},\
  \bibinfo {pages} {437} (\bibinfo {year} {1999})}\BibitemShut {NoStop}%
\bibitem [{\citenamefont {Mann}\ \emph {et~al.}(1998)\citenamefont {Mann},
  \citenamefont {Peacock},\ and\ \citenamefont {Heavens}}]{linear_bias}%
  \BibitemOpen
  \bibfield  {author} {\bibinfo {author} {\bibfnamefont {R.~G.}\ \bibnamefont
  {Mann}}, \bibinfo {author} {\bibfnamefont {J.~A.}\ \bibnamefont {Peacock}}, \
  and\ \bibinfo {author} {\bibfnamefont {A.~F.}\ \bibnamefont {Heavens}},\
  }\href {\doibase 10.1046/j.1365-8711.1998.01053.x} {\bibfield  {journal}
  {\bibinfo  {journal} {Monthly Notices of the Royal Astronomical Society}\
  }\textbf {\bibinfo {volume} {293}},\ \bibinfo {pages} {209} (\bibinfo {year}
  {1998})},\ \Eprint
  {http://arxiv.org/abs/http://oup.prod.sis.lan/mnras/article-pdf/293/3/209/3881609/293-3-209.pdf}
  {http://oup.prod.sis.lan/mnras/article-pdf/293/3/209/3881609/293-3-209.pdf}
  \BibitemShut {NoStop}%
\bibitem [{\citenamefont {Ahrens}\ \emph {et~al.}(2005)\citenamefont {Ahrens},
  \citenamefont {Geveci},\ and\ \citenamefont {Law}}]{paraview}%
  \BibitemOpen
  \bibfield  {author} {\bibinfo {author} {\bibfnamefont {J.}~\bibnamefont
  {Ahrens}}, \bibinfo {author} {\bibfnamefont {B.}~\bibnamefont {Geveci}}, \
  and\ \bibinfo {author} {\bibfnamefont {C.}~\bibnamefont {Law}},\ }\href@noop
  {} {\emph {\bibinfo {title} {ParaView: An End-User Tool for Large Data
  Visualization.}}}\ (\bibinfo  {publisher} {Visualization Handbook,
  Elsevier},\ \bibinfo {year} {2005})\BibitemShut {NoStop}%
\bibitem [{\citenamefont {Seery}\ \emph {et~al.}(2017)\citenamefont {Seery},
  \citenamefont {Smith}, \citenamefont {Eggemeier}, \citenamefont {Regan},\
  and\ \citenamefont {Byun}}]{sussex}%
  \BibitemOpen
  \bibfield  {author} {\bibinfo {author} {\bibfnamefont {D.}~\bibnamefont
  {Seery}}, \bibinfo {author} {\bibfnamefont {R.~E.}\ \bibnamefont {Smith}},
  \bibinfo {author} {\bibfnamefont {A.}~\bibnamefont {Eggemeier}}, \bibinfo
  {author} {\bibfnamefont {D.}~\bibnamefont {Regan}}, \ and\ \bibinfo {author}
  {\bibfnamefont {J.}~\bibnamefont {Byun}},\ }\href {\doibase
  10.1093/mnras/stx1681} {\bibfield  {journal} {\bibinfo  {journal} {Monthly
  Notices of the Royal Astronomical Society}\ }\textbf {\bibinfo {volume}
  {471}},\ \bibinfo {pages} {1581} (\bibinfo {year} {2017})},\ \Eprint
  {http://arxiv.org/abs/http://oup.prod.sis.lan/mnras/article-pdf/471/2/1581/19407307/stx1681.pdf}
  {http://oup.prod.sis.lan/mnras/article-pdf/471/2/1581/19407307/stx1681.pdf}
  \BibitemShut {NoStop}%
\bibitem [{\citenamefont {Lazanu}\ \emph {et~al.}(2017)\citenamefont {Lazanu},
  \citenamefont {Giannantonio}, \citenamefont {Schmittfull},\ and\
  \citenamefont {Shellard}}]{Andrei2}%
  \BibitemOpen
  \bibfield  {author} {\bibinfo {author} {\bibfnamefont {A.}~\bibnamefont
  {Lazanu}}, \bibinfo {author} {\bibfnamefont {T.}~\bibnamefont
  {Giannantonio}}, \bibinfo {author} {\bibfnamefont {M.}~\bibnamefont
  {Schmittfull}}, \ and\ \bibinfo {author} {\bibfnamefont {E.~P.~S.}\
  \bibnamefont {Shellard}},\ }\href {\doibase 10.1103/PhysRevD.95.083511}
  {\bibfield  {journal} {\bibinfo  {journal} {Phys. Rev. D}\ }\textbf {\bibinfo
  {volume} {95}},\ \bibinfo {pages} {083511} (\bibinfo {year}
  {2017})}\BibitemShut {NoStop}%
\bibitem [{\citenamefont {{Bouchet}}\ \emph {et~al.}(1995)\citenamefont
  {{Bouchet}}, \citenamefont {{Colombi}}, \citenamefont {{Hivon}},\ and\
  \citenamefont {{Juszkiewicz}}}]{bouchet}%
  \BibitemOpen
  \bibfield  {author} {\bibinfo {author} {\bibfnamefont {F.~R.}\ \bibnamefont
  {{Bouchet}}}, \bibinfo {author} {\bibfnamefont {S.}~\bibnamefont
  {{Colombi}}}, \bibinfo {author} {\bibfnamefont {E.}~\bibnamefont {{Hivon}}},
  \ and\ \bibinfo {author} {\bibfnamefont {R.}~\bibnamefont {{Juszkiewicz}}},\
  }\href@noop {} {\bibfield  {journal} {\bibinfo  {journal} {A \& A}\ }\textbf
  {\bibinfo {volume} {296}},\ \bibinfo {pages} {575} (\bibinfo {year}
  {1995})},\ \Eprint {http://arxiv.org/abs/astro-ph/9406013} {astro-ph/9406013}
  \BibitemShut {NoStop}%
\bibitem [{\citenamefont {Desjacques}\ \emph {et~al.}(2018)\citenamefont
  {Desjacques}, \citenamefont {Jeong},\ and\ \citenamefont
  {Schmidt}}]{Desjacques}%
  \BibitemOpen
  \bibfield  {author} {\bibinfo {author} {\bibfnamefont {V.}~\bibnamefont
  {Desjacques}}, \bibinfo {author} {\bibfnamefont {D.}~\bibnamefont {Jeong}}, \
  and\ \bibinfo {author} {\bibfnamefont {F.}~\bibnamefont {Schmidt}},\ }\href
  {\doibase https://doi.org/10.1016/j.physrep.2017.12.002} {\bibfield
  {journal} {\bibinfo  {journal} {Physics Reports}\ }\textbf {\bibinfo {volume}
  {733}},\ \bibinfo {pages} {1 } (\bibinfo {year} {2018})},\ \bibinfo {note}
  {large-scale galaxy bias}\BibitemShut {NoStop}%
\bibitem [{\citenamefont {Pace}\ \emph {et~al.}(2015)\citenamefont {Pace},
  \citenamefont {Manera}, \citenamefont {Bacon}, \citenamefont {Crittenden},\
  and\ \citenamefont {Percival}}]{pace}%
  \BibitemOpen
  \bibfield  {author} {\bibinfo {author} {\bibfnamefont {F.}~\bibnamefont
  {Pace}}, \bibinfo {author} {\bibfnamefont {M.}~\bibnamefont {Manera}},
  \bibinfo {author} {\bibfnamefont {D.~J.}\ \bibnamefont {Bacon}}, \bibinfo
  {author} {\bibfnamefont {R.}~\bibnamefont {Crittenden}}, \ and\ \bibinfo
  {author} {\bibfnamefont {W.~J.}\ \bibnamefont {Percival}},\ }\href {\doibase
  10.1093/mnras/stv2019} {\bibfield  {journal} {\bibinfo  {journal} {Monthly
  Notices of the Royal Astronomical Society}\ }\textbf {\bibinfo {volume}
  {454}},\ \bibinfo {pages} {708} (\bibinfo {year} {2015})},\ \Eprint
  {http://arxiv.org/abs/http://oup.prod.sis.lan/mnras/article-pdf/454/1/708/3938653/stv2019.pdf}
  {http://oup.prod.sis.lan/mnras/article-pdf/454/1/708/3938653/stv2019.pdf}
  \BibitemShut {NoStop}%
\bibitem [{\citenamefont {Smith}\ \emph {et~al.}(2008)\citenamefont {Smith},
  \citenamefont {Sheth},\ and\ \citenamefont {Scoccimarro}}]{triaxial}%
  \BibitemOpen
  \bibfield  {author} {\bibinfo {author} {\bibfnamefont {R.~E.}\ \bibnamefont
  {Smith}}, \bibinfo {author} {\bibfnamefont {R.~K.}\ \bibnamefont {Sheth}}, \
  and\ \bibinfo {author} {\bibfnamefont {R.}~\bibnamefont {Scoccimarro}},\
  }\href {\doibase 10.1103/PhysRevD.78.023523} {\bibfield  {journal} {\bibinfo
  {journal} {Phys. Rev. D}\ }\textbf {\bibinfo {volume} {78}},\ \bibinfo
  {pages} {023523} (\bibinfo {year} {2008})}\BibitemShut {NoStop}%
\bibitem [{\citenamefont {Vakili}\ and\ \citenamefont
  {Hahn}(2019)}]{assembly_bias}%
  \BibitemOpen
  \bibfield  {author} {\bibinfo {author} {\bibfnamefont {M.}~\bibnamefont
  {Vakili}}\ and\ \bibinfo {author} {\bibfnamefont {C.}~\bibnamefont {Hahn}},\
  }\href {\doibase 10.3847/1538-4357/aaf1a1} {\bibfield  {journal} {\bibinfo
  {journal} {The Astrophysical Journal}\ }\textbf {\bibinfo {volume} {872}},\
  \bibinfo {pages} {115} (\bibinfo {year} {2019})}\BibitemShut {NoStop}%
\bibitem [{\citenamefont {{Gao}}\ \emph {et~al.}(2005)\citenamefont {{Gao}},
  \citenamefont {{Springel}},\ and\ \citenamefont {{White}}}]{Gao2005}%
  \BibitemOpen
  \bibfield  {author} {\bibinfo {author} {\bibfnamefont {L.}~\bibnamefont
  {{Gao}}}, \bibinfo {author} {\bibfnamefont {V.}~\bibnamefont {{Springel}}}, \
  and\ \bibinfo {author} {\bibfnamefont {S.~D.~M.}\ \bibnamefont {{White}}},\
  }\href {\doibase 10.1111/j.1745-3933.2005.00084.x} {\bibfield  {journal}
  {\bibinfo  {journal} {Monthly Notices of the Royal Astronomical Society}\
  }\textbf {\bibinfo {volume} {363}},\ \bibinfo {pages} {L66} (\bibinfo {year}
  {2005})},\ \Eprint {http://arxiv.org/abs/astro-ph/0506510} {astro-ph/0506510}
  \BibitemShut {NoStop}%
\bibitem [{\citenamefont {{Sunayama}}\ \emph {et~al.}(2016)\citenamefont
  {{Sunayama}}, \citenamefont {{Hearin}}, \citenamefont {{Padmanabhan}},\ and\
  \citenamefont {{Leauthaud}}}]{Sunayama2016}%
  \BibitemOpen
  \bibfield  {author} {\bibinfo {author} {\bibfnamefont {T.}~\bibnamefont
  {{Sunayama}}}, \bibinfo {author} {\bibfnamefont {A.~P.}\ \bibnamefont
  {{Hearin}}}, \bibinfo {author} {\bibfnamefont {N.}~\bibnamefont
  {{Padmanabhan}}}, \ and\ \bibinfo {author} {\bibfnamefont {A.}~\bibnamefont
  {{Leauthaud}}},\ }\href {\doibase 10.1093/mnras/stw332} {\bibfield  {journal}
  {\bibinfo  {journal} {Monthly Notices of the Royal Astronomical Society}\
  }\textbf {\bibinfo {volume} {458}},\ \bibinfo {pages} {1510} (\bibinfo {year}
  {2016})},\ \Eprint {http://arxiv.org/abs/1509.06417} {arXiv:1509.06417}
  \BibitemShut {NoStop}%
\bibitem [{\citenamefont {{Hearin}}\ \emph {et~al.}(2016)\citenamefont
  {{Hearin}}, \citenamefont {{Zentner}}, \citenamefont {{van den Bosch}},
  \citenamefont {{Campbell}},\ and\ \citenamefont {{Tollerud}}}]{Hearin2016}%
  \BibitemOpen
  \bibfield  {author} {\bibinfo {author} {\bibfnamefont {A.~P.}\ \bibnamefont
  {{Hearin}}}, \bibinfo {author} {\bibfnamefont {A.~R.}\ \bibnamefont
  {{Zentner}}}, \bibinfo {author} {\bibfnamefont {F.~C.}\ \bibnamefont {{van
  den Bosch}}}, \bibinfo {author} {\bibfnamefont {D.}~\bibnamefont
  {{Campbell}}}, \ and\ \bibinfo {author} {\bibfnamefont {E.}~\bibnamefont
  {{Tollerud}}},\ }\href {\doibase 10.1093/mnras/stw840} {\bibfield  {journal}
  {\bibinfo  {journal} {Monthly Notices of the Royal Astronomical Society}\
  }\textbf {\bibinfo {volume} {460}},\ \bibinfo {pages} {2552} (\bibinfo {year}
  {2016})},\ \Eprint {http://arxiv.org/abs/1512.03050} {arXiv:1512.03050}
  \BibitemShut {NoStop}%
\bibitem [{\citenamefont {Wechsler}\ \emph {et~al.}(2006)\citenamefont
  {Wechsler}, \citenamefont {Zentner}, \citenamefont {Bullock}, \citenamefont
  {Kravtsov},\ and\ \citenamefont {Allgood}}]{Wechsler2006}%
  \BibitemOpen
  \bibfield  {author} {\bibinfo {author} {\bibfnamefont {R.~H.}\ \bibnamefont
  {Wechsler}}, \bibinfo {author} {\bibfnamefont {A.~R.}\ \bibnamefont
  {Zentner}}, \bibinfo {author} {\bibfnamefont {J.~S.}\ \bibnamefont
  {Bullock}}, \bibinfo {author} {\bibfnamefont {A.~V.}\ \bibnamefont
  {Kravtsov}}, \ and\ \bibinfo {author} {\bibfnamefont {B.}~\bibnamefont
  {Allgood}},\ }\href {\doibase 10.1086/507120} {\bibfield  {journal} {\bibinfo
   {journal} {The Astrophysical Journal}\ }\textbf {\bibinfo {volume} {652}},\
  \bibinfo {pages} {71} (\bibinfo {year} {2006})}\BibitemShut {NoStop}%
\bibitem [{\citenamefont {{Zhao}}\ \emph {et~al.}(2003)\citenamefont {{Zhao}},
  \citenamefont {{Mo}}, \citenamefont {{Jing}},\ and\ \citenamefont
  {{B{\"o}rner}}}]{Zhao2003}%
  \BibitemOpen
  \bibfield  {author} {\bibinfo {author} {\bibfnamefont {D.~H.}\ \bibnamefont
  {{Zhao}}}, \bibinfo {author} {\bibfnamefont {H.~J.}\ \bibnamefont {{Mo}}},
  \bibinfo {author} {\bibfnamefont {Y.~P.}\ \bibnamefont {{Jing}}}, \ and\
  \bibinfo {author} {\bibfnamefont {G.}~\bibnamefont {{B{\"o}rner}}},\ }\href
  {\doibase 10.1046/j.1365-8711.2003.06135.x} {\bibfield  {journal} {\bibinfo
  {journal} {Monthly Notices of the Royal Astronomical Society}\ }\textbf
  {\bibinfo {volume} {339}},\ \bibinfo {pages} {12} (\bibinfo {year} {2003})},\
  \Eprint {http://arxiv.org/abs/astro-ph/0204108} {astro-ph/0204108}
  \BibitemShut {NoStop}%
\bibitem [{\citenamefont {{Zhao}}\ \emph {et~al.}(2009)\citenamefont {{Zhao}},
  \citenamefont {{Jing}}, \citenamefont {{Mo}},\ and\ \citenamefont
  {{B{\"o}rner}}}]{Zhao2009}%
  \BibitemOpen
  \bibfield  {author} {\bibinfo {author} {\bibfnamefont {D.~H.}\ \bibnamefont
  {{Zhao}}}, \bibinfo {author} {\bibfnamefont {Y.~P.}\ \bibnamefont {{Jing}}},
  \bibinfo {author} {\bibfnamefont {H.~J.}\ \bibnamefont {{Mo}}}, \ and\
  \bibinfo {author} {\bibfnamefont {G.}~\bibnamefont {{B{\"o}rner}}},\ }\href
  {\doibase 10.1088/0004-637X/707/1/354} {\bibfield  {journal} {\bibinfo
  {journal} {The Astrophysical Journal}\ }\textbf {\bibinfo {volume} {707}},\
  \bibinfo {pages} {354} (\bibinfo {year} {2009})},\ \Eprint
  {http://arxiv.org/abs/0811.0828} {arXiv:0811.0828} \BibitemShut {NoStop}%
\bibitem [{\citenamefont {{Villarreal}}\ \emph {et~al.}(2017)\citenamefont
  {{Villarreal}}, \citenamefont {{Zentner}}, \citenamefont {{Mao}},
  \citenamefont {{Purcell}}, \citenamefont {{van den Bosch}}, \citenamefont
  {{Diemer}}, \citenamefont {{Lange}}, \citenamefont {{Wang}},\ and\
  \citenamefont {{Campbell}}}]{Villarreal2017}%
  \BibitemOpen
  \bibfield  {author} {\bibinfo {author} {\bibfnamefont {A.~S.}\ \bibnamefont
  {{Villarreal}}}, \bibinfo {author} {\bibfnamefont {A.~R.}\ \bibnamefont
  {{Zentner}}}, \bibinfo {author} {\bibfnamefont {Y.-Y.}\ \bibnamefont
  {{Mao}}}, \bibinfo {author} {\bibfnamefont {C.~W.}\ \bibnamefont
  {{Purcell}}}, \bibinfo {author} {\bibfnamefont {F.~C.}\ \bibnamefont {{van
  den Bosch}}}, \bibinfo {author} {\bibfnamefont {B.}~\bibnamefont {{Diemer}}},
  \bibinfo {author} {\bibfnamefont {J.~U.}\ \bibnamefont {{Lange}}}, \bibinfo
  {author} {\bibfnamefont {K.}~\bibnamefont {{Wang}}}, \ and\ \bibinfo {author}
  {\bibfnamefont {D.}~\bibnamefont {{Campbell}}},\ }\href {\doibase
  10.1093/mnras/stx2045} {\bibfield  {journal} {\bibinfo  {journal} {Monthly
  Notices of the Royal Astronomical Society}\ }\textbf {\bibinfo {volume}
  {472}},\ \bibinfo {pages} {1088} (\bibinfo {year} {2017})},\ \Eprint
  {http://arxiv.org/abs/1705.04327} {arXiv:1705.04327} \BibitemShut {NoStop}%
\bibitem [{\citenamefont {Wechsler}\ \emph {et~al.}(2002)\citenamefont
  {Wechsler}, \citenamefont {Bullock}, \citenamefont {Primack}, \citenamefont
  {Kravtsov},\ and\ \citenamefont {Dekel}}]{Wechsler2002}%
  \BibitemOpen
  \bibfield  {author} {\bibinfo {author} {\bibfnamefont {R.~H.}\ \bibnamefont
  {Wechsler}}, \bibinfo {author} {\bibfnamefont {J.~S.}\ \bibnamefont
  {Bullock}}, \bibinfo {author} {\bibfnamefont {J.~R.}\ \bibnamefont
  {Primack}}, \bibinfo {author} {\bibfnamefont {A.~V.}\ \bibnamefont
  {Kravtsov}}, \ and\ \bibinfo {author} {\bibfnamefont {A.}~\bibnamefont
  {Dekel}},\ }\href {\doibase 10.1086/338765} {\bibfield  {journal} {\bibinfo
  {journal} {The Astrophysical Journal}\ }\textbf {\bibinfo {volume} {568}},\
  \bibinfo {pages} {52} (\bibinfo {year} {2002})}\BibitemShut {NoStop}%
\bibitem [{\citenamefont {Mao}\ \emph {et~al.}(2017)\citenamefont {Mao},
  \citenamefont {Zentner},\ and\ \citenamefont {Wechsler}}]{Wechsler2017}%
  \BibitemOpen
  \bibfield  {author} {\bibinfo {author} {\bibfnamefont {Y.-Y.}\ \bibnamefont
  {Mao}}, \bibinfo {author} {\bibfnamefont {A.~R.}\ \bibnamefont {Zentner}}, \
  and\ \bibinfo {author} {\bibfnamefont {R.~H.}\ \bibnamefont {Wechsler}},\
  }\href {\doibase 10.1093/mnras/stx3111} {\bibfield  {journal} {\bibinfo
  {journal} {Monthly Notices of the Royal Astronomical Society}\ }\textbf
  {\bibinfo {volume} {474}},\ \bibinfo {pages} {5143} (\bibinfo {year}
  {2017})},\ \Eprint
  {http://arxiv.org/abs/http://oup.prod.sis.lan/mnras/article-pdf/474/4/5143/23126779/stx3111.pdf}
  {http://oup.prod.sis.lan/mnras/article-pdf/474/4/5143/23126779/stx3111.pdf}
  \BibitemShut {NoStop}%
\bibitem [{\citenamefont {Stewart}\ \emph {et~al.}(2009)\citenamefont
  {Stewart}, \citenamefont {Bullock}, \citenamefont {Barton},\ and\
  \citenamefont {Wechsler}}]{merger}%
  \BibitemOpen
  \bibfield  {author} {\bibinfo {author} {\bibfnamefont {K.~R.}\ \bibnamefont
  {Stewart}}, \bibinfo {author} {\bibfnamefont {J.~S.}\ \bibnamefont
  {Bullock}}, \bibinfo {author} {\bibfnamefont {E.~J.}\ \bibnamefont {Barton}},
  \ and\ \bibinfo {author} {\bibfnamefont {R.~H.}\ \bibnamefont {Wechsler}},\
  }\href {\doibase 10.1088/0004-637x/702/2/1005} {\bibfield  {journal}
  {\bibinfo  {journal} {The Astrophysical Journal}\ }\textbf {\bibinfo {volume}
  {702}},\ \bibinfo {pages} {1005} (\bibinfo {year} {2009})}\BibitemShut
  {NoStop}%
\bibitem [{\citenamefont {Bayes}\ and\ \citenamefont {Price}(1763)}]{bayes}%
  \BibitemOpen
  \bibfield  {author} {\bibinfo {author} {\bibfnamefont {M.}~\bibnamefont
  {Bayes}}\ and\ \bibinfo {author} {\bibfnamefont {M.}~\bibnamefont {Price}},\
  }\href {\doibase 10.1098/rstl.1763.0053} {\bibfield  {journal} {\bibinfo
  {journal} {Philosophical Transactions}\ }\textbf {\bibinfo {volume} {53}},\
  \bibinfo {pages} {370} (\bibinfo {year} {1763})}\BibitemShut {NoStop}%
\bibitem [{\citenamefont {Eaton}(1983)}]{conditional_gaussian}%
  \BibitemOpen
  \bibfield  {author} {\bibinfo {author} {\bibfnamefont {M.~L.}\ \bibnamefont
  {Eaton}},\ }\href@noop {} {\emph {\bibinfo {title} {Multivariate statistics :
  a vector space approach}}},\ Wiley series in probability and mathematical
  statistics\ (\bibinfo  {publisher} {Wiley},\ \bibinfo {address} {New York ;
  Chichester},\ \bibinfo {year} {1983})\BibitemShut {NoStop}%
\bibitem [{\citenamefont {Monaco}\ \emph {et~al.}(2013)\citenamefont {Monaco},
  \citenamefont {Sefusatti}, \citenamefont {Borgani}, \citenamefont {Crocce},
  \citenamefont {Fosalba}, \citenamefont {Sheth},\ and\ \citenamefont
  {Theuns}}]{pinocchio}%
  \BibitemOpen
  \bibfield  {author} {\bibinfo {author} {\bibfnamefont {P.}~\bibnamefont
  {Monaco}}, \bibinfo {author} {\bibfnamefont {E.}~\bibnamefont {Sefusatti}},
  \bibinfo {author} {\bibfnamefont {S.}~\bibnamefont {Borgani}}, \bibinfo
  {author} {\bibfnamefont {M.}~\bibnamefont {Crocce}}, \bibinfo {author}
  {\bibfnamefont {P.}~\bibnamefont {Fosalba}}, \bibinfo {author} {\bibfnamefont
  {R.~K.}\ \bibnamefont {Sheth}}, \ and\ \bibinfo {author} {\bibfnamefont
  {T.}~\bibnamefont {Theuns}},\ }\href {\doibase 10.1093/mnras/stt907}
  {\bibfield  {journal} {\bibinfo  {journal} {Monthly Notices of the Royal
  Astronomical Society}\ }\textbf {\bibinfo {volume} {433}},\ \bibinfo {pages}
  {2389} (\bibinfo {year} {2013})},\ \Eprint
  {http://arxiv.org/abs/http://oup.prod.sis.lan/mnras/article-pdf/433/3/2389/4061631/stt907.pdf}
  {http://oup.prod.sis.lan/mnras/article-pdf/433/3/2389/4061631/stt907.pdf}
  \BibitemShut {NoStop}%
\bibitem [{\citenamefont {Ludlow}\ \emph {et~al.}(2014)\citenamefont {Ludlow},
  \citenamefont {Navarro}, \citenamefont {Angulo}, \citenamefont
  {Boylan-Kolchin}, \citenamefont {Springel}, \citenamefont {Frenk},\ and\
  \citenamefont {White}}]{merger_history_1}%
  \BibitemOpen
  \bibfield  {author} {\bibinfo {author} {\bibfnamefont {A.~D.}\ \bibnamefont
  {Ludlow}}, \bibinfo {author} {\bibfnamefont {J.~F.}\ \bibnamefont {Navarro}},
  \bibinfo {author} {\bibfnamefont {R.~E.}\ \bibnamefont {Angulo}}, \bibinfo
  {author} {\bibfnamefont {M.}~\bibnamefont {Boylan-Kolchin}}, \bibinfo
  {author} {\bibfnamefont {V.}~\bibnamefont {Springel}}, \bibinfo {author}
  {\bibfnamefont {C.}~\bibnamefont {Frenk}}, \ and\ \bibinfo {author}
  {\bibfnamefont {S.~D.~M.}\ \bibnamefont {White}},\ }\href {\doibase
  10.1093/mnras/stu483} {\bibfield  {journal} {\bibinfo  {journal} {Monthly
  Notices of the Royal Astronomical Society}\ }\textbf {\bibinfo {volume}
  {441}},\ \bibinfo {pages} {378} (\bibinfo {year} {2014})},\ \Eprint
  {http://arxiv.org/abs/http://oup.prod.sis.lan/mnras/article-pdf/441/1/378/2996980/stu483.pdf}
  {http://oup.prod.sis.lan/mnras/article-pdf/441/1/378/2996980/stu483.pdf}
  \BibitemShut {NoStop}%
\bibitem [{\citenamefont {Correa}\ \emph {et~al.}(2015)\citenamefont {Correa},
  \citenamefont {Wyithe}, \citenamefont {Schaye},\ and\ \citenamefont
  {Duffy}}]{merger_history_2}%
  \BibitemOpen
  \bibfield  {author} {\bibinfo {author} {\bibfnamefont {C.~A.}\ \bibnamefont
  {Correa}}, \bibinfo {author} {\bibfnamefont {J.~S.~B.}\ \bibnamefont
  {Wyithe}}, \bibinfo {author} {\bibfnamefont {J.}~\bibnamefont {Schaye}}, \
  and\ \bibinfo {author} {\bibfnamefont {A.~R.}\ \bibnamefont {Duffy}},\ }\href
  {\doibase 10.1093/mnras/stv1363} {\bibfield  {journal} {\bibinfo  {journal}
  {Monthly Notices of the Royal Astronomical Society}\ }\textbf {\bibinfo
  {volume} {452}},\ \bibinfo {pages} {1217} (\bibinfo {year} {2015})},\ \Eprint
  {http://arxiv.org/abs/http://oup.prod.sis.lan/mnras/article-pdf/452/2/1217/18505290/stv1363.pdf}
  {http://oup.prod.sis.lan/mnras/article-pdf/452/2/1217/18505290/stv1363.pdf}
  \BibitemShut {NoStop}%
\bibitem [{\citenamefont {Ludlow}\ \emph {et~al.}(2016)\citenamefont {Ludlow},
  \citenamefont {Bose}, \citenamefont {Angulo}, \citenamefont {Wang},
  \citenamefont {Hellwing}, \citenamefont {Navarro}, \citenamefont {Cole},\
  and\ \citenamefont {Frenk}}]{merger_history_3}%
  \BibitemOpen
  \bibfield  {author} {\bibinfo {author} {\bibfnamefont {A.~D.}\ \bibnamefont
  {Ludlow}}, \bibinfo {author} {\bibfnamefont {S.}~\bibnamefont {Bose}},
  \bibinfo {author} {\bibfnamefont {R.~E.}\ \bibnamefont {Angulo}}, \bibinfo
  {author} {\bibfnamefont {L.}~\bibnamefont {Wang}}, \bibinfo {author}
  {\bibfnamefont {W.~A.}\ \bibnamefont {Hellwing}}, \bibinfo {author}
  {\bibfnamefont {J.~F.}\ \bibnamefont {Navarro}}, \bibinfo {author}
  {\bibfnamefont {S.}~\bibnamefont {Cole}}, \ and\ \bibinfo {author}
  {\bibfnamefont {C.~S.}\ \bibnamefont {Frenk}},\ }\href {\doibase
  10.1093/mnras/stw1046} {\bibfield  {journal} {\bibinfo  {journal} {Monthly
  Notices of the Royal Astronomical Society}\ }\textbf {\bibinfo {volume}
  {460}},\ \bibinfo {pages} {1214} (\bibinfo {year} {2016})},\ \Eprint
  {http://arxiv.org/abs/http://oup.prod.sis.lan/mnras/article-pdf/460/2/1214/8115724/stw1046.pdf}
  {http://oup.prod.sis.lan/mnras/article-pdf/460/2/1214/8115724/stw1046.pdf}
  \BibitemShut {NoStop}%
\bibitem [{\citenamefont {Benson}\ \emph {et~al.}(2019)\citenamefont {Benson},
  \citenamefont {Ludlow},\ and\ \citenamefont {Cole}}]{merger_history_4}%
  \BibitemOpen
  \bibfield  {author} {\bibinfo {author} {\bibfnamefont {A.~J.}\ \bibnamefont
  {Benson}}, \bibinfo {author} {\bibfnamefont {A.}~\bibnamefont {Ludlow}}, \
  and\ \bibinfo {author} {\bibfnamefont {S.}~\bibnamefont {Cole}},\ }\href
  {\doibase 10.1093/mnras/stz695} {\bibfield  {journal} {\bibinfo  {journal}
  {Monthly Notices of the Royal Astronomical Society}\ }\textbf {\bibinfo
  {volume} {485}},\ \bibinfo {pages} {5010} (\bibinfo {year} {2019})},\ \Eprint
  {http://arxiv.org/abs/http://oup.prod.sis.lan/mnras/article-pdf/485/4/5010/28249774/stz695.pdf}
  {http://oup.prod.sis.lan/mnras/article-pdf/485/4/5010/28249774/stz695.pdf}
  \BibitemShut {NoStop}%
\end{thebibliography}%

\end{document}